\documentclass[onecolumn,prd,floats,aps,amsmath,amssymb,nofootinbib, superscriptaddress,preprintnumbers,10pt]{revtex4-2}
\usepackage{comment}
\usepackage[utf8]{inputenc}
\usepackage{mathtools}
\usepackage{bbm}
\usepackage{cancel} 
\usepackage{bm}
\usepackage{blindtext}
\usepackage{xcolor}
\usepackage{hyperref}
\usepackage{graphicx}
\usepackage{amsthm}
\usepackage{bm}
\usepackage{verbatim}
\usepackage{amsmath}
\usepackage{siunitx}
\usepackage{amssymb}
\usepackage{slashed}
\usepackage{tikz}
\usetikzlibrary{positioning, shapes, arrows.meta, decorations.pathmorphing}
\pgfkeys{
  /simplerwick/.cd,
  arrows/.store in=\LstWickArrows,
  arrows={-,-,-,-,-,-,-,-,-},
  arrows/.initial={-,-,-,-,-,-,-,-,-}, 
  below/.code={\WickBelowtrue},
}

\newcommand{\Romatre}{Dipartimento di Matematica e Fisica, Universit\`a  Roma Tre and INFN, Sezione di Roma Tre,\\ Via della Vasca Navale 84, I-00146 Rome, Italy}
\newcommand{\RomatreINFN}{Istituto Nazionale di Fisica Nucleare, Sezione di Roma Tre,\\ Via della Vasca Navale 84, I-00146 Rome, Italy}
\newcommand{\Romadue}{Dipartimento di Fisica and INFN, Universit\`a di Roma ``Tor Vergata",\\ Via della Ricerca Scientifica 1, I-00133 Roma, Italy}
\newcommand{\LaSapienza}{Physics Department and INFN Sezione di Roma La Sapienza,\\ Piazzale Aldo Moro 5, 00185 Roma, Italy}
\newcommand{\soton}{Department of Physics and Astronomy, University of Southampton,\\ Southampton SO17 1BJ, UK}
\newcommand{\be}{\begin{equation}}
\newcommand{\ee}{\end{equation}}
\newcommand{\bs}[1]{\bm{#1}}
\renewcommand{\vec}[1]{\bm{#1}}

\usepackage{hyperref}
\usepackage{physics}
\usepackage{graphicx}

\usepackage{simpler-wick}
\usepackage{comment}
\usetikzlibrary{decorations.markings}
\tikzset{W->-/.style={decoration={
  markings,
  mark=at position 0.5*\pgfdecoratedpathlength+2pt with
  {\draw[-latex] (-2pt,0pt) -- (1pt,0pt);}},postaction={decorate}},
  W-<-/.style={decoration={
  markings,
  mark=at position 0.5*\pgfdecoratedpathlength with
  {\draw[latex-] (-2pt,0pt) -- (1pt,0pt);}},postaction={decorate}}
  }
\newif\ifWickBelow
\WickBelowfalse
\pgfkeys{
  /simplerwick/.cd,
  arrows/.store in=\LstWickArrows,
  arrows={-,-,-,-,-,-,-,-,-},
  arrows/.initial={-,-,-,-,-,-,-,-,-}, 
  below/.code={\WickBelowtrue},
}

\makeatletter
\def\swick@end#1#2{
  \swick@setfalse@#1
  \tikzexternaldisable
  \begin{tikzpicture}[remember picture, baseline=(swick-close#1.base)]
    \node[use as bounding box, inner sep=0pt, outer sep=0pt] (swick-close#1) {$\displaystyle #2$};
  \end{tikzpicture}
  \tikz[remember picture, overlay]
{
\foreach \W@X[count=\W@C] in \LstWickArrows
{\ifnum\W@C=#1
\xdef\myW@style{\W@X}
\fi}
\ifWickBelow
    \draw[\myW@style] ($(swick-open#1.south) + (0, -3pt)$) 
          -- ($(swick-open#1.base) + (0, -\swick@offset) + #1*(0, -\swick@sep)$) 
          -- ($(swick-close#1.base) + (0, -\swick@offset) + #1*(0, -\swick@sep)$) 
          -- ($(swick-close#1.south) + (0, -3pt)$);
\else
    \draw[\myW@style] ($(swick-open#1.north) + (0, 3pt)$) 
          -- ($(swick-open#1.base) + (0, \swick@offset) + #1*(0, \swick@sep)$) 
          -- ($(swick-close#1.base) + (0, \swick@offset) + #1*(0, \swick@sep)$) 
          -- ($(swick-close#1.north) + (0, 3pt)$);
\fi}
  \tikzexternalenable}
\makeatother

\begin{document}
\title{Complete lattice QCD calculation of $K^{-}\to \ell^{-}\bar{\nu}_{\ell}\ell^{'+}\ell^{'-}$ form factors}
\date{\today}
\author{R.\,Di\,Palma}\affiliation{\Romatre}
\author{R.\,Frezzotti}\affiliation{\Romadue} 
\author{G.\,Gagliardi}\affiliation{\RomatreINFN}
\author{V.\,Lubicz}\affiliation{\Romatre} 
\author{G.\,Martinelli}\affiliation{\LaSapienza}
\author{C.T.\,Sachrajda}\affiliation{\soton}
\author{F.\,Sanfilippo}\affiliation{\RomatreINFN}
\author{S.\,Simula}\affiliation{\RomatreINFN}
\author{N.\,Tantalo}\affiliation{\Romadue}
\begin{abstract}
We present the first complete lattice QCD calculation of the four structure-dependent form factors governing the rare charged kaon decay 
$K^-  \to \ell^- \bar{\nu}_\ell \ell'^+ \ell'^-$, with fully controlled statistical and systematic uncertainties.
Our calculation is based on gauge ensembles generated by the Extended Twisted Mass Collaboration (ETMC) with $N_f = 2+1+1$ flavors of Wilson--clover twisted-mass fermions. Simulations are performed directly at the physical values of the light and strange quark masses, and include an estimate of the quark-disconnected contributions in which the virtual photon couples to sea quarks. All four form factors are determined across the kinematical region probed by experiments. The Spectral Function Reconstruction (SFR) method of Ref.~\cite{Frezzotti:2023nun} is employed to overcome the analytic continuation problem for dilepton invariant masses above the two-pion threshold. Finite-volume effects are investigated using ensembles with spatial extents $L\simeq [3.8,7.6]~\mathrm{fm}$, while the continuum limit is obtained from three lattice spacings in the range $a\in[0.057, 0.08]~\mathrm{fm}$. Our results for the form factors enable the evaluation of decay rates and differential observables for all four channels,
$K^- \to e^- \bar{\nu}_e e^+ e^-$,
$K^- \to e^- \bar{\nu}_e \mu^+ \mu^-$,
$K^- \to \mu^- \bar{\nu}_\mu e^+ e^-$,
and
$K^- \to \mu^- \bar{\nu}_\mu \mu^+ \mu^-$,
thereby providing first-principles Standard Model predictions against which existing and upcoming measurements  can be directly compared.
A detailed phenomenological analysis of the decay rates and associated  observables is presented in a companion paper~\cite{DiPalma:pheno}.
\end{abstract}
\maketitle

\section{Introduction}
The decays of  charged pseudoscalar mesons into light leptons with the emission of an off-shell photon,
\be
\label{eq:intro_decay}
P^- 
\to \ell^-\bar{\nu}_\ell \ell'^+\ell'^-, \qquad
\ell, \ell' = e , \mu,
\ee
represent\footnote{Throughout this work, all considerations made for the decay
$P^- \to \ell^-\bar{\nu}_\ell \gamma^*$ are understood to apply also to the
charge-conjugated process
$P^+ \to \ell^+\nu_\ell \gamma^*$.}
powerful probes of the flavour sector of the Standard Model (SM) and of possible
New Physics (NP) effects.
The decay channels in Eq.~(\ref{eq:intro_decay}) allow for alternative
determinations of the relevant Cabibbo--Kobayashi--Maskawa (CKM) matrix elements%
~\cite{Cabibbo:1963yz,Kobayashi:1973fv}, as well as for tests of lepton flavour
universality (LFU), complementing analogous studies involving purely leptonic
decays ($P^- \to \ell^-\bar{\nu}_\ell$) and radiative decays
($P^- \to \ell^-\bar{\nu}_\ell \gamma$), for which we have recently presented the results of a
dedicated lattice calculation~\cite{DiPalma:2025iud}.
Unlike the latter two processes, however, the presence of an intermediate
off-shell photon suppresses the decay rate of
$P^- \to \ell^-\bar{\nu}_\ell \ell'^+\ell'^-$ by two powers of the fine-structure
constant, $\alpha_{\mathrm{em}}^2$, rendering these channels rare in the SM (i.e. the rates start at $O(G_F^2\alpha_{\mathrm{em}}^2)$).
As a consequence, even small NP effects could give rise to sizeable
contributions to experimental observables, potentially revealing
themselves through deviations from corresponding SM predictions.

The determination of the decay amplitude of
$P^- \to \ell^-\bar{\nu}_\ell \ell'^+\ell'^-$
requires knowledge of the pseudoscalar decay constant $f_P$, whose contribution
is commonly referred to as \emph{point-like}, together with four
structure-dependent form factors, $F_V$, $H_1$, $F_A$, and $H_2$ (see Eq.~(\ref{eq:hmunusd}) below).
These form factors are functions of two Lorentz invariants  and encode the
interaction of a hard off-shell photon with the internal constituents of the
pseudoscalar meson. We take these invariants to be the photon virtuality $k^{2}$ and the photon three-momentum $\bs{k}$ in the kaon rest frame. When $\ell \neq \ell'$, both $k^2$ and $\bs{k}$ can  be experimentally determined by measuring the invariant mass and the three-momentum of the charged lepton pair $\ell'^+\ell'^-$.

In this work, we perform a first-principles lattice QCD calculation of the four
structure-dependent form factors governing the kaon leptonic decay
\be
\label{eq:kaon_decay}
K^- \to \ell^- \bar{\nu}_\ell \ell'^+\ell'^-.
\ee
Our analysis is based on gauge ensembles generated by the Extended Twisted Mass
Collaboration (ETMC), with $N_f = 2+1+1$ flavours of Wilson--Clover
twisted-mass fermions. Throughout this work, for convenience, we will refer to the decay in Eq.~(\ref{eq:kaon_decay}) using the shorthand $K_{\ell2\ell'}$.

Prior to this work two lattice QCD studies of $K_{\ell2\ell'}$ decays had been performed: one by Tuo \textit{et al.} using gauge configurations from the RBC/UKQCD collaboration~\cite{Tuo:2021ewr}, and the other by our group, Gagliardi \textit{et al.}, using gauge ensembles from the Extended Twisted Mass Collaboration~\cite{Gagliardi:2022szw}.
The present study however, is the first
lattice QCD calculation performed at the physical point and including a
systematic assessment of all relevant sources of uncertainty.
In particular:
\begin{enumerate}\vspace{-10pt}
    \item simulations are carried out at the physical values of the light and
    strange quark masses;
    \item we include an estimate of the 
    quark-disconnected contributions in which the photon is emitted by sea
    quarks;
    \item all four form factors are determined over the full kinematically
    allowed region in the infinite-volume and continuum limits;
    \item finite-size effects are investigated using ensembles with spatial
    extents ranging from $L \simeq 3.8$\,fm to $L \simeq 7.6$\,fm, while the
    continuum extrapolation is performed using three lattice spacings in the
    range $a \in [0.08,\,0.057]$\,fm.
\end{enumerate}
All these features were absent in the previous exploratory lattice studies,
which were performed at a single lattice spacing, on a single volume and at unphysical
pion masses, $m_\pi > 0.3~\mathrm{GeV}$.

\begin{figure}[]
    \centering
    \includegraphics[width=0.4\columnwidth]{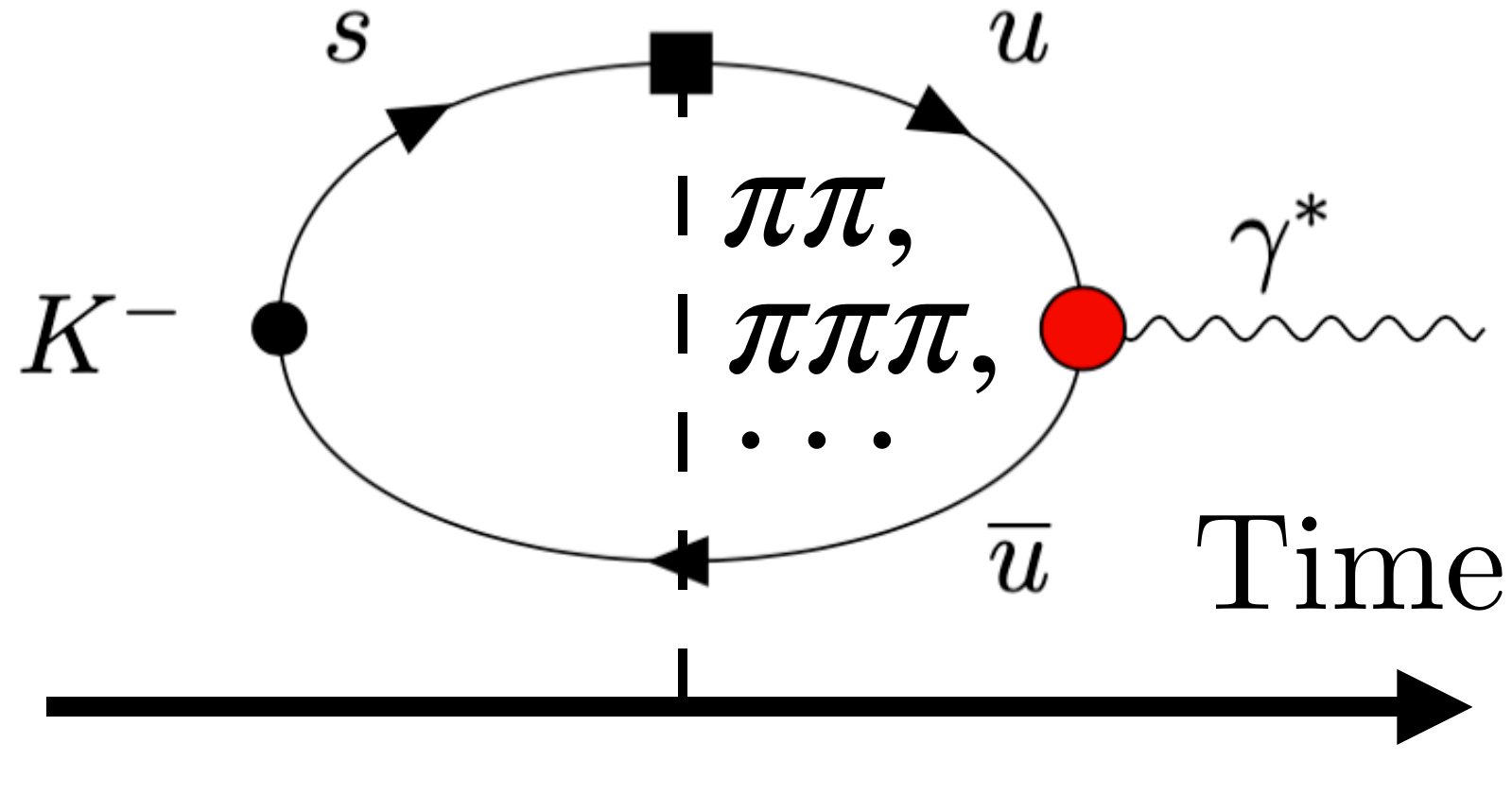}
    \caption{Schematic representation of photon emission from the valence up quark line. The black square denotes the insertion of the weak current, while the
    red circle indicates the electromagnetic vertex. Time flows from left to right and the black dashed line represents the time of insertion of the weak current. Therefore, the diagram corresponds to the time-ordering where the weak current acts before the electromagnetic current. Between the two currents, unflavoured $J^{PC}=1^{--}$ states, including 
    on-shell two-pion states, propagate.}
    \label{fig:Intro_Diagrams}
\end{figure}

Working at the physical pion mass is essential for a reliable determination of
the form factors. To illustrate this point, consider the diagram shown in
Fig.~\ref{fig:Intro_Diagrams}, where the photon is emitted from the valence up quark line in the time-ordering where the weak current (black square) acts before the electromagnetic current (red circle).
In this case, unflavoured hadronic vector states can propagate
between the two currents.
In particular, light two-pion states contribute.
For photon virtualities satisfying
\begin{align}
\label{eq:Intro_condition}
2 m_\pi < \sqrt{k^2} < m_K~,
\end{align}
these states can be produced on shell.
If instead $2m_\pi > m_{K}$, as in earlier lattice studies, the
condition in Eq.~(\ref{eq:Intro_condition}) is never realized, and the contribution of the intermediate two-pion states to the form factors can be therefore significantly
distorted. One obvious consequence is that while for $2m_{\pi} > m_{K} $ all four structure-dependent form factors are real, at the physical point they acquire an imaginary part.

The presence of on-shell two-pion intermediate states has important
consequences for the determination of the form factors from lattice QCD correlation
functions.
As it will be discussed in Sec.~\ref{Sec:Analytic_properties}, such states prevent the
direct extraction of the form factors by integrating Euclidean correlation
functions over Euclidean time in the kinematical region above the two-pion production
threshold, $\sqrt{k^2} > 2 m_\pi$.
In Sec.~\ref{Sec:FFfromCorr}, we show how this limitation can be overcome by
employing the Spectral Function Reconstruction (SFR) method introduced in
Ref.~\cite{Frezzotti:2023nun}, in combination with the
Hansen--Lupo--Tantalo (HLT) algorithm for spectral density reconstruction%
~\cite{Hansen:2019idp}.
Above the two-pion threshold, the form factors become complex, and in this work
we determine, for the first time, both their real and imaginary parts.
This study represents the first application of the SFR method to a complete
lattice calculation of form factors describing exclusive decay channels,
providing a further confirmation of the effectiveness of spectral reconstruction methods, such as the HLT in
addressing inverse Laplace-transform problems in lattice QCD. 

Although the spectral density becomes non-zero at the two-pion production threshold, for photon virtualities $x_k > 3m_\pi/m_K\simeq 0.85$ three-pion states can also go on shell. 
The SFR/HLT approach, however, properly accounts for all such contributions and allows us to handle an arbitrary number of intermediate states with energies lower than those of the external states. This feature becomes particularly advantageous for future applications to heavier mesons, where many states with energies below those of the external states are present.
Recent notable applications of the HLT include the calculation of the energy-smeared R-ratio of $e^{+}e^{-}\to$ hadrons ~\cite{ExtendedTwistedMassCollaborationETMC:2022sta}, the inclusive $\tau$ decays%
~\cite{Evangelista:2023fmt, ExtendedTwistedMass:2024nyi}, the rare decay
$B_s \to \mu^+ \mu^- \gamma$~\cite{Frezzotti:2024kqk}, inclusive semileptonic $D_s$ decays%
~\cite{DeSantis:2025qbb, DeSantis:2025yfm}, and very recently an application to the calculation of long-distance contributions in Flavour Changing Neutral Current (FCNC) $B$-decays~\cite{Frezzotti:2025hif}. 

\textbf{For readers primarily interested in the final results}, we refer to Sec.~\ref{Sec:4L_finalFF}, where our  determinations of the real and imaginary parts of the form factors are presented in Figs.~\ref{fig:4L_ReFinalFF},~\ref{fig:4L_ImFinalFF} and~\ref{fig:4L_ReFinalFFyk}. A convenient polynomial parametrization of the form factors is given in Eqs.~(\ref{eq:4L_ansatzFF}) and~(\ref{eq:4L_ansatzImFF}) for the real and imaginary parts, respectively; the corresponding fit parameters can be found in Tab.~\ref{tab:4L_fitpars}, while their correlation matrices are reported in Tabs.~\ref{tab:4L_Recorr} and~\ref{tab:4L_Imcorr}. A phenomenological analysis based on these results is presented in the companion paper~\cite{DiPalma:pheno}.

The remainder of this paper is organized as follows.
In Sec.~\ref{sec:definitions}, we define the $K_{\ell2\ell'}$ form factors.
In Sec.~\ref{Sec:Analytic_properties}, we discuss the analytic properties of the
correlation functions and the role of on-shell two-pion intermediate states.
Sec.~\ref{Sec:FFfromCorr} introduces the strategy used to evaluate the form
factors below and above the two-pion threshold.
Details of the numerical setup are given in Sec.~\ref{sec:Lattice_setup}.
The analysis below the two-pion threshold is presented in
Sec.~\ref{Sec:belowth}, while Sec.~\ref{Sec:4L_aboveth} collects the results above the two-pion threshold, including a description of the specific HLT implementation which we have adopted.
All the contributions are combined in Sec.~\ref{Sec:4L_finalFF}, and our conclusions are
presented in Sec.~\ref{sec:conclusions}.
Apps.~\ref{App:rhomunupipi_details},~\ref{app:4L_Wick}, and~\ref{app:kaon_radius} contain further details relevant for the evaluation of the form factors, which are explicitly recalled in the text when needed. App.~\ref{app:4L_Corr} contains the correlation matrix of the parameters entering our effective polynomial parametrization of the four form factors. 

\section{Definitions of the form factors}
\label{sec:definitions}
\begin{figure}[]
    \centering
\includegraphics[width=0.7\columnwidth]{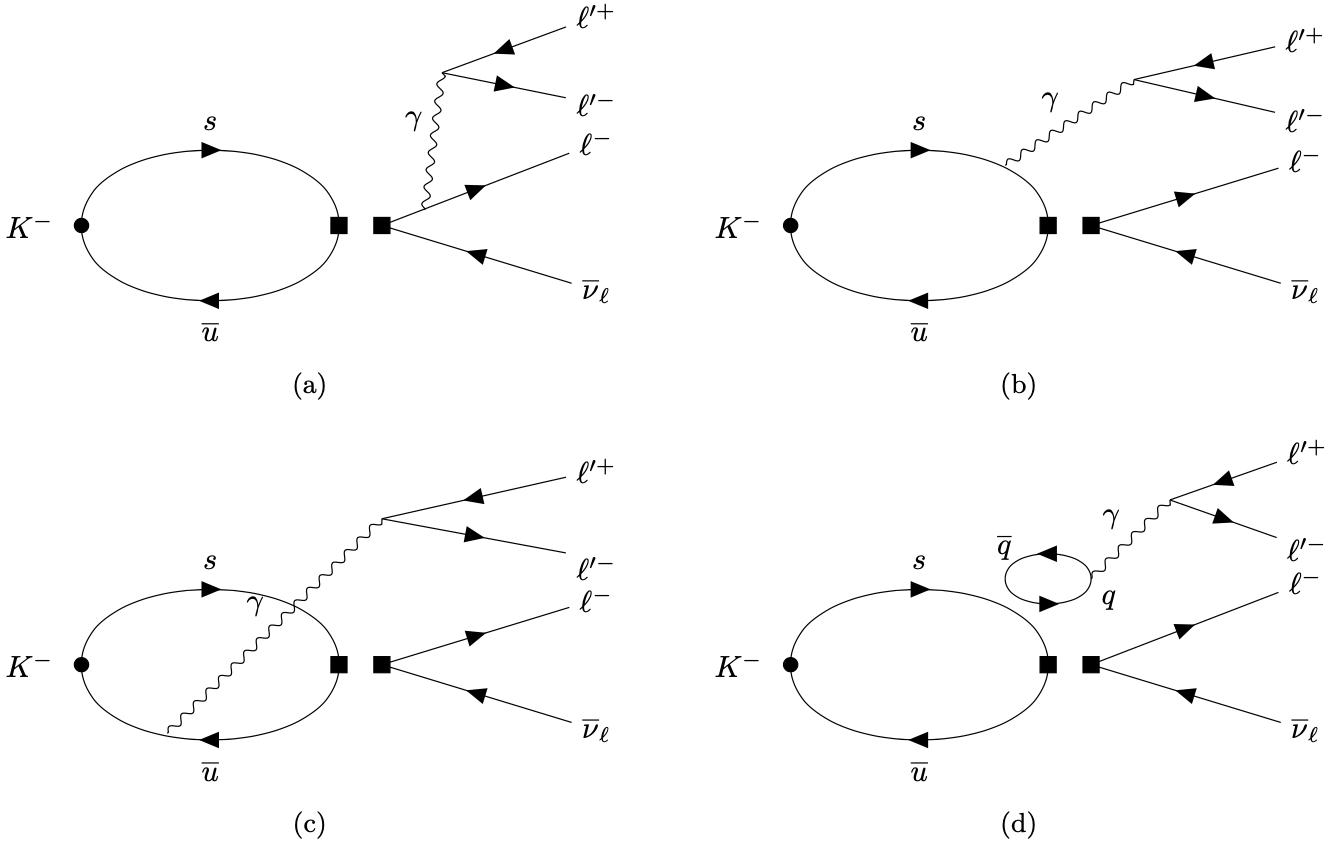}
\caption{Diagrams contributing to the decay $K^{-} \to \ell^{-}\bar{\nu}_{\ell} \ell'^+\ell'^-$ at leading order in $\alpha_{\rm em}$ and $G_F$, for $\ell'^- \neq \ell^-$. The double square represents the 4-fermion weak operator in the Fermi effective theory. Diagram (a) corresponds to the final-state radiation contribution where the photon is emitted from a lepton leg. Diagrams (b) and (c) correspond to the quark-connected contributions, with the photon emitted by the valence strange and anti-up quarks, respectively. Diagram (d) represents the case where the photon is emitted by a sea quark ($q=u,d,s,c$), which we refer to as the quark-disconnected contribution.}
\label{Fig:Diagrams}
\end{figure}

In the low-energy effective description of the weak interactions (Fermi theory), the relevant diagrams~\footnote{In the case of identical charged leptons in the final state ($\ell^- = \ell'^-$), four additional diagrams must be considered, which are obtained by exchanging the momenta of $\ell^-$ and $\ell'^-$ in the diagrams of Fig.~\ref{Fig:Diagrams}.} contributing to the $K_{\ell2\ell'}$ decay at leading-order in the electroweak interactions  are shown in Fig.~\ref{Fig:Diagrams}. 
Diagram~(a) represents the so-called final-state radiation  contribution, where the photon is emitted by one of the charged leptons in the final state, and the only non-perturbative input required for its evaluation is the kaon decay constant, $ f_{K} $. Diagrams~(b) and~(c) represent the emission of the photon from the valence strange and anti-up quarks, respectively, and we refer to them as \textit{quark-connected} contributions, while diagram~(d) corresponds to the emission of the photon from the sea quarks, and we refer to it as \textit{quark-disconnected} contribution. 

The non-perturbative contribution entering  from diagrams~(b),~(c), and~(d) is encoded in the following hadronic tensor: 
\be
H_{W}^{\mu\nu}(E_{\gamma}, \bs{k}, \bs{p})=  \int dt_\gamma \ e^{iE_\gamma t_\gamma} \int  d\bs{y}_{\gamma}\, e^{-i\bs{k}\cdot \bs{y}_{\gamma}}\, \bra{0} \hat{\mathrm{T}}[\,j_W^\nu(0) j^\mu_\mathrm{em}(y_{\gamma})]\ket{K^-(\bs{p})}~,
\label{Eq:Def_hadronictensor}
\ee
where $y_\gamma=(t_\gamma, \bs{y}_\gamma)$, $k=(E_{\gamma} = \sqrt{k^2 + \vert \bs{k} \vert^2}, \bs{k})$ is the four-momentum of the photon with virtuality $\sqrt{k^2}$. The three-momentum of the decaying kaon is indicated with $\bs{p}$, corresponding to a four-momentum $p = (\sqrt{m_K^2 + \vert \bs{p} \vert ^2}, \bs{p})$, where $m_K$ is the mass of the kaon. The $\hat{\mathrm{T}}$ operator in Eq.~(\ref{Eq:Def_hadronictensor}) stands for the \textit{time-ordered-product}. The electromagnetic current $j_{\rm em}^{\mu}(x)$ and the hadronic weak current $j_{W}^{\nu}(x)$ are given by 
\begin{align}
\label{Eq:Def_em_current}
j^{\mu}_{\mathrm{em}}(x) &=  \sum_{f} ~ j^{\mu}_{f}(x) =  \sum_f e_f~ \overline{q}_f(x) \gamma^{\mu} q_f(x)~,\\[8pt]
\label{Eq:Def_weak_current}
j^{\nu}_W(x) &= j^{\nu}_{V}(x) - j^{\nu}_A(x) = \overline{q}_u(x) (\gamma^{\nu} - \gamma^{\nu} \gamma_5) q_s(x)~,
\end{align}
where $q_{f}(x)$ is the quark-field of the flavour $f$, and we have indicated by $e_f$ its electric charge in units of the charge of the positron. Throughout this paper, whenever the subscript $V$ (or $A$) is used in place of $W$, it indicates that the vector (or axial) component of the weak hadronic current in 
Eq.~(\ref{Eq:Def_weak_current}) is being considered. 

The hadronic tensor in Eq.~(\ref{Eq:Def_hadronictensor}) can be decomposed in terms of a point-like (pt) and a structure-dependent (SD) component as
\be
H^{\mu\nu}_W(E_{\gamma}, \bs{k}, \bs{p}) =  \  H^{\mu\nu}_{\mathrm{pt}}(E_{\gamma}, \bs{k}, \bs{p}) +   H^{\mu\nu}_{\mathrm{SD}}(E_{\gamma}, \bs{k}, \bs{p}), \\[10pt]
\label{Eq:Hmunu_decomposition}
\ee
where the first term on the right-hand side is proportional to the kaon decay constant 
\be
\label{eq:Hmunupt}
H^{\mu \nu}_{\mathrm{pt}}(E_{\gamma}, \bs{k}, \bs{p}) = \ f_{K} \bigg[g^{\mu\nu} + \frac{(2p-k)^\mu(p-k)^\nu}{2p\cdot k - k^2}\bigg],
\ee
while the latter depends on four structure-dependent (SD) form factors $F_V$, $H_1$, $F_A$ and $H_2$:\footnote{The SD form factors are functions of two Lorentz-invariants. In this work we parameterize them in terms of $\sqrt{k^{2}}$ and the magnitude of the photon's three-momentum in the kaon's rest frame (see Eq.~(\ref{eq:xkyk})).} 
\begin{align}
H^{\mu\nu}_{\mathrm{SD}}(E_{\gamma}, \bs{k}, \bs{p}) =&  -i \frac{F_V}{m_K} \varepsilon^{\mu\nu\alpha\beta}k_\alpha p_\beta + \frac{F_A}{m_K}[(p\cdot k -k^2)g^{\mu\nu} - (p-k)^\mu k^\nu] \label{eq:hmunusd}
 \\ \nonumber & \qquad \qquad \qquad \quad +\frac{H_1}{m_K} (k^2 g^{\mu \nu} - k^\mu k^\nu) + \frac{H_2}{m_K} \frac{(p\cdot k - k^2)k^\mu - k^2(p-k)^\mu}{(p-k)^2 - m_K^2} (p-k)^\nu. 
\end{align}
With the decomposition of the hadronic tensor into a point-like term 
$H^{\mu\nu}_{\mathrm{pt}}(E_\gamma, \bs{k}, \bs{p})$ and a structure-dependent term 
$H^{\mu\nu}_{\mathrm{SD}}(E_\gamma, \bs{k}, \bs{p})$, the Ward identity satisfied by 
$H^{\mu\nu}_W(E_{\gamma}, \bs{k}, \bs{p})$ reads
\be
k_{\mu}H^{\mu\nu}_{\mathrm{pt}}(E_{\gamma}, \bs{k},\bs{p}) = i \bra{0}j^{\nu}_W(0)\ket{K^-(\bs{p})} = f_Kp^\nu, \qquad k_\mu H^{\mu\nu}_{\mathrm{SD}}(E_{\gamma},\bs{k}, \bs{p}) = 0.
\label{Eq:WI}
\ee 
In the following, we present the calculation of the form factors $F_V$, $H_1$, $F_A$, and $H_2$ over the kinematical range probed by experiments, evaluating all the sources of systematic errors: finite-size effects, discretization effects, and including the quark-disconnected contribution. 
The calculation is performed using $N_{f}=2+1+1$ dynamical quark flavours at the physical point. 

This analysis is carried out in the kaon rest frame ($p = (m_K,\bs{0})$), with the photon momentum chosen to be along the $z$-axis ($\bs{k} = (0,0,k_z)$). For simplicity, we denote the hadronic tensor in this frame as  $H^{\mu\nu}_W(E_\gamma, \bs{k})\equiv H^{\mu\nu}_W(E_\gamma, \bs{k}, \bs{p} = \bs{0})$.  
We parametrize the form factors in terms of two dimensionless Lorentz scalars $x_k$ and $y_k$, defined as
\be
x_k = \frac{\sqrt{k^2}}{m_K}\,, \qquad \qquad y_{k} = \frac{2 \vert \bs{k} \vert}{m_K}\,. 
\label{eq:xkyk}
\ee
The phase space of the decay is then bounded by the following kinematic limits:
\be
\frac{2 m_{\ell'}}{m_K} \leq x_k \leq  1 - \frac{m_\ell}{m_K}, \qquad \qquad 0 \leq y_k \leq \sqrt{ \bigg[ 1 + x_k^2 - \frac{m_\ell^2}{m_K^2} \bigg]^2 - 4 x_k^2} \ ,
\label{eq:xkyklimits}
\ee
where $m_\ell$ and $m_{\ell'}$ are the masses of the charged leptons produced in the weak and virtual photon decays, respectively.
\section{Analytic structure of the hadronic tensor and Wick rotation to Euclidean time}
\label{Sec:Analytic_properties}
As shown in Eq.~(\ref{Eq:Def_hadronictensor}), the elements of the hadronic tensor are obtained from time integrals of suitable Minkowskian correlation functions. Since lattice QCD simulations provide access only to correlation functions evaluated at Euclidean time separations, it is necessary to determine under which conditions the Minkowskian time integral can be related to an integral along imaginary (Euclidean) time via the analytic continuation $t_\gamma \to - i t_\gamma$.

\begin{figure}[]
    \centering
\includegraphics[width=0.4\columnwidth]{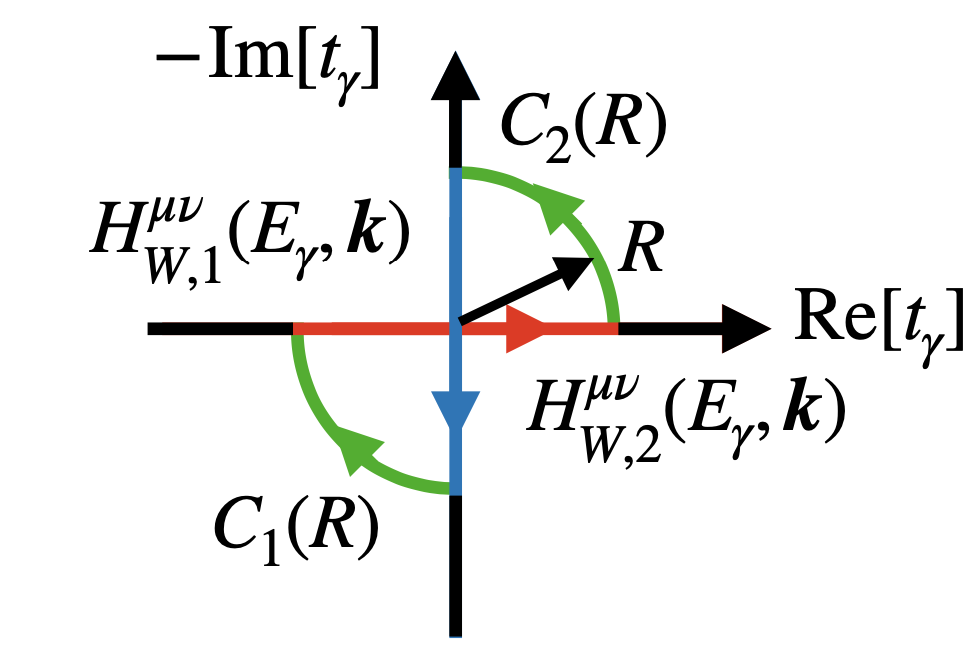}
\caption{Schematic illustration of the Wick rotation relating Minkowskian and Euclidean time integrals. The red line represents the integration along real (Minkowskian) time defining the hadronic tensor, while the blue line corresponds to the Euclidean-time contour accessible in lattice QCD calculations. The deformation of the integration contour is justified by Cauchy's theorem provided that the contributions from the arcs at infinity (green contours) vanish. }
\label{Fig:Wick}
\end{figure}

This situation is illustrated schematically in Fig.~\ref{Fig:Wick}. The goal is to ensure that the integral along the real-time contour (red line), defining the hadronic tensor, can be deformed into the Euclidean contour (blue line), where the correlation functions can be computed on the lattice. According to Cauchy's theorem, this contour deformation is valid provided that the contributions from the arcs at infinity (green contours) vanish.

In this section we analyze the possibility of Wick rotating to Euclidean spacetime by examining separately the contributions from the two contour arcs shown in Fig.~\ref{Fig:Wick}. A crucial role in this analysis is played by the intermediate states propagating between the electromagnetic and weak currents. Since the set of intermediate states depends on the order of the operators, we decompose
$H^{\mu\nu}_W(E_\gamma, \bs{k})$
into contributions associated with negative and positive values of $t_\gamma$. We refer to the time ordering $t_\gamma < 0$ (i.e.\ when the electromagnetic current acts before the weak current) as the \emph{first time ordering}, and to $t_\gamma > 0$ (i.e.\ when the electromagnetic current acts after the weak current) as the \emph{second time ordering}. Accordingly, we write
\be
H^{\mu\nu}_W(E_\gamma, \bs{k}) =
H^{\mu\nu}_{W,1}(E_\gamma, \bs{k}) +
H^{\mu\nu}_{W,2}(E_\gamma, \bs{k}),
\label{eq:Hmunu12}
\ee
where $H^{\mu\nu}_{W,1}$ is given by the integral over negative times
\be
H_{W,1}^{\mu\nu}(E_{\gamma}, \bs{k})=
\int_{-\infty}^{0} dt_\gamma \ e^{iE_\gamma t_\gamma}
\int d\bs{y}_{\gamma}\, e^{-i\bs{k}\cdot \bs{y}_{\gamma}}\,
\bra{0} j_W^\nu(0) j^\mu_\mathrm{em}(y_{\gamma}) \ket{K^-(\bs{0})},
\label{eq:HmunuW1formal}
\ee
while the integral over positive times yields
\be
H_{W,2}^{\mu\nu}(E_{\gamma}, \bs{k})=
\int_0^\infty dt_\gamma \ e^{iE_\gamma t_\gamma}
\int d\bs{y}_{\gamma}\, e^{-i\bs{k}\cdot \bs{y}_{\gamma}}\,
\bra{0} j^\mu_\mathrm{em}(y_{\gamma}) j_W^\nu(0) \ket{K^-(\bs{0})}.
\label{eq:HmunuW2formal}
\ee

We begin by analysing the first time ordering ($t_{\gamma}<0$). Inserting a complete set of states $|n_1(-\bs{k})\rangle$ between the weak and electromagnetic currents in Eq.~(\ref{eq:HmunuW1formal}) and by using
\begin{align}
\label{eq:J_evo}
J^{\mu}_{\rm em}(y_{\gamma}) = e^{i\mathbb{P}y_{\gamma}} J^{\mu}_{\rm em}(0) e^{-i\mathbb{P}y_{\gamma}}~,
\end{align}
where $\mathbb{P}^{\mu}$ is the QCD 4-momentum operator, it is easy to show that 
the contribution from the lower-left arc $C_1(R)$ in Fig.~\ref{Fig:Wick} is given by
\be
H^{\mu\nu}_{W,C_{1}(R)}(E_{\gamma},\bs{k}) =  \sum_{n_1} c_{n_1}^{\mu\nu}(-\bs{k}) \int_{C_{1}(R)} dt_{\gamma}\,
e^{it_{\gamma} (E_\gamma + E_{n_1} - m_K)  },
\label{eq:C1}
\ee
where $E_{n_1}$ is the energy of the intermediate state $|n_{1}(-\bs{k})\rangle$, and
\be
c_{n_1}^{\mu\nu}(\bs{-k}) \equiv
\frac{\bra{0} j^\nu_W(0) \ket{n_1(-\bs{k})}
\bra{n_1(-\bs{k})} j^{\mu}_{\mathrm{em}}(0)\ket{K^-(\bs{0})}}
{2 E_{n_1}}.
\label{eq:cn1}
\ee
It follows that  $\lim\limits_{R\to \infty}H^{\mu\nu}_{W,C_{1}(R)}(E_{\gamma},\bs{k})$ vanishes if and only if
\be
E_\gamma+ E_{n_1} - m_K > 0.
\label{eq:ConvCond1}
\ee
Given the flavour quantum numbers  of the two currents in Eq.~(\ref{eq:HmunuW1formal}), the intermediate states $|n_{1}(-\bs{k})\rangle$ that can propagate between them have strangeness $S=1$. The lightest state is therefore a kaon (with three-momentum $-\bs{k}$), for which the condition in Eq.~(\ref{eq:ConvCond1}) is always satisfied. It follows that the rotation of the integral to Euclidean spacetime is always possible for the contribution from the first time-ordering.

For the second time ordering, we again insert a complete set of states
$|n_2(\bs{k})\rangle$ between the two currents. Using Eq.~(\ref{eq:J_evo}),
the contribution from the upper-right arc $C_{2}(R)$ is given by
\be
H^{\mu\nu}_{W,C_{2}(R)}(E_{\gamma},\bs{k})
=
\sum_{n_2} c_{n_2}^{\mu\nu}(\bs{k}) \int_{C_{2}(R)} dt_{\gamma}\,   e^{-it_{\gamma} (E_{n_2} - E_\gamma) },
\label{eq:C2}
\ee
where $E_{n_{2}}$ is the energy of the intermediate state
$|n_{2}(\bs{k})\rangle$, and
\be
c_{n_2}^{\mu\nu}(\bs{k}) \equiv
\frac{\bra{0} j^{\mu}_{\mathrm{em}}(0) \ket{n_2(\bs{k})}
\bra{n_2(\bs{k})} j^\nu_W(0) \ket{K^-(\bs{0})}}
{2E_{n_2}}.
\ee
It follows that
$\lim\limits_{R\to\infty}H^{\mu\nu}_{W,C_{2}(R)}(E_{\gamma},\bs{k})=0$
provided that the following condition holds for each intermediate state:
\be
E_{n_2} > E_\gamma~.
\label{eq:ConvCond2}
\ee
This condition corresponds to the requirement that no on-shell intermediate
states propagate between the two currents. Whenever it is violated, the Wick
rotation cannot be performed and the hadronic tensor cannot be obtained from a
simple Euclidean-time integral.

To assess whether this condition is satisfied, we identify the lightest
intermediate state allowed by the quantum numbers of the currents.
Since the intermediate states propagating between the two currents are
unflavoured vector states, the lightest contribution to
$H^{\mu\nu}_{W,2}(E_{\gamma},\bs{k})$ is a collinear two-pion state with total
momentum $\bs{k}$ and energy
\be
\label{eq:4L_2pi}
E_{n_2} = \sqrt{4m_\pi^2 + |\bs{k}|^2}.
\ee
Consequently, the Wick rotation of
$H^{\mu\nu}_{W,2}(E_\gamma,\bs{k})$ is not possible whenever
\be
2 m_\pi < \sqrt{k^2} < m_K,
\label{eq:4L_conditions_analyticC}
\ee
i.e.\ above the two-pion production threshold.

It is worth mentioning that if one considers only the quark-connected
contribution corresponding to photon emission from the strange quark
(diagram (b) in Fig.~\ref{Fig:Diagrams}), the lightest contributing intermediate state is a
collinear $K^{+}K^{-}$ state with momentum $\bs{k}/2$. Since for
$\sqrt{k^{2}}\leq m_{K}$ this state  has always an energy larger than $E_{\gamma}$,
this contribution alone satisfies the criterion for analytic continuation to
Euclidean spacetime. 

In the kinematical region above the two-pion threshold in
Eq.~(\ref{eq:4L_conditions_analyticC}), the Wick rotation is thus not possible.
As will be discussed in the next section, we employ the  spectral function
reconstruction (SFR) method introduced in Ref.~\cite{Frezzotti:2023nun} together with the
Hansen--Lupo--Tantalo (HLT) method from Ref.\,\cite{Hansen:2019idp}, to determine  the form factors for $k^{2} > 4m_{\pi}^{2}$. 

Note that the two previous exploratory lattice calculations~\cite{Gagliardi:2022szw, Tuo:2021ewr} did not encounter this issue, as they
employed heavier pions with masses $m_\pi \simeq 0.3~\mathrm{GeV}$, such that
the two-pion production threshold lay outside the physical phase space,
$2m_\pi > m_K$.

\section{Form factors from Euclidean correlation functions}
\label{Sec:FFfromCorr}
In this section we outline the strategy used to determine the four form factors
$F_V$, $F_A$, $H_1$, and $H_2$ over the kinematical range probed by experiments.
As already mentioned, we parametrize the form factors in terms of the normalized
photon virtuality $x_k$ and the normalized photon momentum $y_k$ defined in
Eq.~(\ref{eq:xkyk}).

In Sec.\,\ref{sec:4L_below2pi}, we discuss the region below the two-pion threshold
($\sqrt{k^2} < 2m_\pi$, or equivalently $x_k < 2m_\pi/m_K \simeq 0.55$), where the
hadronic matrix elements can be obtained through a suitable integration of time-dependent
Euclidean correlation functions computed on the lattice.

In Sec.~\ref{sec:4L_strategyFFAbove} we then turn to the region above the
two-pion threshold ($\sqrt{k^2} > 2m_\pi$, or equivalently
$x_k > 2m_\pi/m_K$), where the lack of a analytic continuation
to Euclidean spacetime requires the use of the SFR/HLT method to evaluate the form factors.
The specific implementation of this strategy, including the treatment of the
different diagrams in Fig.~\ref{Fig:Diagrams} will be discussed in details.

\subsection{Strategy for the numerical evaluation of the form factors below the $\pi\pi$ threshold}
\label{sec:4L_below2pi}
The starting point of our calculation are the following Euclidean three-point correlation functions, evaluated on a lattice with finite temporal ($T$) and spatial ($L$) extents:\footnote{Compared to Ref.~\cite{DiPalma:2025iud}, an additional overall factor of $(-i)$ is now included in the definition of $C^{\mu\nu}_{3,W}(t_\gamma, \bs{k}; t_W)$ so as to ensure consistency with the convention used for the hadronic tensor in Eq.~(\ref{Eq:Def_hadronictensor}).} 
\be
C_{3,W}^{\mu \nu}(t_{\gamma},  \bs{k}; t_W) =  (-i) \sum_{\bs{x}} \sum_{\bs{y}_{\gamma}}  \ e^{- i \bs{k}\cdot \bs{y}_{\gamma}} \bra{0}\hat{\mathrm{T}}[j^\nu_{W}(t_W, \bs{0})j^{\mu}_{\mathrm{em}}(t_{\gamma}+t_{W}, \bs{y}_{\gamma}) P^{\dagger}_K(0, \bs{x})] \ket{0},
\label{Eq:Def3pt}
\ee 
where $P_K^\dagger(0, \bs{x})$ is an interpolating operator with the quantum numbers of the kaon\footnote{The variable $t_{\gamma}$ (as well as $t_{W}$) appearing in Eq.~(\ref{Eq:Def3pt}) denotes Euclidean time and should not be confused with the Minkowskian time variable $t_{\gamma}$ introduced in the previous section.}. Note that the correlation functions in Eq.~(\ref{Eq:Def3pt}) are computed by inserting the kaon interpolating operator at the origin of time. Consequently, the time of insertion of the weak and the electromagnetic currents are shifted by a fixed amount $t_W$ with respect to the definition adopted in Eq.~(\ref{Eq:Def_hadronictensor}).
Following Ref.~\cite{DiPalma:2025iud}, we define the following amputated correlation function: 
\be
C_{W}^{\mu\nu}(t_\gamma, E_\gamma,  \bs{k}; t_W) = e^{E_\gamma t_\gamma} \mathcal{N}(t_{W}) ~C_{3, W}^{\mu\nu}(t_\gamma, \bs{k}; t_W )~,\qquad  \mathcal{N}(t_{W}) \equiv e^{m_K t_{W}} \frac{2m_K}{\bra{K^{-}(\bs{0})}P^{\dagger}_K(0)\ket{0}}.
\label{Eq:Def_correlationfunction}
\ee
Below the two-pion threshold, both time orderings satisfy the conditions needed to perform a Wick rotation to Euclidean spacetime, which are given in Eqs.~(\ref{eq:ConvCond1}) and~(\ref{eq:ConvCond2}). As a result, the hadronic tensor can be obtained directly from the time integral of $C_{W}^{\mu}(t_{\gamma}, E_{\gamma}, \bs{k}; t_{W})$, through
\be
H^{\mu\nu}_W(E_\gamma, \bs{k}; t_W)  =  \int_{-\infty}^{\infty} d t_\gamma \   C_{W}^{\mu\nu}(t_\gamma, E_\gamma, \bs{k}; t_W ) = H^{\mu \nu}_W(E_\gamma, \bs{k}) + \cdot \cdot \cdot \ ,
\label{Eq:HmunufromCmunu}
\ee
where the ellipses on the right-hand side stand for the sub-leading exponentials, vanishing in the limit $t_W \to \infty$ in which the kaon is fully isolated. On a 
lattice of temporal extent $T$, the time-integration over $t_{\gamma}$ in Eq.~(\ref{Eq:HmunufromCmunu}) is necessarily 
restricted to the finite interval $t_{\gamma} \in [-t_{W}, t_{\rm max}]$, where 
$t_{\rm max} \ll T - t_{W}$ in order to avoid around-the-world contributions from 
unwanted time orderings. A thorough analysis of the systematic effects arising from the truncation of the integral will be presented in Sec.~\ref{Sec:4L_tgammaintegral}.

As in our previous work on radiative kaon decays~\cite{DiPalma:2025iud}, we adopt, for the computation of $C_{3,W}^{\mu\nu}(t_\gamma, \bs{k}; t_W)$, the 3d method proposed in Refs.~\cite{Tuo:2021ewr} and~\cite{Giusti:2023pot}. The method consists of computing $C_{3,W}^{\mu\nu}(t_\gamma, \bs{k}; t_W)$ at all values of $t_{\gamma}$ but for a fixed $t_W$, at the cost of a single sequential propagator per quark flavour and three-momentum $\bs{k}$. A key advantage of this strategy is that the hadronic tensor can be evaluated at arbitrary photon virtualities without additional numerical costs~\cite{Frezzotti:2023nun,DiPalma:2025iud,Giusti:2025ibe}\,\footnote{For fixed photon momentum $\bs{k}$, changing the photon virtuality corresponds only to modifying the factor $e^{E_{\gamma} t_{\gamma}}$ in Eq.~(\ref{Eq:Def_correlationfunction}). The nonperturbative input $C_{3,W}^{\mu\nu}(t_{\gamma},\bs{k};t_{W})$ is independent of $E_\gamma$.}. Moreover, as discussed in Ref.~\cite{Frezzotti:2023ygt}, the 3d  method mitigates the exponential signal-to-noise degradation that occurs for $E_\gamma > m_\pi$. Note that this strategy differs from that adopted in our preliminary study of $K_{\ell 2\ell'}$ decays in Ref.~\cite{Gagliardi:2022szw}, where we used the so-called 4d method, in which the $t_{\gamma}$ integration was performed at simulation time and the correlation function accessed at all values of $t_{W}$, at the cost of a single sequential propagator for photon momentum $\bs{k}$ and photon energy $E_{\gamma}$.

We now describe our strategy to isolate the four form factors $F_{V},H_{1},H_{2}$ and $F_{A}$ from the nonperturbative input $C_{W}^{\mu\nu}(t_{\gamma},E_{\gamma},\bs{k};t_{W})$. A standard approach consists in first determining the hadronic tensor and then constructing suitable linear combinations of its components that project onto the desired form factors~\cite{DiPalma:2025iud, Gagliardi:2022szw}. In this work, however, we adopt a different strategy and introduce specific linear combinations of the Euclidean correlators $C_{W}^{\mu\nu}(t_{\gamma}, E_{\gamma},\bs{k};t_{W})$, which we refer to as \emph{differential form factors}. They will be defined in such a way that their $t_\gamma$-integral  yields the form factors whenever the Wick rotation is possible (i.e. below the two-pion threshold). This formulation allows us to extract the four form factors without explicitly reconstructing the hadronic tensor. Moreover, this framework proves particularly useful when discussing in Sec.~\ref{sec:4L_strategyFFAbove} the determination of the form factors above the two-pion threshold using the SFR/HLT method.

All the relations derived below are valid in the kaon rest frame, with the photon momentum set along the z-axis ($\bs{k} = (0,0,k_z)$).
We start by discussing the particularly simple case of the vector form factor $F_V$. From the definition of the structure-dependent part of the hadronic tensor in Eq.~(\ref{eq:hmunusd}), it follows that the vector form factor at fixed normalized photon virtuality and photon momentum, $F_{V}(x_k, y_k)$, can be obtained in our kinematic setup from the large-$t_W$ limit of the estimator
\be
F_V(x_k, y_k; t_W)
= \frac{i}{k_z}\, H_V^{21}(E_\gamma, \boldsymbol{k}; t_W)\, .
\label{eq:4L_FV}
\ee
It follows immediately that the following differential vector form factor
\be
\label{eq:4L_dFV}
\delta F_V(t_\gamma, x_k, y_k; t_W)
\equiv \frac{i}{k_z}\, C_V^{21}(t_\gamma, E_\gamma, \boldsymbol{k}; t_W)
\ee
is such that
\be
F_V(x_k, y_k; t_W)
= \int_{-\infty}^{\infty} d t_\gamma \,
\delta F_V(t_\gamma, x_k, y_k; t_W)
= F_V(x_k,y_k) + \cdots \, ,
\ee
where the ellipses denote contributions that vanish exponentially in the limit $t_W\to\infty$.

\vspace{0.2cm}

In the axial channel the construction of the differential form factors is more involved, since the axial correlators contain both structure-dependent contributions and a point-like term proportional to $f_K$. The latter must be removed before isolating the axial form factors.

For convenience we collect the three structure-dependent axial form factors into the vector
\be
A^i(x_k,y_k) = \bigl(H_1,\,H_2,\,F_A\bigr)(x_k,y_k).
\label{eq:4L_Aidef}
\ee

In Ref.~\cite{Gagliardi:2022szw}, in order to isolate the structure-dependent contributions, the following subtracted combinations of axial components of the hadronic tensor had been introduced
\be
\tilde H_A^i(E_\gamma,\bs{k})
\equiv
\left(
\begin{array}{c}
H_A^{30}(E_{\gamma},\bs{k})-H_A^{03}(E_{\gamma},\bs{k})\frac{m_K-E_\gamma}{2m_K-E_\gamma} \\[5pt]
H_A^{33}(E_{\gamma},\bs{k})+H_A^{11}(E_{\gamma},\bs{k})
-\frac{2m_K E_\gamma - E_\gamma^2}{2m_K E_\gamma-k^2}H_A^{33}(0,\bs0)
-H_A^{11}(0,\bs0) \\[5pt]
H_A^{33}(E_{\gamma},\bs{k})-H_A^{11}(E_{\gamma},\bs{k})
-\frac{2m_K E_\gamma - E_\gamma^2}{2m_K E_\gamma-k^2}H_A^{33}(0,\bs0)
+H_A^{11}(0,\bs0)
\end{array}
\right).
\ee
In our kinematic setup, the $\tilde{H}_{A}^{i}(E_{\gamma},\bs{k})$ only depend on the structure-dependent axial form factors which are then obtained as
\be
A^i(x_k,y_k)=M^{ij}\,\tilde H_A^j(E_\gamma,\bs{k}),
\label{eq:4L_AifromH}
\ee
where $M^{ij}$ is the following $3\times3$ matrix of known kinematical factors:
\be
M^{ij} =
\left(
\begin{array}{ccc}
-\frac{E_\gamma k_z}{2 m_K-E_\gamma} & -\frac{k_z (m_K-E_\gamma)}{2 m_K-E_\gamma} & \frac{k_z m_K}{2 m_K-E_\gamma} \\[5pt]
 -\frac{E_\gamma^2+k^2}{m_K} & \frac{E_\gamma k_z^2}{2 E_\gamma m_K-k^2} & \frac{E_\gamma^2-2 E_\gamma m_K+k^2}{m_K} \\[5pt]
  \frac{k^2-E_\gamma^2}{m_K} & \frac{E_\gamma k_z^2}{2 E_\gamma m_K-k^2} & \frac{E_\gamma^2-k^2}{m_K}
\end{array}
\right)^{-1}.
\label{eq:4L_Mij}
\ee

To construct the axial differential form factors, it is sufficient to reproduce the same subtraction directly at the level of the Euclidean correlation functions $C_A^{\mu\nu}(t_\gamma,E_\gamma,\bs{k}; t_W)$. To this end,  we first define
\begin{align}
\label{eq:axial_corr}
 C_A^{[3,0]}(t_\gamma, E_\gamma, \bs{k}; t_W)
&\equiv
 C_A^{30}(t_\gamma, E_\gamma, \bs{k}; t_W)
 - C_A^{03}(t_\gamma, E_\gamma, \bs{k}; t_W)
 \frac{m_K - E_\gamma}{2 m_K - E_\gamma},
\\
\tilde{C}_A^{11}(t_\gamma, E_\gamma, \bs{k}; t_W)
&\equiv
C_A^{11}(t_\gamma, E_\gamma, \bs{k}; t_W)
- C_A^{11}(t_\gamma, 0, \bs{0}; t_W),
\\
\tilde{C}_A^{33}(t_\gamma, E_\gamma, \bs{k}; t_W)
&\equiv
C_A^{33}(t_\gamma, E_\gamma, \bs{k}; t_W)
- C_A^{33}(t_\gamma, 0, \bs{0}; t_W)
\frac{2m_K E_\gamma - E_\gamma^2}{2m_K E_\gamma - k^2},
\end{align}
we then combine them into the following vector
\be
\tilde{C}^i_A(t_\gamma, E_\gamma, \bs{k}; t_W)
\equiv  \left(
\begin{array}{c}
C^{[3,0]}_{A}(t_\gamma, E_\gamma, \bs{k}; t_W) \\[5pt]
(\tilde{C}^{33}_{A} + \tilde{C}^{11}_{A})(t_\gamma, E_\gamma, \bs{k}; t_W) \\[5pt]
(\tilde{C}^{33}_{A} - \tilde{C}^{11}_{A})(t_\gamma, E_\gamma, \bs{k}; t_W)
\end{array}
\right) = \overline{C}_{A}^{\,i}(t_\gamma, E_\gamma, \bs{k}; t_W) - C_{pt}^{i}(t_\gamma, E_\gamma, \bs{k}; t_W),
\label{eq:def_CAi}
\ee
where on the rightmost side of Eq.~(\ref{eq:def_CAi}), we formally separate the combination of correlation functions computed at $(E_\gamma, \bs{k})$, which we group into $\overline{C}_{A}^{\,i}(t_\gamma, E_\gamma, \bs{k}; t_W)$, from those computed at zero photon momentum and energy, which correspond to the point-like contribution $C_{pt}^{i}(t_\gamma, E_\gamma, \bs{k}; t_W)$. Explicitly, we have 
\begin{align}
\label{eq:axial_pt_corr}
\overline{C}^{\,i}_A(t_\gamma, E_\gamma, \bs{k}; t_W) \equiv 
\left(
\begin{array}{c}
C^{[3,0]}_{A}(t_\gamma, E_\gamma, \bs{k}; t_W)  \\[5pt]
(C^{33}_{A} + C^{11}_{A})(t_\gamma, E_\gamma, \bs{k}; t_W) \\[5pt]
(C^{33}_{A}- C^{11}_{A})(t_\gamma, E_\gamma, \bs{k}; t_W)
\end{array}
\right), \quad 
C^i_{pt}(t_\gamma, E_\gamma, \bs{k}; t_W) \equiv 
\left(
\begin{array}{c}
0 \\[5pt]
(C^{33}_{A}\frac{2m_K E_\gamma - E_\gamma^2}{2m_K E_\gamma - k^2} + C^{11}_{A})(t_\gamma, 0, \bs{0}; t_W) \\[5pt]
(C^{33}_{A}\frac{2m_K E_\gamma - E_\gamma^2}{2m_K E_\gamma - k^2} - C^{11}_{A})(t_\gamma, 0, \bs{0}; t_W)
\end{array}
\right).
\end{align}
Finally, we introduce the differential axial form factors through
\be
\delta A^i(t_\gamma, x_k, y_k; t_W)
\equiv 
M^{ij}\,\tilde{C}^j_A(t_\gamma, E_\gamma, \bs{k}; t_W) \equiv (\delta \overline{A}^{\,i} -\delta A^{i}_{pt})(t_\gamma, x_k, y_k; t_W)~,
\label{eq:4L_dFFA}
\ee
where $\delta \overline{A}^{\,i}$ and $\delta A^{i}_{pt}$ are defined as 
\be
\delta \overline{A}^{\,i}(t_\gamma, x_k, y_k; t_W) \equiv M^{ij} \overline{C}^{\,j}_A(t_\gamma, E_\gamma, \bs{k}; t_W), \qquad \delta A^{i}_{pt}(t_\gamma, x_k, y_k; t_W) \equiv M^{ij} C^j_{pt}(t_\gamma, E_\gamma, \bs{k}; t_W).
\ee
By construction, the $t_\gamma$-integrals of the differential axial form factors define estimators of the corresponding axial form factors,
\be
A^i(x_k, y_k; t_W)
= \int_{-\infty}^{\infty} d t_\gamma \,
\delta A^i(t_\gamma, x_k, y_k; t_W)
= A^i(x_k,y_k) + \ldots \, ,
\label{eq:FFAfromdFFA}
\ee
where the ellipses denote contributions exponentially suppressed in $t_W$. For future convenience, we also define contributions to the form factors from the first and second time-orderings as
\begin{align}
\label{eq:TO_separation}
F_{1}(x_{k},y_{k};t_{W}) &= \int_{-\infty}^{0} dt_{\gamma} \, \delta F(t_\gamma, x_k, y_k; t_W)
= F_{1}(x_k,y_k) + \ldots~,\qquad F=\{F_{V},A^{i}\}~, \nonumber \\
F_{2}(x_{k},y_{k};t_{W}) &= \int^{\infty}_{0} dt_{\gamma} \, \delta F(t_\gamma, x_k, y_k; t_W)
= F_{2}(x_k,y_k) + \ldots~,\qquad F=\{F_{V}, A^{i}\}~, 
\end{align}
so that $F(x_{k},y_{k}) = F_{1}(x_k,y_k) + F_{2}(x_k,y_k)$~.

In Sec.~\ref{Sec:belowth} we present the numerical results for the differential form factors defined in Eqs.~(\ref{eq:4L_dFV}) and~(\ref{eq:4L_dFFA}). We analyse in detail the systematic uncertainties associated with finite-$t_W$ effects, as well as those arising from the reconstruction of the $t_\gamma$ integral on a lattice with finite temporal extent. In particular, we discuss several strategies aimed at accelerating the convergence of the time integral. The introduction of differential form factors proves especially advantageous in this context, as it allows these effects to be investigated directly at the level of the integrand defining the physical observables. We then present our final results for the form factors in the kinematical region below the two-pion threshold. The strategy adopted to determine the form factors in the complementary region $x_k \geq 2m_\pi/m_K$ is described in the following subsection.

\subsection{Strategy for the numerical evaluation of the form factors above the $\pi\pi$ threshold}
\label{sec:4L_strategyFFAbove}
We now describe the strategy adopted to determine the form factors in the kinematical region $x_k \geq \frac{2m_\pi}{m_K}$, i.e.\ above the two-pion production threshold. As discussed in Sec.~\ref{Sec:Analytic_properties}, in this region the Wick rotation of the second time-ordering contribution to the hadronic tensor cannot be performed. To overcome this issue we employ the SFR method introduced in Ref.~\cite{Frezzotti:2023nun}.

To illustrate the basic idea behind the SFR approach, consider a generic Minkowskian correlator of the form
\be
C(t) = \langle 0 | J_{1}(t,\bs{k})\,J_{2}(0) | P\rangle, \qquad t>0,
\ee
where $J_{1}$ and $J_{2}$ are external currents (in our case the weak and electromagnetic currents), and $P$ is a single-hadron state, stable under the strong interactions (in our case a charged kaon). We introduce the associated amplitude
\be
H(E_{\gamma}) = \int_0^\infty dt\, e^{iE_{\gamma}t}\, C(t),
\label{eq:SFR_genericH}
\ee
where $E_{\gamma}$ is an external energy (the photon energy in the present case). Introducing the spectral density $\rho(E)$ through
\be
C(t)=\int_{0}^{\infty}\frac{dE}{2\pi}\, e^{-iE t}\,\rho(E),
\ee
one obtains the spectral representation
\be
H(E_{\gamma})=\lim_{\varepsilon\to0^+}\int_{0}^{\infty}\frac{dE}{2\pi}
\frac{\rho(E)}{E-E_{\gamma}-i\varepsilon} = {\rm P.V.} \int_{0}^{\infty}\frac{dE}{2\pi}
\frac{\rho(E)}{E-E_{\gamma}}  + \frac{i}{2}\rho(E_{\gamma})
\label{eq:SFR_dispersion}
\ee
When $\rho(E)$ has support at $E = E_{\gamma}$, the $i\varepsilon$ prescription generates an imaginary part for the physical amplitude.

On the lattice, as already explained, one has access instead to the Euclidean correlator
\be
C_E(\tau)= C(t=-i\tau)
=\int_{0}^{\infty}\frac{dE}{2\pi}\, e^{-E\tau}\,\rho(E),
\label{eq:SFR_EuclGeneric}
\ee
and a Wick rotation of Eq.~(\ref{eq:SFR_genericH}) would lead, for the amplitude $H(E_{\gamma})$, to the integral
\be
\int_0^\infty d\tau\, e^{E_{\gamma}\tau}\,C_E(\tau)~,
\ee
which converges only if the spectral density vanishes at and below the external energy $E_{\gamma}$; this corresponds to the standard Euclidean-time integration formula used below the two-pion threshold. Whenever intermediate states with energies $E\leq E_{\gamma}$ contribute, the Euclidean integral diverges, signaling that the Minkowskian amplitude cannot be reconstructed directly from a simple Euclidean-time integration in that kinematical regime~\cite{Frezzotti:2023nun}. This is precisely what occurs in our case (see  Sec.~\ref{Sec:Analytic_properties}) for $k^{2} > 4m_{\pi}^{2}$ in the second time ordering $t_{\gamma} > 0$. In principle, if one could invert Eq.~(\ref{eq:SFR_EuclGeneric}), the spectral density would be directly accessible from Euclidean lattice data, and $H(E_{\gamma})$ determined performing directly the integral in Eq.~(\ref{eq:SFR_dispersion}).  However, extracting $\rho(E)$ directly from $C_{E}(\tau)$ is ill-posed, as it requires to perform an inverse Laplace transform of the Euclidean correlator computed on the lattice, which is known only on a finite number of time separations and is affected by statistical errors. 

The SFR method overcomes this problem by keeping the regulator $\varepsilon$ finite and computing instead the smeared amplitude
\be
H(E_{\gamma};\varepsilon)=
\int_{0}^{\infty}\frac{dE}{2\pi}
\rho(E)\,K(E-E_{\gamma};\varepsilon)~,\qquad K(x;\varepsilon) \equiv \frac{1}{x-i\varepsilon}~.
\label{eq:SFR_smeared_generic}
\ee
Crucially, the smeared amplitude at non-zero $\varepsilon$ can be reconstructed from knowledge of the Euclidean correlator using spectral-density reconstruction algorithms such as the Hansen--Lupo--Tantalo (HLT) method~\cite{Hansen:2019idp}. The HLT method is nowadays a standard tool in lattice QCD, and it has been applied in many cases where analytic continuation issues arise. We will discuss the technical details of the HLT implementation in Sec.~\ref{Sec:4L_aboveth}. The physical amplitude is then obtained by computing the smeared amplitude for several non-zero values of $\varepsilon$, and then extrapolating the smeared results to $\varepsilon\to0^+$. 

The practical applicability of this strategy requires working in a window where two conditions can be simultaneously satisfied~\cite{Frezzotti:2023nun}. First, at finite spatial extent $L$, the energy spectrum is always discrete (the spectral density is always a collection of delta functions, even if the infinite-volume spectral density contains continuum multiparticle states) and the smearing induced by $\varepsilon$ must be larger than the typical level spacing of the finite-volume states in order to keep finite-size effects manageable; this corresponds parametrically to the condition
\be
\frac{1}{L}\ll \varepsilon.
\label{eq:SFR_epsL}
\ee
Second, the $\varepsilon\to0$ extrapolation, which must be carried out after performing the infinite-volume limit $L\to\infty$, requires $\varepsilon$ to be small compared to the characteristic energy scale over which the physical amplitude varies significantly around $E_{\gamma}$,
\be
\varepsilon \ll \Delta(E_{\gamma}),
\label{eq:SFR_epsDelta}
\ee
so that a controlled extrapolation to $\varepsilon\to0^+$ can be carried out. In particular, if the spectral density presents a resonance peak located at $M$ and of width $\Gamma$, Ref.~\cite{Frezzotti:2023nun} showed that, assuming a Breit-Wigner parameterization, one has 
\be
\Delta(E_{\gamma})\simeq \sqrt{(E_{\gamma}-M)^2 + (\Gamma/2)^2}.
\label{eq:SFR_DeltaRes}
\ee
Equation\,(\ref{eq:SFR_DeltaRes}) indicates that the extrapolation becomes more delicate when $E_{\gamma}$ lies close to the peak of a narrow resonance (small $\Gamma$), whereas the extrapolation is smoother when the relevant energies are sufficiently far from the resonance region and/or when the resonance is broad. When the condition in Eq.~(\ref{eq:SFR_DeltaRes}) is satisfied, the $\varepsilon$-dependence of the smeared amplitude can be accurately described by a low-order polynomial in $\varepsilon$ (see Ref.~\cite{Frezzotti:2023nun} for further details). Because the evaluation of the smeared amplitude becomes increasingly challenging as $\varepsilon$ decreases, the feasibility of a controlled $\varepsilon \to 0$ extrapolation depends critically on the specific problem under consideration (namely, the range of $\varepsilon$ over which the previous conditions are satisfied) as well as on the achievable statistical precision of the correlator $C_E(\tau)$. The latter strongly affects the smallest value of $\varepsilon$ that can be reliably determined.

We now relate the general discussion of the SFR/HLT method to the present calculation. In our case, the role of the Euclidean correlator $C_E(\tau)$ is played by
$e^{-E_{\gamma}t_{\gamma}}C_{W}^{\mu\nu}(t_\gamma,E_{\gamma},\bs{k};t_W)$,
defined in Eq.~(\ref{Eq:Def_correlationfunction}).
As already discussed, only a subset of the contributions to the form factors is affected by analytic-continuation issues, and therefore only this part requires the use of the SFR/HLT method. In particular, the contribution from the first time-ordering contribution presents no difficulty and is computed using the standard Euclidean-time integration described in the previous section. In the notation introduced in Eq.~(\ref{eq:TO_separation}), this corresponds to the contribution $F_{1}$, where $F=(A^{i},F_{V})$. Moreover, as discussed in Sec.~\ref{Sec:Analytic_properties}, diagram (b) in Fig.~\ref{Fig:Diagrams}, corresponding to the quark-connected contribution in which the photon is emitted by the strange quark, does not suffer from analytic-continuation problems even in the second time ordering, and can be computed using standard Euclidean-time integration.

Then, it is useful to decompose the Euclidean correlation function in Eq.~(\ref{Eq:Def_correlationfunction}) as
\begin{align}
C^{\mu\nu}_{W}(t_\gamma, E_\gamma, \boldsymbol{k}; t_W)
= C^{\mu\nu}_{W;s}(t_\gamma, E_\gamma, \boldsymbol{k}; t_W)
+ C^{\mu\nu}_{W;u}(t_\gamma, E_\gamma, \boldsymbol{k}; t_W)\, ,
\label{eq:4L_Cels}
\end{align}
where $C_{W;s}^{\mu\nu}$ corresponds to the contribution of diagram (b) in Fig.~\ref{Fig:Diagrams}, while $C_{W;u}^{\mu\nu}$ contains the remaining contributions (diagram (c) together with the quark-disconnected diagram (d)). We denote their contribution to $F_{2}(x_k,y_k)$ by $F_{2,s}(x_{k},y_{k})$ and $F_{2,u}(x_k,y_k)$ (with $F=(A^{i},F_{V})$), respectively. In general, we will use the subscripts ``$s$'' and ``$u$'' to indicate quantities originating from the $C_{W;s}^{\mu\nu}$ and $C_{W;u}^{\mu\nu}$ components of the correlation function, respectively.

The contribution from the second time-ordering from diagrams (c) and (d) is evaluated using the SFR/HLT method. In analogy with the general discussion presented above, we introduce the spectral density associated with $C_{W;u}^{\mu\nu}$ through
\begin{equation}
\label{eq:4L_CErho}
C^{\mu\nu}_{W;u}(t_\gamma,E_\gamma,\bs{k};t_W)
=
\int_{0}^{\infty}\frac{dE}{2\pi}\,
e^{-t_\gamma(E-E_\gamma)}\,\rho^{\mu\nu}_{W;u}(E,\bs{k};t_W)\, .
\end{equation}
The corresponding contribution to the smeared hadronic tensor in terms of this spectral density is given by 
\begin{equation}
H^{\mu\nu}_{W,2;u}(\varepsilon,E_\gamma,\bs{k};t_W)
=
\int_{0}^{\infty}\frac{dE}{2\pi}\,
K(E-E_\gamma;\varepsilon)\,\rho^{\mu\nu}_{W;u}(E,\bs{k};t_W)\, ,
\end{equation}
so that 
\begin{equation}
\label{eq:4L_Hrho}
H^{\mu\nu}_{W,2;u}(E_\gamma,\bs{k};t_W)
=
\lim_{\varepsilon\to0^+}
H^{\mu\nu}_{W,2;u}(\varepsilon,E_\gamma,\bs{k};t_W)~.
\end{equation}
The smeared hadronic tensor $H^{\mu\nu}_{W,2;u}(\varepsilon,E_\gamma,\bs{k};t_W)$ can be reconstructed using the HLT method directly from the Euclidean correlators $C^{\mu\nu}_{W;u}(t_\gamma,E_\gamma,\bs{k};t_W)$.

We now construct an analogous spectral representation directly at the level of the form factors. For the vector form factor this proceeds straightforwardly. We introduce the spectral density $\rho_{V;u}$ associated with the differential form factor $\delta F_{V;u}$ via
\begin{equation}
\delta F_{V;u}(t_\gamma,x_k,y_k;t_W)
=
\int_{0}^{\infty}\frac{dE}{2\pi}
e^{-t_\gamma(E-E_\gamma)}\rho_{V;u}(E,x_{k},y_{k};t_W)\, .
\label{eq:diff_V_HLT}
\end{equation}
The full vector form factor above the two-pion threshold is decomposed as
\begin{align}
\label{eq:FV_split}
F_{V}(x_k,y_k;t_W)
=
F_{V,\mathrm{Wick}}(x_k,y_k;t_W)
+
F_{V,\mathrm{SFR}}(x_k,y_k;t_W)\, .
\end{align}
The two terms on the right-hand side are defined as follows.

The Wick contribution collects all terms that can be computed from Euclidean-time integration of the corresponding differential form factors,
\begin{align}
\label{eq:FV_Wick_def}
F_{V,\mathrm{Wick}}(x_k,y_k;t_W)
&=
F_{V,1}(x_k,y_k;t_W)
+F_{V,2;s}(x_k,y_k;t_W)
+\int_{0}^{\infty}dt_\gamma\,e^{-E_\gamma t_\gamma}\,
\delta F_{V;u}(t_\gamma,x_k,y_k;t_W)\, ,
\end{align}
where we remind that $F_{V,1}$ denotes the full first time-ordering contribution, i.e. $F_{V,1}=F_{V,1;u}+F_{V,1;s}$.

The remaining contribution is evaluated using the SFR/HLT method. The SFR form factor is defined as the $\varepsilon\to0^+$ limit of a smeared quantity,
\begin{equation}
\label{eq:FV_SFR_def}
F_{V,\mathrm{SFR}}(x_k,y_k;t_W)
=
\lim_{\varepsilon\to0^+}
F_{V,\mathrm{SFR}}(\varepsilon,x_k,y_k;t_W)\, ,
\end{equation}
where the smeared SFR term is given by
\begin{equation}
\label{eq:FF_V_HLT}
F_{V,\mathrm{SFR}}(\varepsilon,x_k,y_k;t_W)
=
\int_{0}^{\infty}\frac{dE}{2\pi}
\rho_{V;u}(E,x_{k},y_{k};t_W)
\bigl[K(E-E_\gamma;\varepsilon)-K(E;\varepsilon)\bigr]\, .
\end{equation}

Note that the Wick contribution includes an additional term (the last term in Eq.~\eqref{eq:FV_Wick_def}) that is subtracted in the SFR part through the modified kernel
\begin{align}
\label{eq:Ktilde_def}
K_{{\rm sub}}(E-E_\gamma;\varepsilon)
=
K(E-E_\gamma;\varepsilon)-K(E;\varepsilon)\, .
\end{align}
This rearrangement leaves the total form factor unchanged, but allows the Wick and SFR contributions to be separately $O(a)$-improved, as discussed in Sec.~IV-A of Ref.~\cite{Frezzotti:2024kqk} for $B_s\to\mu^+\mu^-\gamma$.\footnote{A naive separation into first- and second-time-ordering contributions leads to an $O(a)$ ambiguity associated with the contact term ($t_{\gamma}=0$, which does not belong to either time ordering). In the present decomposition, the SFR term is free of contact terms due to the zero-energy subtraction, while the Wick term contains both time orderings, for which the treatment of the contact term is unambiguous.}
In the following, we analyze and perform continuum extrapolations separately for $F_{V,\mathrm{Wick}}$ and $F_{V,\mathrm{SFR}}$.

The axial form factors admit a decomposition analogous to the one introduced for the vector case. The only additional ingredient concerns the treatment of the point-like contribution. Since the latter enters the differential form factors (Eqs.~(\ref{eq:axial_pt_corr})--(\ref{eq:4L_dFFA})) only through the axial correlation function $C_{A}^{\mu\nu}$ evaluated at $E_{\gamma}=0$ and $\bs{k}=0$, it can be obtained directly by Euclidean-time integration and therefore does not require the use of the SFR/HLT method.

For this reason, we introduce a spectral-density representation for the unsubtracted differential form factors $\delta\overline{A}^{\,i}$ in Eq.~(\ref{eq:4L_dFFA}), defining the spectral densities $\rho_{A^{i};u}$ through
\begin{align}
\label{eq:diff_A_HLT}
\delta \overline{A}^{\,i}_{u}(t_{\gamma},x_{k},y_{k};t_{W}) =
\int_{0}^{\infty}\frac{dE}{2\pi}
e^{-t_\gamma(E-E_\gamma)}\,\rho_{A^{i};u}(E,x_{k},y_{k};t_W)\, .
\end{align}
As in the vector case, we introduce the smeared SFR contribution as
\begin{equation}
\label{eq:FF_A_HLT}
A^{i}_{\mathrm{SFR}}(\varepsilon,x_k,y_k;t_W)
=
\int_{0}^{\infty}\frac{dE}{2\pi}
\rho_{A^{i};u}(E,x_{k},y_{k};t_W)\,K_{\rm sub}(E-E_\gamma;\varepsilon)\, ,
\end{equation}
and write the full axial form factors above the two-pion threshold as
\begin{align}
A^{i}(x_{k},y_{k};t_{W}) =
A^{i}_{{\rm Wick}}(x_{k},y_{k};t_{W})
+
A^{i}_{\rm SFR}(x_{k},y_{k};t_{W}) ,
\end{align}
where $A^{i}_{\rm SFR}(x_{k},y_{k};t_{W})=\lim \limits _{ \varepsilon\to 0^{+}} A^{i}_{\rm SFR}(\varepsilon,x_{k},y_{k};t_{W})$.

The ``Wick'' contribution, obtained through Euclidean-time integration, is defined as
\begin{align}
\label{eq:FA_wick}
A^{i}_{{\rm Wick}}(x_{k},y_{k};t_{W}) \equiv
A^{i}_{1}(x_{k},y_{k};t_{W})
+
A^{i}_{2;s}(x_{k},y_{k};t_{W})
+
\int_{0}^{\infty} dt_{\gamma}
\left(
e^{-E_{\gamma}t_{\gamma}}\,
\delta \overline{A}^{\,i}_{u}(t_{\gamma},x_{k},y_{k};t_{W})
-
\delta A^{i}_{{\rm pt};u}(t_{\gamma}, x_{k},y_{k};t_{W})
\right) ,
\end{align}
which differs from the definition used for the vector form factor in Eq.~(\ref{eq:FV_Wick_def}) only through the explicit subtraction of the point-like contribution (last term in Eq.~(\ref{eq:FA_wick})).

In the $t_{W}\to \infty$ limit, one has for each of the four form factors, $F=\{F_{V},A^{i}\}$,
\begin{align}
 F_{\rm SFR}(x_{k},y_{k})\equiv \lim_{t_{W}\to \infty} F_{\rm SFR}(x_{k},y_{k},t_{W})~, \quad F_{\rm Wick}(x_{k},y_{k})\equiv \lim_{t_{W}\to \infty} F_{\rm Wick}(x_{k},y_{k},t_{W})~,
\end{align}
and the full form factors are given by
\begin{align}
F(x_{k},y_{k}) = F_{{\rm SFR}}(x_{k},y_{k}) + F_{\rm Wick}(x_{k},y_{k})~.
\end{align}

In Sec.~\ref{Sec:4L_aboveth}, we present our results for $F_{V,{\rm SFR}}(x_{k},y_{k})$ and $A^{i}_{\rm SFR}(x_{k},y_{k})$, while the Wick contributions $F_{V,{\rm Wick}}(x_{k},y_{k})$ and $A^{i}_{\rm Wick}(x_{k},y_{k})$, which are obtained following the same strategy used below the two-pion threshold, are reported in App.~\ref{app:4L_Wick}.

To accelerate the convergence of the smeared amplitudes towards the $\varepsilon \to 0$ limit, we further employ in the HLT reconstruction a modified kernel that suppresses the leading finite-$\varepsilon$ effects. Specifically, instead of the naive kernel, for which the leading correction starts at $O(\varepsilon)$~\cite{Frezzotti:2023nun}, we use the improved kernel
\begin{align}
\label{eq:ImprovedKernel}
\tilde{K}(x;\varepsilon) \;\to\;
\frac{
\varepsilon^{+}\,K_{\rm sub}(x;\varepsilon^{-})
-
\varepsilon^{-}\,K_{\rm sub}(x;\varepsilon^{+})
}{
\varepsilon^{+}-\varepsilon^{-}
},
\qquad
\varepsilon^{\pm} = \varepsilon(1 \pm\delta),
\end{align}
with $\delta \in (0,1)$. One can readily verify that if the unimproved kernel induces corrections starting at $O(\varepsilon)$, the modified kernel generates only $O(\varepsilon^{2})$ effects. In the following we will use this improved kernel in the HLT procedure. After preliminary studies, we adopt $\delta=0.3$. The results at non-zero $\varepsilon$ show only a mild dependence on this parameter; in particular, varying $\delta$ within the range $\delta \in [0.05,0.4]$ produces negligible changes within statistical uncertainties.

Before closing this section, we comment on the range of $\varepsilon$ for which a controlled polynomial extrapolation to $\varepsilon \to 0^{+}$ can be expected. In the infinite-volume limit, the spectral densities $\rho_{V}$ and $\rho_{A^{i}}$ are dominated by the two-pion continuum. On physical grounds one expects these spectral densities to be relatively smooth functions near threshold, with a peak associated with the $\rho$ resonance. Assuming a Breit--Wigner description of the resonance and applying the criterion of Eq.~(\ref{eq:SFR_DeltaRes}), one obtains
\begin{align}
\Delta_{\rho}(E_{\gamma},\bs{k}) =
\sqrt{ (E_{\gamma}-E_{\rho}(\bs{k}))^{2} + (\Gamma_{\rho}/2)^{2}},
\end{align}
with $E_{\rho}(\bs{k}) = \sqrt{ m_{\rho}^{2}+|\bs{k}|^{2}}$. Since in the physical region of interest $k^{2}< m_{K}^{2}$ and $|\bs{k}|\leq m_{K}/2$, implying $E_{\gamma} \leq \sqrt{m_{K}^{2} + |\bs{k}|^{2}}$, one finds
\begin{align}
\label{eq:rho_max}
\Delta_{\rho}(E_{\gamma},\bs{k}) \gtrsim \Delta_{\rho}^{\rm max} = 270~{\rm MeV}~,
\end{align}
using $m_{\rho}=0.775~{\rm GeV}$ and $\Gamma_{\rho}=0.150~{\rm GeV}$. This channel thus provides an ideal arena for the application of the SFR method, as moderately large values of $\varepsilon$ already allow one to reach the asymptotic regime in which corrections to the $\varepsilon\to0$ limit are described by a low-order polynomial in $\varepsilon$.

\section{Simulation details}
\label{sec:Lattice_setup}
This work uses the gauge configurations produced by the Extended Twisted Mass Collaboration (ETMC) with $N_{f}=2+1+1$ dynamical Wilson-clover twisted mass fermions~\cite{Frezzotti:2000nk,Frezzotti:2003xj} at the physical point\footnote{The simulated pion masses differ from the charged pion mass $m_{\pi^{\pm}} \simeq 139~{\rm MeV}$ by only about $\pm 2~{\rm MeV}$. The resulting effects are expected to be of the same order of the neglected isospin-breaking corrections, which are parametrically of order $O(\alpha_{\rm em}) \simeq O( \frac{m_d - m_u}{\Lambda_{QCD}}) \simeq O(1\%)$, and therefore smaller than the errors of the results presented in Secs.~\ref{Sec:belowth} and~\ref{Sec:4L_aboveth}.}. This framework guarantees automatic $\mathcal{O}(a)$ improvement of parity-even observables~\cite{Frezzotti:2003ni,Frezzotti:2004wz}. Essential information on the ensembles that we use for the present computation are collected in Table~\ref{tab:simudetails}, and we refer the reader to Ref.~\cite{ExtendedTwistedMassCollaborationETMC:2024xdf} for additional details. 
At each lattice spacing, the mass $m_{s}$ of the strange quark has been set in order to reproduce $m_K= 494.6~{\rm MeV}$, following the prescription recommended for isospin-symmetric QCD in the latest FLAG review~\cite{FlavourLatticeAveragingGroupFLAG:2024oxs}. 
We employ the mixed-action lattice setup introduced in~Ref.~\cite{Frezzotti:2004wz}, and described in the appendices of Ref.~\cite{ExtendedTwistedMassCollaborationETMC:2024xdf}. In this setup, the action of the valence quarks is discretized in the so-called Osterwalder--Seiler (OS) regularization, namely
\begin{align}
\label{eq:tm_action}
S = \sum_{f=u,d,s,c}\sum_{x} \bar{q}_{f}(x) \left[ \gamma_{\mu}\bar{\nabla}_{\mu}[U] -ir_{f}\gamma^{5}(W^{\rm cl}[U] + m_{\rm cr}) + m_{f}  \right] q_{f}(x)~,
\end{align}
where $W^{\rm cl}[U]$ is the Wilson-clover term~\cite{Sheikholeslami:1985ij}, $m_{\rm cr}$ is the critical mass, $m_{f}$ the quark mass of the flavour $f$ (with $m_{u}=m_{d}=m_{l}$), and $r_{f}=\pm 1$ is the sign of the twisted-Wilson parameter  for the flavour $f$ ($r_{u,c}= -r_{d,s}=1$). 

In the OS regularization, the weak hadronic current is written as 
\begin{align}
j_{W}^{\nu}(x) = \bar{q}_{u}(x)\;\Gamma^{\nu}_{W}\;q_{s}(x) = \bar{q}_{u}(x)\;( Z_{A}\gamma^{\nu} - Z_{V}\gamma^{\nu}\gamma_{5})\; q_{s}(x)~, 
\end{align}
where $Z_{A/V}$ are the renormalization factors ensuring that the Ward identities are satisfied\,\footnote{Note that the renormalisation factors to be used in Twisted-Mass at maximal twist are chirally-rotated with respect to the ones of standard Wilson fermions. This is a consequence of the fact that the up-type and down-type quark fields in the action of Eq.~(\ref{eq:tm_action}) are discretised
with opposite values of the Wilson parameter.}, and which are reported in Tab.~\ref{tab:simudetails}. For this work, we employ the local version of the  electromagnetic current, which renormalizes multiplicatively with $Z_V$.
The interpolating operator appearing in Eq.~(\ref{Eq:Def3pt}) is given by 
\begin{align}
\label{eq:interpolators}
P_{K}^{\dag}(t,\vec{x}) = \sum_{\vec{y}}\bar{q}_{s}(t,\vec{x}) G^{N}_{t}(\vec{x},\vec{y}) \gamma_5 q_{u}(t,\vec{y}) \;, 
\end{align}
where $G_{t}
(\vec{x},\vec{y})$ is the Gaussian smearing operator
\begin{align}
G_{t}(\vec{x},\vec{y}) = \frac{1}{1+ 6\kappa}\left( \delta_{\vec{x},\vec{y}} + \kappa H_{t}(\vec{x},\vec{y})    \right)~,
\end{align}
with
\begin{flalign}
H_{t}(\vec{x}, \vec{y}) = \sum_{\mu=1}^{3}\left( U^{\star}_{\mu}(t,\vec{x})\delta_{\vec{x}+\hat{\mu},\vec{y}} + U^{\star\dagger}_{\mu}(t,\vec{x}-\hat{\mu})\delta_{\vec{x}-\hat{\mu},\vec{y}}    \right)~,
\end{flalign}
and we have indicated by $U^{\star}_{\mu}(x)$ the APE-smeared links, defined as in Ref.~\cite{Becirevic:2012dc}.
We set $\kappa=0.4$, and fix on each ensemble the number of smearing-steps $N$ in Eq.~(\ref{eq:interpolators}) to obtain a smearing radius $r_{0}= a\sqrt{N}/\sqrt{\kappa^{-1}+6} \simeq 0.4~{\rm fm}$. We have found that this choice of the smearing radius provides the optimal overlap with the ground-state charged kaon~\cite{DiPalma:2024jsp}. The mass of the kaon $m_K$ and the matrix element $\langle K^{-}(\vec{0}) | P_{K}^{\dagger}(0) | 0 \rangle$ appearing in Eq.~(\ref{Eq:Def_correlationfunction}) are extracted from the following two-point correlation function:
\begin{align}
C_{\rm 2}(t) = \sum_{\vec{x}}  \langle 0 | \hat{\mathrm{T}} [P_{K}(t,\vec{x}) P_{K}^{\dag}(0)] | 0 \rangle~.  
\end{align}

\begin{figure}[]
    \centering
    \includegraphics[width=0.45
    \linewidth]{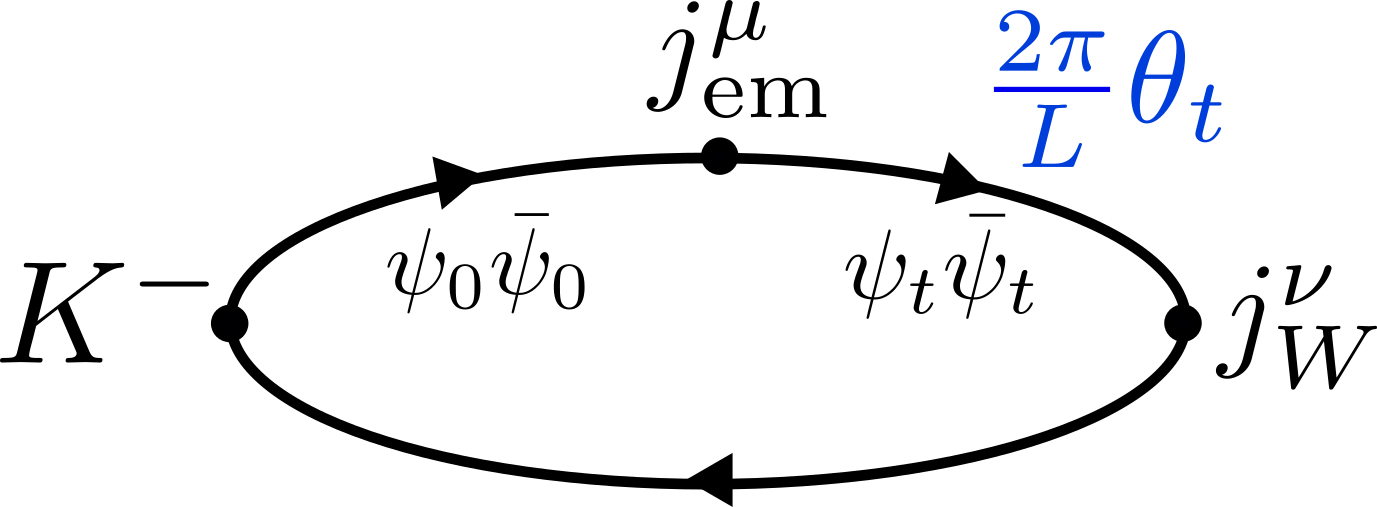}
    \caption{The diagram represents the quark-line connected contribution and illustrates our choice of the spatial boundary conditions, which allows us to set arbitrary values for the meson and photon spatial momenta.}
    \label{fig:twisted}
\end{figure}

We now describe the strategy adopted to compute the dominant quark-connected contribution. We have determined the three-point correlation function $C_{3,W}^{\mu\nu}(t_\gamma, \bs{k}; t_W)$ in Eq.~(\ref{Eq:Def3pt}) on all the five ensembles of Tab.~\ref{tab:simudetails}. For each gauge configuration, we employ 72 (36) time-wall (spin-diluted) stochastic sources to evaluate the Wick contraction corresponding to photon emission from the anti-up quark (strange quark). We use the so-called twisted boundary conditions~\cite{Sachrajda:2004mi,deDivitiis:2004kq} to inject arbitrary  values of the photon and kaon spatial momenta in addition to those proportional to $2\pi/L$ obtained with periodic boundary conditions. Specifically, the two propagators, among which  the electromagnetic current is inserted, are computed from the contraction of two different quark fields that we call $\psi_0(x)$ and $\psi_t(x)$. These fields have identical masses and quantum numbers but satisfy different spatial boundary conditions, given by\footnote{We corrected a transcription typo in the definition of the phase implementing the boundary condition of $\psi_t(x)$, which was present in~\cite{Desiderio:2020oej} and in subsequent works.}
\be
\psi_{0}(x + \bs{n}L) = \psi_{0}(x), \qquad \psi_{t}(x + \bs{n}L) = \exp [ i 2\pi \bs{n}\cdot \bs{\theta}_t ] \psi_{t}(x), \qquad \bs{\theta}_t=(0,0,\theta_t)~,
\label{Eq:Boundaryconditions}
\ee
where $\bs{n}$ is an integer valued three-vector, and $\theta_{t}$ is related to the lattice three-momentum of the photon by
\be
k_z = -\frac{2}{a} \sin \bigg[ \frac{a \pi}{L} \theta_t \bigg].
\label{Eq:Momenta}
\ee
Fig.~\ref{fig:twisted} shows the connected Feynman diagram with the quark field $\psi_{t}$ which satisfies non-periodic spatial boundary conditions\footnote{An alternative choice of twisted-boundary condition will be discussed in Sec.~\ref{sec:4L_FVsfrandAisfr}.}. As discussed in Refs.~\cite{Sachrajda:2004mi,deDivitiis:2004kq}, the use of twisted-boundary conditions induces small violations of unitarity, which are however exponentially vanishing with the lattice spatial extent $L$. 

\begin{figure}[t]
    \centering
    \includegraphics[width=0.8\linewidth]{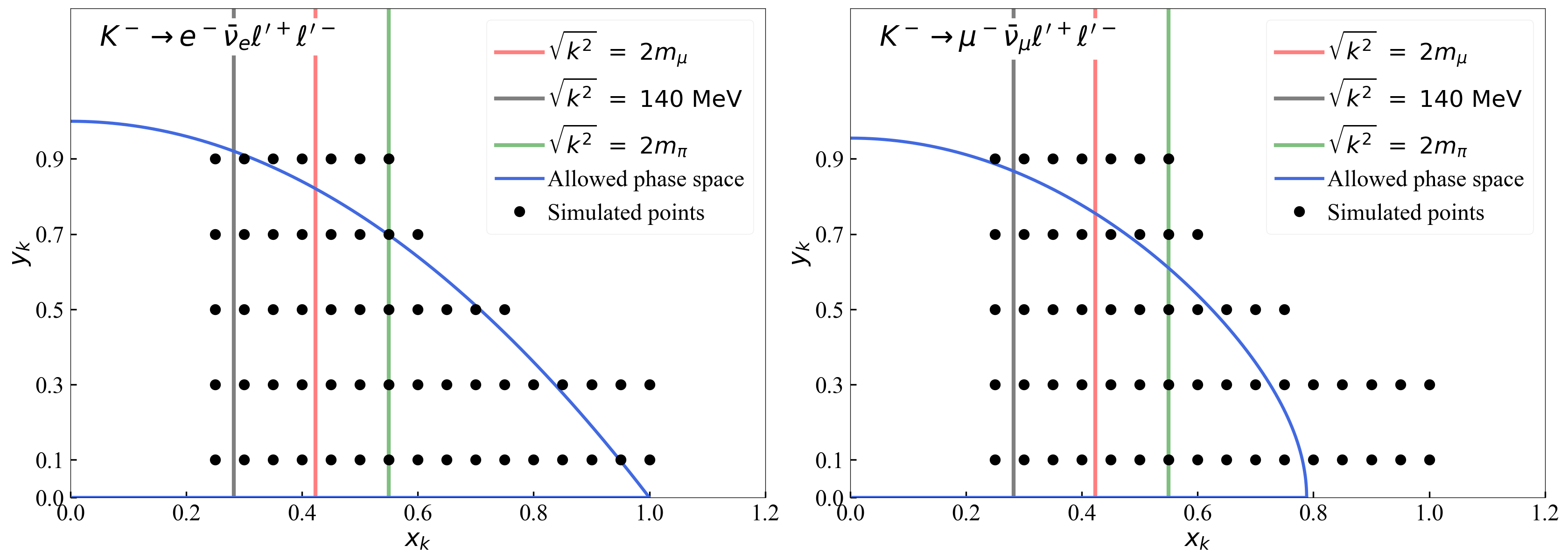}
    \caption{ The blue lines show the largest allowed value of $y_{k}$ as a function of $x_{k}$ in the case of the  $K^- \to e^- \bar{\nu}_e \ell'^+\ell'^-$ (left panel) and $K^- \to \mu^- \bar{\nu}_\mu \ell'^+\ell'^-$ (right panel) decays. The black points correspond to the simulated values of $x_{k}$ and $y_{k}$. The black (red) vertical line shows the minimum value of $x_k$ when the final states contain an $e^{+}e^{-}$ ($\mu^{+}\mu^{-}$) pair. Finally, the green line indicates the position of the two-pion threshold.}
    \label{fig:4L_Ps}
\end{figure}

For the quark-connected contribution, the form factors are computed at five equally-spaced values of $y_k$, namely
\be
\label{eq:ykchoice}
y_k = 0.1,\ 0.3,\ 0.5,\ 0.7,\ 0.9,
\ee
together with $y_k = 0$, which is required to perform the subtraction of the point-like contribution proportional to $f_{K}$.
 \begin{table}[]
\begin{ruledtabular}
\begin{tabular}{lcccccc}
\textrm{ID} & $L/a$ & $m_{\pi}$ [\textrm{GeV}] & $a$ [\textrm{fm}] & $Z_{V}$ & $Z_{A}$ & $N_{g}$ \\
\colrule
\textrm{B48} & $48$ & $140.3(3)$ & $0.07948(11)$ &  $0.706354(54)$ & $0.74296(19)$  & $436$ \\
\textrm{B64} & $64$ & $140.2(3)$ & $0.07948(11)$ & $0.706354(54)$ & $0.74296(19)$  & $189$ \\
\textrm{B96} & $96$ & $140.1(3)$ & $0.07948(11)$ & $0.706406(52)$ & 
$0.74261(19)$  & $66$ \\
\textrm{C80} & $80$ & $136.7(3)$ & $0.06819(14)$ & $0.725440(33)$ &  $0.75814(13)$ & $169$  \\
\textrm{D96} & $96$ & $140.8(3)$ & $0.056850(90)$ & $0.744132(31)$ &  $0.77367(10)$  &  $211$
\end{tabular}
\end{ruledtabular}
\caption{$N_{f}=2+1+1$ ETMC gauge ensembles used for the calculation of the quark-connected contribution. We give the spatial extent in lattice units $L/a$, the pion mass $m_{\pi}$, the lattice spacing $a$, the vector ($Z_{V}$) and axial ($Z_{A}$) renormalization constants, and the number of gauge configurations $N_{g}$ analyzed. 
The quark-disconnected contribution is computed exclusively on the B96 ensemble employing 300 gauge configurations.
\label{tab:simudetails}}
\end{table}
For fixed photon momentum (i.e.\ fixed $y_k$), the 3d method~\cite{Tuo:2021ewr,Giusti:2023pot} allows us to access arbitrary values of the photon virtuality using the same three-point correlation function as input. We evaluate the form factors at the following set of values
\be
x_k = 0.25 + (n-1)\,\Delta x_k, \qquad n=1,\dots,16,
\ee
with step size $\Delta x_k = 0.05$. In practice, for a given value of $y_k$, only the subset of $x_k$ values needed to cover the relevant kinematical region is computed. The actual simulated points in the $(x_k,y_k)$ plane are indicated by the black markers in Fig.~\ref{fig:4L_Ps}, which therefore specify explicitly the set of $(x_k,y_k)$ values used in the lattice calculation.

In the figure, the blue curves indicate the largest possible value of $y_{k}$ for fixed $x_{k}$ for the decays $K^- \to e^- \bar{\nu}_e \ell'^+\ell'^-$ (left panel) and $K^- \to \mu^- \bar{\nu}_\mu \ell'^+\ell'^-$ (right panel). Moreover, the red vertical lines correspond to the lowest possible value of $x_{k}$ when the final-state $\ell'^{+}\ell'^{-}$ is a dimuon ($x_{k}^{\rm min}= 2m_{\mu}/m_{K}$), while the black vertical lines correspond to the lower $x_{k}$ cut adopted in experimental analyses when $\ell'^{+}\ell'^{-}=e^{+}e^{-}$~\cite{Poblaguev:2002ug}.\footnote{In the case of $e^{+}e^{-}$ final states, experimental measurements below $x_{k}=m_{\pi^{0}}/m_{K}$ are hindered
by the large $K^{-}\to\pi^{0}\ell^{-}\bar{\nu}_{\ell}$ background.} The distribution of the simulated points allows us to basically cover the entire phase space covered by experiments. The green line marks the position of the two-pion threshold; above this line, the form factors are computed using the strategy discussed in Sec.~\ref{sec:4L_strategyFFAbove}.

The quark-disconnected contribution is treated separately from the quark-connected one.
As in the case of real-photon emission studied in Ref.\,\cite{DiPalma:2025iud}, the quark-disconnected contribution is expected to be (and indeed will turn out to be) numerically small compared to the quark-connected one. For this reason, and in order to optimize the use of computational resources, we adopt a simplified numerical setup for its evaluation. The disconnected diagrams are computed on a single gauge ensemble with $L\simeq7.6~\mathrm{fm}$ (the B96 ensemble) and for a single non-zero value of the photon momentum, $\bs{k}=(0,0,2\pi/L)$, corresponding to $y_k\simeq 0.7$, together with $y_k=0$. The resulting contribution is included in the final analysis by assigning a conservative $100\%$ uncertainty, intended to account for the absence of a dedicated continuum and infinite-volume extrapolation as well as for the limited kinematic coverage. This strategy follows closely the approach adopted in Ref.\,\cite{DiPalma:2025iud}.

The quark-disconnected contribution $C^{\mu\nu}_{3,W;\mathrm{disc}}(t_\gamma,\bs{k};t_W)$ to the three-point correlation function $C^{\mu\nu}_{3,W}(t_\gamma,\bs{k};t_W)$ in Eq.\,(\ref{Eq:Def3pt}) reads
\begin{align}
C^{\mu\nu}_{3,W;\mathrm{disc}}(t_{\gamma}, \vec{k}; t_{W})
&= Z_{V}\,\sum_{\vec{x}}\sum_{\vec{y}_{\gamma}}\,
   e^{-i\,\vec{k}\cdot\vec{y}_{\gamma}}\,
  {\rm \hat{T}} \langle 0 | \!\Bigl[
   \wick[arrows={W->-,W-<-}, below]{%
     \c1 {\bar{q}_u}(x_W)\, \Gamma_{W}^{\nu}\,\c2 q_s(x_W)\;
     \c2 {\bar{q}_s}(x)\,\gamma^5\,\c1 q_u(x)
     {\sum_{q'=u,d,s,c}} \!\! e_{q'}\;
     \c1 {\bar{q}_{q'}}(y_\gamma)\,\gamma^\mu\,\c1 q_{q'}(y_\gamma)
   }
   \Bigr] | 0 \rangle \nonumber \\[12pt]
&= Z_{V}\,\sum_{\vec{x}}\sum_{\vec{y}_{\gamma}}
    e^{-i\vec{k}\cdot\vec{y}_{\gamma}}\,
    \biggl\langle
      \mathrm{Tr}\Bigl[
        \Gamma_{W}^{\nu}
        S_{s}(x,x_{W})\gamma^{5}S_{u}(x_{W},x)\Bigr]\times\!\!\!\!\sum_{q'=u,d,s,c}\!\!e_{q'}\;\mathrm{Tr}\Bigl[\gamma^{\mu}\,S_{q'}(y_{\gamma},y_{\gamma})
      \Bigr]
    \biggr\rangle_{U},
\label{eq:disco_corr}
\end{align}
where $S_f$ denotes the quark propagator of flavour $f$, and the traces are taken over colour and Dirac indices.

In practice we only include the contributions from the $u$, $d$ and $s$ quarks in the vector loop, neglecting the charm contribution which is expected to be suppressed due to its heavy mass. Using $m_u=m_d\equiv m_{ud}$, the flavour sum can be rewritten as
\begin{align}
\sum_{q'=u,d,s} e_{q'} \Tr \left[ \gamma^{\mu}  S_{q'} (y_{\gamma},y_{\gamma})\right]
= \frac{1}{3} \Tr \left[ \gamma^{\mu} \left(  S_{u} (y_{\gamma},y_{\gamma}) - S_{s}(y_{\gamma},y_{\gamma}) \right)\right],
\label{eq:Tracesumquarks}
\end{align}
which shows explicitly that the disconnected term vanishes in the $\mathrm{SU}(3)$-symmetric limit $m_s=m_{ud}$.

The trace on the right-hand side of Eq.~(\ref{eq:Tracesumquarks}) is evaluated using the one-end trick introduced in Ref.\,\cite{DiPalma:2025iud},
\begin{align}
\Tr \left[ \gamma^{\mu} \left(  S_{u} (y_{\gamma},y_{\gamma}) - S_{s}(y_{\gamma},y_{\gamma}) \right)\right]
=  \frac{(m_{l}+m_{s})}{2} \sum_{z} \Tr\left[    
 \gamma^{5} S_{u}(z,y_{\gamma})\gamma^{5}\gamma^{\mu} S_{s}(y_{\gamma},z) - \gamma^{5}S_{s}(z,y_{\gamma})\gamma^{5}\gamma^{\mu} S_{u}(y_{\gamma},z) \right],
\label{eq:OET}
\end{align}
where the sum over $z$ is estimated stochastically using $100$ volume sources.

For the quark-disconnected contribution, we have access to all $t_W \in [0, T)$ and $t_W + t_\gamma \in [0, T)$. We exploit this fact together with the following properties of the correlation function under time-reversal
\begin{align}
\label{eq:symmantisymmC}
C^{\mu\nu}_{3,A}(t_{\gamma}, \vec{k}; t_W) = C^{\mu\nu}_{3,A}(-t_{\gamma},\vec{k}; T-t_{W})~, \quad
C^{\mu\nu}_{3,V}(t_{\gamma}, \vec{k}; t_W) = -C^{\mu\nu}_{3,V}(-t_{\gamma},\vec{k}; T-t_{W})
\end{align}
to either symmetrize or anti-symmetrize $C_{3,W;{\rm disc}}^{\mu \nu}(t_{\gamma}, \bs{k}; t_W)$, leading to a reduction of the statistical errors by a factor of about $\sqrt{2}$.

\section{Results below the $\mathbf{\pi\pi}$ threshold}
\label{Sec:belowth}
In this section we present our results in the kinematic region $x_k < \frac{2 m_\pi}{m_K}$~, where, as discussed in Sec.~\ref{Sec:FFfromCorr}, the form factors can be obtained directly from the time integral of the Euclidean correlators, or equivalently from the differential form factors defined in Eqs.~(\ref{eq:4L_dFV}) and~(\ref{eq:4L_dFFA}). The analysis closely follows the strategy adopted in Ref.~\cite{DiPalma:2025iud} for the radiative decay
$K^- \to \ell^- \bar{\nu}_\ell \gamma$.

We first illustrate the impact of several improvements applied to the differential form factors aimed at reducing statistical noise. We then discuss the dominant sources of systematic uncertainty, including the truncation of the $t_\gamma$ integral, finite-$t_W$ effects, and finite-volume corrections. Finally, we present the form factors extrapolated to the infinite-volume and continuum limits.

\begin{figure}[]
    \centering
\includegraphics[width=1.\columnwidth]{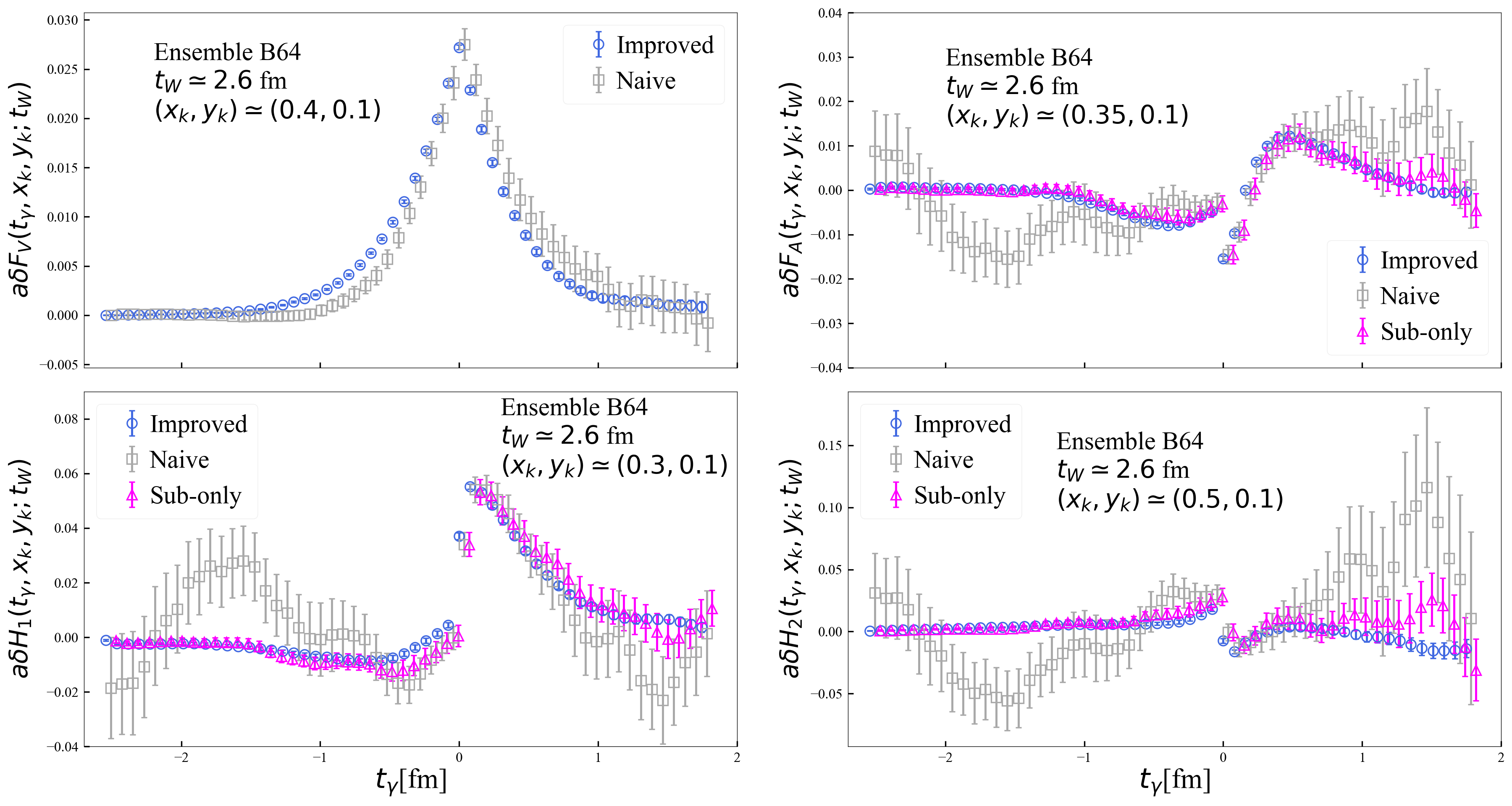}
\caption{Quark-connected contributions to the differential form factors, as functions of $t_\gamma$, corresponding to the photon emission from the anti-up quark. The results are obtained on the B64 ensemble with $t_W \simeq 2.6~\mathrm{fm}$ for several values of the photon virtuality $x_k$, and fixed momentum $y_k \simeq 0.1$. The blue circles represent the improved estimators, while the gray squares and the magenta triangles correspond to naive estimators (see text for details). Data-points are slightly shifted horizontally for clarity.}
\label{fig:4L_dFF_imp}
\end{figure}

Fig.~\ref{fig:4L_dFF_imp} shows the quark-connected contributions to the differential form factors as functions of $t_\gamma$, corresponding to photon emission from the anti-up quark. The results are obtained on the B64 ensemble with $t_W \simeq 2.6~\mathrm{fm}$, at the smallest photon momentum considered ($y_k \simeq 0.1$), and for several values of the photon virtuality $x_k$. The gray squares represent the naive estimators defined in Eqs.~(\ref{eq:4L_dFV}) and~(\ref{eq:4L_dFFA}).

The blue circles in the upper-left panel correspond to an improved estimator of the differential vector form factor,
\begin{equation}
\delta F_V(t_\gamma, x_k, y_k; t_W)
= \frac{1}{k_z} \bigg[
C_V^{21}(t_\gamma, E_\gamma, \bs{k}; t_W)
- e^{E_\gamma t_\gamma} C_V^{21}(t_\gamma, 0, \bs{0}; t_W)
\bigg],
\label{eq:dFV_impr}
\end{equation}
where we have used that $C_V^{21}(t_\gamma,0,\bs0;t_W)$ is zero up to statistical noise.

For the axial differential form factors (upper-right and bottom panels of Fig.~\ref{fig:4L_dFF_imp}), the improved estimators are obtained by modifying the correlation functions entering Eq.~(\ref{eq:axial_corr}) as follows:
\begin{align}
\label{eq:impCAij}
C_A^{ij}(t_\gamma, E_\gamma, \bs{k}; t_W) \to \ &  C_A^{ij}(t_\gamma, E_\gamma, \bs{k}; t_W) - e^{E_\gamma t_\gamma} C_A^{ij}(t_\gamma, 0, \bs{0}; t_W)~,\quad ij=30,03 \\[10pt]
 \label{eq:impCAii}
  \tilde{C}_A^{ii}(t_\gamma, E_\gamma, \bs{k}; t_W) \to \ & \frac{1}{2} \bigg[ \tilde{C}_A^{ii}(t_\gamma, E_\gamma, \bs{k}; t_W) + \tilde{C}_A^{ii}(t_\gamma, E_\gamma, -\bs{k}; t_W)  \bigg]~,\qquad i=1,3~. 
\end{align}
Additionally, we perform the subtraction
\begin{equation}
\tilde C_A^3(t_\gamma,E_\gamma,\bs k;t_W)
\to
\tilde C_A^3(t_\gamma,E_\gamma,\bs k;t_W)
-
\tilde C_A^3(t_\gamma,\sqrt{k^2},\bs0;t_W),
\label{eq:vecCtildeA3}
\end{equation}
where we recall that 
$\tilde C_A^3 =\tilde C_A^{33}-\tilde C_A^{11}$ (see Eq.~(\ref{eq:def_CAi})),
which therefore vanishes for $|\bs{k}|=0$. The correlated subtraction of the $\bs{k}=\bs{0}$ contribution thus removes the momentum-independent noise components from $\tilde{C}_{A}^{3}$.

The improved estimators exhibit significantly reduced statistical fluctuations compared to the naive ones, as illustrated in Fig.~\ref{fig:4L_dFF_imp}. In the remainder of this section we analyse the origin of these improvements.

\begin{figure}[]
    \centering
\includegraphics[width=1.\columnwidth]{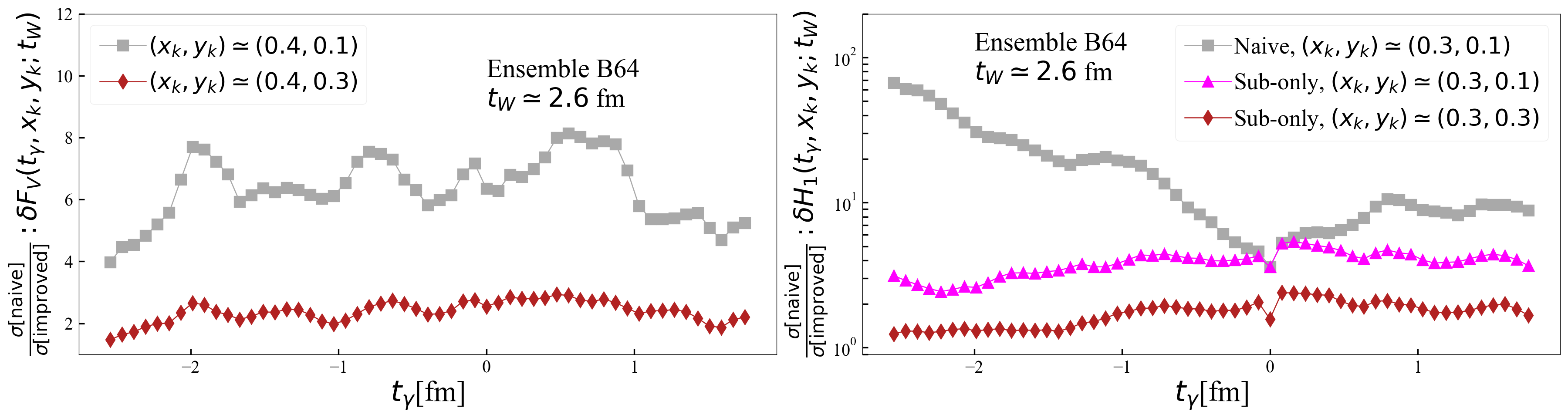}
\caption{Ratios between the uncertainties of the naive and improved estimators (see text for details) of the quark-connected contributions to the differential form factors $\delta F_V(t_\gamma, x_k, y_k; t_W)$ (left panel) and $\delta H_1(t_\gamma, x_k, y_k; t_W)$ (right panel). The data-points correspond to the contribution of the photon emission by the anti-up quark. The results are obtained on the B64 ensemble with $t_W \simeq 2.6~\mathrm{fm}$ and different values of the photon virtuality $x_k$ and momentum $y_k$. Note that for the right panel, we employed the logarithmic scale.}
\label{fig:4L_dFF_errs}
\end{figure}

For the vector form factor, the zero-momentum subtraction in Eq.~(\ref{eq:dFV_impr}) reduces the statistical uncertainty by nearly an order of magnitude at small photon momentum, due to the correlated zero-momentum subtraction, which removes momentum-independent noise components. The left panel of Fig.~\ref{fig:4L_dFF_errs} shows the ratio between the uncertainties of the improved and naive estimators in the vector channel as a function of $t_\gamma$ for two values of $y_k$. The effectiveness of the subtraction decreases with increasing photon momentum, becoming negligible for $y_k \gtrsim 0.7$.

For the axial form factors, the overall improvement results from the combined effect of the modifications in Eqs.~(\ref{eq:impCAij})–(\ref{eq:vecCtildeA3}). The subtraction in Eq.~(\ref{eq:vecCtildeA3}) provides the dominant reduction in statistical uncertainty, as illustrated by the ``sub-only'' estimators (magenta triangles in Fig.~\ref{fig:4L_dFF_imp}). The right panel of Fig.~\ref{fig:4L_dFF_errs} quantifies this effect in the case of the form factor $H_{1}$: at $t_\gamma \simeq -1~\mathrm{fm}$ the fully improved estimator reduces the uncertainty by roughly a factor of $20$ compared to the naive one.

The remaining improvements in the axial channel arise from removing momentum-independent noise component in the odd-in-$\bs{k}$ correlators $C_{A}^{30}$ and $C_{A}^{03}$ (see Eq.~(\ref{eq:impCAij})), and from averaging over opposite momenta the even-in-$\bs{k}$ correlators $C_{A}^{11}$ and $C_{A}^{33}$ (Eq.~(\ref{eq:impCAii})). Averaging over opposite momenta correlators that are even in $\bs{k}$ cancels noise components that are odd under $\bs k \to -\bs k$, which can be sizeable in twisted-mass QCD due to parity-breaking effects, as already observed in Ref.~\cite{DiPalma:2025iud}. As in the case of $F_{V}$, the impact of all these improvements decreases with increasing photon momentum.

All results presented in the remainder of this paper are obtained using the improved estimators.

\begin{figure}[]
    \centering
\includegraphics[width=1.\columnwidth]{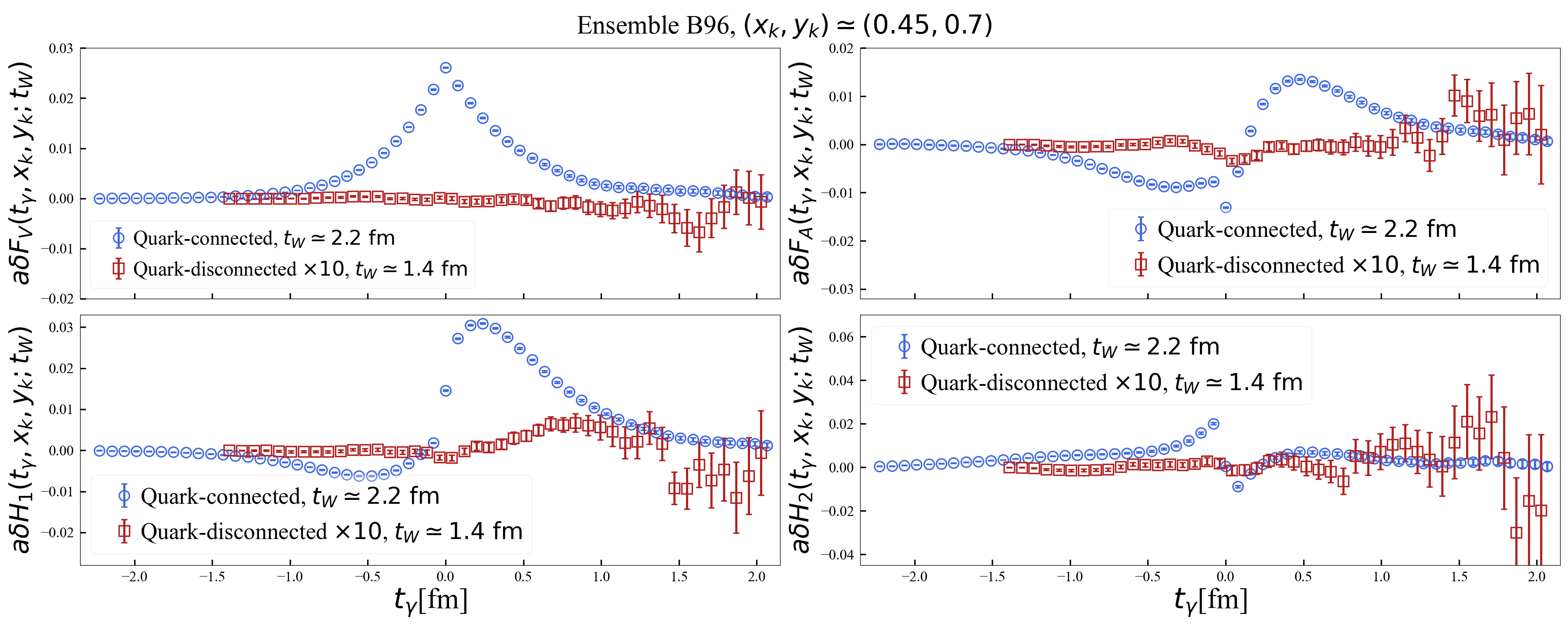}
\caption{Comparison between quark-connected (blue circles) and quark-disconnected (red squares) contributions to the differential form factors. Results are presented for the B96 ensemble, with $(x_k, y_k) \simeq (0.45, 0.7)$. The quark-connected contribution corresponds to the photon emission by the anti-up quark. The quark-disconnected contribution is multiplied by a factor of 10 for visualization purpose. }
\label{fig:4L_dFF_disc}
\end{figure}

Finally, Fig.~\ref{fig:4L_dFF_disc} compares quark-connected and quark-disconnected contributions to the differential form factors for representative kinematics. The disconnected contributions are significantly smaller in the regions that dominate the $t_\gamma$ integral and exhibit larger relative uncertainties.

In the following subsections we discuss the main sources of systematic uncertainty:
(i) truncation of the $t_\gamma$ integral,
(ii) finite-$t_W$ effects, and
(iii) finite-volume effects.
We then present the extrapolation to the continuum-limit and the final results for the form factors below the two-pion threshold.

\subsection{Truncation effects in the $t_{\gamma}$ integration}
\label{Sec:4L_tgammaintegral}

In this section we address the systematic uncertainty associated with the fact that, on a lattice with finite temporal extent $T$, the integrals over $t_\gamma$ of the differential form factors cannot be performed over the full range $t_\gamma \in (-\infty,+\infty)$.
At fixed $t_W$, the differential form factors are available only for
\begin{equation}
t_\gamma \in [-t_W,\; t_\gamma^{\rm max}]\,,\qquad t_\gamma^{\rm max}\ll T,
\end{equation}
where the upper bound is required to suppress wrap-around effects.
As a consequence, the contributions from the missing ``tails'' of the integrals,
\begin{equation}
t_\gamma<-t_W\quad \text{(first time ordering)}\,,\qquad
t_\gamma>t_\gamma^{\rm max}\quad \text{(second time ordering)}\,,
\end{equation}
must be controlled and shown to be negligible at the level of precision of the present analysis.

Our strategy follows the reconstruction method introduced in Ref.~\cite{Tuo:2021ewr} and used extensively in Ref.~\cite{DiPalma:2025iud} for the calculation of the form factors for the radiative decay, $K\to\ell\nu_{\ell}\gamma$.
The basic idea is to introduce an intermediate ``switching time'' $t_\gamma^*$ and:
(i) integrate the differential form factors directly using lattice data up to $t_\gamma^*$, and
(ii) approximate the remaining tail by assuming an asymptotic approximation for the corresponding Euclidean correlator.
The stability of the resulting estimator as a function of $t_\gamma^*$ provides a direct, data-driven test of convergence.
This strategy is very effective in accelerating the convergence compared to a naive truncation of the time integral, as already observed in Ref.~\cite{DiPalma:2025iud}.

Since the intermediate states propagating between the two currents differ in the two time orderings, we discuss the corresponding truncation effects separately.
We recall that Eq.~(\ref{eq:TO_separation}) defines the first- and second-time-ordering contributions to the form factors.
Compared to Ref.~\cite{DiPalma:2025iud}, which focused on real-photon emission, the off-shell case considered here introduces an additional complication in the axial channel:
in the first time ordering, a single-kaon intermediate state with momentum $-\bs{k}$ contributes to the form factors.
At small photon momentum and virtuality, this kaon pole produces a very slowly decaying contribution to the axial differential form factors, thereby significantly worsening the convergence of the $t_\gamma$ integral.

To overcome this problem, we exploit the fact that the kaon-pole contribution can be computed nonperturbatively on the lattice from the knowledge of the kaon electromagnetic form factor and the kaon-to-vacuum axial-current matrix element.
We subtract this contribution directly at the level of the differential form factors, so that the remainder is governed by heavier intermediate states and decays more rapidly.
The kaon-pole contribution is then added back after integrating it analytically in $t_\gamma$.
We now discuss the strategy to estimate the $t_\gamma$ integrals, starting from the vector form factor.

\subsection*{Vector channel}
We start by discussing the determination of the first time-ordering contribution, $F_{V,1}$.
For a switching time $t_\gamma^*\in[0,t_W]$ we define
\begin{equation}
F_{V,1}(t_\gamma^*,x_k,y_k;t_W)
=
\int_{-t_\gamma^*}^{0} dt_\gamma\;\delta F_V(t_\gamma,x_k,y_k;t_W)
+\int_{-\infty}^{-t_\gamma^*} dt_\gamma\;\delta F_{V,1;{\rm asy}}(t_\gamma,x_k,y_k;t_W),
\label{eq:4L_contFV1split_new}
\end{equation}
where the first integral is computed from lattice data, while the tail is approximated by 
\begin{equation}
\delta F_{V,1;{\rm asy}}(t_\gamma,x_k,y_k;t_W)
\equiv
\exp\!\Big[(\mathcal{E}_{V,1}+E_\gamma-m_K)\,(t_\gamma+t_\gamma^*)\Big]\;
\delta F_{V}(-t_\gamma^*,x_k,y_k;t_W),
\qquad t_\gamma\le -t_\gamma^*.
\label{eq:4L_contdFVasy_new}
\end{equation}
Here $\mathcal{E}_{V,1}$ is the energy of the lightest state (with three-momentum $-\bs{k}$) propagating between the two currents in the first time ordering.
Integrating explicitly the contribution of $\delta F_{V,1;{\rm asy}}$ in Eq.~(\ref{eq:4L_contFV1split_new}) yields the estimator
\begin{equation}
F_{V,1}(t_\gamma^*,x_k,y_k;t_W)
=
\int_{-t_\gamma^*}^{0} dt_\gamma\;\delta F_V(t_\gamma,x_k,y_k;t_W)
+\frac{\delta F_V(-t_\gamma^*,x_k,y_k;t_W)}{E_\gamma+\mathcal{E}_{V,1}-m_K}\,.
\label{eq:4L_contFV1_new}
\end{equation}
We then study $F_{V,1}(t_\gamma^*,x_k,y_k;t_W)$ as a function of $t_\gamma^*$ and extract
$F_{V,1}(x_k,y_k;t_W)$ from a constant fit in a region where a plateau is observed.

For the second time ordering we proceed analogously.
We introduce a switching time $t_\gamma^*\in[0,t_\gamma^{\rm max}]$ and define
\begin{equation}
F_{V,2}(t_\gamma^*,x_k,y_k;t_W)
=
\int_{0}^{t_\gamma^*} dt_\gamma\;\delta F_V(t_\gamma,x_k,y_k;t_W)
+\int_{t_\gamma^*}^{\infty} dt_\gamma\;\delta F_{V,2;{\rm asy}}(t_\gamma,x_k,y_k;t_W),
\label{eq:4L_contFV2split_new}
\end{equation}
with the asymptotic approximation for $t_\gamma\ge t_\gamma^*$,
\begin{equation}
\delta F_{V,2;{\rm asy}}(t_\gamma,x_k,y_k;t_W)
\equiv
\exp\!\Big[-(\mathcal{E}_2-E_\gamma)\,(t_\gamma-t_\gamma^*)\Big]\;
\delta F_V(t_\gamma^*,x_k,y_k;t_W),
\label{eq:4L_dFV2asy_new}
\end{equation}
where $\mathcal{E}_2$ is the energy of the lightest state with momentum $+\bs{k}$
propagating between the two currents in the second time ordering.
This yields
\begin{equation}
F_{V,2}(t_\gamma^*,x_k,y_k;t_W)
=
\int_{0}^{t_\gamma^*} dt_\gamma\;\delta F_V(t_\gamma,x_k,y_k;t_W)
+\frac{\delta F_V(t_\gamma^*,x_k,y_k;t_W)}{\mathcal{E}_2-E_\gamma}.
\label{eq:FV2_new}
\end{equation}

As discussed in Ref.~\cite{DiPalma:2025iud} for $K^- \to \ell^-\bar{\nu}_\ell \gamma$, the optimal choices for $\mathcal{E}_{V,1}$ and $\mathcal{E}_{2}$ in our setup are given by\footnote{
In principle, one could choose $\mathcal{E}_{V,1}$ and $\mathcal{E}_{2}$ equal to the energies of the lightest finite-volume multiparticle states, such as $K\pi$ or $\pi\pi$.
We have tested this alternative and found that, although it leads to compatible plateau values within uncertainties, the resulting plateaux are significantly shorter (see also Ref.~\cite{DiPalma:2025iud}).
This behavior is expected, since in finite volume the multiparticle spectrum consists of a tower of energy levels whose combined contribution effectively reproduces the resonance.
Using instead the resonance energies provides a more performant effective description of the tail and leads to faster convergence.}
\be
\mathcal{E}_{V,1} = \sqrt{m_{K^*}^2 + \vert \bs{k} \vert^2}, \qquad
\mathcal{E}_{2} = \sqrt{m_{\rho}^2 + \vert \bs{k} \vert^2},\ \sqrt{m_{\phi}^2 + \vert \bs{k} \vert^2}\,,
\label{eq:4L_eVA}
\ee
where $m_{K^{*}}, m_{\rho}$ and $m_{\phi}$ are the masses of the $K^{*}$-, $\rho$-, and $\phi$-resonances.
In the second time ordering, we use the energy of the $\phi$ resonance when considering the contribution from diagram (b) in Fig.~\ref{Fig:Diagrams}, i.e.\ the quark-connected contribution with photon emission from the strange quark, and the energy of the $\rho$ for the remaining diagrams.
\subsection*{Axial channel}
For the differential axial form factors, $\delta A^i$, in the first time ordering, a sizable and slowly decaying contribution from the elastic scattering of the initial kaon off the (virtual) photon is present.
This contribution becomes very slowly decaying when $|\bs{k}|/m_K\ll 1$ and $\sqrt{k^2}$ is small, because the energy gap
$E_\gamma+E_K-m_K$ becomes small,
\begin{equation}
\label{eq:Egap}
E_\gamma+E_K-m_K \simeq E_\gamma+\frac{|\bs{k}|^2}{2m_K}\,,
\qquad
E_K=\sqrt{m_K^2+|\bs{k}|^2}\,.
\end{equation}
In this situation the reconstruction method adopted for the vector form factor can still be applied, but convergence can be slow.
Since the intermediate kaon corresponds to an isolated pole, its matrix elements with the weak and electromagnetic currents can be determined straightforwardly, enabling an exact reconstruction of its contribution.
Specifically, its contribution to the axial part of the Euclidean correlator $C_{W}^{\mu\nu}(t_{\gamma},E_{\gamma},\bs{k};t_{W})$ is given by\footnote{Since the kaon-pole contribution does not depend on $t_{W}$ we have dropped the $t_{W}$ dependence when defining $C_{A,K}^{\mu\nu}$.}
\begin{equation}
C^{\mu\nu}_{A;K}(t_\gamma,E_\gamma,\bs{k})
\equiv
-i\frac{e^{(E_\gamma+E_K-m_K)t_\gamma}}{2E_K}\,
\bra{0} j^\nu_A(0) \ket{K^-(\!-\bs{k})}\,
\bra{K^-(\!-\bs{k})} j^\mu_{\rm em}(0) \ket{K^-(\bs{0})}\,,
\label{eq:4L_CmunuAK_new}
\end{equation}
with
\begin{align}
\bra{0} j^\nu_A(0) \ket{K^-(\!-\bs{k})}
&= -i f_K\, p_K^\nu,
\qquad\quad\qquad\quad\,\, p_K=(E_K,-\bs{k}),
\\[8pt]
\bra{K^-(\!-\bs{k})} j^\mu_{\rm em}(0) \ket{K^-(\bs{0})}
&=
i\,F^{K}_{\rm em}\!\big(q_{K}^2\big)\,(p+p_K)^\mu,
\quad\,\,\,\, q_{K}=p-p_{K},
\label{eq:4L_emKme_new}
\end{align}
where $F^{K}_{\rm em}(q_{K}^{2})$ is the kaon electromagnetic form factor, and we recall that our calculation is carried out in the kaon rest frame $p = (m_K, \bs{0})$.

To reconstruct $C_{A;K}^{\mu\nu}$ at fixed lattice spacing, we evaluate on the same gauge ensembles in Tab.~\ref{tab:simudetails} the kaon decay constant $f_{K}$ from the $\nu=0$ component of the axial-current kaon-to-vacuum matrix element. In the same way, we determine the form factor $F^{K}_{\rm em}(q_{K}^{2})$ from the $\mu=0$ component of the electromagnetic matrix element in Eq.~(\ref{eq:4L_emKme_new}).\footnote{In order to check for possible hypercubic artifacts associated with the breaking of $O(4)$ symmetry at finite lattice spacing, we have also evaluated $f_K$ and $F^{K}_{\rm em}(q_{K}^{2})$ from the corresponding spatial components ($\nu=3$ and $\mu=3$). The results agree with those obtained from the temporal components within percent-level accuracy, indicating that hypercubic effects are negligible at our present precision.}

The electromagnetic kaon form factor can be expanded as
\begin{align}
F_{\rm em}^{K}(q_{K}^{2})
=
1 + \frac{\langle r_{K}^{2}\rangle}{6}\, q_{K}^{2}
+ c_{K}^{(4)}(q_{K}^{2})^{2}
+ O\!\big((q_{K}^{2})^{3}\big)\,,
\label{eq:FK_expanded}
\end{align}
where $\langle r^{2}_{K}\rangle$ is the kaon electromagnetic mean square radius.
Assuming the vector-meson-dominance (VMD) approximation, i.e.\ $F_{\rm em}^{K}(q_{K}^{2}) = 1/(1-q_{K}^{2}/m_{V}^{2})$, the $O((q_{K}^{2})^{2})$ coefficient is expected to satisfy $c_{K}^{(4)} \sim (\langle r_{K}^{2}\rangle/6)^{2}$.
Since the largest value (in modulus) of $q_{K}^{2}$ in our analysis is $q_{K}^{2} \simeq -|\bs{k}|^{2} \simeq -(0.22~{\rm GeV})^{2}$, the $O((q_{K}^{2})^{2})$ contribution is expected to be very small (at the level of $\sim 0.5\%$ relative to the point-like term at the largest momentum considered), using the PDG value~\cite{ParticleDataGroup:2024cfk} $\langle r_{K}^{2}\rangle \simeq 0.31~{\rm fm^{2}}$.
For this reason, in order to limit computational costs, we restrict our calculation to the $O(q_{K}^{2})$ term and determine $\langle r_{K}^{2}\rangle$ on the lattice.
\begin{table}[t]
\begin{ruledtabular}
\begin{tabular}{lccc}
\textrm{ID} & $a~[\textrm{fm}]$ & $\langle r^2_K \rangle~[\textrm{fm}^2]$ & $Z_{{\rm{em}}}\equiv 1 + \delta Z_{{\rm{em}}}$   \\
\colrule
\textrm{B48}, B64, B96 & $0.07948(11)$ & $0.334(14)$ & $0.994(5)$   \\
\textrm{C80} & $0.06819(14)$ & $0.339(13)$ & $1.005(5)$   \\
\textrm{D96} & $0.056850(90)$ & $0.335(10)$ & $0.996(4)$ 
\end{tabular}
\end{ruledtabular}
\caption{Kaon mean squared charge radius $\langle r^2_K \rangle$ and $Z_{{\rm{em}}}\equiv 1+\delta Z_{{\rm em}}$, determined on each lattice spacing employed in our analysis. 
\label{tab:rsqKZem}}
\end{table}
Moreover, the point-like normalization $F_{\rm em}^{K}(0)=1$ is guaranteed at finite lattice spacing only when employing the conserved vector current. 
Since we use a local vector current, renormalized through a condition that does not enforce $F_{\rm em}^{K}(0)=1$, the relation becomes $F_{\rm em}^{K}(0)=1+\delta Z_{\rm em}$, where $\delta Z_{\rm em}$ is a pure $O(a^{2})$ lattice artifact. 
Because our goal is to determine $C_{A;K}^{\mu\nu}$ at fixed lattice spacing, including its discretization effects, we evaluate $\delta Z_{\rm em}$ explicitly for each ensemble rather than imposing the continuum normalization condition. 
The resulting values of $\delta Z_{\rm em}$ are reported in Tab.~\ref{tab:rsqKZem} and are compatible with zero within uncertainties. 
Our determination of the kaon mean square radius is also given in Tab.~\ref{tab:rsqKZem}.

In order to account conservatively for possible $O((q_{K}^{2})^{2})$ effects, when evaluating $C_{A;K}^{\mu\nu}$ we inflate the uncertainty on $\langle r_{K}^{2}\rangle$ by a factor of two, bringing it to the level of about $6$--$10\%$. This, together with the $\sim 0.5\%$ uncertainty on $Z_{\rm em}=1+\delta Z_{\rm em}$ (see Tab.~\ref{tab:rsqKZem}), is expected to conservatively cover the residual impact of neglected higher-order terms. Further details of the analysis of the kaon charge radius are provided in App.~\ref{app:kaon_radius}.

In this way, we build $C_{A;K}^{\mu\nu}(t_{\gamma}, E_{\gamma},\bs{k})$ on each gauge ensemble and compute from it the corresponding differential axial form factors, denoted by $\delta A_{K}^{i}(t_{\gamma},x_{k},y_{k})$, which take the pure exponential form
\begin{align}
\label{eq:kaonpole_contribution}
\delta A_{K}^{i}(t_{\gamma},x_{k},y_{k})
=
A^{i}_{K}(x_{k},y_{k})\,
\exp\!\left[(E_{\gamma}+E_{K}-m_{K})\,t_{\gamma}\right].
\end{align}
We then subtract this contribution from the full differential form factors $\delta A^{i}$ in the first time ordering ($t_{\gamma}<0$), defining the kaon-subtracted differential form factors
\begin{align}
\label{eq:kaon_pole_subtracted}
\delta A^i_{K\text{-}{\rm sub}}(t_\gamma,x_k,y_k;t_W)
\equiv
\delta A^i(t_{\gamma},x_k,y_k;t_W)
-\delta A^{i}_{K}(t_\gamma,x_k,y_k),
\qquad t_{\gamma} < 0.
\end{align}
The kaon-subtracted differential form factor now contains only contributions from heavier axial-vector states, and therefore converges more rapidly.

\begin{figure}[]
    \centering
\includegraphics[width=1.\columnwidth]{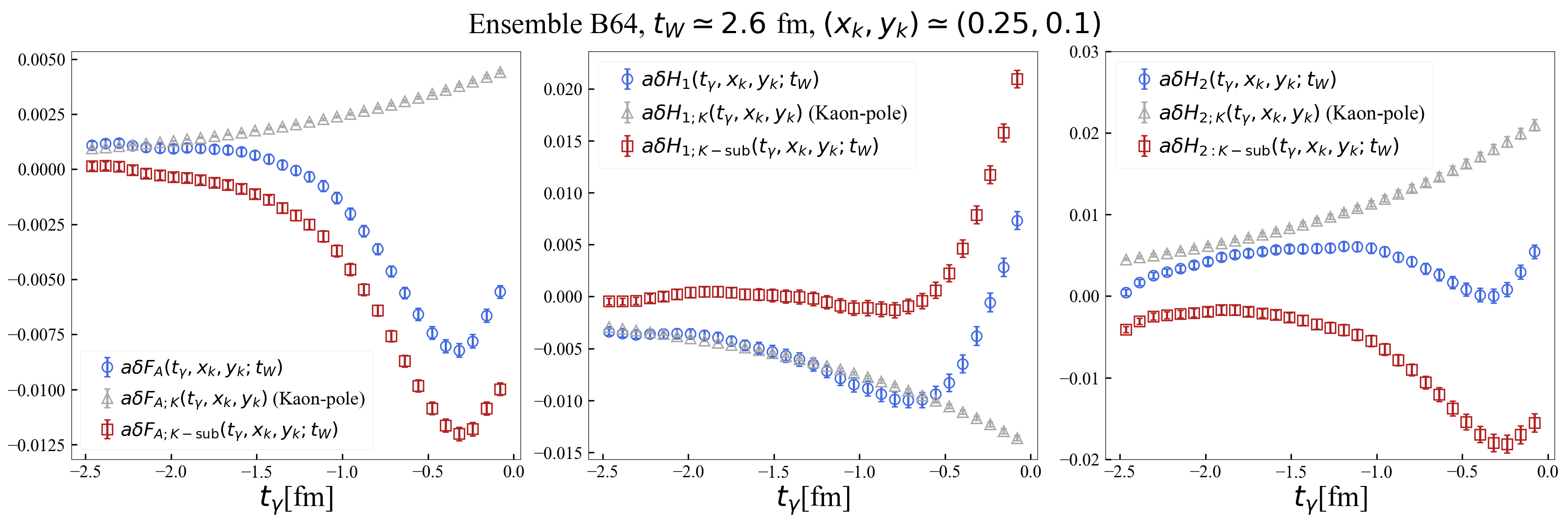}
    \caption{Quark-connected contribution to the differential axial form factors, determined on the B64 ensemble:
$\delta F_A(t_\gamma, x_k, y_k; t_W)$ (left panel),
$\delta H_1(t_\gamma, x_k, y_k; t_W)$ (center panel), and
$\delta H_2(t_\gamma, x_k, y_k; t_W)$ (right panel),
with $t_W \simeq 2.6~\mathrm{fm}$ and $(x_k, y_k) \simeq (0.25, 0.1)$.
In each panel we show the differential axial form factor $\delta A^{i}$ (blue circles), the kaon-pole contribution $\delta A^{i}_{K}$ (gray triangles), and the kaon-pole-subtracted differential form factor $\delta A^{i}_{K\text{-}{\rm sub}}$ (red squares) defined in Eq.~(\ref{eq:kaon_pole_subtracted}).}
    \label{fig:4L_dFF_tails}
\end{figure}

Fig.~\ref{fig:4L_dFF_tails} shows the quark-connected contribution in the first time ordering to the differential axial form factors $\delta A^{i}$ (blue circles), the kaon-pole contribution $\delta A_{K}^{i}$ (gray triangles), and the kaon-pole-subtracted differential form factors $\delta A^{i}_{K\text{-}{\rm sub}}$ (red squares), obtained on the B64 ensemble with $t_W \simeq 2.6~\mathrm{fm}$.

The data points shown in the figure correspond to the smallest values of the photon virtuality and momentum used in the present analysis, $(x_k, y_k) \simeq (0.25, 0.1)$. This is the kinematic configuration for which the integral of the differential form factors converges most slowly to its plateau value, since the energy gap in Eq.~(\ref{eq:Egap}) reaches its lowest value among those considered. This can be clearly seen from the large negative-time behaviour of $\delta A^i$, which approaches slowly zero without reaching it within the available lattice time slices. This highlights the importance of the kaon-pole subtraction strategy described above in controlling the systematic effects associated with the missing tails of the differential axial form factors.

The differential form factors $\delta H_1$ and $\delta F_A$ behave as expected: for $t_\gamma < -0.8~\mathrm{fm}$ and $t_\gamma < -1.5~\mathrm{fm}$, respectively, they are dominated by the kaon-pole contribution. As a consequence, the kaon-pole-subtracted differential form factors $\delta H_{1,K\text{-}\mathrm{sub}}$ and $\delta F_{A,K\text{-}\mathrm{sub}}$ decay more rapidly to zero, thereby improving the convergence of the corresponding time-integrals.

On the other hand, for $\delta H_2$, we observe that around $t_\gamma \simeq -1.5~\mathrm{fm}$ the kaon-pole contribution approaches $\delta H_2$, only to deviate from it again as $t_\gamma$ becomes more negative. We attribute this behaviour to contamination from excited states different from the kaon in $\delta H_2$, arising when the electromagnetic current approaches the interpolating operator $P^\dagger_K$, placed at $t_\gamma = -t_W \simeq -2.6~\mathrm{fm}$. In the following paragraph, we describe how we investigated this hypothesis in detail.

Following Ref.~\cite{Giusti:2023pot}, we perform an additional simulation on the B64 ensemble at the two smallest values of the photon momentum ($y_k \simeq 0.1,\,0.3$), employing a different setup to the present one, obtained by exchanging the roles of the electromagnetic and weak currents. Specifically, we fix the insertion time of the electromagnetic current at $t_{\mathrm{em}} \simeq 2~\mathrm{fm}$ to keep a suitable time-separation between the interpolating operator and the currents, and compute the amputated correlation functions as functions of the insertion time of the weak current, as follows:
\be
M^{\mu\nu}_W(t_\gamma,  \bs{k}; t_{\mathrm{em}}) = e^{ -(E_\gamma - m_K) t_\gamma}  \mathcal{N}(t_{\mathrm{em}})  (-i) \sum_{\bs{x}} \sum_{\bs{y}}  \ e^{- i \bs{k}\cdot \bs{y}} \bra{0}\hat{\mathrm{T}}[j^\nu_{W}(t_\gamma+t_{\mathrm{em}}, \bs{y})j^{\mu}_{\mathrm{em}}(t_{\mathrm{em}}, \bs{0}) P^{\dagger}_K(0, \bs{x})] \ket{0},
\ee
with $\mathcal{N}$ being the amputation factor defined in Eq.~(\ref{Eq:Def3pt}).
The latter are related to the amputated correlation functions in the fixed-$t_W$ setup (see Eq.~(\ref{Eq:Def_correlationfunction})) as
\be
C^{\mu\nu}_W(t_\gamma, \bs{k}; t_W) = M^{\mu\nu}_W(-t_\gamma, -\bs{k}; t_{\mathrm{em}}), \qquad t_W, t_\mathrm{em} \to \infty
\ee
The fixed-$t_{\mathrm{em}}$ setup has the practical advantage that, in the first time ordering, the time separation between the kaon interpolating operator and both currents is always at least $t_{\mathrm{em}}$, thereby suppressing contamination from excited states other than the kaon and allowing us to evaluate the differential form factors also for $t_\gamma < -2.6~\mathrm{fm}$. Although this fixed-$t_{\mathrm{em}}$ setup is well suited for the determination of the contribution to the form factors from the first time ordering, difficulties analogous to those described above arise in the second time ordering, for which the fixed-$t_W$ setup is instead better suited.

\begin{figure}[]
    \centering
\includegraphics[width=1.\columnwidth]{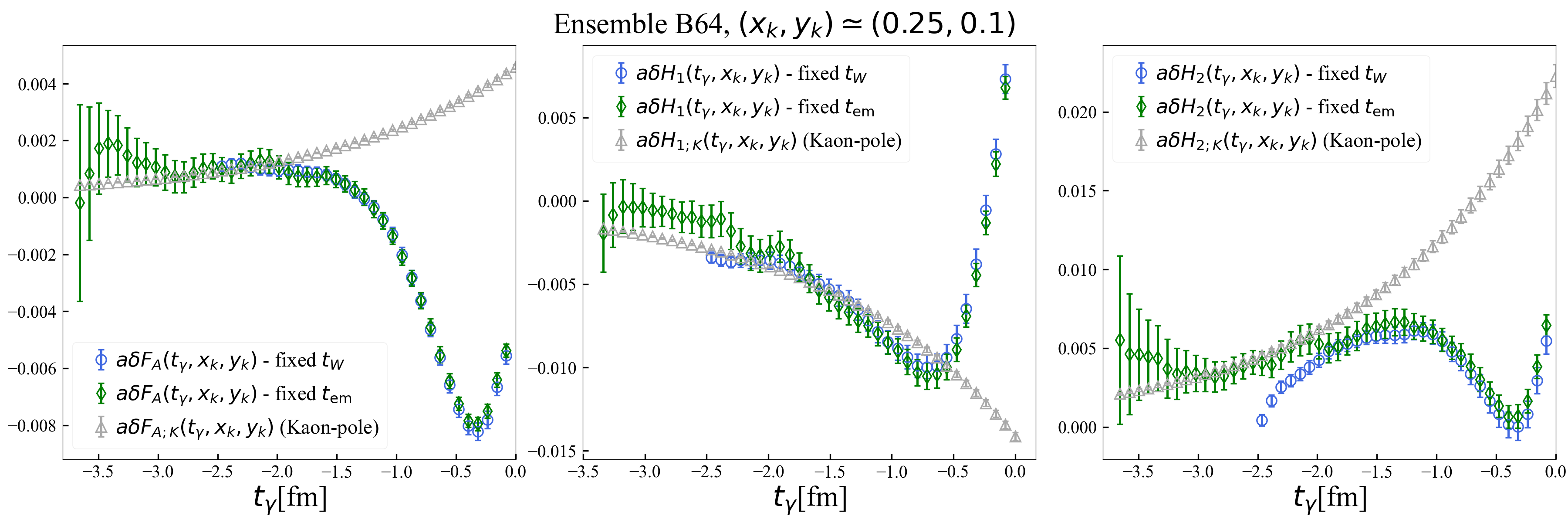}
    \caption{Quark-connected contribution to the differential axial form factors, determined on the B64 ensemble:
$\delta F_A(t_\gamma, x_k, y_k; t_W)$ (left panel),
$\delta H_1(t_\gamma, x_k, y_k; t_W)$ (center panel), and
$\delta H_2(t_\gamma, x_k, y_k; t_W)$ (right panel).
In each panel, we show the differential axial form factor $\delta A^{i}$, obtained by keeping fixed the time of insertion of the weak current $t_W \simeq 2.6~\mathrm{fm}$ (blue circles) or fixing the time of insertion of the electromagnetic current $t_{\mathrm{em}} \simeq 2~\mathrm{fm}$ (green diamond), as well as  the kaon-pole contribution $\delta A^{i}_{K}$ (gray triangles).}
    \label{fig:4L_dFF_tails2}
\end{figure}

Fig.~\ref{fig:4L_dFF_tails2} shows a comparison of the differential axial form factors in the first time ordering, obtained either by fixing the insertion time of the weak current, $t_W \simeq 2.6~\mathrm{fm}$ (blue circles), or of the electromagnetic current, $t_{\mathrm{em}} \simeq 2~\mathrm{fm}$ (green diamonds), together with the kaon-pole contribution (gray triangles). The setup with fixed $t_{\mathrm{em}}$ allows us to study the behaviour of the differential form factors also in the region $t_\gamma < -2.6~\mathrm{fm}$. As expected, we observe that all differential form factors decay to zero at large negative times. Crucially, the figure shows that in the fixed-$t_{\mathrm{em}}$ setup, the kaon-pole contribution provides an effective description of the tails of all the differential form factors, including $\delta H_2$, for $t_\gamma < -2~\mathrm{fm}$. Therefore, we ascribe the observed deviation in the fixed-$t_W$ setup to the presence of excited states created by the kaon interpolating operator, whose effects grow as the electromagnetic current approaches it. In the following, we take this effect into account by assigning a systematic uncertainty when evaluating the first time-ordering contribution to $H_2$.

To determine the first time-ordering contribution to the axial form factors we proceed as follows:
we apply the reconstruction strategy discussed above for the vector form factor to the kaon-pole-subtracted differential form factors $\delta A^{i}_{K\text{-}{\rm sub}}$, and then add the kaon-pole contribution separately as an analytic term.
This is done by introducing a switching time $t_{\gamma}^{*}\in[0,t_{W}]$ and defining
\begin{align}
A^{i}_{1}(t_{\gamma}^{*}, x_{k}, y_{k};t_{W})
&=
\frac{A^i_K(x_k,y_k)}{E_\gamma+E_K-m_K}
+\int_{-t_\gamma^*}^{0} dt_\gamma\;\delta A^{i}_{K\text{-}{\rm sub}}(t_\gamma,x_k,y_k;t_W)
+\frac{\delta A^{i}_{K\text{-}{\rm sub}}(-t_\gamma^*,x_k,y_k;t_W)}{E_\gamma+\mathcal{E}_{A,1}-m_K}\,,
\label{eq:esti_impro_axial}
\end{align}
where the first term is the kaon-pole contribution, integrated analytically over $t_{\gamma}\in(-\infty,0]$ using Eq.~(\ref{eq:kaonpole_contribution}), and $\mathcal{E}_{A,1}$ is the energy of the lowest-lying state contributing to the kaon-subtracted differential form factor (an axial-vector state).
As in the vector case, we study $A^{i}_{1}(t_\gamma^*,x_k,y_k;t_W)$ as a function of $t_\gamma^*$ and extract
$A^{i}_{1}(x_k,y_k;t_W)$ from a constant fit in a region where a plateau is observed.
For $\mathcal{E}_{A,1}$ we take the energy of the $K_{1}$ resonance,
\begin{align}
\mathcal{E}_{A,1}= \sqrt{m_{K_{1}}^{2} + |\bs{k}|^{2}}\,,
\end{align}
where $m_{K_{1}}$ is the mass of the $K_{1}$.
We have also tested using for $\mathcal{E}_{A,1}$ the energy of a single finite-volume $K\pi\pi$ state, and found that, while the plateau value remains unchanged within uncertainties, the convergence at small virtuality and three-momentum is substantially faster when using the $K_1$ resonance energy.

We will furthermore compare this strategy with the one in which we do not perform the explicit subtraction of the kaon pole from the axial differential form factors, and instead apply the reconstruction method directly to $\delta A^{i}$ exactly as in the vector case.
In this alternative strategy we set $\mathcal{E}_{A,1}=E_{K}$.

\begin{figure}[]
    \centering
\includegraphics[width=1.\columnwidth]{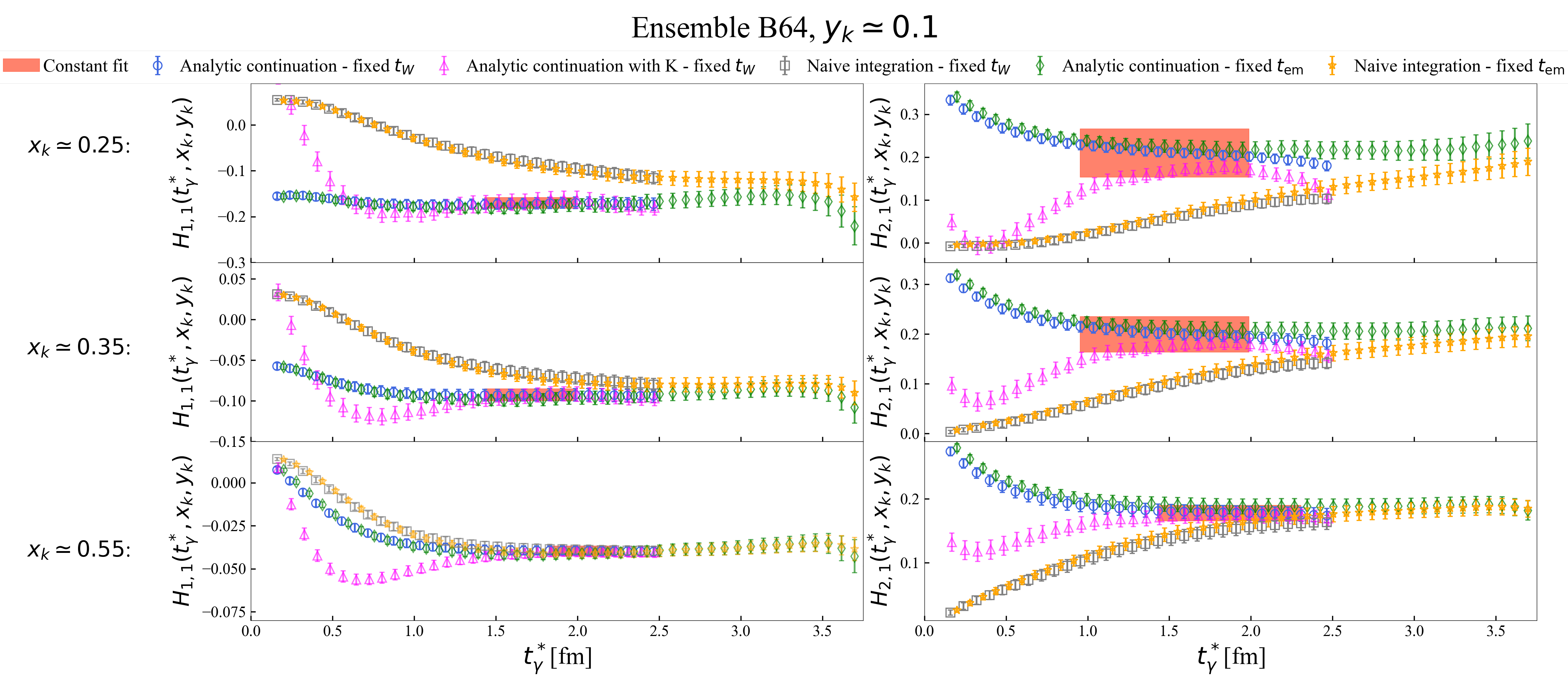}
    \caption{Quark-connected contribution to $H_1$ (left panels) and $H_2$ (right panels), obtained on the B64 ensemble with $y_k \simeq 0.1$, and increasing values of $x_k$ (from top to bottom). Blue circles, gray squares, and magenta triangles correspond to different estimators, obtained using the fixed $t_W$-setup (see text for details), with $t_W \simeq 2.6~\mathrm{fm}$. Green diamonds and orange stars correspond to analogous estimators, obtained in the fixed-$t_{\mathrm{em}}$ setup, with $t_{\mathrm{em}} \simeq 2~\mathrm{fm}$. Red bands represent our estimates of the contribution to the form factors from the first time-ordering. Data points are slightly shifted horizontally to improve readability.}
    \label{fig:4L_FFpltxKaon}
\end{figure}

Let us now discuss the effect of the kaon-pole contribution in the first time-ordering on the estimators of the axial form factors. 
Fig.~\ref{fig:4L_FFpltxKaon} shows the contribution from the first time-ordering to $H_1$ (left panels) and $H_2$ (right panels), computed on the B64 ensemble at the smallest simulated value of the photon momentum, $y_k \simeq 0.1$.

We first focus on the estimators obtained in the fixed-$t_W$ setup: blue circles, gray squares, and magenta triangles. Blue circles correspond to our preferred estimators in Eq.~(\ref{eq:esti_impro_axial}), while gray squares are determined by integrating the corresponding differential form factors up to $t_\gamma^*$ (naive integration). Magenta triangles represent an additional estimator obtained by not performing the kaon-pole subtraction and by applying the reconstruction method directly to $\delta A^{i}$, using $\mathcal{E}_{A,1}=E_{K}$. 

At the lowest value of the photon virtuality, $x_k \simeq 0.25$ (top panels), we observe that the naive integration for both form factors does not reach convergence within the available time slices (in the fixed-$t_W$ setup, $t_\gamma^* \in [0,t_W \simeq 2.6~\mathrm{fm}]$). This is precisely the effect of the slowly decaying kaon-pole contribution, already observed when discussing the differential form factors in Fig.~\ref{fig:4L_dFF_tails}.

The figure clearly shows the importance of using improved estimators (blue circles and magenta triangles) to reliably determine the form factors at small values of the photon virtuality. For $H_{1,1}$, both the improved estimators reach convergence to compatible results already at $t_\gamma^* \simeq 1.2~\mathrm{fm}$, even for $x_k \simeq 0.25$. Although not shown, we observe the same behavior also for $F_{A,1}$. 

On the other hand, contamination from excited states other than the kaon, which becomes relevant when the electromagnetic current approaches the kaon interpolating operator in the fixed-$t_W$ setup (for $t_\gamma^* > 2~\mathrm{fm}$), distorts the signal of $H_{2,1}$, most significantly at the lowest value of the photon virtuality, $x_k \simeq 0.25$. We account for these effects by assigning a systematic uncertainty to $H_{2,1}$ corresponding to half the difference between the results of the two improved estimators. Our final results for $H_{1,1}$ and $H_{2,1}$ are shown as red bands.

As anticipated above, to further investigate the effects of the kaon interpolating operator on $H_{2,1}$ and to assess the reliability of our systematic uncertainty, we performed an additional simulation in a setup where the roles of the two currents are exchanged, keeping fixed the insertion time of the electromagnetic current at $t_{\mathrm{em}} \simeq 2~\mathrm{fm}$. Green diamonds and orange stars correspond to the estimator in Eq.~(\ref{eq:esti_impro_axial}) and to the naive integration, respectively, as obtained in the fixed-$t_{\mathrm{em}}$ setup.  

Note that in the fixed-$t_{\mathrm{em}}$ setup, the kaon interpolating operator is always separated from the currents by at least $t_{\mathrm{em}}$, and possible contaminations from excited states other than the kaon are therefore suppressed. Furthermore, we have access to the form factors for $t_\gamma^* > 2.6~\mathrm{fm}$, which allows us to investigate the behavior of the estimators at large times. Crucially, we observe that: (i) for both form factors, all estimators (including those from naive integration) converge to compatible plateau values; (ii) our final estimate of $H_{2,1}$ (red band) is consistent with the improved estimator obtained in the fixed-$t_{\mathrm{em}}$ setup (green diamonds), supporting our choice of the systematic uncertainty.

Finally, we observe that as the photon virtuality increases (from top to bottom panels), all difficulties associated with the slowly decaying kaon-pole contribution are reduced, since the energy gap in Eq.~(\ref{eq:Egap}) increases, and, as expected, all estimators reach convergence and yield compatible results.

For the contribution from the second time-ordering, we proceed analogously to the vector case.
We define
\begin{equation}
A^{i}_{2}(t_\gamma^*,x_k,y_k;t_W)
=
\int_{0}^{t_\gamma^*} dt_\gamma\;\delta A^{i}(t_\gamma,x_k,y_k;t_W)
+\int_{t_\gamma^*}^{\infty} dt_\gamma\;\delta A^{i}_{2;{\rm asy}}(t_\gamma,x_k,y_k;t_W),
\end{equation}
where for $t_\gamma\ge t_\gamma^*$,
\begin{equation}
\delta A^{i}_{2;{\rm asy}}(t_\gamma,x_k,y_k;t_W)
\equiv
\exp\!\Big[-(\mathcal{E}_2-E_\gamma)\,(t_\gamma-t_\gamma^*)\Big]\;
\delta A^{i}(t_\gamma^*,x_k,y_k;t_W).
\end{equation}
This yields the estimator
\begin{equation}
A^{i}_{2}(t_\gamma^*,x_k,y_k;t_W)
=
\int_{0}^{t_\gamma^*} dt_\gamma\;\delta A^{i}(t_\gamma,x_k,y_k;t_W)
+\frac{\delta A^{i}(t_\gamma^*,x_k,y_k;t_W)}{\mathcal{E}_2-E_\gamma}\,,
\label{eq:FA2_est}
\end{equation}
which we study as a function of $t_{\gamma}^{*}$.
In the region where the estimator displays a plateau we perform a constant fit and extract $A^{i}_{2}(x_{k},y_{k};t_{W})$.
The energy $\mathcal{E}_{2}$ in the previous equations is the same as in the vector case (see Eq.~(\ref{eq:4L_eVA})), since in both channels the intermediate states propagating in the second time ordering are unflavoured vector states.

\begin{figure}[]
    \centering
\includegraphics[width=1.\columnwidth]{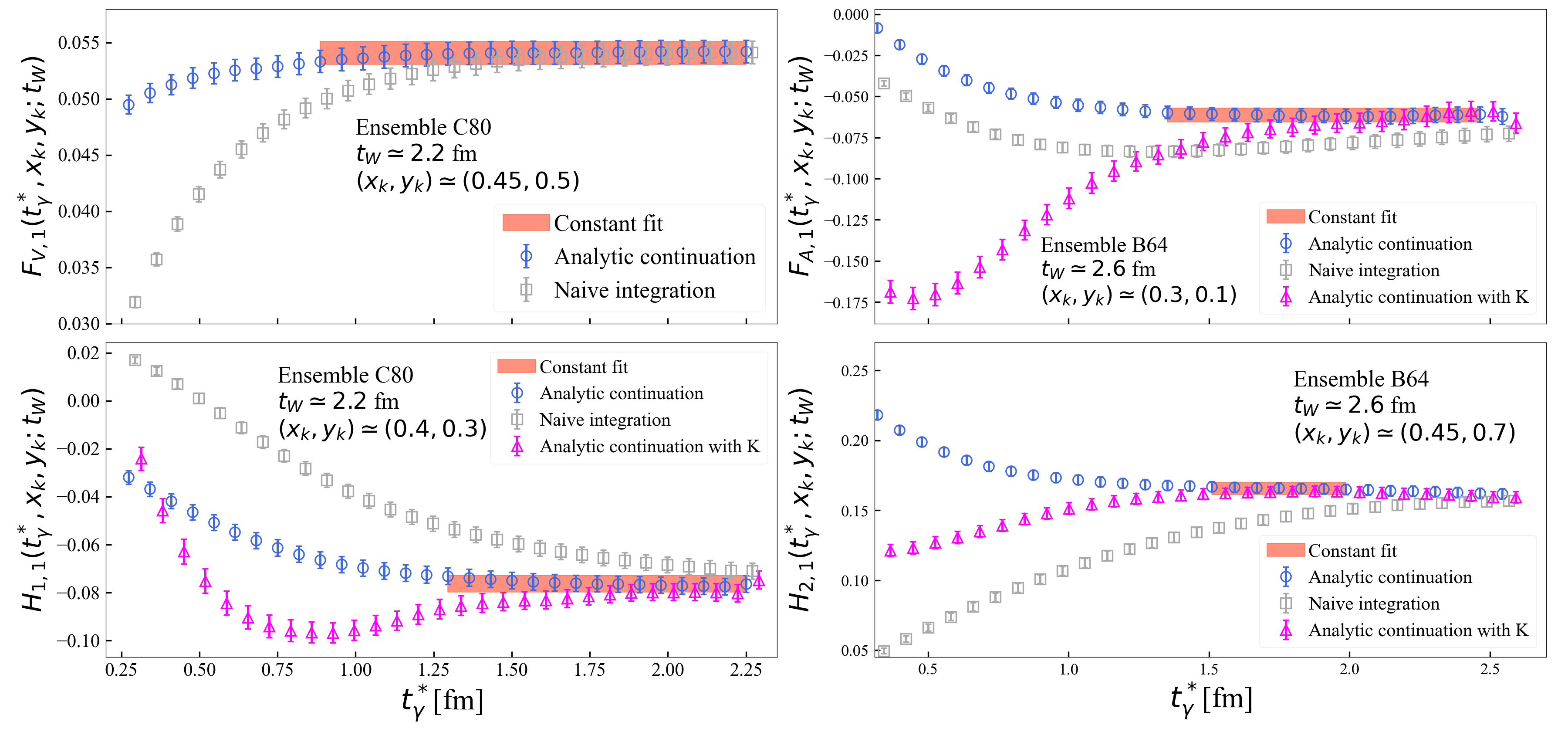}
    \caption{Quark-connected contributions to the form factors from the first time ordering as functions of the switching time $t_\gamma^*$.
The results in the left (right) panels are obtained on the C80 (B64) ensemble with $t_W \simeq 2.2~\mathrm{fm}$ ($t_W \simeq 2.6~\mathrm{fm}$), at different values of the photon virtuality $x_k$ and momentum $y_k$.
Blue circles, gray squares, and magenta triangles correspond to different estimators (see text for details).
The red bands correspond to the results of constant fits to our best estimators (blue circles) in the region where a plateau is visible.
Data points have been slightly shifted horizontally to improve readability.}
    \label{fig:4L_FFfirstpltx}
\end{figure}

In Fig.~\ref{fig:4L_FFfirstpltx} we show the stability plots for the quark-connected contributions to $F_{V,1}(x_{k},y_{k};t_{W})$ and $A^{i}_{1}(x_{k},y_{k};t_{W})$.
The gray squares correspond to the \textit{naive integration}.
The blue circles correspond to our preferred estimators, defined in Eq.~(\ref{eq:4L_contFV1_new}) for the vector channel and in Eq.~(\ref{eq:esti_impro_axial}) for the axial channel.
As anticipated, in the axial channel we also show an additional estimator (magenta triangles), obtained using $\mathcal{E}_{A,1}=E_{K}$ for the analytic continuation, without performing the subtraction of the kaon-pole contribution.

Although, in most cases, all estimators shown in Fig.~\ref{fig:4L_FFfirstpltx} yield compatible results once convergence is reached, our best estimators exhibit a larger plateau region compared to the others.
\begin{figure}[]
    \centering
\includegraphics[width=1.\columnwidth]{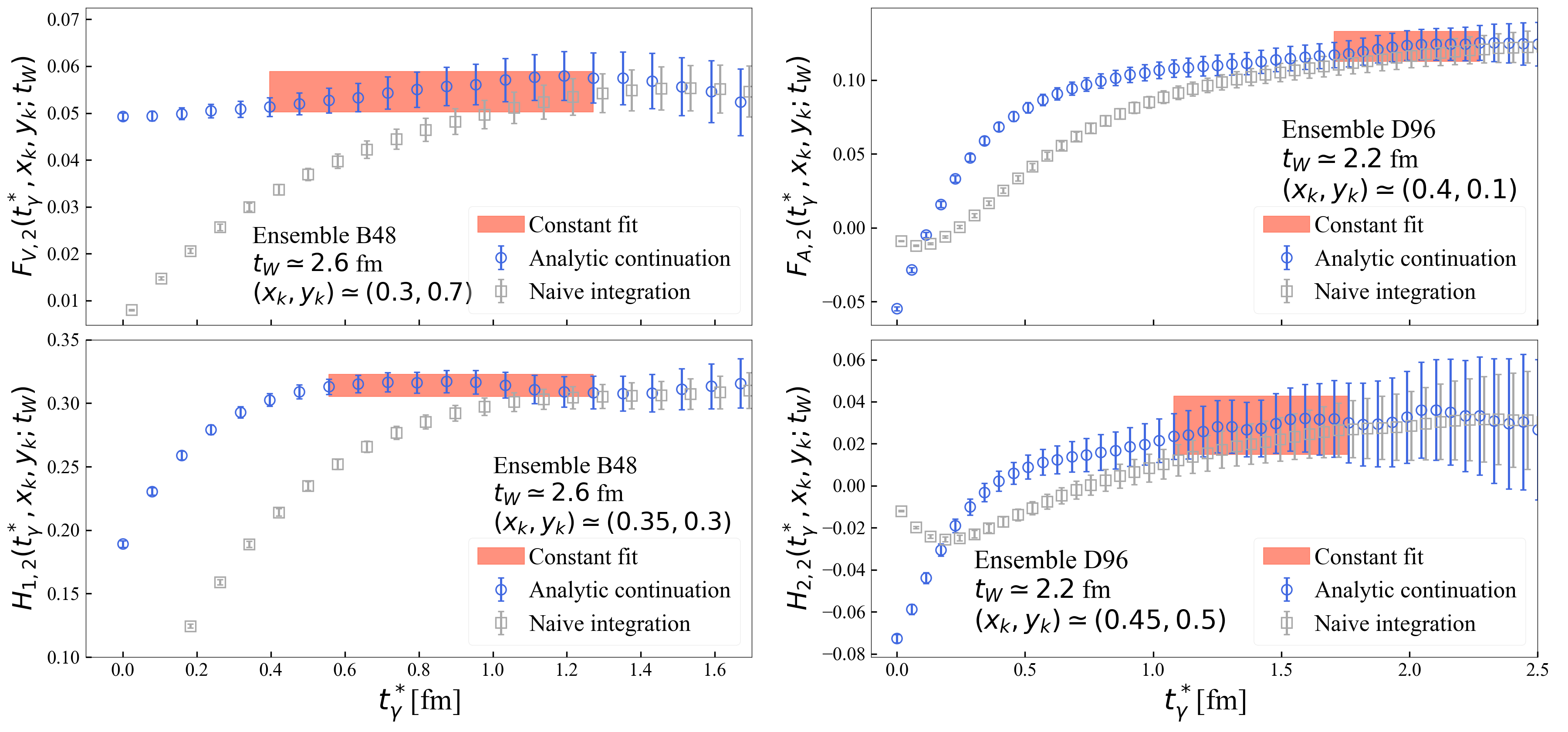}
    \caption{Quark-connected contributions to the form factors from the second time ordering as functions of the switching time $t_\gamma^*$.
The results in the right (left) panels are obtained on the D96 (B48) ensemble with $t_W \simeq 2.2~\mathrm{fm}$ ($t_W \simeq 2.6~\mathrm{fm}$), at different values of the photon virtuality $x_k$ and momentum $y_k$.
Blue circles and gray squares correspond to different estimators (see text for details).
Red bands represent the results of a constant fit to our best estimators (blue circles).
Data points have been slightly shifted horizontally to improve readability.}
    \label{fig:4L_FFsecondpltx}
\end{figure}
This highlights the importance of carefully assessing and controlling the kaon-pole contribution. All results discussed in the following sections are obtained using the estimators defined in Eqs.~(\ref{eq:4L_contFV1_new}) and~(\ref{eq:esti_impro_axial}).

We now discuss the corresponding analysis for the second time-ordering contribution.
An illustrative stability plot for the quark-connected contribution to $F_{V,2}(x_{k},y_{k})$ and $A^{i}_{2}(x_{k},y_{k})$ is shown in Fig.~\ref{fig:4L_FFsecondpltx}.
\begin{figure}[]
    \centering
\includegraphics[width=1.\columnwidth]{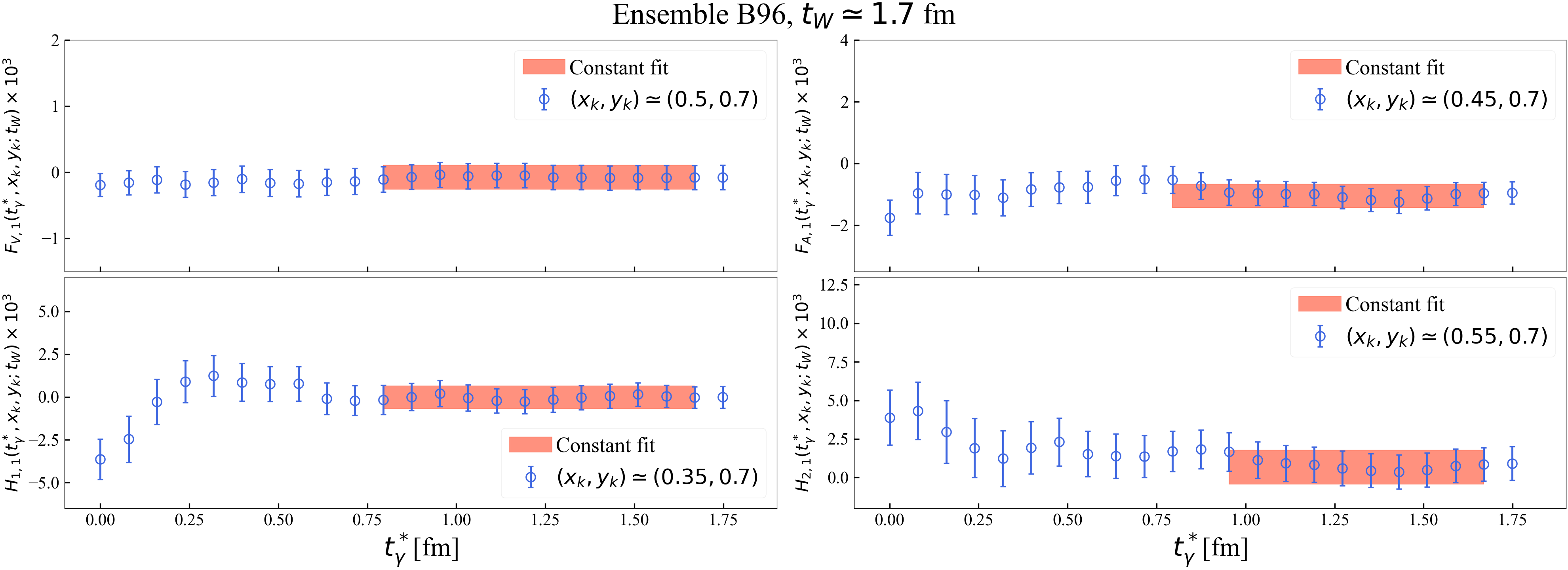}
    \caption{Quark-disconnected contribution to the form factors (multiplied by $10^3$) from the first time ordering.
The results are obtained on the B96 ensemble with $t_W \simeq 1.7~\mathrm{fm}$.
Red bands represent the results of constant fits to the data points (blue circles).}
    \label{fig:4L_DiscFFfirstpltx}
\end{figure}
As before, the gray squares correspond to the naive integration, obtained by integrating the differential form factors in the second time ordering up to $t_{\gamma}^{*}$, while the blue circles correspond to the improved estimators defined in Eqs.~(\ref{eq:FV2_new}) and~(\ref{eq:FA2_est}) for the vector and axial channel, respectively.
As in the case of the first time ordering, the reconstruction method leads to convergence at smaller values of $t_\gamma^*$ and yields more extended plateaux compared to the naive integration.
In this case, however, we always find that the naive integration eventually reaches a plateau consistent with the one obtained using the improved estimator, although convergence occurs at significantly larger Euclidean times. The results presented in the following sections are obtained using the estimators defined in Eqs.~(\ref{eq:FV2_new}) and~(\ref{eq:FA2_est}).

\begin{figure}[]
    \centering
\includegraphics[width=1.\columnwidth]{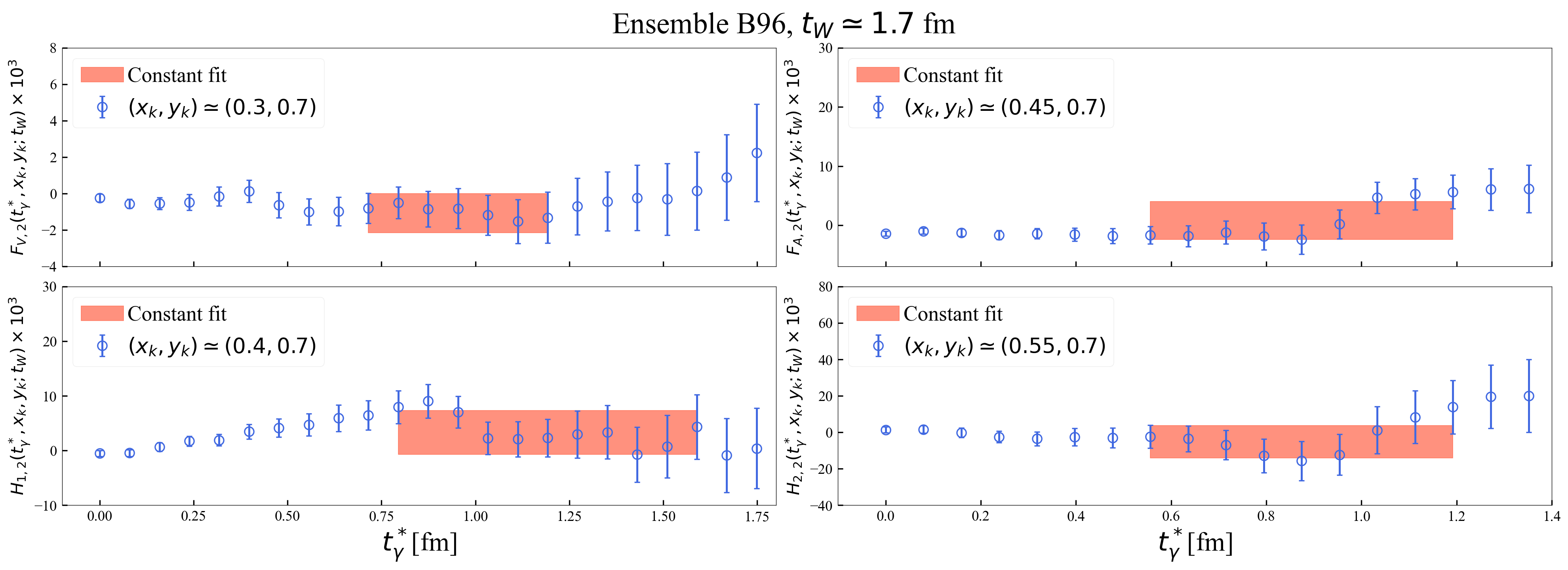}
    \caption{Quark-disconnected contribution to the form factors (multiplied by $10^3$) from the second time ordering.
The results are obtained on the B96 ensemble with $t_W \simeq 1.7~\mathrm{fm}$.
Red bands represent the results of constant fits to the data points (blue circles).}
    \label{fig:4L_DiscFFsecondpltx}
\end{figure}

Finally, Figs.~\ref{fig:4L_DiscFFfirstpltx} and~\ref{fig:4L_DiscFFsecondpltx} show the analogous results for the quark-disconnected contributions from the first and second time orderings, respectively.
In this case, within the large relative uncertainties of the disconnected contribution, we observe clear plateaux already when considering only the naive integral of the quark-disconnected contribution to the differential form factors (which is the one shown in the figures).
The blue circles represent the data obtained on the B96 ensemble with $t_W \simeq 1.7~\mathrm{fm}$ at the single simulated value of the photon momentum, corresponding to $y_{k}\simeq 0.7$.
Similar behaviors are observed for other values of $t_W$, as will be discussed in detail in the next subsection.
The quark-disconnected contributions from both time orderings are compatible with zero (within at most $3\sigma$) for all form factors.

\subsection{Finite-$t_W$ effects}
\label{Sec:4L_tweff}
\begin{figure}[]
    \centering
\includegraphics[width=1.\columnwidth]{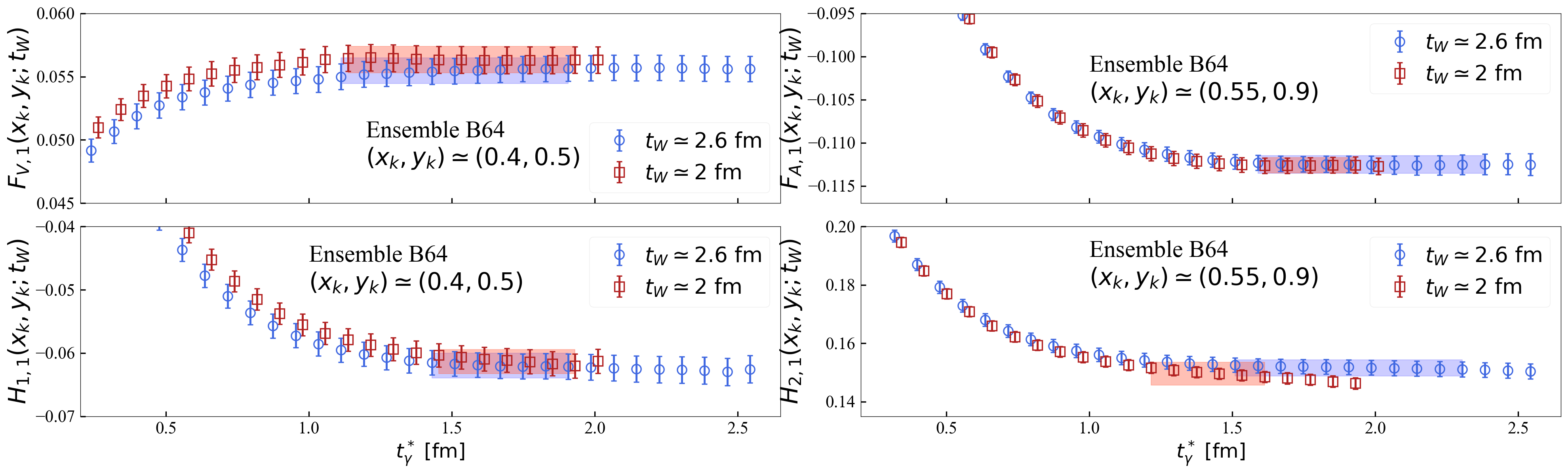}
    \caption{Quark-connected contribution to the form factors from the first time-ordering. Results are shown for the B64 ensemble using $t_W \simeq 2~\mathrm{fm}$ (red squares) and $t_W \simeq 2.6~\mathrm{fm}$ (blue circles). Colored bands represent constant fits to the corresponding data points. Data points have been slightly shifted horizontally for clarity.}
    \label{fig:4L_FFfirstTOtwdep}
\end{figure}

\begin{figure}[]
    \centering
\includegraphics[width=1.\columnwidth]{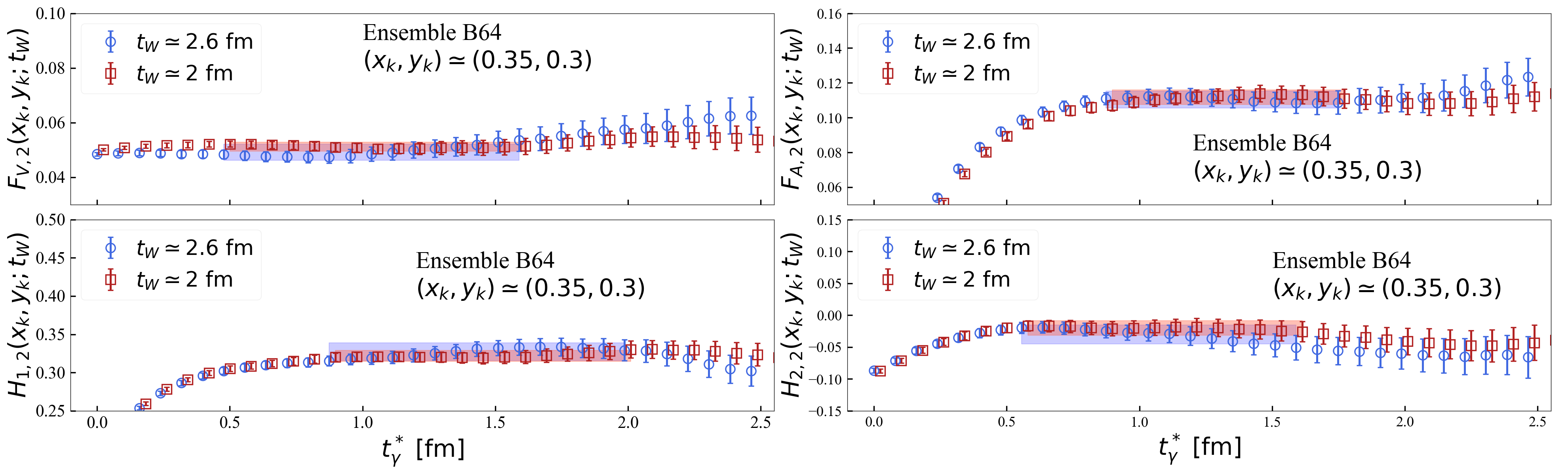}
    \caption{Quark-connected contribution to the form factors from the second time-ordering. Results are shown for the B64 ensemble using $t_W \simeq 2~\mathrm{fm}$ (red squares) and $t_W \simeq 2.6~\mathrm{fm}$ (blue circles). Colored bands represent constant fits to the corresponding data points. Data points have been slightly shifted horizontally for clarity.}
    \label{fig:4L_FFsecondTOtwdep}
\end{figure}

The extraction of the form factors from lattice correlation functions requires taking the limit $ t_W \to \infty $ in order to isolate the initial-state kaon. In this subsection, since $t_W$ is unavoidably finite, we provide a detailed analysis of the systematic uncertainties related to the isolation of the kaon . \\ \\ 
We carry out the analysis using the full-statistics on the B64 ensemble for two values of $t_W$: $t_W \simeq 2\,\mathrm{fm}$ and $t_W \simeq 2.6\,\mathrm{fm}$. 
Figs.~\ref{fig:4L_FFfirstTOtwdep} and~\ref{fig:4L_FFsecondTOtwdep} show the quark-connected contributions to the form factors from the first and second time-orderings, respectively. 
The results for $t_W~\simeq 2~\mathrm{fm}$ (red squares) and $t_W \simeq 2.6~\mathrm{fm}$ (blue circles) are in excellent agreement, with a possible dependence on $t_W$ being significantly smaller than the statistical uncertainties. \\ \\ 
To account for the residual finite-$t_W$ effects, we estimate a systematic uncertainty for each of the form factors, defined as:
\begin{equation}
\label{eq:4L_Systtw}
\Sigma_{F}(x_k, y_k) 
= \Delta_{F}(x_k, y_k) \,\mathrm{erf}\!\Bigg(\frac{\Delta_{F}(x_k, y_k)}{\sqrt{2}\,\sigma_{\Delta_{F}}(x_k, y_k)}\Bigg)~,\qquad F=\{F_{V},A^{i}\}~,
\end{equation}
where $\mathrm{erf}(x)$ is the error function and 
\begin{align}
    \Delta_{F}(x_k, y_k) 
=& \  \bigl|\,
    F(x_k,y_k; t_{W} \simeq 2.6~\mathrm{fm}) 
  - F(x_k, y_k; t_{W} \simeq 2.0~\mathrm{fm}) \,
\bigr|~, \qquad F=\{F_{V},A^{i}\}~,
\end{align}
with $\sigma_{\Delta_{F}}(x_k, y_k)$ representing the statistical uncertainty of $\Delta_{F}(x_k,  y_k)$. For the B64 ensemble, we take as the central value of each form factor the average between the results obtained at $t_W \simeq 2\,\mathrm{fm}$ and $t_W \simeq 2.6\,\mathrm{fm}$. 
For the remaining ensembles, simulations are performed at a single value of $t_W \simeq 2.2\text{--}2.3~\mathrm{fm}$. The corresponding systematic uncertainty is estimated by assigning the same relative systematic error as determined on the B64 ensemble.\\ \\ 
We have compared $\Sigma_{F}(x_{k}, y_k)$ to the statistical uncertainties of the 4 form factors across all 5 ensembles, for every combination of $x_k$ and $y_k$ considered below the two-pion threshold, for a total of 700 comparisons. We then add the systematic uncertainty in Eq.~(\ref{eq:4L_Systtw}) in quadrature to the statistical uncertainty of each form factor.  In nearly all cases, $\Sigma_{F}(x_{k}, y_k)$ is small: in about
77\% of cases, it is smaller than the statistical uncertainty; in 96\% of cases, it is below twice the statistical uncertainty; and it is never larger than $3.5$ times the statistical uncertainty.  \\ \\ 
For the quark-disconnected contribution, we evaluate the correlation functions for all values of $t_\gamma$ and $t_W$.
\begin{figure}[]
    \centering
\includegraphics[width=1.\columnwidth]{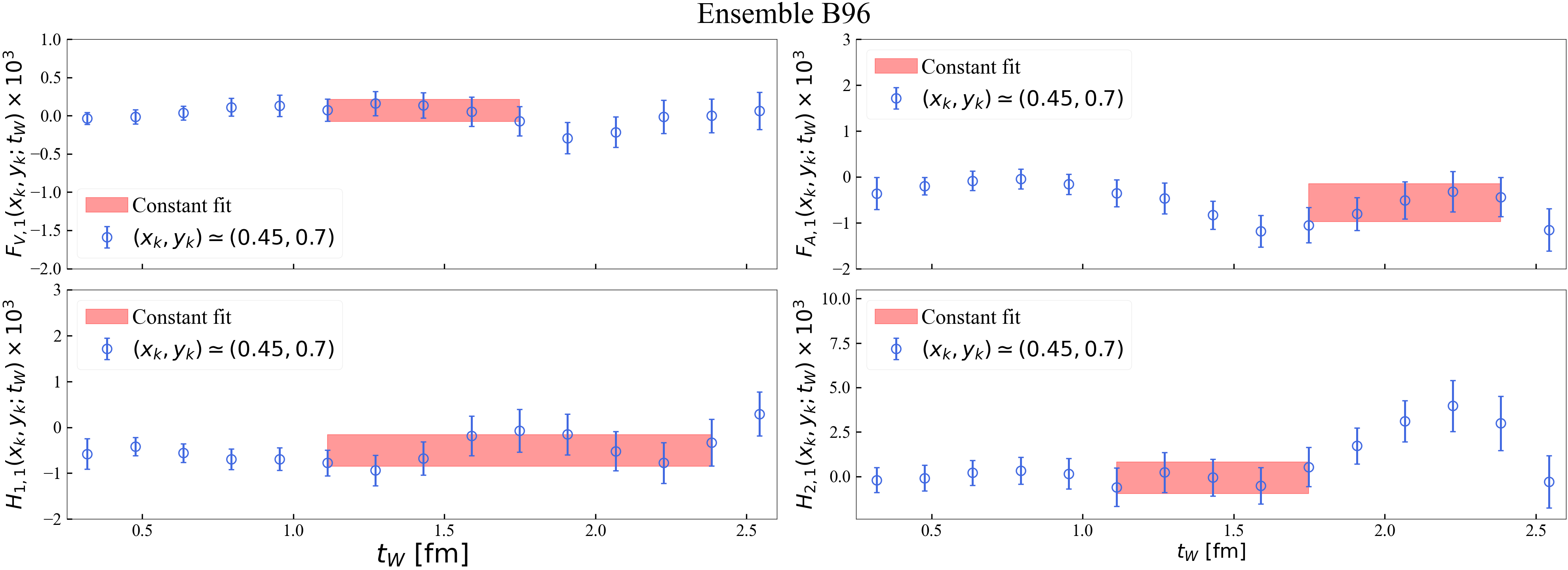}
    \caption{Quark-disconnected contribution to the form factors from the first time-ordering, as functions of $t_W$. Results (blue circles) are obtained on the B96 ensemble. Red bands represent constant fits to the data points.}
    \label{fig:4L_DiscFFfirstTOtwdep}
\end{figure}
\begin{figure}[]
    \centering
\includegraphics[width=1.\columnwidth]{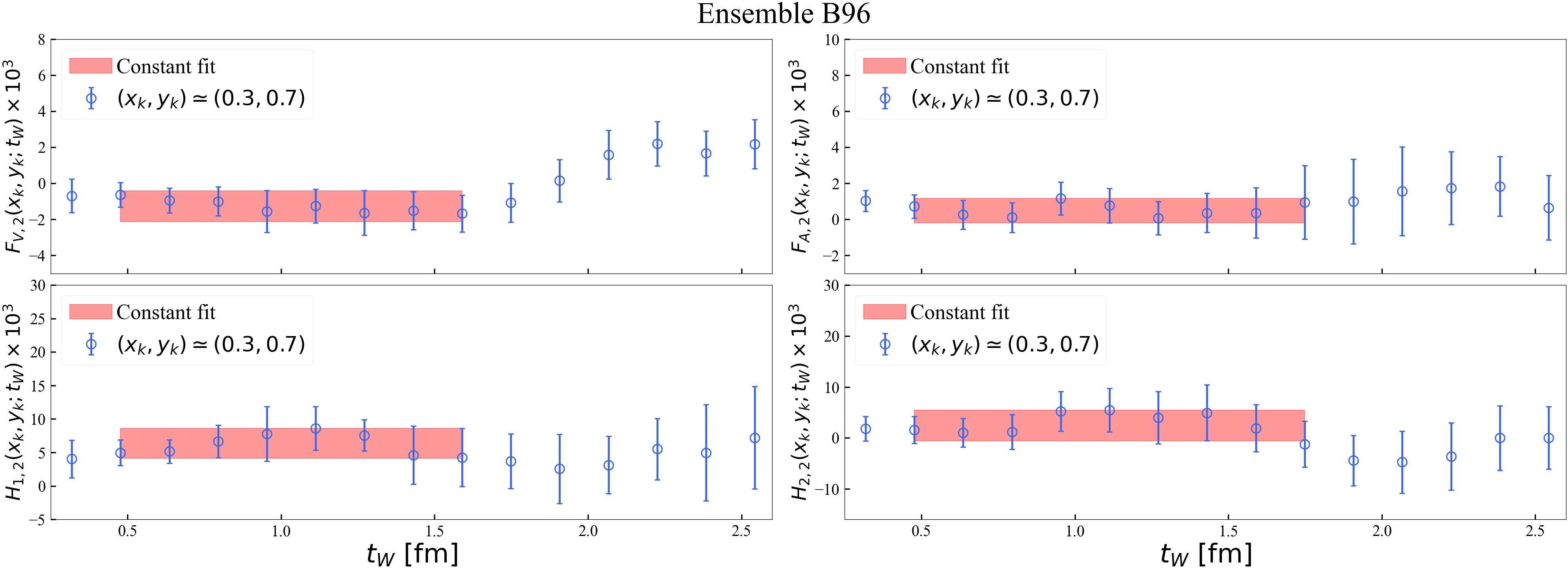}
    \caption{Quark-disconnected contributions to the form factors from the second time-ordering, as functions of $t_W$. The results (blue circles) are obtained on the B96 ensemble. Red bands represent constant fits to the data points.}
    \label{fig:4L_DiscFFsecondTOtwdep}
\end{figure}
Figs.~\ref{fig:4L_DiscFFfirstTOtwdep} and~\ref{fig:4L_DiscFFsecondTOtwdep}  show the contributions to the  form factors from the first and second time-orderings, respectively as functions of $t_W$. Within uncertainties, we observe no significant 
 $t_W$-dependence. We therefore extract our estimators of the form factors by performing constant fits to the data points (blue circles). The results of the fits are indicated by the red bands in the figures.

\subsection{Finite-size effects}
\label{sec:4L_fse}
In this section we describe the procedure we have adopted to account for finite-size effects and to perform the extrapolation to the infinite-volume limit.

We begin by discussing finite-size effects for the dominant quark-connected contribution. For this calculation we have performed simulations on three ensembles (B48, B64, and B96) that share the same lattice spacing and quark masses, differing only in the spatial lattice volume. The spatial extent is $L \simeq 3.8~{\rm fm}$, $5.1~{\rm fm}$, and $7.6~{\rm fm}$ for the B48, B64, and B96 ensembles, respectively. 

To extrapolate to the infinite-volume limit we adopt the following strategy. First, we fit the form factors obtained on the B ensembles using the ansatz
\begin{align}
\label{eq:4L_FSE_ansatz}
F(x_k, y_k, L) \;=\; C_{F}(x_k, y_k ) \;+\; D_{F}(x_k, y_k)\,\exp(-m_{\pi}L), 
\qquad F=\{F_{V},A^{i}\},
\end{align}
where $F(x_k, y_k, L)$ denotes the four form factors evaluated on a lattice with spatial extent $L$.
\begin{figure}[]
    \centering
\includegraphics[width=1.\columnwidth]{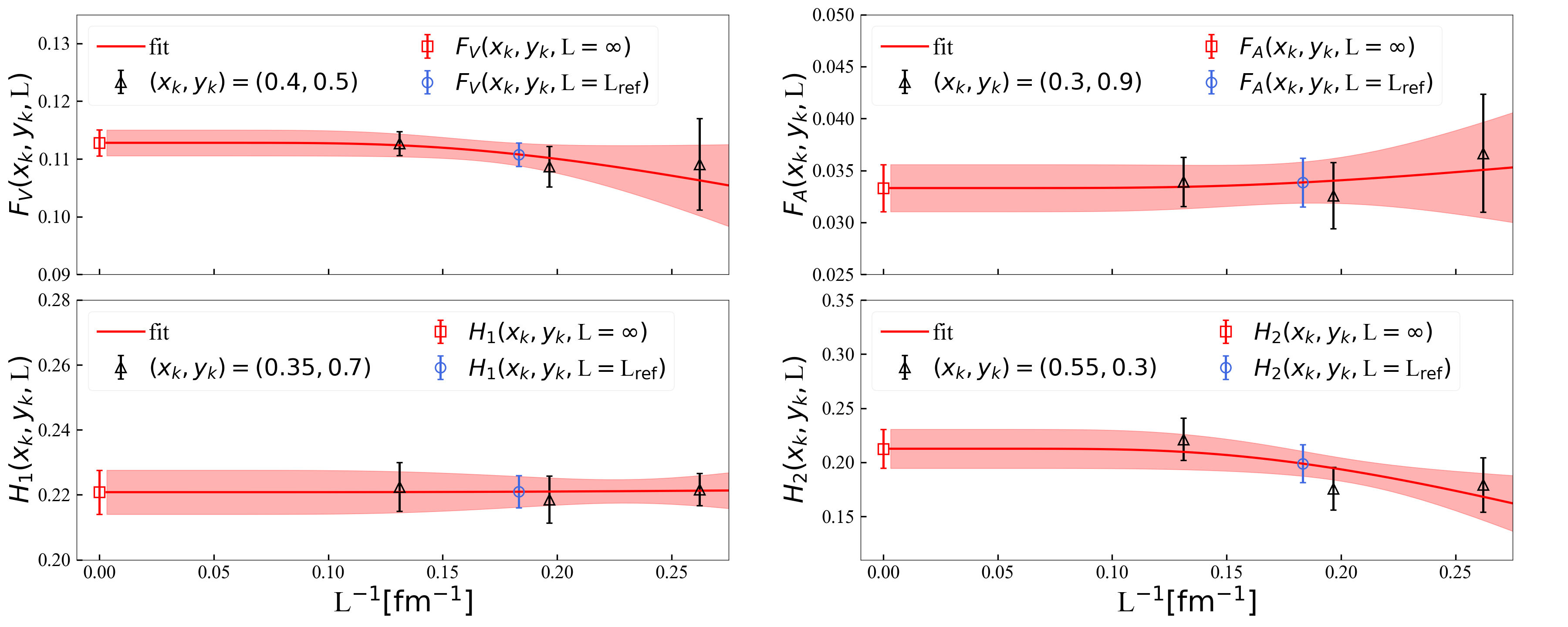}
    \caption{Dependence of the quark-connected contribution to the form factors on the inverse spatial lattice extent, $L^{-1}$. The black triangles show the form factors computed on the B48, B64, and B96 ensembles, ordered from larger to smaller $L^{-1}$. The red bands represent exponential fits to these data points, while the red squares indicate the result of the infinite-volume extrapolation at the lattice spacing $a\simeq 0.795~{\rm fm}$. The blue circles denote the form factors interpolated to a reference spatial extent $L_{\mathrm{ref}} \simeq 5.46~\mathrm{fm}$, matching the volume of the C80 and D96 ensembles.}
    \label{fig:4L_Vfit}
\end{figure}
The coefficients $C_{F}(x_k, y_k)$ and $D_{F}(x_k, y_k)$, for each form factor $F=\{F_{V},A^{i}\}$, are treated as free fit parameters. Separate fits are performed for each value of the photon virtuality and momentum $(x_k,y_k)$.

Once the fit parameters are determined, the form factors are interpolated to the reference volume $L = L_{\mathrm{ref}} \simeq 5.46~\mathrm{fm}$, corresponding to the spatial extent of the ensembles with finer lattice spacings, namely D96 and C80. This choice allows us to carry out the continuum-limit extrapolation at the fixed volume $L_{\mathrm{ref}}$.

In Fig.~\ref{fig:4L_Vfit} we show illustrative results of the volume fits based on Eq.~(\ref{eq:4L_FSE_ansatz}) for selected values of $x_k$ and $y_k$. The blue circles represent the interpolated results at $L=L_{\mathrm{ref}}$. As a conservative choice, we inflate the uncertainties of the interpolated points so that their errors do not become smaller than those of the most precise B-ensemble determination (black triangles). As shown in the figure, finite-volume effects are found to be very small, and the ansatz in Eq.~(\ref{eq:4L_FSE_ansatz}) provides an excellent description of the data.

After performing the continuum-limit extrapolation at fixed spatial extent $L_{\mathrm{ref}}$, the final step consists in extrapolating from $L=L_{\mathrm{ref}}$ to the infinite-volume limit. This is achieved by adding to the continuum results at $L_{\mathrm{ref}}$ the correction
\begin{align}
\label{eq:4L_finite-VCorr}
\Delta_L F(x_k, y_k, L_{\mathrm{ref}}) =  - D_F(x_k, y_k)\,\exp(- m_\pi L_{\mathrm{ref}}),
\qquad F=\{F_{V},A^{i}\},
\end{align}
where $D_{F}(x_k,y_k)$ are the fit parameters defined in Eq.~(\ref{eq:4L_FSE_ansatz}).

The quark-disconnected contribution is evaluated only on the largest available ensemble (B96). Given the current level of precision and the small size of the quark-disconnected contribution relative to the quark-connected one, we treat the results obtained on the B96 ensemble as effectively corresponding to the infinite-volume limit. To remain conservative, we assign a $100\%$ uncertainty to the quark-disconnected contribution which should account for possible finite-size and discretization effects.

\subsection{Continuum-limit extrapolation}
\label{Sec:4L_Continuum}
We now discuss the continuum-limit extrapolation of the form factors, starting from the quark-connected contribution.
In this case, the form factors are determined on ensembles with three different lattice spacings: $a \simeq 0.079~\mathrm{fm}$ (B), $a \simeq 0.068~\mathrm{fm}$ (C80), and $a \simeq 0.057~\mathrm{fm}$ (D96), all sharing the same lattice spatial extent $L_{\mathrm{ref}} \simeq 5.46~\mathrm{fm}$.

To perform the continuum-limit extrapolation we first fit the form factors obtained on all the three lattice spacings using the following linear ansatz in $a^2$: 
\begin{align}
\label{eq:4L_asqfitlin}
F(x_k, y_k, a) =  R_{F}(x_k, y_k) + B_{F}(x_k, y_k) a^2,  \qquad F=\{F_{V},A^{i}\}~,
\end{align}
where we employ distinct parameters $R_{F}(x_k, y_k)$ and $B_{F}(x_k, y_k)$ for each $(x_k, y_k)$ considered. A second fit is then performed using only the data at the two finest lattice spacings (D96 and C80 ensembles), adopting a constant ansatz (i.e. setting $B_{F}=0$).

\begin{figure}[t]
    \centering
\includegraphics[width=1.\columnwidth]{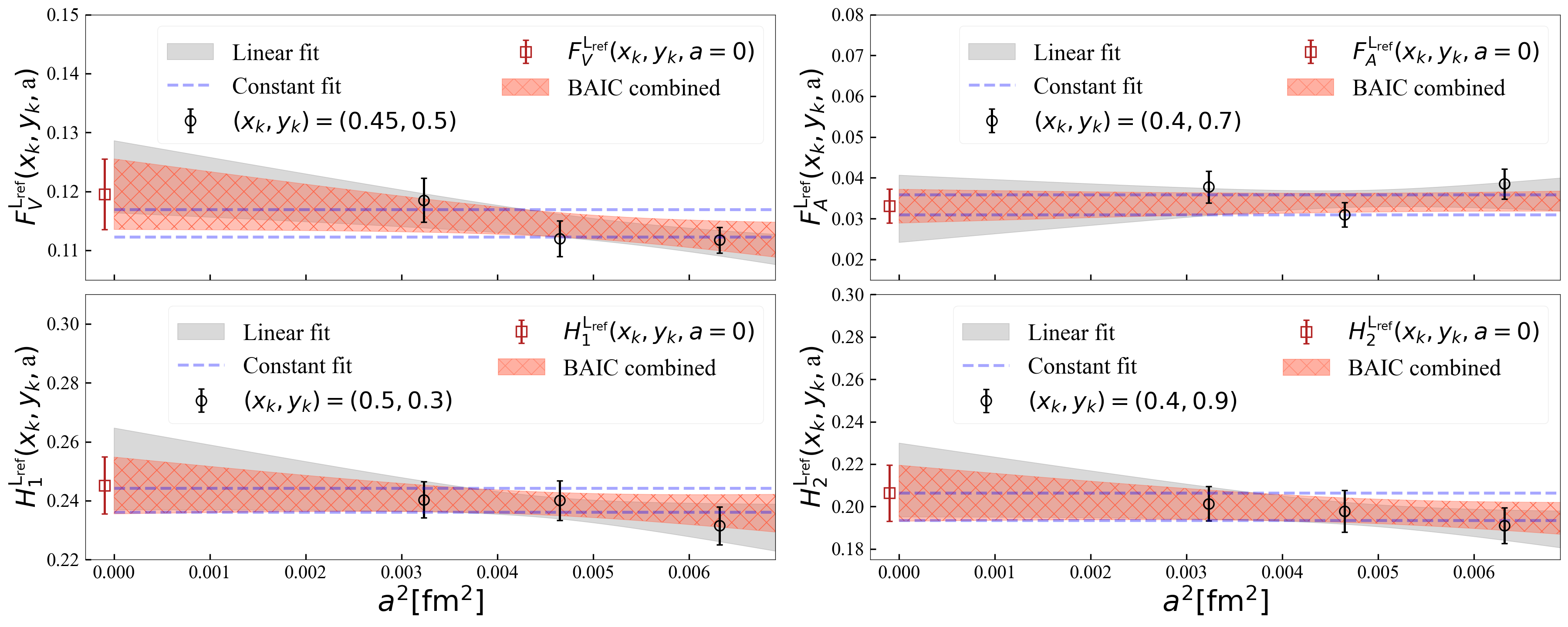}
    \caption{Continuum-limit extrapolation of the quark-connected contribution to the form factors at fixed lattice spatial extent $L_{\mathrm{ref}} \simeq 5.46~\mathrm{fm}$. The black circles correspond, from finer to coarser lattice spacings, to the D96, C80, and B ensemble. Gray bands show the linear fit to all three data points, while the area between the blue dashed lines correspond to a constant fit to the two finest lattice spacings.  The meshed red bands represent the BAIC-weighted combination of the two fits. Red squares mark the continuum-limit extrapolation results at $L_{\mathrm{ref}}$.}
    \label{fig:4L_afit}
\end{figure}

We then combine the two fits via the Bayesian Akaike Criterion (BAIC)~\cite{Neil:2022joj}, which we now briefly describe. Let $x_{1}$ and $x_{2}$ be the outcomes of the two fits. The final central value is given by
\begin{equation}
\bar{x} = w_{1}\,x_{1} + w_{2}\,x_{2},
\end{equation}
and the total error is
\begin{equation}
\sigma^{2} = \sigma_{\mathrm{stat}}^{2} + w_{1}(x_{1}-\bar{x})^{2} + w_{2}(x_{2}-\bar{x})^{2},
\end{equation}
where $\sigma_{\mathrm{stat}}$ is the statistical error of $\bar{x}$, and $w_{1}$ and $w_{2}$ are weights normalized to one and given by
\begin{equation}
w_{i} \propto \exp\Bigl[-\bigl(\chi_{i}^{2} + 2\,N_{i}^{\mathrm{pars}} - 2\,N_{i}^{\mathrm{data}}\bigr)/2\Bigr]~,\qquad i=1,2~.
\end{equation}
Here, $\chi_{i}^{2}$ denotes the chi-squared of the $i$-th fit, while $N_{i}^{\mathrm{pars}}$ and $N_{i}^{\mathrm{data}}$ are its number of free parameters and data points, respectively.  \\ \\ 
Fig.~\ref{fig:4L_afit} shows the continuum-limit extrapolation of the form factors for selected values of $x_{k}$ and $y_{k}$. 
In each panel, the gray band correspond to the linear fit performed on all three lattice spacings (black circles), while the area between the blue dashed lines shows the constant fit restricted to the two finest lattice spacing ensembles.
\begin{figure}[t]
    \centering
\includegraphics[width=1.\columnwidth]{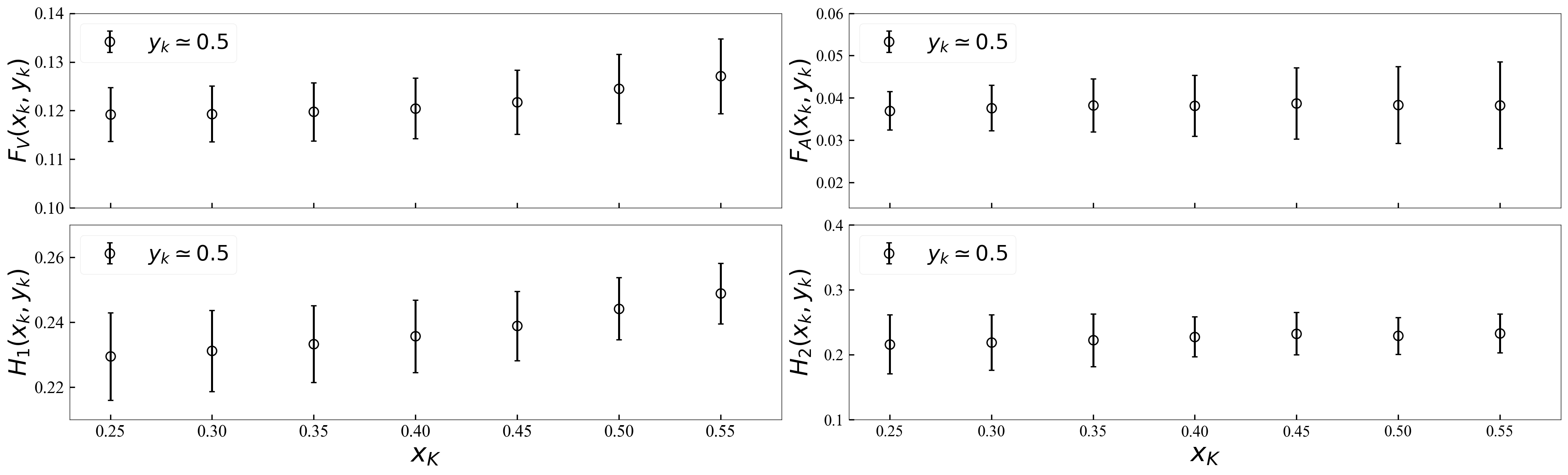}
    \caption{Quark-connected contribution to the form factors, as functions of $x_k$ with $y_k \simeq 0.5$. Black circles represents the results after performing the continuum and infinite-volume limits.}
    \label{fig:4L_FFundervsxk}
\end{figure}
The meshed red band represents the BAIC weighted average of the two fit functions, while the red point in square indicates the final continuum extrapolated value at $L_{\mathrm{ref}}$. Given the current uncertainties, no significant cutoff effects are observed,  which supports the validity of including a constant fit to the two finest lattice spacings.

\begin{figure}[t]
    \centering
\includegraphics[width=1.\columnwidth]{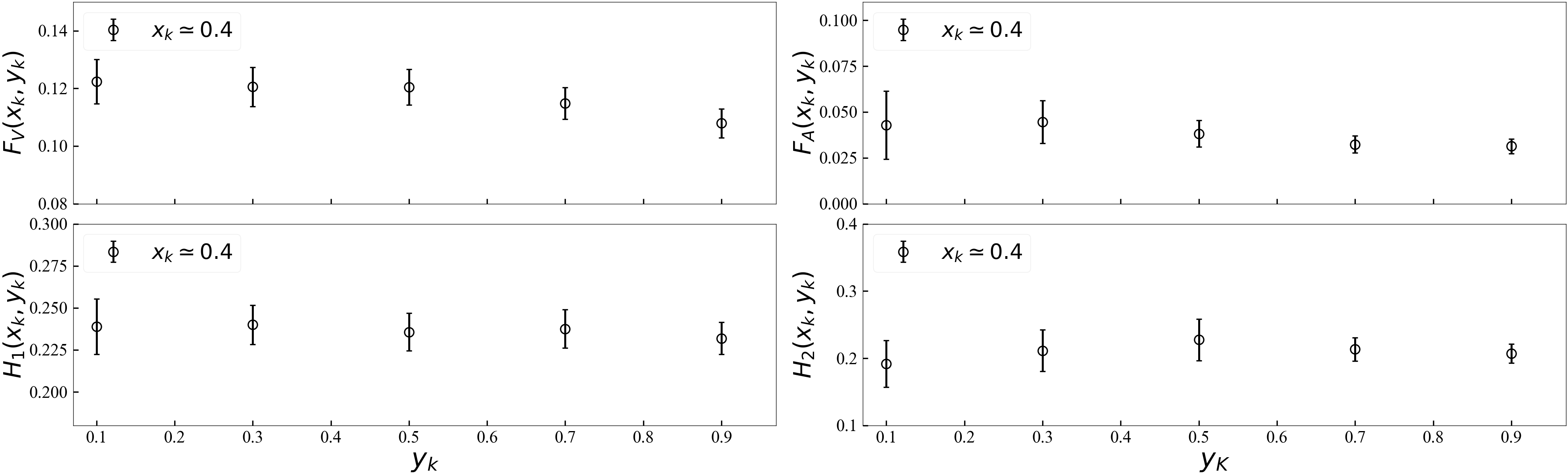}
    \caption{Quark-connected contribution to the form factors, as functions of $y_k$ with $x_k \simeq 0.4$. Black circles represents the results after performing the continuum and infinite-volume limits.}
    \label{fig:4L_FFundervsyk}
\end{figure}

\begin{figure}[]
    \centering
\includegraphics[width=1.\columnwidth]{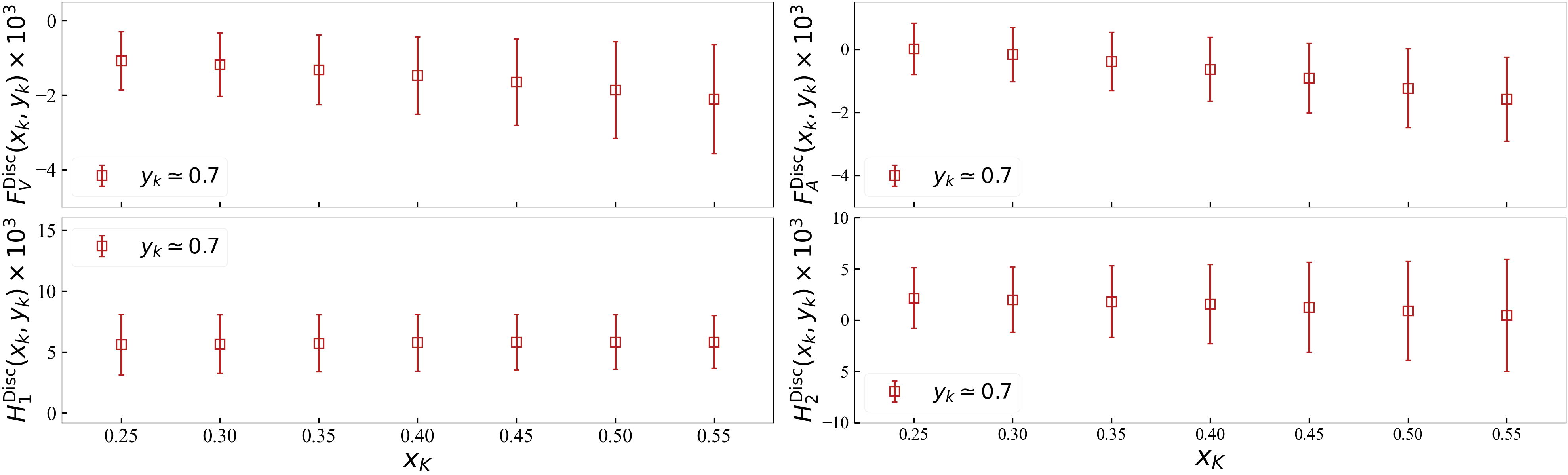}
    \caption{Quark-disconnected contribution to the form factors multiplied by $10^3$, as function of $x_k$ at fixed $y_k \simeq 0.7$. Red squares represent the results obtained on the B96 ensemble. To provide a conservative estimate of the missing continuum and infinite-volume extrapolation for this contribution, we will assign a $100\%$ uncertainty to the disconnected contribution, which is not shown in the figure. }
    \label{fig:4L_DiscFFundervsxk}
\end{figure}

Finally, we extrapolate to the infinite volume limit by adding the finite-volume corrections $\Delta_L F(x_k, y_k, L_{\mathrm{ref}})$, with $F=\{F_{V},A^{i}\}$, in Eq.~(\ref{eq:4L_finite-VCorr}) to the results of the continuum-limit extrapolation at $L_{\mathrm{ref}}$~(red squares in Fig.~\ref{fig:4L_afit}). 
The resulting form factors are shown in Fig.~\ref{fig:4L_FFundervsxk} and Fig.~\ref{fig:4L_FFundervsyk} as functions of $x_k$, at fixed $y_k \simeq 0.5$, and of $y_k$, at fixed $x_k \simeq 0.4$, respectively. 
We observe overall a mild dependence on $x_k$ and $y_k$ below the two-pion threshold. For the quark-disconnected contribution, as already stated,  we have results at a fixed $y_k\simeq 0.7$ on a single ensemble, the B96 ensemble, with  a lattice spacing $a \simeq 0.079~\mathrm{fm}$, and spatial extent $L \simeq 7.6~\mathrm{fm}$. The form factors computed on the B96 ensemble are shown as red squares in Fig.~\ref{fig:4L_DiscFFundervsxk}.
We find that the quark-disconnected contributions are suppressed by more than a factor of $50$ compared to the connected ones and are negligible within the current uncertainties. Consequently, cutoff and finite-volume effects on these contributions are expected to be subleading. 
To conservatively account for these effects, we assign a $100\%$  error to our estimates\footnote{To be more precise we take as final error the larger of a  $100\%$ relative uncertainty on the disconnected contribution and its statistical uncertainty.} of the quark-disconnected contribution. 

In Sec.~\ref{Sec:4L_finalFF} we will provide phenomenological parameterizations of the form factors as a function of $x_{k}$ and $y_{k}$. Before doing so, we discuss the determination of the form factors above the two-pion threshold.  

\section{Results above the two-pion threshold}
\label{Sec:4L_aboveth}
In this section we present our results in the kinematic region
$x_k > 2 m_\pi / m_K$, where intermediate on-shell two-pion states prevent the determination of the form factors through a simple Euclidean-time integration of the differential form factors.

As discussed in Sec.~\ref{sec:4L_strategyFFAbove}, the calculation of the form factors above the two-pion threshold can be organized by decomposing them into two contributions
\begin{align}
\label{eq:4L_FVWickpsfr}
F(x_k, y_k)
   ={}& F_{\mathrm{Wick}}(x_k, y_k)
     + 
       F_{\mathrm{SFR}}(x_k, y_k), \quad F = \{ F_V, A^i \}
\end{align}
The ``Wick'' terms are obtained by integrating the differential form factors over Euclidean time (see Eqs.~(\ref{eq:FV_Wick_def})--(\ref{eq:FA_wick})). Since their analysis proceeds in close analogy with the strategy adopted in the previous section for the determination of the form factors below the two-pion threshold, we do not repeat it here; details of the evaluation of the Wick contribution are provided in App.~\ref{app:4L_Wick}.

The remaining contributions,
$F_{V,\mathrm{SFR}}(\varepsilon, x_k, y_k)$ and
$A^i_{\mathrm{SFR}}(\varepsilon, x_k, y_k)$,
are instead determined using the SFR/HLT method described in Sec.~\ref{sec:4L_strategyFFAbove} and constitute the main focus of the present section. Before presenting the lattice-QCD results, we study
$F_{V,\mathrm{SFR}}(\varepsilon,x_k,y_k)$ and
$A^i_{\mathrm{SFR}}(\varepsilon,x_k,y_k)$
within a simple physically motivated model for the $\pi\pi$ contribution to the spectral density $\rho^{\mu\nu}_{W;u}$ in Eq.~(\ref{eq:4L_CErho}). 
The purpose of this analysis is to develop an intuition for the $\varepsilon$-dependence of the smeared form factors and to guide the construction of a robust strategy for the extrapolation to the physical $\varepsilon \to 0^+$ limit.

\subsection{An effective model for the $\pi^{+}\pi^{-}$ contribution to $F_{V,{\rm SFR}}$ and $A^{i}_{{\rm SFR}}$}
\label{sec:4L_model}

In this subsection we introduce an effective model for the contribution of intermediate
$\pi^+\pi^-$ states to the spectral densities entering the SFR form factors
$F_{V,{\rm SFR}}(\varepsilon,x_k,y_k)$ and $A^{i}_{{\rm SFR}}(\varepsilon,x_k,y_k)$.
The purpose of the model is twofold: i) to provide a framework in which
the qualitative $\varepsilon$-dependence of the smeared form factors can be understood,
and ii) to test and motivate improved strategies for the $\varepsilon\to0^+$ extrapolation
that will be employed on the lattice data in the next subsection.
The comparison with lattice correlators shown below is meant only as a sanity check
that the model captures the expected long-distance behaviour; the model is \emph{not}
used to obtain any quantitative determination of the physical form factors.

Throughout this subsection we work directly in the infinite-volume limit. All smeared quantities are reconstructed using
the \emph{improved} kernel introduced in Sec.~\ref{sec:4L_strategyFFAbove},
so that the leading finite-$\varepsilon$ corrections for both the real and imaginary part of the smeared form factors start at $O(\varepsilon^2)$.

We start from the two-pion contribution to the spectral density,
which arises from the isospin $I=1$ component of the electromagnetic current 
\begin{align}
\label{eq:jmuemIeq1}
j^{\mu}_{I=1} = \frac{1}{2}\left[ \bar{q}_{u}\gamma^{\mu}q_{u} - \bar{q}_{d}\gamma^{\mu}q_{d} \right]~.
\end{align}
The $\pi\pi$ contribution to $\rho_{W;u}^{\mu\nu}$ is given by
\begin{align}
\label{eq:4L_rhotwopions_model_revised}
\rho^{\mu\nu}_{W;\pi\pi}(E,\bs{k})
  \equiv -2\pi  \int 
      \frac{d\bs{p}_\pi}{(2\pi)^3\, 4 E_{\pi^+} E_{\pi^-}}\;
      \delta(E - E_{\pi^+} - E_{\pi^-})
   \bra{0} j^\mu_{I=1}(0) \ket{\pi\pi(\bs{k}, \bs{p}_\pi)}\,
   \bra{\pi\pi(\bs{k}, \bs{p}_\pi)} j^\nu_W(0) \ket{K^-(\bs{0})},
\end{align}
where the relative momentum $\bs{p}_\pi$ is defined such that
\begin{equation}
p_{\pi^\pm}
 = \bigl(E_{\pi^\pm},\, \bs{p}_{\pi^\pm} = \bs{k}/2 \pm \bs{p}_\pi \bigr),\qquad
E_{\pi^\pm}
  = \sqrt{m_\pi^2 + \lvert \bs{k}/2 \pm \bs{p}_\pi \rvert^{\,2}}.
\end{equation}
The spectral density in Eq.~(\ref{eq:4L_rhotwopions_model_revised}) vanishes below the
two-pion threshold, i.e. it has support only for
\begin{equation}
E \ge E^*(\bs{k}) \equiv 
2\sqrt{m_\pi^2+|\bs{k}|^2/4}\,.
\label{eq:threshold_infL}
\end{equation}

The corresponding contribution to the smeared hadronic tensor (and, after projection,
to the smeared form factors) is obtained through
\begin{equation}
H^{\mu\nu}_{W,2;\pi\pi}(\varepsilon,E_\gamma,\bs{k};\pi\pi)
\equiv
\int_0^\infty\frac{dE}{2\pi}\,
\tilde K(E-E_\gamma;\varepsilon)\,\rho^{\mu\nu}_{W;u}(E,\bs{k};\pi\pi),
\label{eq:H_pipi}
\end{equation}
with $\tilde K$ the (improved) kernel defined in Eq.\,(\ref{eq:ImprovedKernel}) in Sec.~\ref{sec:4L_strategyFFAbove}, such that the leading non-zero $\varepsilon$ effects are of order $O(\varepsilon^{2})$. In the spectral representation above we have set $E=0$ as the lower integration limit with the understanding that the integral effectively starts at $E^*(\bs{k})$.

The electromagnetic matrix element is parameterized as
\begin{equation}
\label{eq:4L_modelem_model_revised}
\bra{0} j^\mu_{I=1}(0) \ket{\pi\pi(\bs{k}, \bs{p}_\pi)}
  = i \bigl(p^\mu_{\pi^{+}} - p^\mu_{\pi^{-}}\bigr)\, F_\pi(s),
\end{equation}
where $F_\pi(s)$ is the timelike pion electromagnetic form factor and
\begin{equation}
s \equiv (p_{\pi^{+}} + p_{\pi^{-}})^2
  = (E_{\pi^{+}} + E_{\pi^{-}})^2 - |\bs{k}|^{\,2}.
\end{equation}
We model $F_\pi(s)$ using the Gounaris--Sakurai (GS) parametrization~\cite{Gounaris:1968mw}. More details on the GS parameterization and on the specific input parameters adopted can be found in App.~\ref{App:rhomunupipi_details}.

For the weak matrix elements we follow the standard $K_{\ell4}$ parametrization
(see e.g.\ Ref.~\cite{Bijnens:1994ie}),
\begin{align}
\label{eq:4L_Vmodel_model_revised}
\bra{\pi\pi(\bs{k}, \bs{p}_\pi)} j^\nu_V(0) \ket{K^-(\bs{0})}
   &= -\,\frac{H}{m_K^3}\,
     \varepsilon^{\nu\alpha\rho\sigma}
     p_\alpha\,
     (p_{\pi^{+}} + p_{\pi^{-}})_\rho\,
     (p_{\pi^{+}} - p_{\pi^{-}})_\sigma,
\\[6pt]
\label{eq:4L_Amodel_model_revised}
\bra{\pi\pi(\bs{k}, \bs{p}_\pi)} j^\nu_A(0) \ket{K^-(\bs{0})}
  &= \frac{i}{m_K}
    \Big[
      (p_{\pi^{+}} + p_{\pi^{-}})^\nu F
      + (p_{\pi^{+}} - p_{\pi^{-}})^\nu G
      + \bigl(p - p_{\pi^{+}} - p_{\pi^{-}}\bigr)^\nu R
    \Bigr],
\end{align}
where the $K_{\ell4}$ form factors $F,G,H,R$ are functions of $s$ and of the
Mandelstam variables
\begin{equation}
t = (p_{\pi^{+}} - p)^2,\qquad
u = (p_{\pi^{-}} - p)^2.
\end{equation}
To simplify the model we will assume that the form factors are only $s$-dependent. Within this approximation, one can show that $F$ and $R$ do not contribute to the spectral density, since their contributions to the integrand in Eq.~(\ref{eq:4L_rhotwopions_model_revised}) is odd under $\bs{p}_{\pi}\to -\bs{p}_{\pi}$, and therefore vanish after integration. 

We thus consider only the contribution of the form factors $H$ and $G$, which can be written as~\cite{Cirigliano:2011ny}
\begin{align}
H(s) = h(s)e^{i\delta(s)}~, \quad G(s) = g(s)e^{i\delta(s)}~.
\end{align}
The two form factors are well extracted from the $K_{\ell 4}$ experimental data in the physical region of the decay, namely $4m_{\pi}^{2}\leq s\leq m_{K}^{2}$. Phenomenological analyses ~\cite{Rosselet:1976pu,Amoros:1999qq,Cirigliano:2011ny} indicate the following values for $h(s)$ and $g(s)$ at the two pion threshold
\begin{align}
g(4m_{\pi}^{2}) \simeq 5~, \qquad \frac{h(4m_{\pi}^{2})}{g(4m_{\pi}^{2})} \simeq -0.46~,
\end{align}
and we take these two values as input for our model. Moreover,
according to Watson's theorem~\cite{Watson:1952ji}, in the elastic region the phases of the
$K\to\pi\pi$ amplitudes coincide with the $\pi\pi$ scattering phase shift in the corresponding
channel. Since the electromagnetic matrix element carries the phase of $F_\pi(s)$, consistency
requires the weak form factors to carry the opposite phase so that the phase cancels in the
product entering the spectral density.
There are several possible ways to implement this constraint. In this work we adopt a minimal
choice that simultaneously (i) enforces the elastic-phase cancellation and (ii) provides a
reasonable extension of the weak form factors into the region $s\sim m_\rho^2$, where a $\rho$-driven enhancement can be expected. This region is not
kinematically accessible in physical $K_{\ell4}$ decays, but is probed by the spectral integral in Eq.~(\ref{eq:H_pipi}).
Concretely, we impose
\begin{equation}
\label{eq:4L_FFWatson_model_revised}
G(s)=g(4m_\pi^2)\,\frac{F_\pi^*(s)}{|F_{\pi}(4m_{\pi}^{2})|},\qquad
H(s)=h(4m_\pi^2)\,\frac{F_\pi^*(s)}{|F_{\pi}(4m_{\pi}^{2})|},
\end{equation}
so that $\delta(s)=-\arg[F_\pi(s)]$ and the phase cancels in the product of matrix elements.

Having defined our spectral density model, $\rho^{\mu\nu}_{W;\pi\pi}(E,\bs{k})$, we construct the corresponding model spectral densities associated with the form factors, namely $\rho_{V;u}$ and $\rho^{i}_{A;u}$. For convenience we denote these by $\rho_{V}^{\pi\pi}(E,x_{k},y_{k})$ and $\rho_{A^{i}}^{\pi\pi}(E,x_{k},y_{k})$. From these model spectral densities one obtains the differential form factors entering the SFR/HLT reconstruction (cf.~Eqs.~(\ref{eq:diff_V_HLT}) and (\ref{eq:diff_A_HLT})) through
\begin{align}
\delta F_{V;u}^{\pi\pi}(t_\gamma,x_k,y_k)
&=\int_0^\infty\frac{dE}{2\pi}\,e^{-t_\gamma(E-E_\gamma)}\,\rho_{V}^{\pi\pi}(E,x_k,y_k),
\\[8pt]
\delta\overline{A}^{\,i;\pi\pi}_u(t_\gamma,x_k,y_k)
&=\int_0^\infty\frac{dE}{2\pi}\,e^{-t_\gamma(E-E_\gamma)}\,\rho^{\pi\pi}_{A^{i}}(E,x_k,y_k),
\end{align}
as well as the smeared form factors (cf.~Eqs.~(\ref{eq:FF_V_HLT}) and (\ref{eq:FF_A_HLT}))
\begin{align}
F_{V,{\rm SFR}}^{\pi\pi}(\varepsilon,x_k,y_k)
&=
\int_0^\infty\frac{dE}{2\pi}\,\rho_{V}^{\pi\pi}(E,x_k,y_k)\,
\tilde K(E-E_\gamma;\varepsilon),
\\[8pt]
A^{i;\pi\pi}_{{\rm SFR}}(\varepsilon,x_k,y_k)
&=
\int_0^\infty\frac{dE}{2\pi}\,\rho^{\pi\pi}_{A^{i}}(E,x_k,y_k)\,
\tilde K(E-E_\gamma;\varepsilon).
\end{align}

\begin{figure}[]
    \centering
    \includegraphics[width=1.\columnwidth]{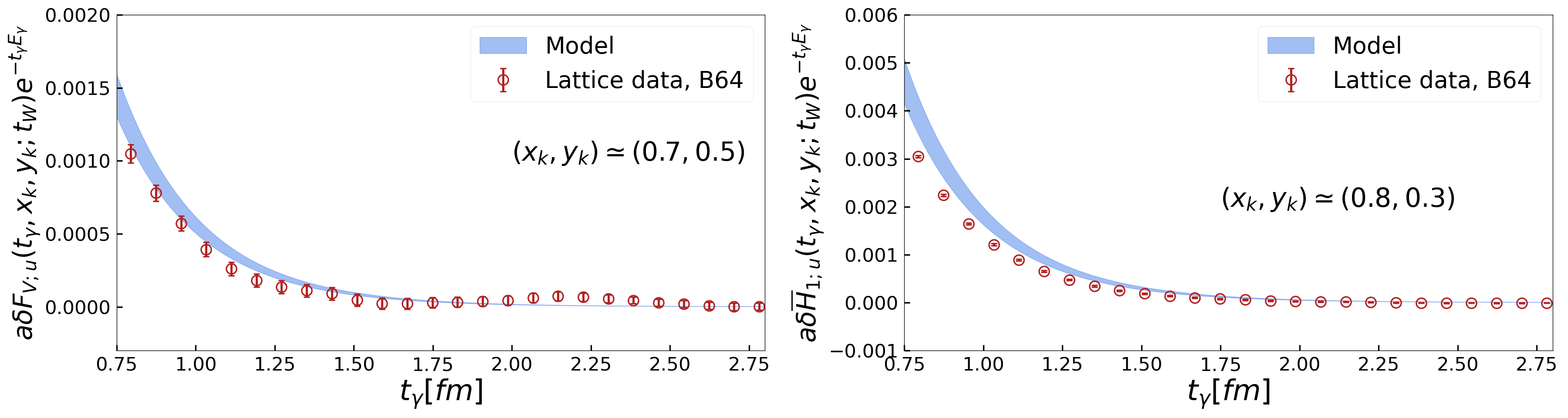}
    \caption{Differential form factors entering the HLT reconstructions of $F_{V,\mathrm{SFR}}(\varepsilon, x_k, y_k; t_W)$ (left panel) and $H_{1,\mathrm{SFR}}(\varepsilon, x_k, y_k; t_W)$ (right panel). The blue bands represent the predictions of our effective model (the width of the band corresponds to a $10\%$ uncertainty),  while the red points correspond to the lattice data, obtained on the B64 ensemble for $t_{w}\simeq 2.2~{\rm fm}$.}
    \label{fig:4L_ModelFF}
\end{figure}

In Fig.~\ref{fig:4L_ModelFF} we compare the lattice results for the quark-connected contributions to $\delta F_{V;u}$ and $\delta \overline{H}_{1;u}$ (diagram (c) in Fig.~\ref{Fig:Diagrams}) with the corresponding model predictions $\delta F_{V;u}^{\pi\pi}$ and $\delta \overline{H}_{1;u}^{\,\pi\pi}$. Since the $\pi\pi$ contribution to the quark-connected part of $\delta F_{V;u}$ and $\delta \overline{H}_{1;u}$ is equal to $2e_u=4/3$ times the total $\pi\pi$ contribution, we include this factor in the model results when performing the comparison with the lattice data. The comparison is shown for representative values of $x_{k}$ and $y_{k}$; for all other values the comparison looks similar. The lattice data (red circles) are obtained on the B64 ensemble at $t_{W}\simeq 2~{\rm fm}$, while the model predictions are shown as coloured bands. The width of the band corresponds to a $10\%$ uncertainty. As can be seen, the model provides a realistic description of the lattice data at large times, where the two-pion states are expected to be the dominant contribution. 

By contrast, analogous comparisons for the other two form factors, $F_{A}$ and $H_{2}$, reveal larger discrepancies. This behaviour can be traced back to the fact that $H_{1}$ is primarily sensitive to the $G$ form factor, which is included in the model and is well constrained by experiment. On the other hand, $H_{2}$ and $F_{A}$ receive sizeable contributions from the $R$ and $F$ form factors, which we have neglected for simplicity and which are also not well determined experimentally.\footnote{This is particularly true for the form factor $R$, whose contribution to the $K^{-}\to \pi^{+}\pi^{-}e^{-}\bar{\nu}_{e}$ decay rate is chirality suppressed.}

\begin{figure}[]
    \centering
    \includegraphics[width=1.\columnwidth]{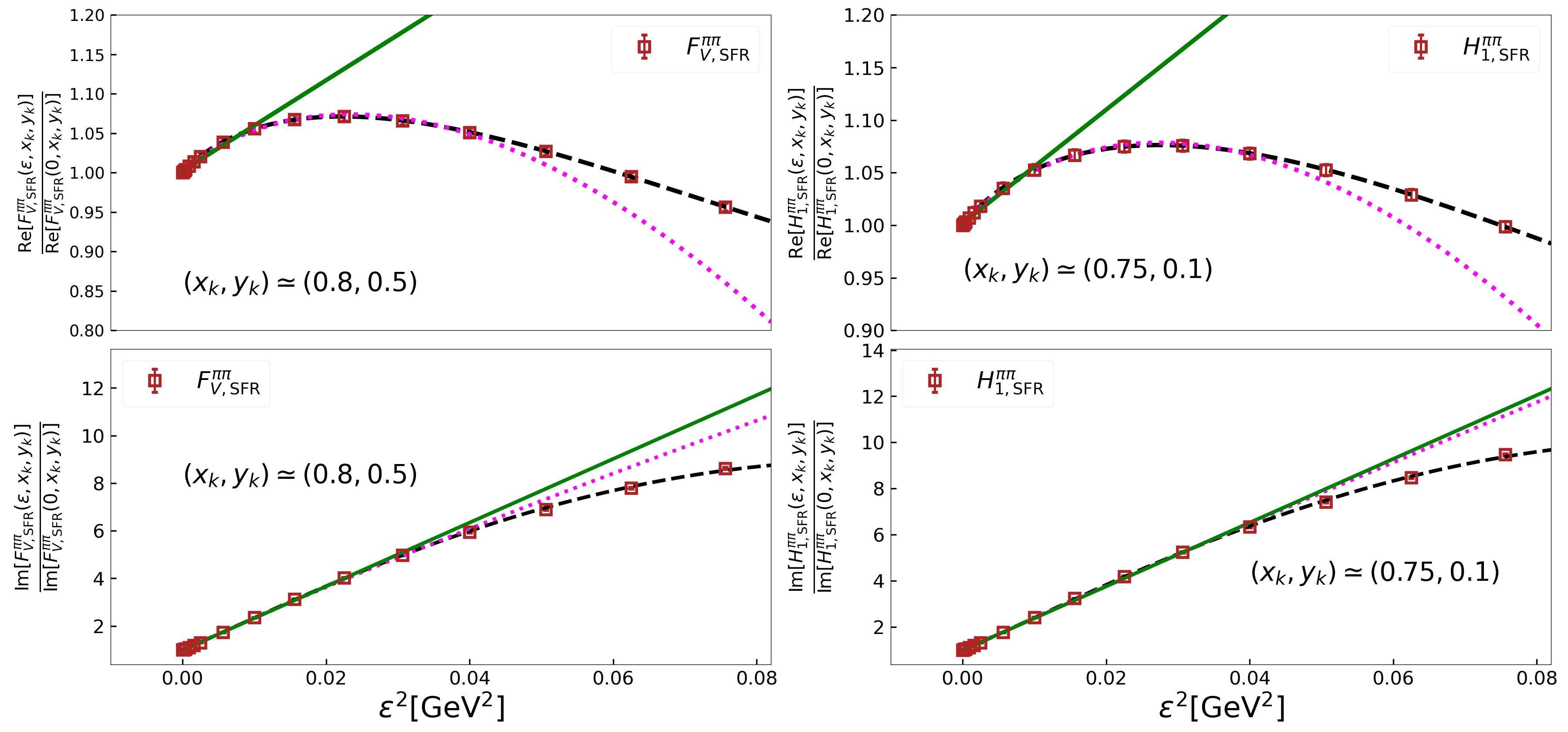}
    \caption{Real (upper panels) and imaginary (lower panels) parts of the SFR form factors $F_{V,\mathrm{SFR}}$ (left panels) and $H_{1,\mathrm{SFR}}$ (right panels), computed within the effective model as functions of $\varepsilon^2$ and normalized to their values at $\varepsilon = 0$, computed within the model. The data are displayed with a $0.5\%$ uncertainty. In each panel, the black dashed line represents a polynomial fit including terms up to $\varepsilon^4$ and performed over the full data set. The magenta dotted curve and the green solid line correspond to polynomial fits including terms up to $\varepsilon^3$ and $\varepsilon^2$, respectively, applied to a restricted subset of data (see text for details).}
    \label{fig:4L_ModelFFeps}
\end{figure}

Given these observations, in the following we restrict our analysis of the $\varepsilon$-dependence to the form factors $F_{V}$ and $H_{1}$. In Fig.~\ref{fig:4L_ModelFFeps} we display the $\varepsilon$-dependence of $F_{V,{\rm SFR}}^{\pi\pi}$ and $H_{1,{\rm SFR}}^{\pi\pi}$ (red squares). The upper panels show the real parts, while the lower panels correspond to the imaginary parts. In all cases the smeared form factors are normalized to their $\varepsilon\to0$ values.

We attempt to describe the $\varepsilon$-dependence of the smeared form factors using a polynomial expansion starting at $O(\varepsilon^{2})$, as dictated by the improved kernel. Considering first the interval $\varepsilon \in [0,275]~{\rm MeV}$, where $275~{\rm MeV}$ is close to our estimate of $\Delta_{\rho}^{\rm max}$ in Eq.~(\ref{eq:rho_max}), we find that an excellent description of the data (even assuming $0.5\%$ uncertainties) is achieved with a polynomial including terms up to $\varepsilon^{4}$. The resulting $O(\varepsilon^{4})$ fits are shown by the black dashed curves. For $\varepsilon \lesssim 0.2~{\rm GeV}$ the $\varepsilon^{4}$ contribution becomes negligible, and a fit including only $\varepsilon^{2}$ and $\varepsilon^{3}$ terms (magenta dotted curves) is sufficient. Restricting further to $\varepsilon \lesssim 0.1~{\rm GeV}$, a pure $O(\varepsilon^{2})$ description (solid green lines) becomes adequate (with a somewhat wider validity range for the imaginary parts). 

Anticipating the discussion of the next subsection, the smallest value for which reliable lattice data are available is $\varepsilon_{\rm min}\simeq 0.1~{\rm GeV}$. The fact that higher-order terms are not negligible in the region where we have lattice results motivates the development of alternative estimators designed to suppress higher-order corrections and thus accelerate convergence toward the $\varepsilon\to0^{+}$ limit.
\begin{figure}[]
    \centering
    \includegraphics[width=1.\columnwidth]{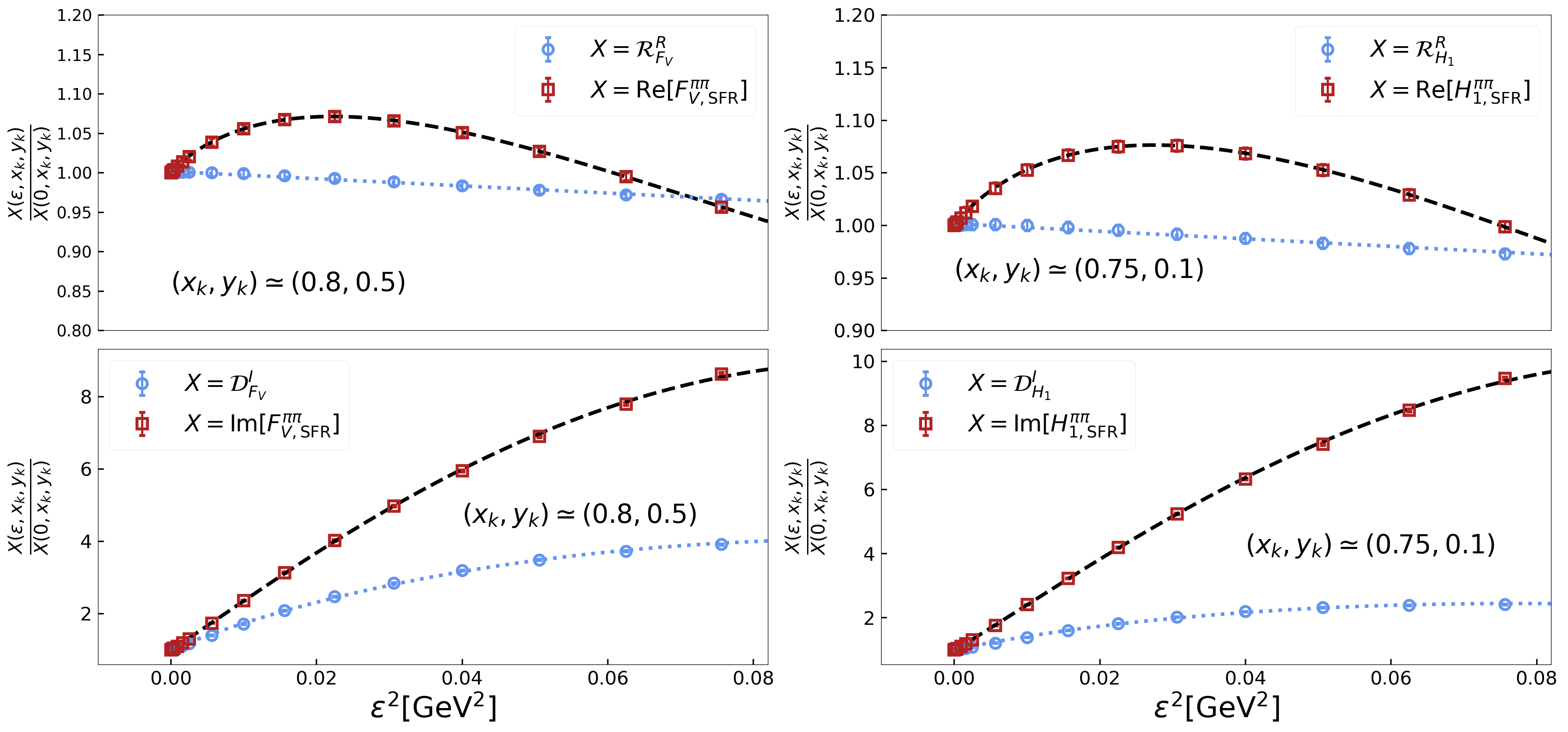}
    \caption{Top panels: comparison of the $\varepsilon$-dependence of the real parts of the smeared form factors with the corresponding ratios $\mathcal{R}_{F}^{R}$, for $F=F_{V}$ (left) and $F=H_{1}$ (right). 
Bottom panels: analogous comparison for the imaginary parts, showing the smeared form factors together with the differences $\mathcal{D}_{F}^{I}$, again for $F=F_{V}$ (left) and $F=H_{1}$ (right).}
    \label{fig:4L_ModelFFeps2}
\end{figure}

This consideration leads us to introduce a strategy which we refer to as ratio-difference preconditioning (RDP). The central idea of the RDP is to exploit the fact that below the two-pion threshold we can determine the form factors directly at $\varepsilon = 0$ (as performed in the previous section using standard Euclidean time integration), but also study their $\varepsilon$-dependence. We then use this information in the analysis above threshold. Although the $\varepsilon$-dependence will not be identical in the two regions, it is reasonable to expect that it varies smoothly as a function of $x_k$. The present effective model provides a controlled setting in which to test this strategy.

In practice, for fixed $y_{k}$, we proceed as follows. We start from the kinematic point immediately below the two-pion threshold (in our case $x_{k}^{0}=0.55$), and reconstruct the form factor at the first point above threshold, $x_{k}=x_{k}^{0}+\Delta x_{k}=0.6$, via
\begin{align}
\label{eq:im1_diff}
{\rm Im }[F_{\rm SFR}(x_{k}^{0}+\Delta x_{k},y_{k})]
&= \lim_{\varepsilon\to 0^+}\mathcal{D}^{I}_{F}(\varepsilon,x_k^{0}+\Delta x_{k},y_k), 
\\[6pt]
\label{eq:re1_diff}
{\rm Re }[F_{\rm SFR}(x_{k}^{0}+\Delta x_{k},y_{k})]
&= {\rm Re}[F_{\rm SFR}(x_{k}^{0}, y_{k})]
   \times\lim_{\varepsilon\to 0^+}\mathcal{R}^{R}_{F}(\varepsilon,x_k^{0}+\Delta x_{k},y_k),
\end{align}
where $F=\{F_{V},A^{i}\}$, and
\begin{align}
\label{eq:im_RDP_sm}
\mathcal{D}_{F}^{I}(\varepsilon,x_{k},y_{k})
&\equiv
\Im[F_{\rm SFR}(\varepsilon,x_k,y_k)]
-
\Im[F_{\rm SFR}(\varepsilon,x_k-\Delta x_k,y_k)],
\\[6pt]
\label{eq:re_RDP_sm}
\mathcal{R}_{F}^{R}(\varepsilon,x_{k},y_{k})
&\equiv
\Re[F_{\rm SFR}(\varepsilon,x_k,y_k)]
/
\Re[F_{\rm SFR}(\varepsilon,x_k-\Delta x_k,y_k)].
\end{align}
Eq.~(\ref{eq:im1_diff}) follows from the fact that the form factors have vanishing imaginary parts below the two-pion threshold, and thus at $x_{k}^{0}$. For the real part we have used the fact that below threshold the form factor $F_{\rm SFR}(x_{k}^{0}, y_{k})$ can be directly computed at $\varepsilon=0$ limit using the Euclidean-time integration technique. Iterating the procedure for $x_{k}=x_{k}^{0}+n\Delta x_{k}$ yields
\begin{align}
\label{eq:im_RDP}
{\rm Im}[F_{\rm SFR}(x_{k},y_{k})]
&=
\sum_{j=1}^{n} 
\mathcal{D}_{F}^{I}(x_{k}^{0}+j\Delta x_{k},y_{k})~,\qquad\qquad\qquad\quad\,\, F=\{F_{V},A^{i}\}~,
\\[6pt]
\label{eq:re_RDP}
{\rm Re}[F_{\rm SFR}(x_{k},y_{k})]
&=
F_{\rm SFR}(x_{k}^{0},y_{k})
\prod_{j=1}^{n} 
\mathcal{R}_{F}^{R}(x_{k}^{0}+j\Delta x_{k},y_{k})~,\qquad F=\{F_{V},A^{i}\}~,
\end{align}
where $\mathcal{D}_{F}^{I}(x_{k},y_{k})$ and $\mathcal{R}_{F}^{R}(x_{k},y_{k})$ denote the $\varepsilon\to0^{+}$ limits of $\mathcal{D}_{F}^{I}(\varepsilon,x_{k},y_{k})$ and $\mathcal{R}_{F}^{R}(\varepsilon,x_{k},y_{k})$, respectively. The expectation, which we now test within the model,  is that $\mathcal{D}_{F}^{I}$ and $R_{F}^{I}$ have a smoother dependence on $\varepsilon$, reducing the impact of higher-order terms. Fig.~\ref{fig:4L_ModelFFeps2} compares the $\varepsilon$-dependence of the original smeared form factors with that of $\mathcal{D}_{F}^{I}$ and $\mathcal{R}_{F}^{R}$. All quantities are normalized to their values at $\varepsilon=0$, computed within the model. Both the differences and the ratios exhibit a markedly flatter $\varepsilon$-dependence, especially the ratios. In the latter case, the data over the full range $\varepsilon \in [0,275]~{\rm MeV}$ are well described by a single linear term in $\varepsilon^{2}$ with better than $0.5\%$ accuracy. For the differences, the behaviour is also significantly improved, though inclusion of an $\varepsilon^{3}$ term is required for a satisfactory description over the full interval. The same behaviour is observed for all kinematic points $(x_{k},y_{k})$ we considered. 

The model study thus provides a strong evidence that the RDP strategy enables significantly more controlled $\varepsilon\to0^{+}$ extrapolations than a direct analysis of the smeared form factors. Guided by these findings, in the next subsection we apply the RDP strategy to the lattice data in order to extract the SFR form factors.

\subsection{Numerical evaluation of $F_{V,{\rm SFR}}(x_{k},y_{k})$ and $A^{i}_{\rm SFR}(x_{k},y_{k})$ from lattice data}
\label{sec:4L_FVsfrandAisfr}
In this subsection we present our numerical results for the SFR form factors obtained using lattice QCD. To determine the form factors $F_{\rm SFR}(x_{k},y_{k})$ above the two-pion threshold we make use of the RDP strategy described in the previous subsection. We therefore evaluate the quantities $\mathcal{R}_{F}^{R}(x_{k},y_{k})$ and $\mathcal{D}_{F}^{I}(x_{k},y_{k})$, from which the real and imaginary parts of the SFR form factors can be reconstructed using Eqs.~(\ref{eq:im_RDP}) and (\ref{eq:re_RDP}).

The SFR form factors presented below are obtained using a single value of $t_W \gtrsim 2~\mathrm{fm}$. Since they receive contributions only from the second time ordering—where the kaon interpolator and the currents are separated by at least $t_W$—finite-$t_W$ effects are negligible, as already observed in Fig.~\ref{fig:4L_FFsecondTOtwdep} for photon virtualities below the two-pion threshold. For this reason, in the following we omit the explicit dependence on $t_W$ in both the differential and the smeared form factors.

To obtain  $\mathcal{R}_{F}^{R}(x_{k},y_{k})$ and $\mathcal{D}_{F}^{I}(x_{k},y_{k})$ for $x_{k} > 2m_{\pi}/m_{K}$, we first evaluate their $\varepsilon$-smeared counterparts defined in Eqs.~(\ref{eq:im_RDP_sm})--(\ref{eq:re_RDP_sm}). We then take their $\varepsilon\to 0^{+}$ limit. This limit must be carried out only after performing the infinite-volume extrapolation.

The discussion in Sec.~\ref{sec:4L_strategyFFAbove}, as well as the model results presented in the previous subsection, have shown that the onset of the polynomial regime, where non-zero $\varepsilon$ effects on smeared quantities can be described by a low-order polynomial in $\varepsilon$, should occur for $\varepsilon \lesssim \Delta_{\rho}^{\rm max} = 270~{\rm MeV}$. We therefore only use values 
of $\varepsilon$ within this range. 
Moreover, in order to perform a controlled infinite-volume extrapolation, the values of $\varepsilon$ should be large compared to $1/L$. The ETMC ensembles employed for the present calculation are such that $1/L$ ranges from about $50~{\rm MeV}$ on the B48 ensemble to about $25~{\rm MeV}$ on the B96 ensemble. For this reason we have decided to evaluate the smeared quantities for the following values of $\varepsilon$:
\be
\label{eq:4L_epsilonused}
\varepsilon = \{0.100,\;0.125,\;0.150,\;0.175,\;0.200,\;0.225,\;0.250,\;0.275\}~\mathrm{GeV}.
\ee
Computations at these values are expected not to be affected by excessively large finite-size effects. Note that the smallest value of $\varepsilon$ in Eq.\,(\ref{eq:4L_epsilonused}) is not much below $m_{\pi} \sim 135\,{\rm MeV}$.

We now discuss the determination of the smeared $\mathcal{D}_{F}^{I}$ and $\mathcal{R}_{F}^{R}$ at fixed $\varepsilon$ and volume. This determination is performed using the HLT method, which we now summarize. We begin by describing how the smeared SFR form factors entering the ratios and differences are obtained.

The main idea behind the HLT is to evaluate the smeared SFR form factors on a given gauge ensemble using only the differential form factors $\delta F_{V;u}$ and $\delta \overline{A}^{\,i}_{u}$ as nonperturbative lattice input, using
\begin{align}
\label{eq:4LFsfr}
F_{V,\mathrm{SFR}}(\varepsilon, x_k, y_k)
 &= \sum_{n=1}^{n_{\mathrm{max}}}
    \bigl(g_{R, F_{V}}(n, E_\gamma, \varepsilon)
       + i\, g_{I,F_{V}}(n, E_\gamma, \varepsilon)\bigr)
   e^{- an E_\gamma}\,
   \delta F_{u}(an, x_k, y_k)~, \nonumber \\[8pt]
 A^{i}_{\mathrm{SFR}}(\varepsilon, x_k, y_k)
 &= \sum_{n=1}^{n_{\mathrm{max}}}
    \bigl(g_{R, A^{i}}(n, E_\gamma, \varepsilon)
       + i\, g_{I,A^{i}}(n, E_\gamma, \varepsilon)\bigr)
   e^{- an E_\gamma}\,
   \delta \overline{A}^{\,i}_{u}(an, x_k, y_k)~,
\end{align}
where $a$ is the lattice spacing. The coefficients $g_{R,F}$ and $g_{I,F}$, with $F=\{ F_{V}, A^{i}\}$, are the basis coefficients of the following approximation to the kernel function $\tilde{K}$ in terms of exponential functions:
\begin{align}
\label{eq:HLT_kernel_approx}
 &\tilde{K}_{R}(E-E_{\gamma};\varepsilon) \equiv {\rm Re}[ \tilde{K}(E-E_{\gamma};\varepsilon)] \simeq \sum_{n =1}^{n_{\rm max}} g_{R,F}(n,E_{\gamma},\varepsilon) e^{-anE}~, \qquad F=\{F_{V},A^{i}\}~, \nonumber \\[8pt]
 &\tilde{K}_{I}(E-E_{\gamma};\varepsilon) \equiv {\rm Im}[ \tilde{K}(E-E_{\gamma};\varepsilon)]\simeq \sum_{n =1}^{n_{\rm max}} g_{I,F}(n,E_{\gamma},\varepsilon) e^{-anE}~, \qquad\, F=\{F_{V},A^{i}\}~. 
\end{align}
The kernel function we consider is the one in Eq.~(\ref{eq:ImprovedKernel}), except that, following Ref.~\cite{Frezzotti:2023nun}, at finite lattice spacing we make the replacement
\begin{align}
\frac{1}{x-i\varepsilon}\,  \longmapsto \,
\frac{ e^{-a(x-i\varepsilon)/2}}
{\frac{2}{a}\sinh\left(\frac{ax-ia\varepsilon}{2}\right)}~,
\end{align}
which reduces to $1/(x-i\varepsilon)$ in the continuum limit.
For this analysis we fix $a\cdot n_{\rm max}\sim 2~{\rm fm}$. Using Eqs.~(\ref{eq:diff_V_HLT})--(\ref{eq:FF_V_HLT}) and Eqs.~(\ref{eq:diff_A_HLT})--(\ref{eq:FF_A_HLT}), it is straightforward to show that, if the r.h.s.\ of Eq.~(\ref{eq:HLT_kernel_approx}) provides a good approximation to the kernel function, then Eq.~(\ref{eq:4LFsfr}) yields the smeared SFR form factors.

Obtaining the coefficients $g_{R,F}$ and $g_{I,F}$ is, however, non-trivial. Any reconstruction of the smeared form factors through Eq.~(\ref{eq:4LFsfr}) is affected simultaneously by systematic uncertainties, due to the imperfect kernel reconstruction, and by statistical uncertainties, arising from fluctuations of the differential form factors. In the HLT approach one finds an optimal compromise between these two sources of uncertainty by obtaining the coefficients $g_{R,F}$ and $g_{I,F}$ through the minimization of the following functional:
\begin{equation}
\label{eq:4L_ABW}
W^{F}_{R(I)}[\boldsymbol{g}]
 = \frac{A_{R(I)}[\boldsymbol{g}]}{A_{R(I)}[\boldsymbol{0}]}
   + \lambda\, B^{F}_{R(I)}[\boldsymbol{g}], \quad F = \{F_{V}, A^i\}.
\end{equation}
The first term in Eq.~(\ref{eq:4L_ABW}), $A_{R(I)}[\boldsymbol{g}]$, provides a measure (in functional space) of the distance between the target kernel function and its reconstruction in terms of a finite number ($n_{\rm max}$) of exponentials. It is given by
\begin{equation}
\label{eq:4L_appAg}
A_{R(I)}[\boldsymbol{g}]
 \equiv \int_{E_{\min}}^\infty dE\, w^{2}(E)\,
   \left|\,\sum_{n=1}^{n_{\max}} g(n,E_\gamma,\varepsilon)\,
   e^{-an E}
   - \tilde K_{R(I)}(E-E_\gamma;\varepsilon)\,\right|^2,
\end{equation}
and we comment below on our choices for the threshold parameter $E_{\rm min}$ as well as for the weight function $w(E)$. The second term in Eq.~(\ref{eq:4L_ABW}), $B^{F}_{R(I)}[\boldsymbol{g}]$, is the so-called error functional and is given by
\be
B^{F}_{R(I)}[\boldsymbol{g}] \equiv
\sum_{n_{1},n_{2}=1}^{n_{\rm max}}
g(n_{1}, E_\gamma, \varepsilon)\, g(n_{2}, E_\gamma, \varepsilon)\,
\mathrm{Cov}^F_{R(I)}(n_{1}, n_{2}), \qquad F = \{F_{V}, A^i\}~,
\ee
where ${\rm Cov}^{F}_{R}(n_{1},n_{2})$ (${\rm Cov}^{F}_{I}(n_{1},n_{2})$) is the covariance of the real (imaginary) part of the differential form factors, $\delta F_{V;u}$ for $F=F_{V}$ and $\delta \overline{A}^{\,i}_u$ for $F=A^{i}$, at Euclidean times $t_{\gamma,1}=n_{1}a$ and $t_{\gamma,2}=n_{2}a$.
The parameter $\lambda$ appearing in Eq.~(\ref{eq:4L_ABW}) is the so-called \emph{trade-off} parameter. Its role is to balance the competing requirements of keeping the statistical uncertainty under control (small $B^{F}_{R(I)}[\bs{g}]$) and achieving an accurate kernel reconstruction (small $A_{R(I)}[\bs{g}]$). The smeared form factors are then obtained through the so-called \emph{stability analysis}, in which one monitors the dependence of the smeared form factors on the value of $\lambda$ employed in the minimization of the functional(s) $W_{R(I)}$. The optimal value, $\lambda^{*}$, is chosen to lie in the statistically dominated regime. In this regime, $\lambda$ is sufficiently small that the systematic error due to the kernel reconstruction is smaller than the statistical error (and therefore the results are stable under variations of $\lambda$ within statistical uncertainties), but still large enough to yield reasonably small statistical uncertainties. This is a standard strategy adopted in several calculations where the HLT reconstruction method has been used.

In our computation we aim to determine the smeared $\mathcal{D}_{F}^{I}(\varepsilon,x_{k},y_{k})$ and $\mathcal{R}_{F}^{R}(\varepsilon,x_{k},y_{k})$. To do so we perform the stability analysis by either subtracting the imaginary parts (for $\mathcal{D}_{F}^{I}$) or dividing the real parts (for $\mathcal{R}_{F}^{R}$) of the smeared form factors at nearby $x_{k}$ values, computed with a common value of $\lambda$. We then monitor the stability of $\mathcal{D}_{F}^{I}(\varepsilon,x_{k},y_{k})$ and $\mathcal{R}_{F}^{R}(\varepsilon,x_{k},y_{k})$ as functions of the common $\lambda$, and determine the optimal value $\lambda^{*}$.

Before showing the results of the corresponding stability analyses, we would like to comment on the choice of the algorithmic parameter $E_{\rm min}$ and of the weight function $w(E)$ entering the norm functional $A_{R(I)}[\bs{g}]$ in Eq.~(\ref{eq:4L_appAg}). We start with the latter. In principle, any choice of $w(E)\neq 0$ would be allowed. For instance, one might choose a uniform weight function $w(E)=1$. However, with this choice the HLT algorithm attempts to reconstruct the kernel with equal accuracy in all energy regions. This includes both regions where the spectral density contributes significantly (and where we would like a better reconstruction of the kernel) and regions where the spectral density is strongly suppressed and the form factors receive negligible contributions. This is clearly suboptimal, since one would instead like to prioritize an accurate reconstruction of the kernel in the energy regions that contribute most to the integral. Choosing a weight function $w(E)$ that reproduces the main qualitative features of the spectral density therefore helps to reduce the systematic error associated with the kernel reconstruction at fixed value of the trade-off parameter $\lambda$.

\begin{figure}[]
    \centering
    \includegraphics[width=1\columnwidth]{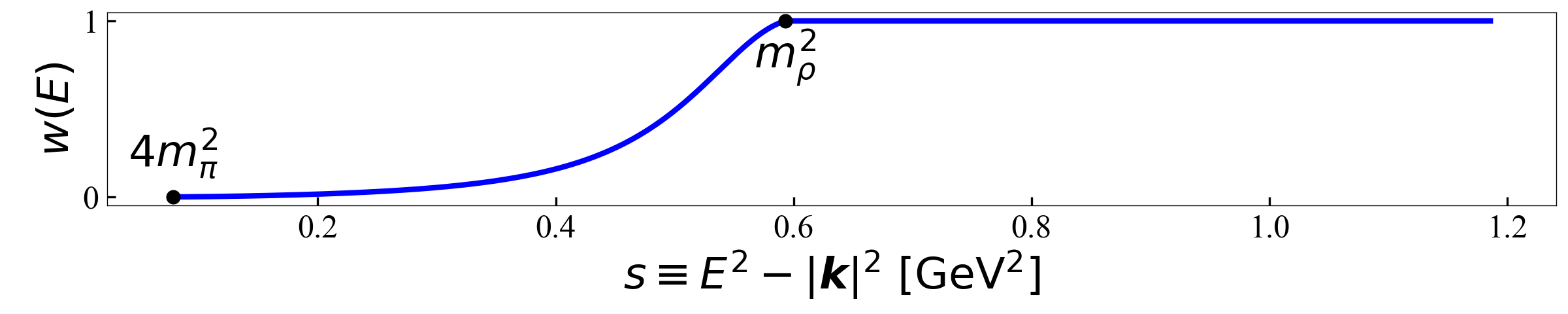}
    \caption{Weight function $w(E)$ shown as a function of $s\equiv E^2 - \vert \bs{k} \vert^2$.}
    \label{fig:4L_appwe}
\end{figure}

In the present case, we have some understanding of the behavior of the spectral density, at least in the low-energy part which is dominated by the $\pi\pi$ contribution. This motivates the use of an improved weight function, which takes into account the $\pi\pi$-threshold behavior and reproduces the resonant behavior at the $\rho$ mass. We take this function to be
\be
\label{eq:4L_appweight_func}
w(E) \equiv
\begin{cases}
\displaystyle
A \,s(E)\left(1 - \frac{4 m_\pi^2}{s(E)} \right)^{\!3/2} \frac{\Gamma_{\rho}^{2}}{(s(E)-m_{\rho}^{2})^{2} + m_{\rho}^{2}\Gamma_{\rho}^{2}},
& s(E) < m_\rho^2, \\[10pt]
1, & s(E) \geq m_\rho^2,
\end{cases}
\ee
where we have introduced $s(E) \equiv E^2 - |\bs{k}|^2$, and $m_\rho = 0.775~\mathrm{GeV}$ ($\Gamma_\rho \simeq 0.15~\mathrm{GeV}$) is the mass (decay width) of the $\rho$ resonance.
The weight function defined in Eq.~(\ref{eq:4L_appweight_func}) is shown in Fig.~\ref{fig:4L_appwe} as a function of $s$.
The first factor,
\be
\label{eq:4L_suppressionfactor}
A\,s(E)\,\left(1 - \frac{4 m_\pi^2}{s(E)} \right)^{\!3/2}\!\!, \,\,\,  {\textrm{with}}\,\,\, A = \left(1 - \frac{4 m_\pi^2}{m_\rho^2} \right)^{-3/2}\!\!,
\ee
reproduces the expected behavior of the spectral density around the two-pion threshold, and the normalization factor $A$ ensures continuity at $s(E_{\rho}) = m_\rho^2$.
The second factor,
\be
 \frac{\Gamma_{\rho}^{2}}{(s(E)-m_{\rho}^{2})^{2} + m_{\rho}^{2}\Gamma_{\rho}^{2}}~,
\ee
describes the $\rho$-resonance contribution in the relativistic Breit--Wigner approximation.
At higher energies, $s(E) \geq m_\rho^2$, we assume no prior knowledge of the spectral density and therefore assign a uniform weight, $w(E)=1$. \\

\begin{figure}
\includegraphics[scale=1.15]{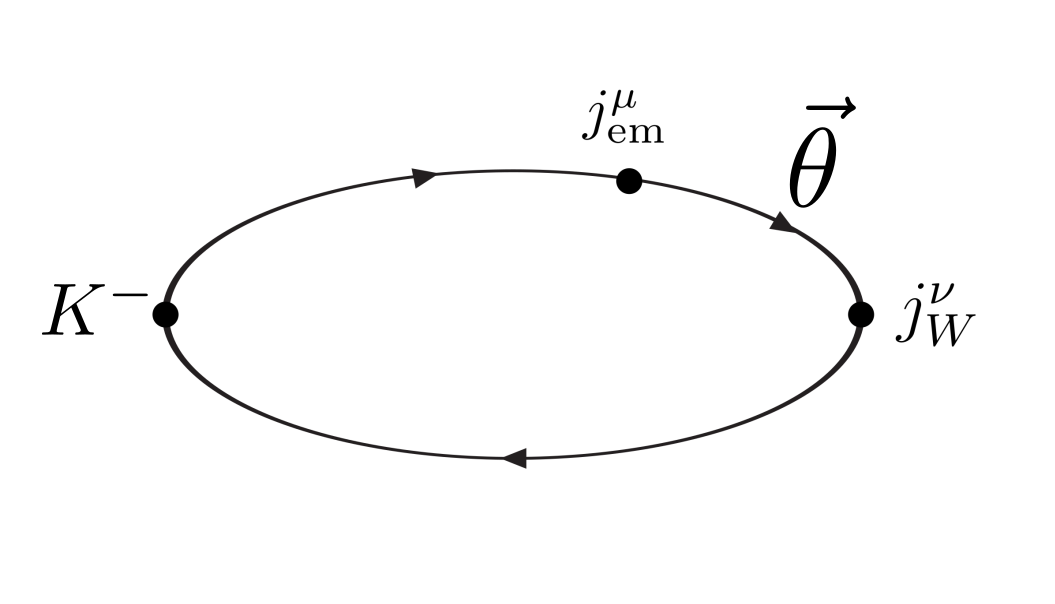}
\includegraphics[scale=1.15]{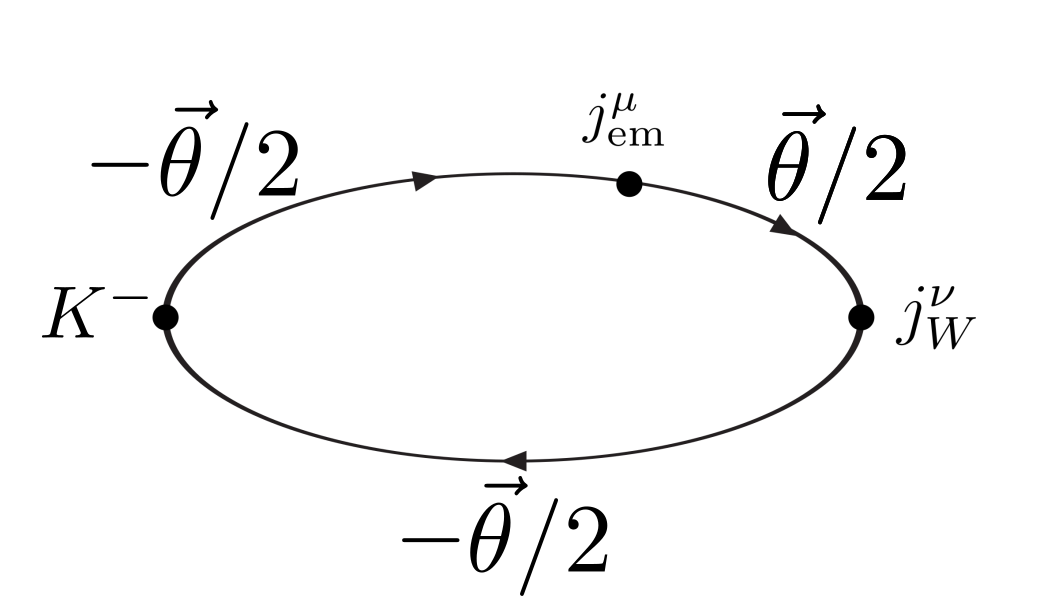}
\caption{{\textit Left:} schematic illustration of the choice of the twisted boundary conditions adopted for the calculation of the three-point function for the determination of the form factors below the $\pi\pi$ threshold. {\textit Right:} additional set of twisted boundary conditions used in the calculation of the SFR form factors. In this case the quark and anti-quark fields at the electromagnetic vertex satisfy the same spatial boundary conditions. Note that this choice of twisted boundary conditions ensures that the interpolated kaon remains at rest. In the figure $|\vec{\theta}\frac{2\pi}{L}| = |\bs{k}|$. }
\label{fig:new_theta}
\end{figure}

As for the choice of the threshold parameter $E_{\rm min}$, this quantity should be smaller than the energy value at which the spectral density on the given ensemble becomes non-zero. Clearly, lowering $E_{\rm min}$ too much into the region where the spectral density is zero deteriorates the reconstruction. The reason is that the norm functional then attempts to reconstruct the kernel in an energy region that does not contribute to the smeared form factors. In our case, in the infinite-volume limit, the spectral density becomes non-zero at the energy value $E^{*}(\bs{k})$ in Eq.~(\ref{eq:threshold_infL}). However, when working at finite volume, the multiparticle $\pi\pi$ continuum becomes discrete, and the requirement that the two pions be in a $P$ wave modifies the threshold value.

In the $\bs{k}\to \bs{0}$ limit, working at finite volume with spatial lattice extent $L$, the lowest two-pion state energy lies slightly below the non-interacting energy
$E^{*}_{L} = 2\sqrt{ m_{\pi}^{2} + (2\pi/L)^{2}}$.
The downward shift is due to the attractive interaction between the two-pion states. For non-zero $\bs{k}$, which we inject using twisted boundary conditions, isotropy is broken. As a consequence, this energy level splits into different energy levels depending on the relative orientation of the lattice quantized momentum ($\bs{n}2\pi/L$) with respect to $\bs{k}$. The values of these energy levels also depend on the specific choice of twisted boundary conditions adopted in the calculation of the three-point function $C_{W}^{\mu\nu}$.

For example, with the choice of twisted boundary conditions adopted for the calculation of the form factors below the two-pion threshold (see Fig.~\ref{fig:twisted}), where a single (valence) quark field carries the full twist angle, these (non-interacting) energy levels become
\begin{align}
\label{eq:pion_lev_tbc1}
E_{L}^{\parallel \pm}(\bs{k}) &= \sqrt{ m_{\pi}^{2} + \left(|\bs{k}| \pm \frac{2\pi}{L}\right)^{2}} + \sqrt{m_{\pi}^{2} + \left(\frac{2\pi}{L}\right)^{2} } \nonumber\\[8pt]
E_{L}^{\perp}(\bs{k}) &= \sqrt{m_{\pi}^{2} + |\bs{k}|^{2} + \left(\frac{2\pi}{L}\right)^{2}}  + \sqrt{ m_{\pi}^{2} + \left(\frac{2\pi}{L}\right)^{2}}~.
\end{align}
Moreover, with that choice of twisted boundary conditions, a lighter two-pion state close to the infinite-volume threshold, which does not carry any lattice quantized momentum, can be present. The energy of this state, neglecting final-state interactions, is simply $m_{\pi}+\sqrt{m_{\pi}^{2} + |\bs{k}|^{2}}$. The contribution of this single finite-volume state is expected to be very suppressed by phase space, and therefore expected to give a very small and probably negligible contribution to the smeared form factors. For this reason we would like to avoid lowering $E_{\rm min}$ below its energy value, as doing so would worsen the kernel reconstruction at higher energies.

For this reason, and only for the calculation of the SFR form factors, we have computed the three-point functions adopting a different choice of boundary conditions, which we refer to as the ``symmetric'' boundary conditions. This second choice is schematically illustrated in Fig.~\ref{fig:new_theta}. Since the quark and antiquark fields entering the electromagnetic-current bilinear now carry the same momentum $\bs{k}/2$, a $\pi^{+}\pi^{-}$ state carrying no lattice quantized momentum cannot be present, since it would not have any $P$-wave component.\footnote{In twisted mass QCD, due to the breaking of isospin symmetry at finite lattice spacing, we cannot rule out the possible presence of an $O(a^{2})$ contribution from a $\pi^{OS}$--$\pi^{0}$ state with energy $E_{\pi^{OS}-\pi^{0}}= \sqrt{m_{\pi^{OS}}^{2} + |\bs{k}|^{2}} + m_{\pi^{0}}$. This state would be present also with the second set of twisted-boundary conditions adopted. Since $m_{\pi}^{OS} \gg m_{\pi}$ at the lattice spacings considered in this calculation, where $m_{\pi}$ is the charged pion mass (i.e.\ the mass of the only true Goldstone boson in the massless limit at finite lattice spacing), this state would be heavier, suppressed by phase space, and correspond to a pure $O(a^{2})$ lattice artifact. For these reasons, we consider this contribution negligible.}
Moreover, with this choice of twisted-boundary conditions the analog of the energies in Eq.~(\ref{eq:pion_lev_tbc1}), which we indicate with the subscript ``(2)'', becomes
\begin{align}
E_{(2),L}^{\parallel}(\bs{k}) &=  \sqrt{ m_{\pi}^{2} + \left(\frac{|\bs{k}|}{2} + \frac{2\pi}{L}\right)^{2}} + \sqrt{ m_{\pi}^{2} + \left(\frac{|\bs{k}|}{2} - \frac{2\pi}{L}\right)^{2}}~,\nonumber\\[8pt]
E_{(2),L}^{\perp}(\bs{k}) &=  2\sqrt{ m_{\pi}^{2}+ \left(\frac{|\bs{k}|}{2}\right)^{2} + \left(\frac{2\pi}{L}\right)^{2}}~.
\end{align}
For the calculation of $\mathcal{D}_{F}^{I}$ and $\mathcal{R}_{F}^{R}$ only, we employ the ``symmetric'' set of boundary conditions. The results shown below correspond to this choice, which will be used to quote our final results. We set $E_{\rm min}=0.98\,E_{(2),L}^{\parallel}(\bs{k})$, where the $2\%$ decrease with respect to the non-interacting energy value is expected to account for the effects of interactions, which for the first $\pi^{+}\pi^{-}$ state are known to be very small. However,  we will also perform the calculation employing the ``asymmetric'' twisted boundary conditions in Fig.~\ref{fig:new_theta}, and compare the two results at the end of the section. The discussion above concerns the quark-connected contribution, for which twisted boundary conditions have been employed in the lattice calculation. The quark-disconnected contribution, for which we will only provide a semi-quantitative estimate above threshold, will be discussed at the end of Sec.~\ref{sec:eps_extr}.

\begin{figure}[]
    \centering
    \includegraphics[width=1.\columnwidth]{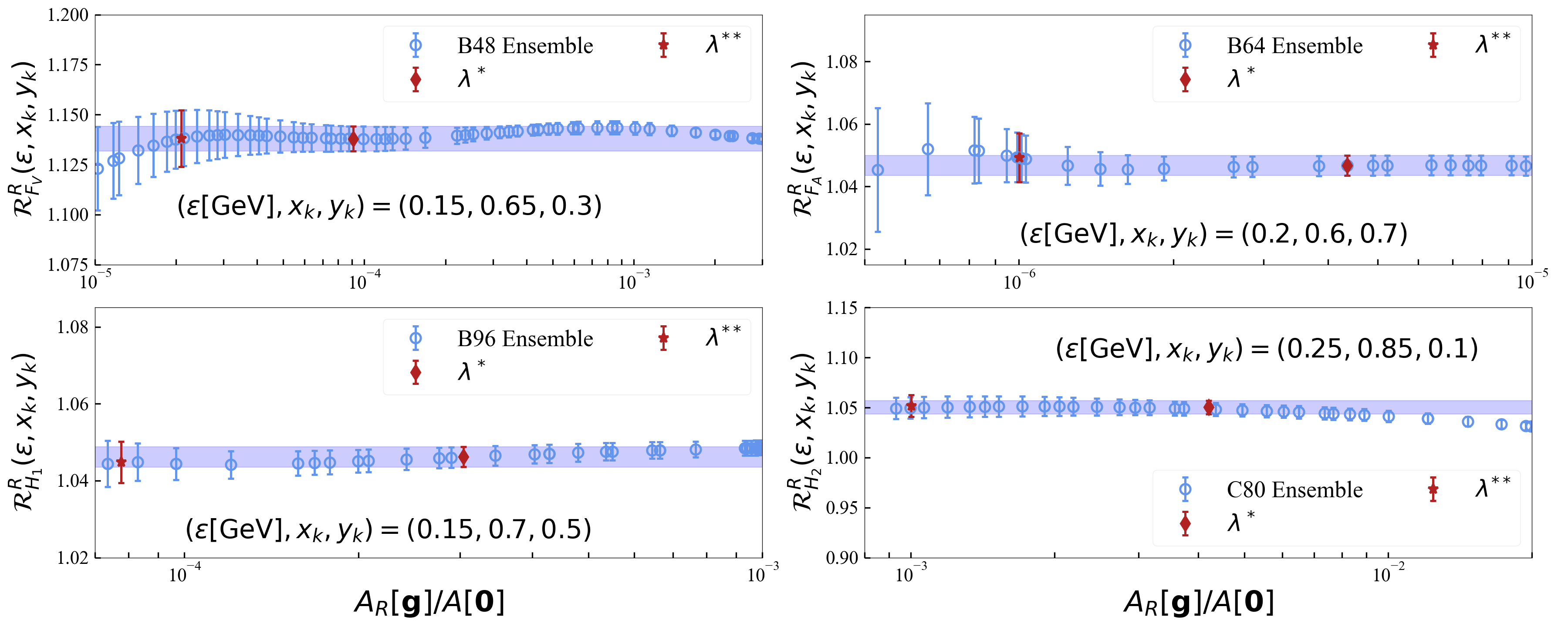}
    \caption{Stability analysis for the quark-connected contribution to $\mathcal{R}^{R}_{F}(\varepsilon, x_k, y_k)$. In each panel, the data points (blue circles) are shown as functions of the squared systematic error associated with the kernel reconstruction, $A_R[\bs{g}]/A_R[\bs{0}]$, on a logarithmic scale. The blue bands indicate our final determinations of the smeared ratios.}
    \label{fig:4L_Restab}
\end{figure}

\begin{figure}[]
    \centering
    \includegraphics[width=1.\columnwidth]{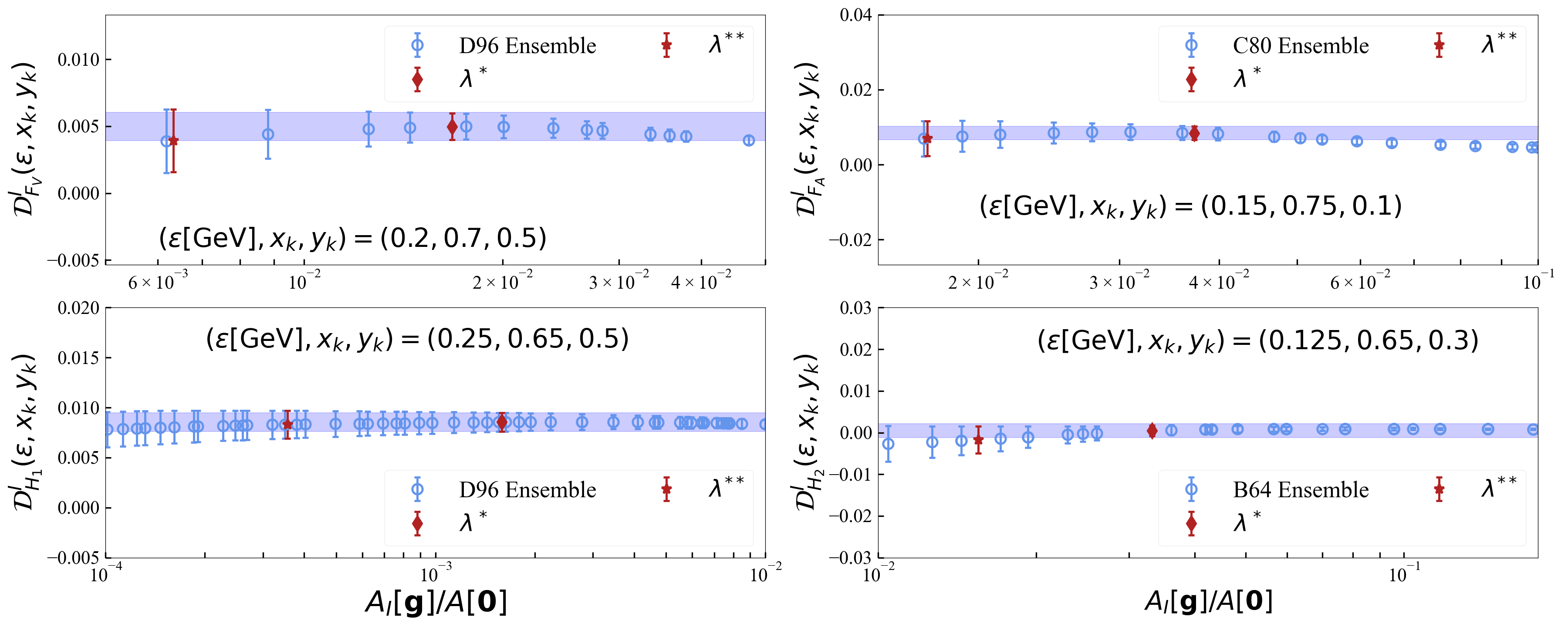}
    \caption{Same as Fig.~\ref{fig:4L_Restab} for $\mathcal{D}_{F}^{I}$.}
    \label{fig:4L_Imstab}
\end{figure}

Having described the details of our HLT implementation, in Figs.~\ref{fig:4L_Restab}--\ref{fig:4L_Imstab} we show illustrative stability-analysis plots for the quark-connected contribution to $\mathcal{D}_{F}^{I}$ and $\mathcal{R}_{F}^{R}$ for selected values of $x_{k}$ and $y_{k}$ and of the smearing parameter $\varepsilon$.

In each panel we display the reconstructed quantities as functions of the trade-off parameter $\lambda$. However, instead of plotting $\lambda$ on the x-axis, we show the value of $A_{R(I)}[\bs{g}]/A_{R(I)}[\bs{0}]$ obtained at that $\lambda$ in the reconstruction of the larger of the two $x_{k}$ values entering $\mathcal{D}_{F}^{I}$ and $\mathcal{R}_{F}^{R}$. Note that $A_{R(I)}[\bs{g}]/A_{R(I)}[\bs{0}]$ provides a measure of the (squared) systematic error due to the imperfect kernel reconstruction. In all cases, a broad interval in which the reconstructed values remain stable over at least one order of magnitude in $A_{R(I)}[\bs{g}]/A_{R(I)}[\bs{0}]$ is observed\,\footnote{In most cases the stability region extends over several orders of magnitude in $A_{R(I)}[\bs{g}]/A_{R(I)}[\bs{0}]$.}.
Within this interval, where the statistical error exceeds the systematic one because the results are stable, we determine our value of $\lambda^{*}$ (the rightmost red diamond). To quantify the systematic error we further select, in each stability plot, a second value of $\lambda$, denoted $\lambda^{**}$, corresponding to an even better kernel reconstruction. The value of $\lambda^{**}$ is determined as
\begin{equation}
\label{eq:lambdastarstar}
\frac{B^{F}_{R(I)}[\boldsymbol{g}]}{A_{R(I)}[\boldsymbol{g}]}\bigg|_{\lambda=\lambda^{\star\star}}
 = \kappa\,
   \frac{B^{F}_{R(I)}[\boldsymbol{g}]}{A_{R(I)}[\boldsymbol{g}]}\bigg|_{\lambda=\lambda^{\star}},
\qquad \kappa = 10, \quad F = \{F_{V}, A^i\}~,
\end{equation}
and it is indicated by the leftmost red star in the figures. Any statistically significant difference between the results obtained at $\lambda^\star$ and $\lambda^{\star\star}$ is included as a systematic error in the final determination, which is shown in the figures as blue bands.

\begin{figure}[]
    \centering
    \includegraphics[width=1\columnwidth]{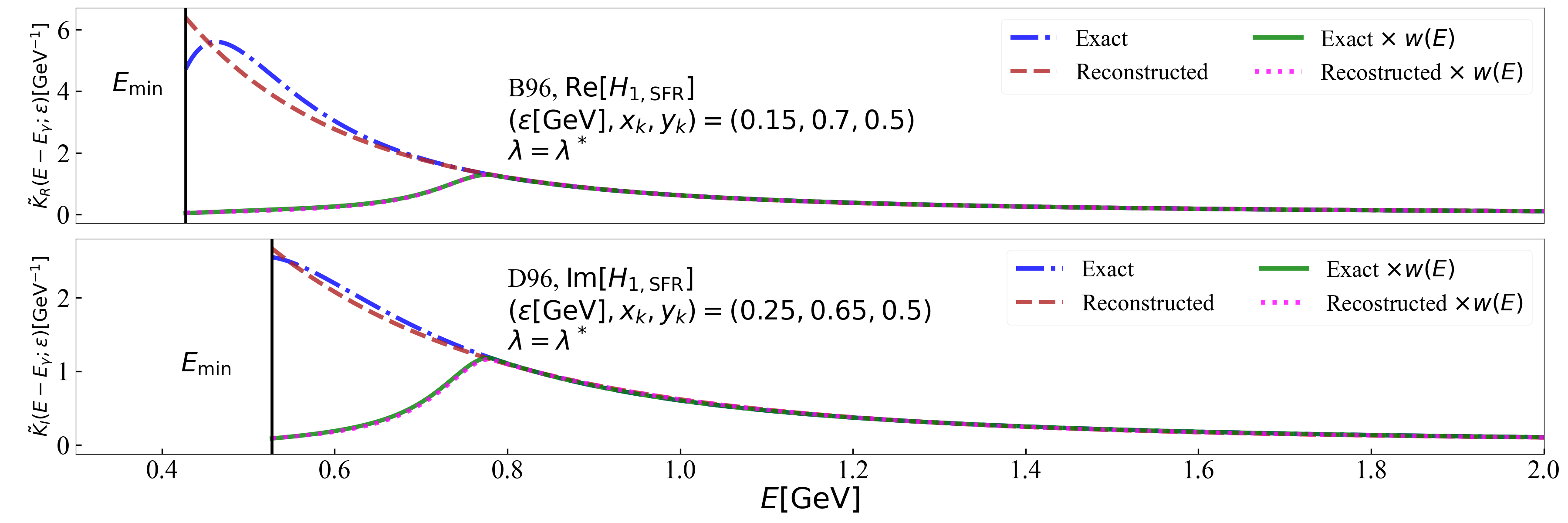}
    \caption{Real (top panel) and imaginary (bottom panel) parts of the improved kernel $\tilde{K}(E - E_\gamma; \varepsilon)$ used for the HLT reconstruction as function of the energy $E$ for $\lambda=\lambda^{*}$. The blue dash-dotted line (red dashed) curves correspond to the exact (reconstructed) kernel. The green solid and magenta dotted curves represent the same quantities multiplied by the weight function $w(E)$. Vertical black lines represent the threshold parameter $E_{\mathrm{min}}$. }
    \label{fig:4L_Kernel}
\end{figure}
Finally, in Fig.~\ref{fig:4L_Kernel} we compare the exact kernel function $\tilde{K}$ with its reconstruction obtained through the HLT method using a finite number of exponentials. The reconstruction shown corresponds to the value $\lambda^{*}$ for the $H_{1}$ form factor, at the same kinematic point and value of $\varepsilon$ considered in Figs.~\ref{fig:4L_Restab}--\ref{fig:4L_Imstab}.
The blue dash-dotted curve represents the exact kernel, while the red dashed curve shows the reconstructed one. We also display the product of the weight function $w(E)$ with the exact kernel (green solid curve) and with the reconstructed kernel (magenta dotted curve).

The figure shows that, for $\lambda=\lambda^*$, the reconstructed kernel weighted with $w(E)$ provides an accurate approximation of the exact counterpart over the full range of energies $E$. It also illustrates the effect of the weight function $w(E)$. The quality of the reconstruction is higher for large values of $E$ than the region around the phase-space threshold, $E \simeq E_{\mathrm{min}}$. This is precisely the desired behavior: to be less stringent where the spectral density is suppressed ($E \simeq E_{\mathrm{min}}$), while accurately reconstructing the kernel in the regions that give the dominant contribution to the integral (large $E$).

\subsection{Infinite-volume and continuum-limit extrapolation of $\mathcal{D}_{F}^{I}(\varepsilon,x_{k},y_{k})$ and $\mathcal{R}_{F}^{R}(\varepsilon,x_{k},y_{k})$}

Using the HLT implementation described in the previous subsection, we determine the quark-connected contribution to $\mathcal{D}_{F}^{I}(\varepsilon,x_{k},y_{k})$ and $\mathcal{R}_{F}^{R}(\varepsilon,x_{k},y_{k})$ on all gauge ensembles.
As discussed above, three different extrapolations are required in order to obtain the physical quantities $\mathcal{D}_{F}^{I}(x_{k},y_{k})$ and $\mathcal{R}_{F}^{R}(x_{k},y_{k})$: the infinite-volume, continuum-limit, and vanishing-$\varepsilon$ extrapolations. While the order of the latter two limits can be interchanged, the infinite-volume extrapolation must be performed before taking the $\varepsilon\to 0^{+}$ limit. Only when the limits are taken in this order does one recover the physical infinite-volume spectral density and, consequently, physical results for the form factors.
In this subsection we discuss the infinite-volume and continuum-limit extrapolations, both performed at fixed value of $\varepsilon$. The extrapolation $\varepsilon\to 0^{+}$ will be addressed in the next subsection.

The strategy closely follows that adopted for the extrapolation of the form factors below the $\pi\pi$ threshold (see Sec.~\ref{sec:4L_fse} and Sec.~\ref{Sec:4L_Continuum}). We begin with the infinite-volume extrapolation. We consider results at fixed $\varepsilon$ obtained on the B48, B64, and B96 ensembles, which differ only in the spatial extent $L$ of the lattice. We denote by $\mathcal{D}_{F}^{I}(\varepsilon,x_{k},y_{k},L)$ and $\mathcal{R}_{F}^{R}(\varepsilon,x_{k},y_{k},L)$ the corresponding quantities evaluated on a lattice with spatial extent $L$.

In Ref.~\cite{Bulava:2021fre} it has been shown that finite-size effects on smeared quantities (i.e. at non-zero $\varepsilon$) are exponentially suppressed. Motivated by these considerations, we extrapolate the results obtained on the three B ensembles, at fixed $\varepsilon$, $x_{k}$, and $y_{k}$, to the infinite-volume limit, adopting the ansatz
\begin{align}
\label{eq:4L_appVfitansatzreim}
\mathcal{R}^R_{F}(\varepsilon, x_k, y_k, L)
 &= \mathcal{A}_{F}^{R}(\varepsilon, x_k, y_k)
    + \mathcal{B}_{F}^R(\varepsilon, x_k, y_k)\, e^{-L\delta},
\\[8pt]
\mathcal{D}^{I}_{F}(\varepsilon, x_k, y_k, L)
 &= \mathcal{A}_{F}^{I}(\varepsilon, x_k, y_k)
    + \mathcal{B}_{F}^I(\varepsilon, x_k, y_k)\, e^{-L\delta},
    \qquad F = \{F_{V}, A^i\}~,
\end{align}
where $\mathcal{A}_{F}^{R(I)}$ and $\mathcal{B}_{F}^{R(I)}$ are free fit parameters. Separate fits are performed for each $F=\{F_{V},A^{i}\}$ and for each simulated value of $x_{k}$, $y_{k}$, and $\varepsilon$. The fit ansatz is analogous to that of Eq.~(\ref{eq:4L_FSE_ansatz}) when setting $\delta = m_{\pi}$. In the present case, however, the exponential suppression of finite-size effects may be governed either by the lowest hadronic scale in the problem, $m_{\pi}$, or by the smearing parameter $\varepsilon$, which constitutes an additional infrared scale. For this reason we perform two independent infinite-volume extrapolations adopting either $\delta=m_{\pi}$ or $\delta=\varepsilon$. In the explored range of $\varepsilon$, both ansätze provide a good description of the data and yield compatible infinite-volume limits.

\begin{figure}[]
    \centering
    \includegraphics[width=1.\columnwidth]{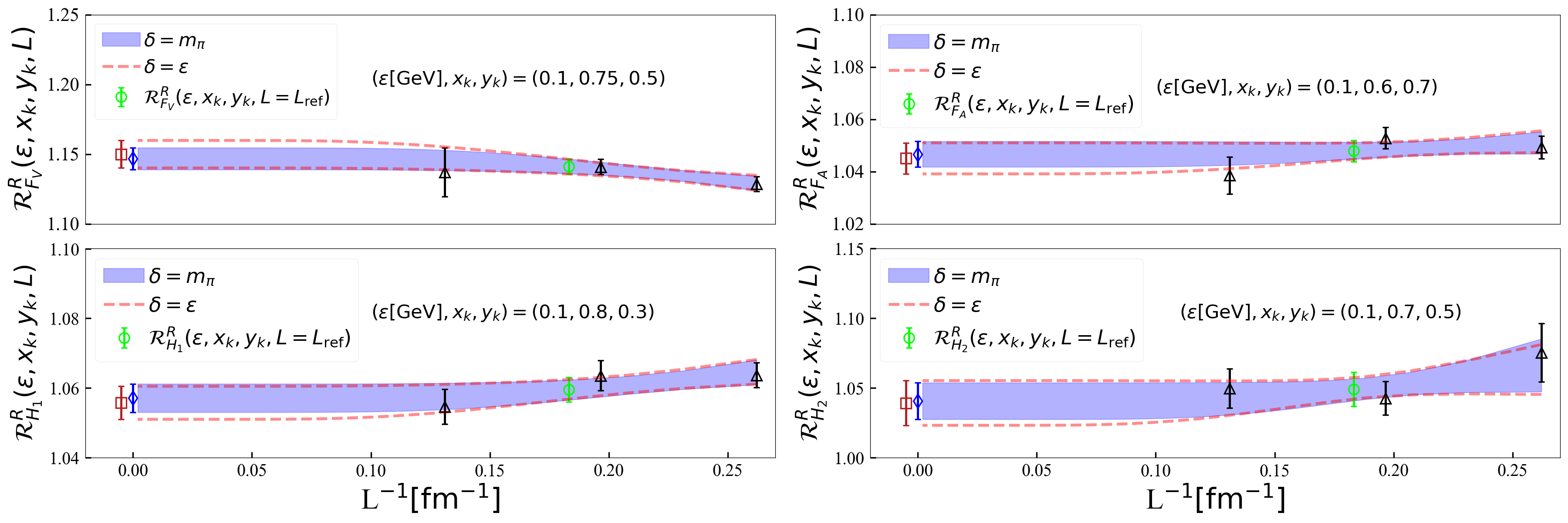}
    \caption{Volume dependence of $\mathcal{R}_{F}^{R}(\varepsilon,x_{k},y_{k})$, $F=\{F_{V},A^{i}\}$, for selected values of $x_{k}$ and $y_{k}$ at $\varepsilon=100~{\rm MeV}$ (black triangles). Blue bands and the area between red dashed lines correspond to fits with $\delta = m_\pi$ and $\delta = \varepsilon$, respectively. Green circles indicate the interpolation to $L = L_{\mathrm{ref}}$, while blue diamonds and red squares correspond to the infinite-volume extrapolation at $a \simeq 0.079~\mathrm{fm}$ obtained with $\delta=m_{\pi}$ and $\delta=\varepsilon$, respectively.}
    \label{fig:4L_Above_Re_Vfitepsp1}
\end{figure}

\begin{figure}[]
    \centering
\includegraphics[width=1.\columnwidth]{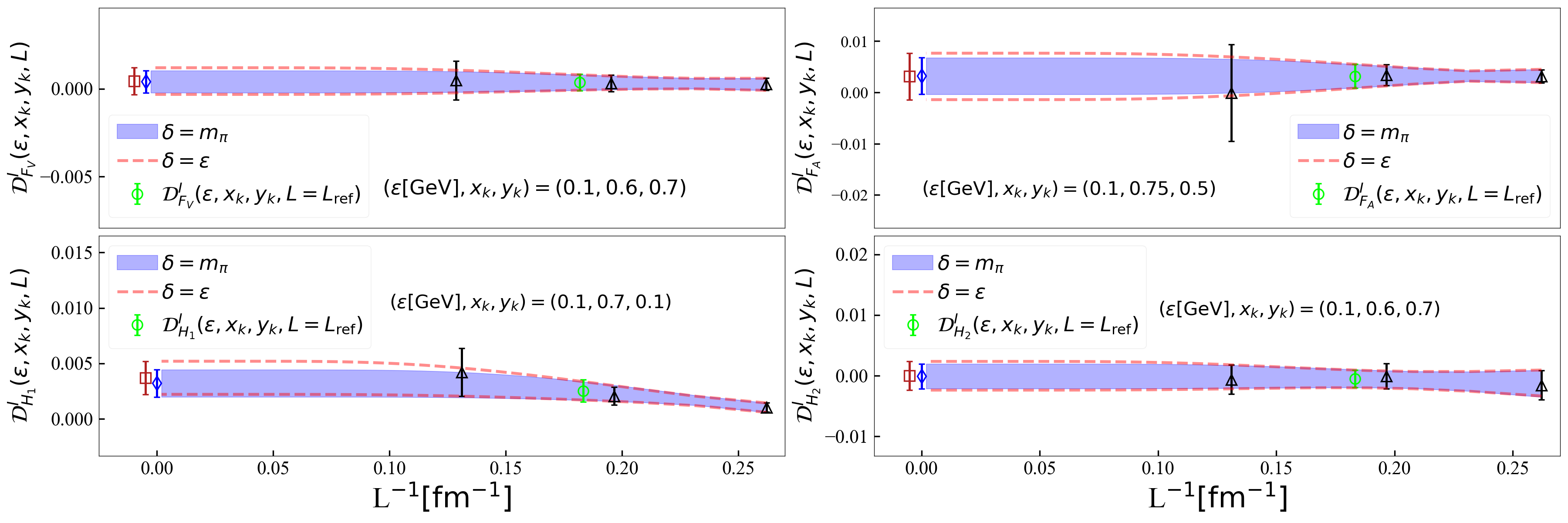}
    \caption{Volume dependence of $\mathcal{D}_{F}^{I}(\varepsilon,x_{k},y_{k})$, $F=\{F_{V},A^{i}\}$, for selected values of $x_{k}$ and $y_{k}$ at $\varepsilon=100~{\rm MeV}$ (black triangles). Blue bands and the area between red dashed lines correspond to fits with $\delta = m_\pi$ and $\delta = \varepsilon$, respectively. Green circles indicate the interpolation to $L = L_{\mathrm{ref}}$, while blue diamonds and red squares correspond to the infinite-volume extrapolation at $a \simeq 0.079~\mathrm{fm}$ obtained with $\delta=m_{\pi}$ and $\delta=\varepsilon$, respectively.}
    \label{fig:4L_Above_Im_Vfitepsp1}
\end{figure}

Figs.~\ref{fig:4L_Above_Re_Vfitepsp1}-~\ref{fig:4L_Above_Im_Vfitepsp1} show representative infinite-volume fits for example values of $x_{k}$ and $y_{k}$ at the smallest smearing parameter considered, namely $\varepsilon=100~{\rm MeV}$, where finite-volume effects might be more significant. As the figures show, the magnitude of finite-size effects is reasonably small, and the two ansätze ($\delta=m_{\pi}$ and $\delta=\varepsilon$) yield essentially identical results in the infinite-volume limit. Similar behaviors are observed for all kinematic points and values of $\varepsilon$, which are not shown explicitly.

Our final result is obtained by averaging the results obtained with the two fit ansätze. Even though the two extrapolations give very similar results, we include the difference between them as an additional systematic uncertainty.

\begin{figure}[]
    \centering
    \includegraphics[width=1.\columnwidth]{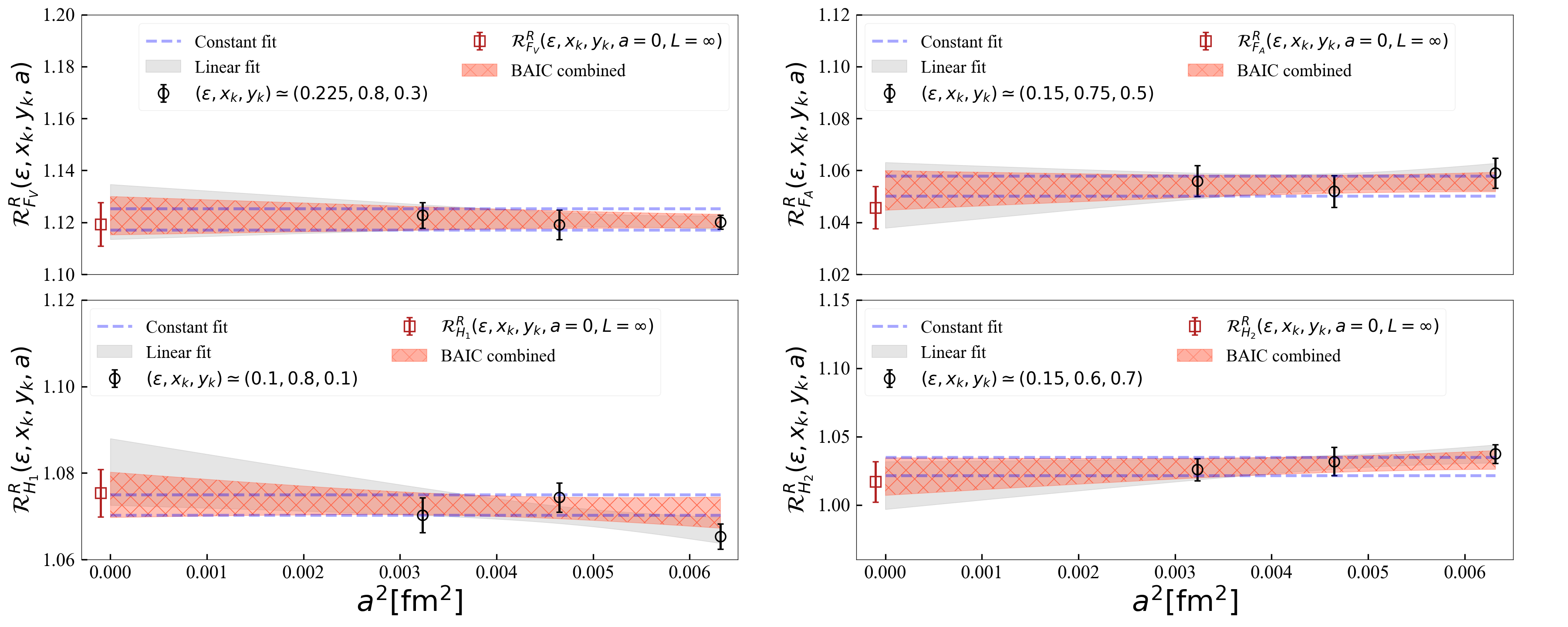}
    \caption{Ratios $\mathcal{R}_{F}^{R}(\varepsilon,x_{k},y_{k})$ for selected values of $x_{k}$, $y_{k}$ and $\varepsilon$ as a function of $a^2$ at $L=L_{\rm ref}$ (black circles). The gray band shows the result of a linear fit to the data at the three lattice spacings; the area between the blue dashed lines corresponds to a constant fit to the two finest lattice spacings, while the meshed red band represents the BAIC average of the two fits. The red squares denote the continuum and infinite-volume limit after applying the finite-volume correction.}
    \label{fig:4L_Above_Re_afit}
\end{figure}

\begin{figure}[]
    \centering
    \includegraphics[width=1.\columnwidth]{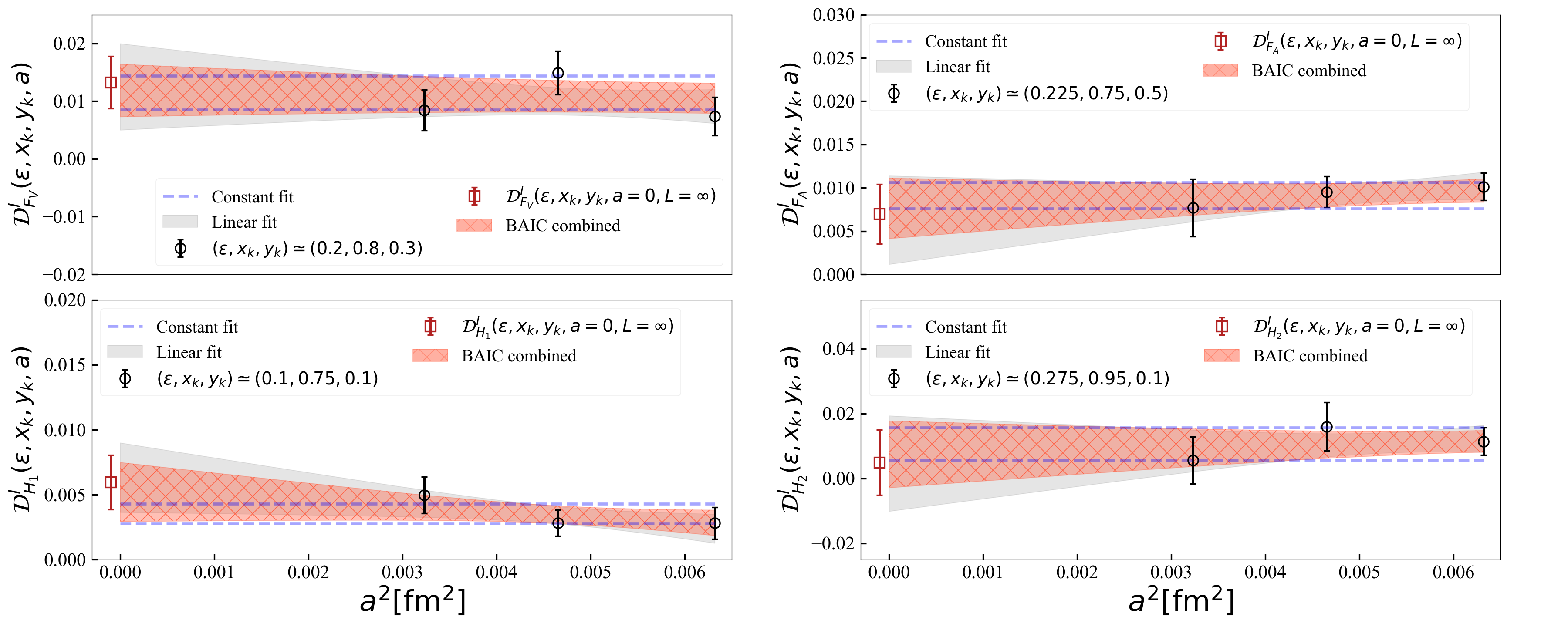}
    \caption{Differences $\mathcal{D}_{F}^{I}(\varepsilon,x_{k},y_{k})$ for selected values of $x_{k}$, $y_{k}$ and $\varepsilon$ as a function of $a^2$ at $L=L_{\rm ref}$ (black circles). The gray band shows the result of a linear fit to the data at the three lattice spacings; the area between the blue dashed lines corresponds to a constant fit to the two finest lattice spacings, while the meshed red band represents the BAIC average of the two fits. The red squares denote the continuum and infinite-volume limit after applying the finite-volume correction.}
    \label{fig:4L_Above_Im_afit}
\end{figure}

In analogy with Sec.~\ref{sec:4L_fse} and Sec.~\ref{Sec:4L_Continuum}, we then perform the continuum-limit extrapolation at fixed $\varepsilon$ and spatial volume $L_{\rm ref}^{3}$, with $L_{\rm ref}\simeq 5.46~{\rm fm}$ corresponding to the spatial volume of the C80 and D96 ensembles. For the B-type ensembles we use the information from the infinite-volume fits to interpolate the results obtained on the B48, B64, and B96 ensembles to $L_{\rm ref}$ (green circles in Figs.~\ref{fig:4L_Above_Re_Vfitepsp1}--\ref{fig:4L_Above_Im_Vfitepsp1}). After performing the continuum-limit extrapolation at $L=L_{\rm ref}$, we add the finite-volume correction required to extrapolate to $L\to \infty$, as done for the form factors below threshold.

The continuum-limit extrapolation of $\mathcal{D}_{F}^{I}(\varepsilon,x_{k},y_{k})$ and $\mathcal{R}_{F}^{R}(\varepsilon,x_{k},y_{k})$ is performed in the same way as for the form factors below threshold. Separate continuum extrapolations are carried out for each $x_{k}$, $y_{k}$, and $\varepsilon$, and for each $F=\{F_{V},A^{i}\}$. We perform both a linear fit in $a^{2}$ using data at the three lattice spacings and a constant fit using only the two finest lattice spacings. The two fits are then combined using the BAIC. As in the case of the form factors below the $\pi\pi$ threshold, both $\mathcal{D}_{F}^{I}(\varepsilon,x_{k},y_{k})$ and $\mathcal{R}_{F}^{R}(\varepsilon,x_{k},y_{k})$ display very mild cutoff effects, with reduced $\chi^{2}$ values close to or below unity for all fits performed. Representative continuum extrapolations are shown in Figs.~\ref{fig:4L_Above_Re_afit} and~\ref{fig:4L_Above_Im_afit}.

\subsection{Vanishing-$\varepsilon$ extrapolation and final results for  $F_{V,{\rm SFR}}(x_{k},y_{k})$ and $A^{i}_{\rm SFR}(x_{k},y_{k})$}
\label{sec:eps_extr}

Having obtained the quark-connected contribution to the smeared quantities
$\mathcal{R}^{R}_{F}(\varepsilon, x_k, y_k)$ and
$\mathcal{D}^{I}_{F}(\varepsilon, x_k, y_k)$ in the infinite-volume and
continuum limit, the final step consists in extrapolating the results to
$\varepsilon \to 0^{+}$. The available data span the range
$\varepsilon \in [100,275]~{\rm MeV}$. Owing to the improved kernel of
Eq.~(\ref{eq:ImprovedKernel}), finite-$\varepsilon$ corrections start at
$O(\varepsilon^2)$.

As discussed in Sec.~\ref{sec:4L_model}, the effective model indicates that
the simulated values of $\varepsilon$ already lie in the regime where the
$\varepsilon$-dependence is well described by a low-order polynomial,
with higher-order $O(\varepsilon^{3})$ effects remaining small within the explored range.
Motivated by these considerations, we adopt the polynomial ansätze
\begin{align}
\label{eq:4L_FVsfrPOLYimp_new}
\mathcal{R}^{R}_{F}(\varepsilon, x_k, y_k)
  &= \mathcal{R}^{R}_{F}(x_k, y_k)
     + \varepsilon^{2} Q_{F}^R
     + \varepsilon^{3} N_{F}^R,
     \qquad F=\{F_{V},A^{i}\},\\[6pt]
\label{eq:4L_AisfrPOLYimp_new}
\mathcal{D}^{I}_{F}(\varepsilon, x_k, y_k)
  &= \mathcal{D}^{I}_{F}(x_k, y_k)
     + \varepsilon^{2} Q^I_{F}
     + \varepsilon^{3} N^I_{F},
     \qquad F=\{F_{V},A^{i}\}.
\end{align}

To assess the stability of the extrapolation and estimate systematic
uncertainties, three different fits are performed:\footnote{ 
The $\varepsilon \to 0$ extrapolations of the real (imaginary) parts of the form factors for $x_k \geq 0.9$ ($x_k \geq 0.8$) are carried out excluding the two (three) smallest values of $\varepsilon$, since the corresponding HLT reconstructions become noisier. Consequently, for these large values of the photon virtuality, the constant fit is not performed. }
\begin{enumerate}
\item[(i)] a cubic fit in $\varepsilon$ (including terms up to
$\varepsilon^{3}$) using the full set of simulated values;
\item[(ii)] a quadratic fit, obtained by setting the cubic terms to zero,
restricted to the four smallest values of $\varepsilon$;
\item[(iii)] a constant fit using only the two smallest values of
$\varepsilon$.
\end{enumerate}

All fits are performed jackknife-by-jackknife, ensuring a correct
propagation of statistical correlations among results obtained at values of different $\varepsilon$. The final result is obtained as the average of the three fits. We do not employ information-criterion weighting (such as BAIC). Although correlations among data at different $\varepsilon$ are
properly propagated through the jackknife analysis, the covariance matrix
is not included explicitly in the definition of the fit $\chi^{2}$, since we find that it is poorly conditioned.
Consequently, the absolute normalization of $\chi^{2}$ is not meaningful,
and criteria relying directly on $\chi^{2}$ values are avoided.

The systematic uncertainty associated with the extrapolation is estimated
from the spread of the three fits used to define the average. Let
$X^{(i)}$, with $i=1,2,3$, denote the values extrapolated to $\epsilon=0^{+}$ obtained from
the three fit ansätze. We define
\begin{align}
\Delta_{\varepsilon}
  = \max_{i,j}\, \bigl|X^{(i)}-X^{(j)}\bigr|,
\end{align}
i.e.\ the largest difference between any pair of extrapolated central
values. This difference is weighted by
\begin{align}
\erf\!\left(
\frac{\Delta_{\varepsilon}}
{\sqrt{2(\sigma_i^{2}+\sigma_j^{2})}}
\right),
\end{align}
where $\sigma_i$ and $\sigma_j$ are the statistical uncertainties of the
two fits giving the largest difference, and assigned as a systematic error.

\begin{figure}[]
    \centering
    \includegraphics[width=1.\columnwidth]{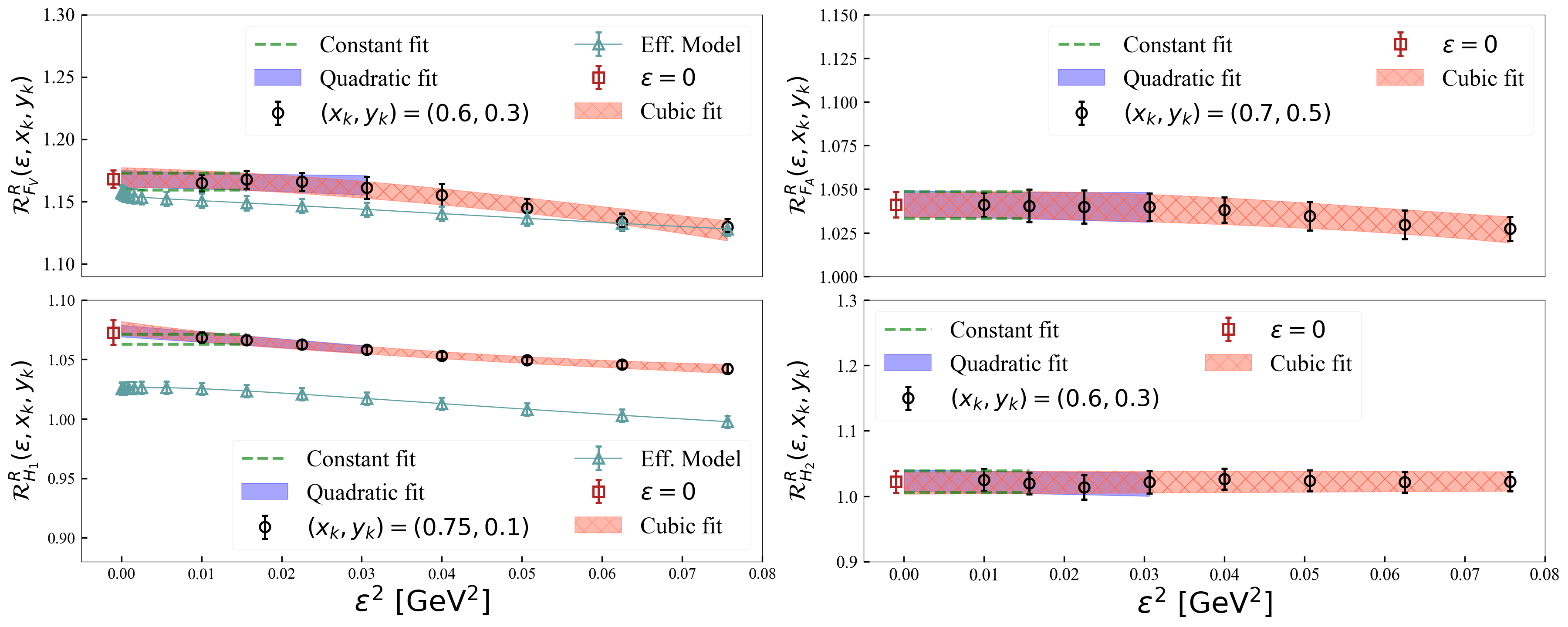}
    \caption{Vanishing-$\varepsilon$ extrapolation of
    $\mathcal{R}_{F}^{R}(\varepsilon,x_k,y_k)$.
    Black circles: lattice data as functions of $\varepsilon^{2}$.
    Light-blue triangles: effective-model prediction for $F=F_V$
    and $F=H_1$.
    Red band: cubic fit to the full $\varepsilon$ range.
    Blue band: quadratic fit to the four smallest $\varepsilon$.
    Area between green dashed lines: constant fit to the two smallest $\varepsilon$.
    Red square: final result at $\varepsilon=0$, including the
    systematic uncertainty described in the text.}
    \label{fig:4L_FFfinalSFR}
\end{figure}

\begin{figure}[]
    \centering
    \includegraphics[width=1.\columnwidth]{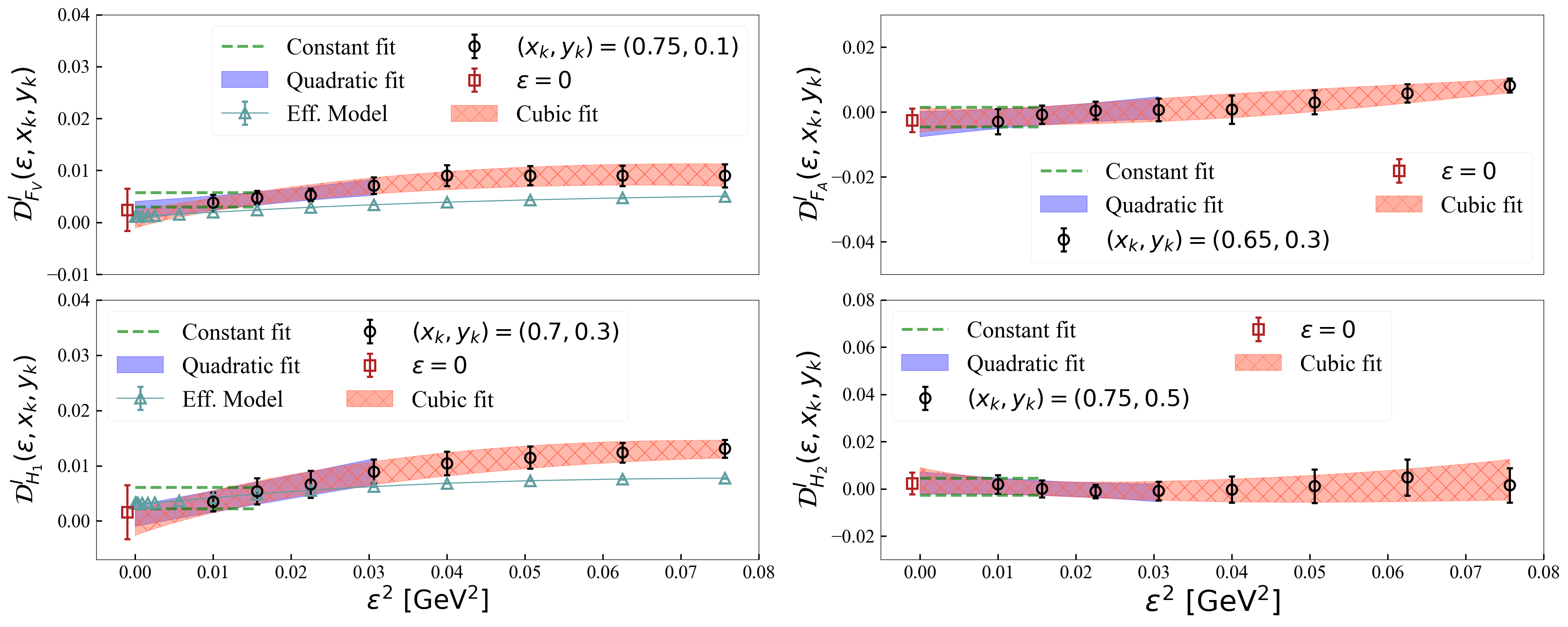}
    \caption{Same as Fig.~\ref{fig:4L_FFfinalSFR}, but for
    $\mathcal{D}_{F}^{I}(\varepsilon,x_k,y_k)$.}
    \label{fig:4L_Imepsextr}
\end{figure}

Fig.~\ref{fig:4L_FFfinalSFR} shows representative extrapolations of
$\mathcal{R}_{F}^{R}(\varepsilon,x_k,y_k)$. In the region where lattice
data are available, the dependence on $\varepsilon$ is extremely mild and
qualitatively consistent with the behaviour observed in the effective
model for $F_V$ and $H_1$. While the normalization differ—as expected,
since the model includes only the two-pion contribution—the similarity of
the $\varepsilon$ dependence provides additional support for the adopted
extrapolation strategy.

An identical analysis is performed for
$\mathcal{D}_{F}^{I}(\varepsilon,x_k,y_k)$, shown in
Fig.~\ref{fig:4L_Imepsextr}. Also in this case the weak $\varepsilon$
dependence leads to stable and controlled extrapolations to the
$\varepsilon\to0$ limit.

\begin{figure}[]
    \centering
    \includegraphics[width=1.\columnwidth]{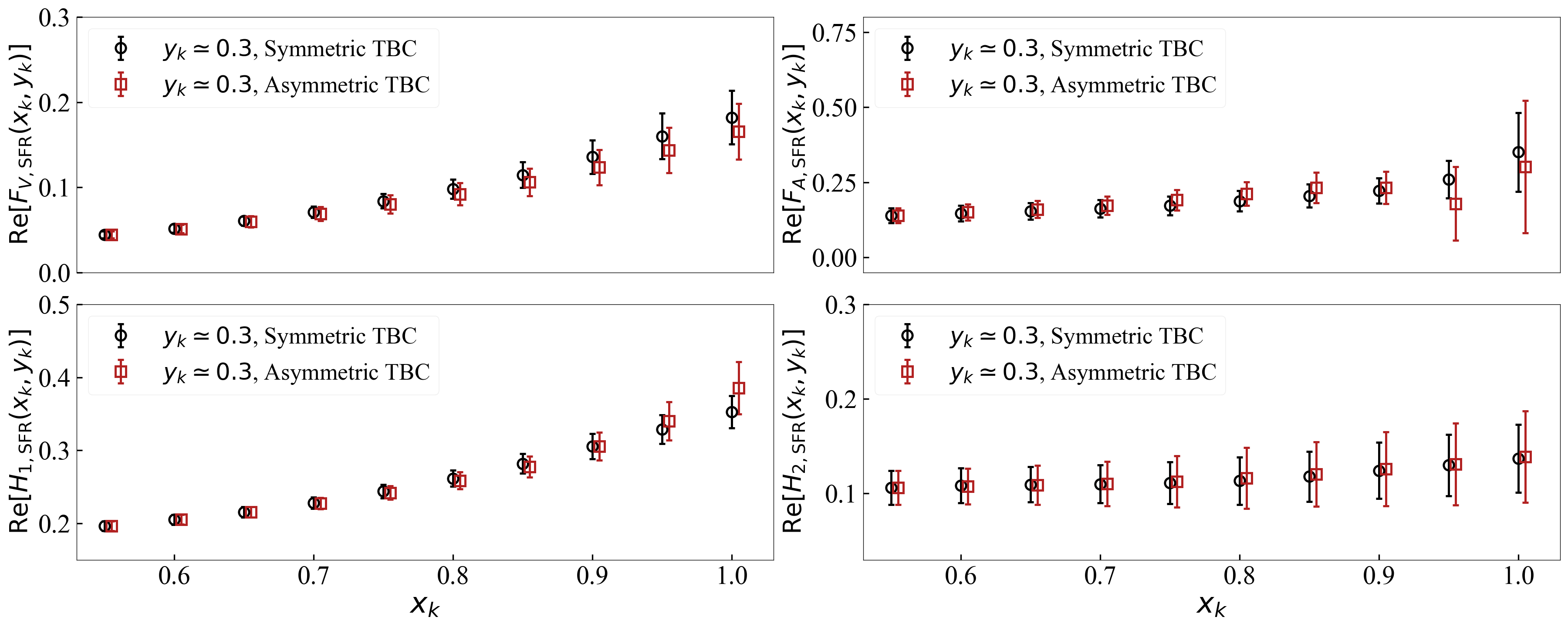}
    \caption{Our final results for the real part of the SFR form factors $F_{V}(x_{k},y_{k})$ and $A^{i}(x_{k},y_{k})$ above the two-pion threshold, for $y_{k}=0.3$ (black circles). For comparison, we show in the figure the corresponding results obtained using the asymmetric boundary conditions in which a single valence quark satisfies non-periodic spatial boundary conditions (red squares). }
    \label{fig:comp_RE_SFR}
\end{figure}

\begin{figure}[]
    \centering
    \includegraphics[width=1.\columnwidth]{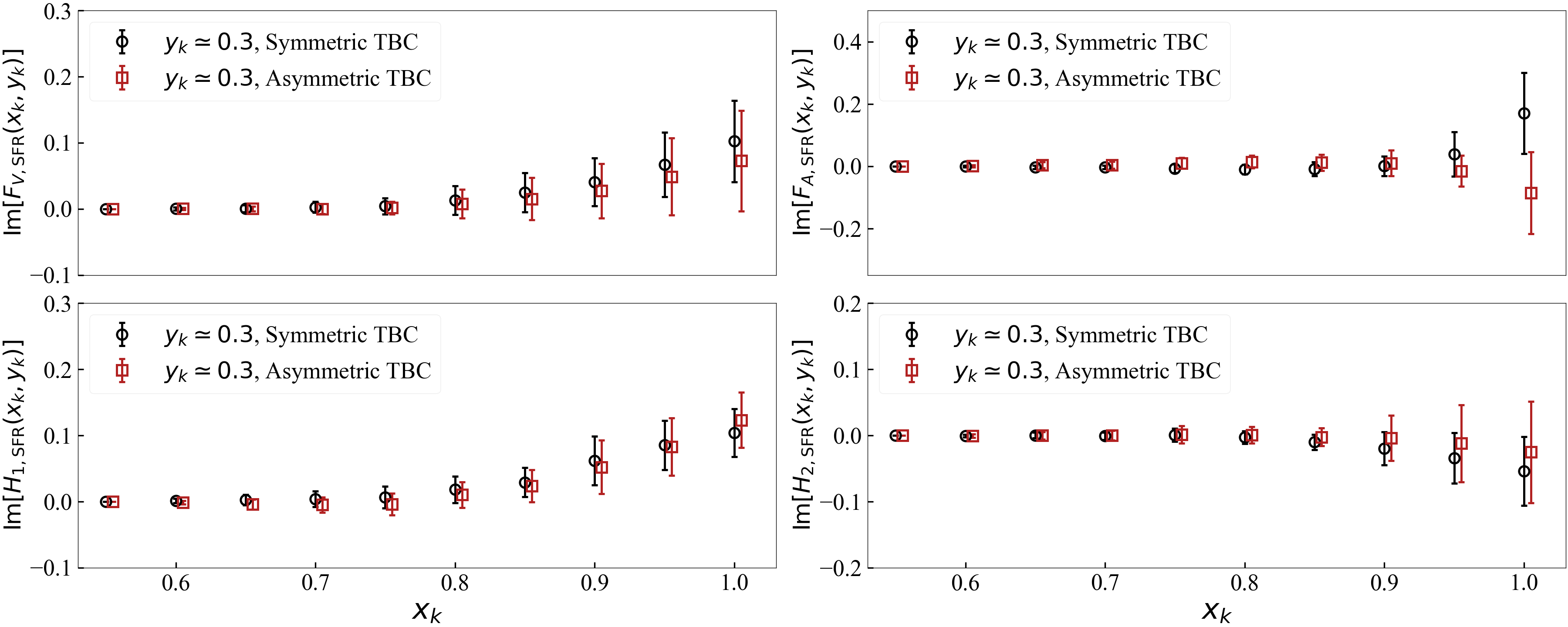}
    \caption{Same as in Fig.~\ref{fig:comp_RE_SFR} for the imaginary part of the SFR form factors $F_{V}(x_{k},y_{k})$ and $A^{i}(x_{k},y_{k})$. }
    \label{fig:comp_IM_SFR}
\end{figure}

Having determined the quark-connected contributions to the ratios
$\mathcal{R}_{F}^{R}(x_{k},y_{k})$ and differences
$\mathcal{D}_{F}^{I}(x_{k},y_{k})$, we reconstruct the corresponding
quark-connected contributions to the real and imaginary parts of the SFR
form factors, $F_{V,{\rm SFR}}(x_{k},y_{k})$ and
$A^{i}_{\rm SFR}(x_{k},y_{k})$, using
Eqs.~(\ref{eq:im_RDP})--(\ref{eq:re_RDP}). The results are shown as black circles in
Fig.~\ref{fig:comp_RE_SFR} for the real part and in
Fig.~\ref{fig:comp_IM_SFR} for the imaginary part, at fixed
$y_{k}=0.3$.

In the figures we also include (red square points) the results obtained
using asymmetric twisted boundary conditions, in which a single valence
quark satisfies non-periodic spatial boundary conditions. As the figures
show, the results obtained with the two choices of twisted boundary
conditions are in excellent agreement\footnote{The correlation functions
computed with the two sets of twisted boundary conditions have been
evaluated on the same gauge configurations but using independent sets of
stochastic sources. Since the statistical uncertainty is dominated by
stochastic noise, the two determinations are effectively uncorrelated,
which explains the small differences—always within uncertainties—between
their central values.}. This observation indicates that the light
finite-volume $\pi^{+}\pi^{-}$ state in which the pions carry no lattice
momentum gives a negligible contribution.

As expected, the real parts of most form factors increase with increasing
$x_{k}$, and the statistical uncertainties also grow toward
larger values of $x_{k}$. For the imaginary parts, the present
uncertainties are of order $100\%$ and increase further in absolute size
as $x_{k}$ grows. However, since the imaginary parts are substantially
smaller in magnitude than the corresponding real parts, the present
precision is already sufficient for a reliable determination of the
branching fractions of $K_{\ell2\ell'}$ decays.

We conclude this section with a discussion of the quark-disconnected
contribution to the SFR form factors, which has not been included so far.
We recall that this contribution has been computed only on a single gauge
ensemble (B96) at fixed $y_{k}\simeq0.7$, where it was found that below
the $\pi\pi$ threshold it is smaller than the current uncertainty on the
connected contribution. We now provide an estimate of its contribution above the threshold.

We begin with the imaginary part. The starting point is the isospin
decomposition of the (charmless) electromagnetic current,
\begin{align}
\label{eq:JemIeq0}
J_{I=0}^{\mu}
 = \frac{1}{6}[ \bar{q}_{u}\gamma^{\mu}q_{u}
 + \bar{q}_{d}\gamma^{\mu}q_{d}
 -2\bar{q}_{s}\gamma^{\mu}q_{s}]~,\qquad
J_{\rm em}^{\mu}=J_{I=1}^{\mu}+J_{I=0}^{\mu}.
\end{align}
The isotriplet ($I=1$) component introduced in
Eq.~(\ref{eq:jmuemIeq1}) is proportional to the up-quark connected
contribution, with proportionality factor $(2e_{u})^{-1}=3/4$, and contains the
$\pi\pi$ intermediate states. In contrast, in the isospin-symmetric limit  $m_u = m_d$, the isosinglet current does not couple to two-pion states; the lightest state it creates is instead a three-pion state (it is the $\omega$ channel).

Performing the relevant Wick contractions between $J^\mu_{I=0}$ in Eq.~(\ref{eq:JemIeq0}), the weak current, and the kaon interpolating operator, the contribution to the form factors in the isospin-symmetric limit obtained from the iso-singlet component of the electromagnetic current, denoted by $F_{I=0}(x_k,y_k)$, can be written as
\be
\label{eq:FIeq0}
F_{I=0}(x_k,y_k)=F_s^{\mathrm{Conn}}(x_k,y_k)+\frac{1}{4}F_u^{\mathrm{Conn}}(x_k,y_k)+F^{\mathrm{Disc}}(x_k,y_k),
\qquad
F=\{F_V,A^i\}.
\ee
The first two terms on the right-hand side of Eq.~(\ref{eq:FIeq0}) correspond to the quark-connected (Conn) contributions\footnote{We recall that our definitions of $F_s^{\mathrm{Conn}}$ and $F_u^{\mathrm{Conn}}$ already include the electric charges of the quarks. Therefore, the coefficients appearing in Eq.~(\ref{eq:FIeq0}) are given by the product of the weights with which the quark fields enter the definition of $J^\mu_{I=0}$ and the inverse quark charges, namely $1/3\,e_s^{-1}=1$ and $1/6\,e_u^{-1}=1/4$ for strange and up-type quarks, respectively.} from strange and up-type quarks (Diags.~(b) and~(c) in Fig.~\ref{Fig:Diagrams}, respectively), while the last term represents the quark-disconnected (Disc) contribution (Diag.~(d) in Fig.~\ref{Fig:Diagrams}).

Let us first consider the imaginary parts of the form factors. Below the three-pion threshold,
$2m_\pi/m_K<x_k<3m_\pi/m_K\simeq 0.85$,
only two-pion intermediate states contribute to the imaginary parts of the form factors, which are proportional to the spectral density (see Eq.~(\ref{eq:SFR_dispersion})). Since the strange-quark connected term has a spectral density starting only at the $K^+K^-$ threshold, and since there is no $\pi\pi$ contribution in the $I=0$ channel, one has
\be
\mathrm{Im}[F_s^{\mathrm{Conn}}(x_k,y_k)]=0,
\qquad
\mathrm{Im}[F_{I=0}(x_k,y_k)]=0.
\ee
Substituting these relations into Eq.~(\ref{eq:FIeq0}) gives
\be
\label{eq:ImFDiscIeq0}
\mathrm{Im}[F^{\mathrm{Disc}}(x_k,y_k)]
=
-\frac{1}{4}\mathrm{Im}[F_u^{\mathrm{Conn}}(x_k,y_k)],
\qquad
\frac{2m_\pi}{m_K}<x_k<\frac{3m_\pi}{m_K}.
\ee
Note that Eq.~(\ref{eq:ImFDiscIeq0}) is exact and provides a practical way to determine $\mathrm{Im}[F^{\mathrm{Disc}}]$ indirectly.
Given the $O(50-100\%)$ uncertainties affecting our determination of the connected contribution to the imaginary parts of the form factors (Fig.~\ref{fig:comp_IM_SFR}), the quoted errors already conservatively account for the disconnected contribution.

The situation is different for the real parts of the form factors. In this case, even in the $\varepsilon \to 0^{+}$ limit, the real parts receive contributions from all intermediate states propagating between the two currents, so that the argument used for the imaginary parts cannot be directly applied. In principle, the form factors could be determined by employing the SFR-HLT strategy with the quark-disconnected differential form factors as input. However, since the latter are affected by large statistical uncertainties, such an extraction would be extremely challenging in practice. We therefore adopt a different strategy and write the real part of the quark-disconnected contribution as
\begin{align}
\nonumber
\mathrm{Re}[F^{\mathrm{Disc}}(x_k,y_k)]
=&\,
-\mathrm{Re}[F_s^{\mathrm{Conn}}(x_k,y_k)]
-\frac{1}{4}\,\mathrm{Re}[F_u^{\mathrm{Conn}}(x_k,y_k)]
\\ \label{eq:disctoconnandIeq0}
&\,
+\bigg(
\mathrm{Re}[F^{\mathrm{Disc}}(x_k,y_k)]
+\frac{1}{4}\,\mathrm{Re}[F_u^{\mathrm{Conn}}(x_k,y_k)]
+\mathrm{Re}[F_s^{\mathrm{Conn}}(x_k,y_k)]
\bigg),
\end{align}
where we have added and subtracted the strange-quark connected contribution together with one quarter of the up-quark connected one.

We evaluate $\mathrm{Re}[F_s^{\mathrm{Conn}}]$ through standard Euclidean-time integration, while $\mathrm{Re}[F_u^{\mathrm{Conn}}]$ is determined using the SFR-HLT method, following the same procedure described in the previous sections. The term in parentheses in the second line of Eq.~(\ref{eq:disctoconnandIeq0}) coincides exactly with $\mathrm{Re}[F_{I=0}]$. As noted above, the $I=0$ component of the electromagnetic current does not couple to $\pi\pi$ states, but only to states containing at least three pions. Therefore, for $x_k < 3m_\pi/m_K$, we determine $\mathrm{Re}[F_{I=0}(x_k,y_k)]$ by integrating over Euclidean time the corresponding differential form factors.

\begin{figure}[]
    \centering
\includegraphics[width=1.\columnwidth]{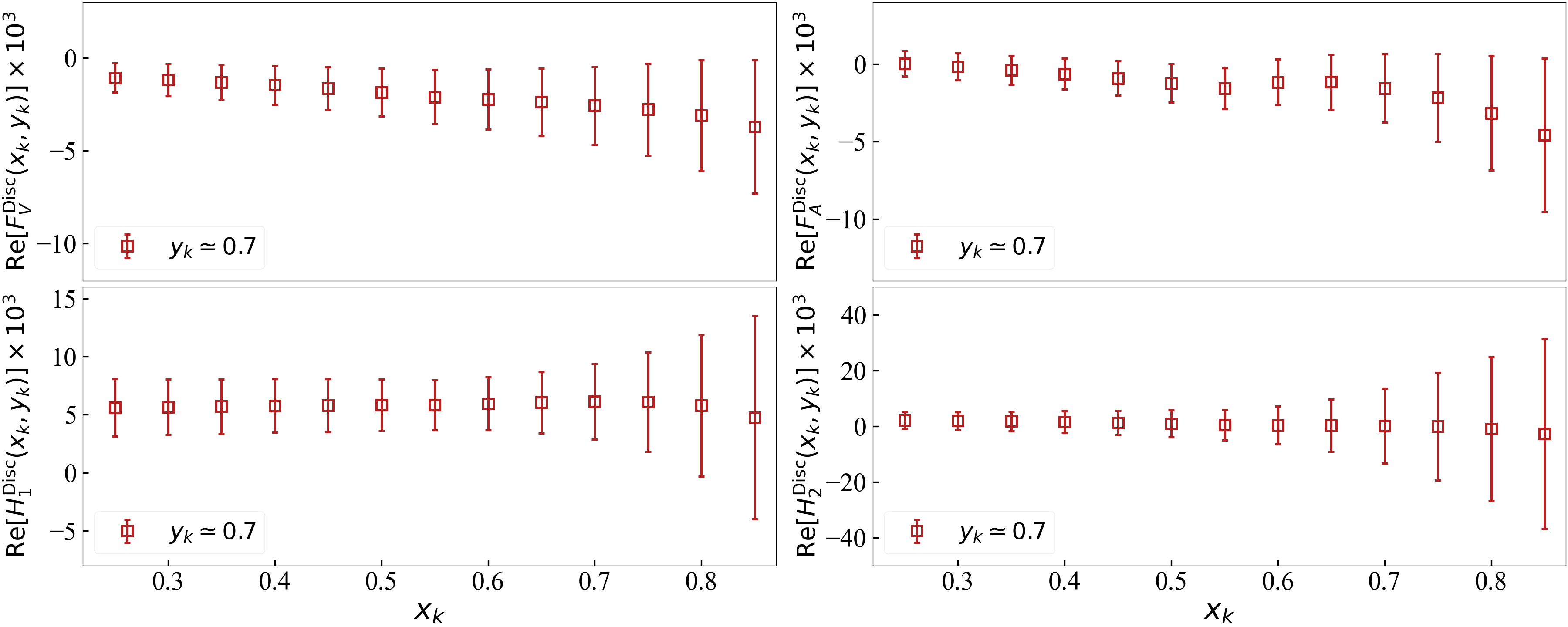}
    \caption{Real parts of the quark-disconnected contribution to the form factors multiplied by $10^3$, as function of $x_k$ at fixed $y_k \simeq 0.7$. Red squares represent the results obtained on the B96 ensemble. To provide a conservative estimate of the missing continuum and infinite-volume extrapolation for this contribution, we will assign a $100\%$ uncertainty to the disconnected contribution, which is not shown in the figure. }
    \label{fig:4L_FinalDisc}
\end{figure}

Fig.~\ref{fig:4L_FinalDisc} shows our final results for the real parts of the quark-disconnected contributions to the form factors as functions of the photon virtuality in the region $x_k < 3m_\pi/m_K$. To conservatively account for the absence of continuum and infinite-volume extrapolations, the uncertainties will be inflated to $100\%$ whenever smaller. Since no significant dependence on $x_k$ is observed, for $x_k > 0.85$ we estimate the quark-disconnected contribution to the form factors in the interval $0.85 < x_k \leq 1$, reasonably assuming that the ratio between disconnected and connected contributions remains approximately equal to that found at $x_k \simeq 0.85$.
Let us stress that the kinematical region $0.85 < x_k \leq 1$ is very marginal, contributing only to the $K_{e2\ell'}$ decays, where an electron is produced in the weak decay, and only for the photon momentum $y_k \simeq 0.1$, as can be seen from Fig.~\ref{fig:4L_Ps}.

\section{Final results for the form factors $F_V$, $F_A$, $H_1$, and $H_2$}
\label{Sec:4L_finalFF}

In the previous sections, we have expressed the form factors above the
two-pion threshold as the sum of two contributions:
\begin{align}
\label{eq:4L_FWickpsfrfinal}
F(x_k, y_k)
   ={}& F_{\mathrm{Wick}}(x_k, y_k)
     +
       F_{\mathrm{SFR}}(x_k, y_k)~,\qquad F=\{ F_{V},A^{i}\}~.
\end{align}
The lattice results for the SFR contribution were presented in the previous section, while those for $F_{V,{\rm Wick}}(x_{k},y_{k})$ and $A^{i}_{\rm Wick}(x_{k},y_{k})$, obtained using the standard Euclidean-time integration technique, are discussed in App.~\ref{app:4L_Wick}, as their analysis closely follows that adopted for evaluating the form factors below the two-pion threshold. Combining the two contributions, including the
estimate of the quark-disconnected diagrams above threshold described at
the end of the previous section, and incorporating the results obtained
below the two-pion threshold
($x_{k} < 2m_{\pi}/m_{K}$) in Sec.~\ref{Sec:belowth}, we obtain our final determinations of the
form factors.

\begin{figure}[t]
    \centering
    \includegraphics[width=1.\columnwidth]{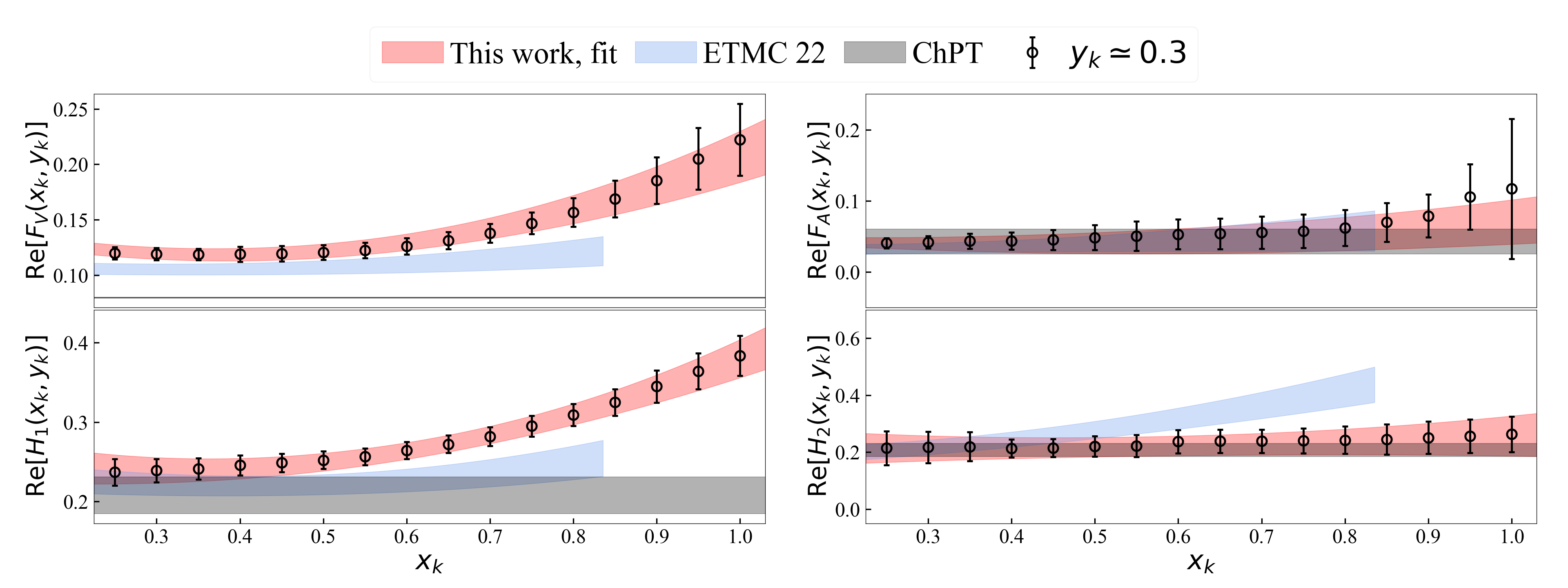}
    \caption{Final results for the real parts of the form factors (black
    circles) as functions of the photon virtuality at fixed photon
    momentum $y_k \simeq 0.3$. The red bands correspond to our
    phenomenological parameterization of the form factors
    (see Eq.~(\ref{eq:4L_ansatzFF})), the blue bands to the results of
    our previous exploratory study~\cite{Gagliardi:2022szw}, and the
    black bands to the $O(p^{4})$ ChPT predictions.}
    \label{fig:4L_ReFinalFF}
\end{figure}

\begin{figure}[]
    \centering
    \includegraphics[width=1.\columnwidth]{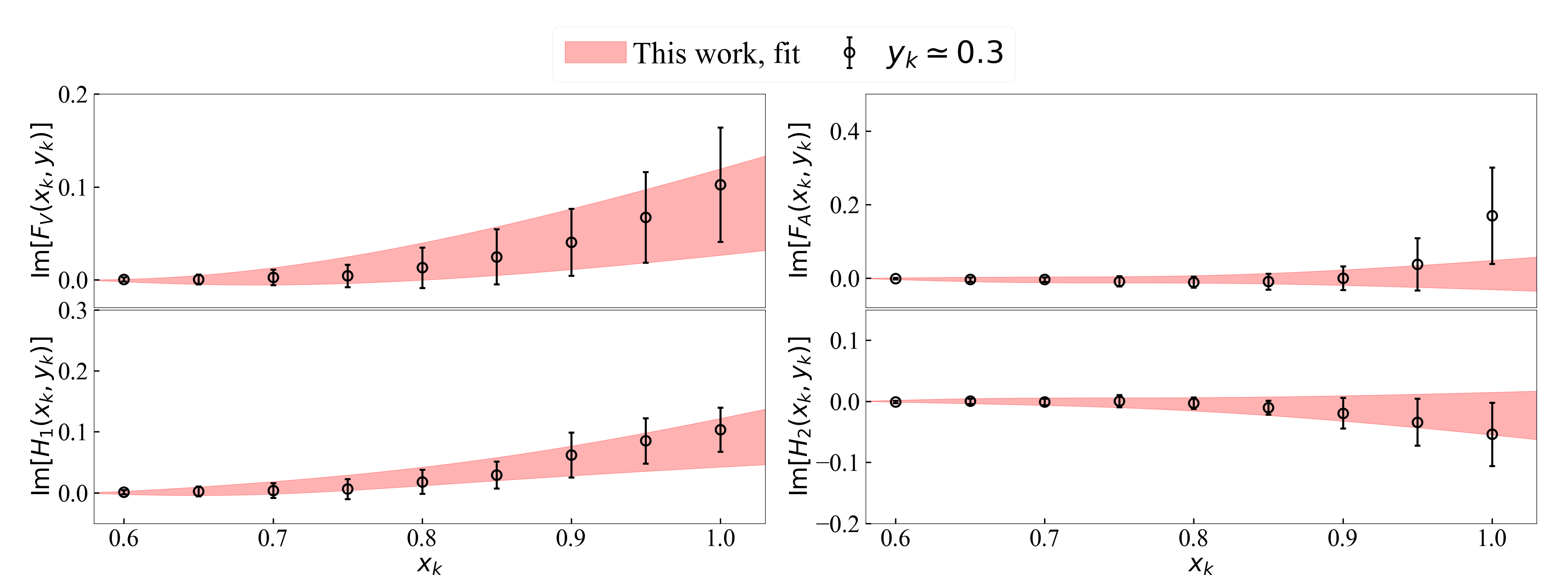}
    \caption{Final results for the imaginary parts of the form factors
    (black circles) as functions of the photon virtuality at fixed photon
    momentum $y_k \simeq 0.3$. The red bands represent the results of
    our phenomenological fit.}
    \label{fig:4L_ImFinalFF}
\end{figure}

The final results are shown in Fig.~\ref{fig:4L_ReFinalFF} for the real
parts and in Fig.~\ref{fig:4L_ImFinalFF} for the imaginary parts, for a
representative value $y_{k}\simeq 0.3$. As already observed previously,
the real parts of some of the form factors (in particular $H_{1}$ and
$F_{V}$) increase with increasing $x_{k}$, while the imaginary parts are
currently determined with uncertainties of order $50-100\%$. The latter are
nevertheless substantially smaller in magnitude than the real parts. The
final values at the simulated $x_{k}$ points are displayed in the figures as black circles.

The real parts of the form factors in  Fig.~\ref{fig:4L_ReFinalFF} are also  compared with the
results obtained in our earlier exploratory ETMC
determination~\cite{Gagliardi:2022szw} (blue bands), which employed a
single gauge ensemble at an unphysical pion mass of about
$320~{\rm MeV}$, as well as with next-to-leading-order ChPT predictions
at $O(p^{4})$ (black bands), given by
\begin{align}
\mathrm{Re}[F_V(x_k, y_k)] &= \frac{m_K}{4\pi^2 f_K}, \qquad
\mathrm{Re}[F_A(x_k, y_k)] = \frac{8 m_K}{f_K} (L_9^r + L_{10}^r), \\[6pt]
\mathrm{Re}[H_1(x_k, y_k)] &= \mathrm{Re}[H_2(x_k, y_k)]
= 2 f_K m_K \frac{F_K(k^2)-1}{k^2}
\simeq \frac{f_{K}m_{K}}{3}\langle r_{K}^{2}\rangle~,
\end{align}
for which we use the inputs~\cite{ParticleDataGroup:2024cfk}
\begin{align}
m_K &= 493.677(13)\,\mathrm{MeV}, \qquad
\langle r_K^2 \rangle = 0.31(3)\,\mathrm{fm}^2, \qquad
f_K = 156.4(6)\,\mathrm{MeV},
\end{align}
and $L_9^r + L_{10}^r = 0.0017(7)$~\cite{Bijnens:2014lea}.

Our results and those of Ref.~\cite{Gagliardi:2022szw} are compatible
within $1\sigma$ for $F_A$, while differences exceeding $2\sigma$ are
observed for the remaining form factors. For $H_1$ and $H_2$, the
difference increases with photon virtuality and becomes progressively
more pronounced above the two-pion threshold ($x_k \gtrsim 0.55$). In
particular, the difference between our determination of $H_1$ and that
of ETMC~2022 amounts to approximately $13\%$ at threshold and increases
to about $25\%$ at $x_k \simeq 0.8$. Since $H_{1}$ provides the dominant
contribution to the $K_{\ell2\ell'}$ decay rates, this enhancement
has a significant impact on the predicted branching fractions, which
will be discussed in a companion paper~\cite{DiPalma:pheno}.

\begin{figure}[]
    \centering
    \includegraphics[width=1.\columnwidth]{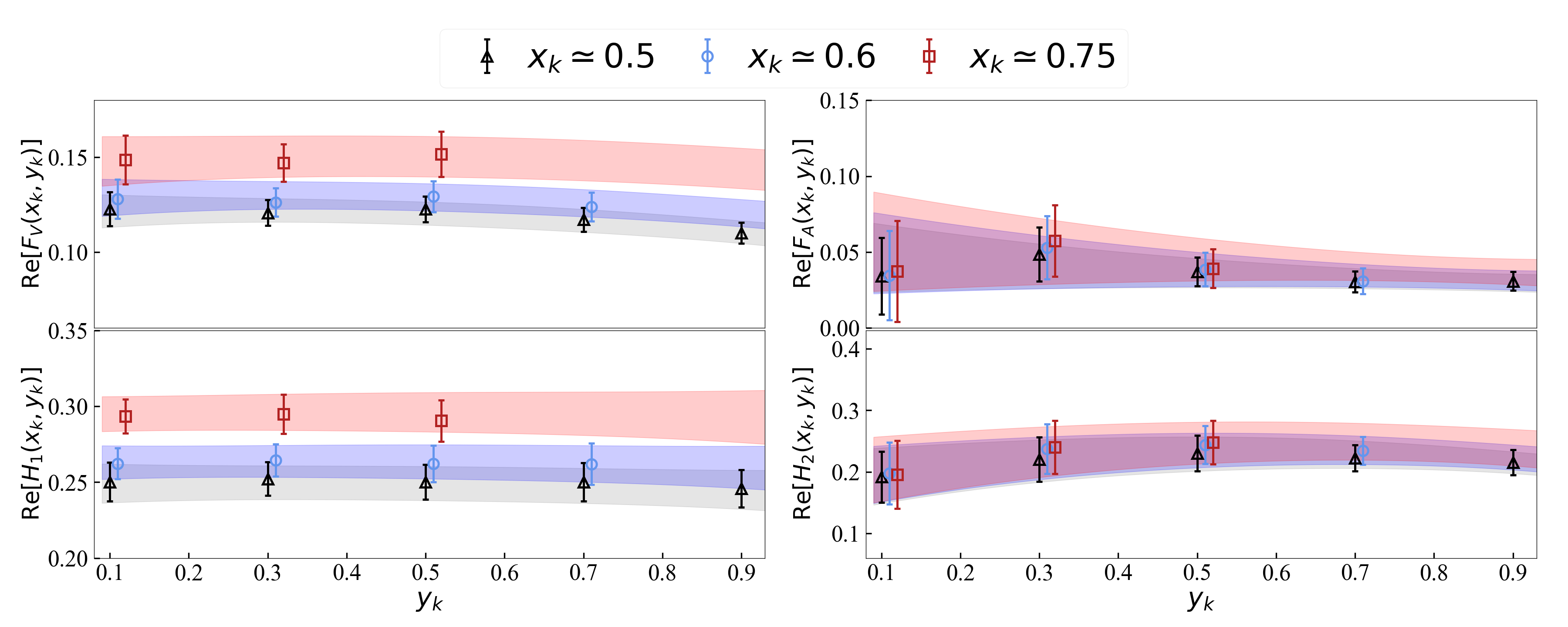}
    \caption{Final results for the real parts of the form factors as
    functions of the photon momentum for
    $x_k \simeq 0.5$ (black triangles),
    $x_k \simeq 0.6$ (blue circles), and
    $x_k \simeq 0.75$ (red squares). Data points are slightly shifted
    horizontally for clarity. The coloured bands correspond to the
    phenomenological parameterization of
    Eq.~(\ref{eq:4L_ansatzFF}).}
    \label{fig:4L_ReFinalFFyk}
\end{figure}

Fig.~\ref{fig:4L_ReFinalFFyk} shows the real parts of the form factors as
functions of the photon momentum for three representative values of the
photon virtuality:
$x_k \simeq 0.5$ (black triangles),
$x_k \simeq 0.6$ (blue circles), and
$x_k \simeq 0.75$ (red squares).
Similarly to what we observed for $x_{k}<2m_{\pi}/m_{K}$
(Fig.~\ref{fig:4L_FFundervsyk}), the form factors exhibit only a very mild
dependence on the photon momentum.

Before concluding, we provide phenomenological parameterizations of the
real and imaginary parts of the four form factors, suitable for direct
use in phenomenological analyses. We start from the real part, for which we assume the following polynomial parameterization
\begin{equation}
\label{eq:4L_ansatzFF}
{\rm{Re}}[F(x_k, y_k)] = a_F + b_F x_k + c_F y_k + d_F x_k^2 + e_F y_k^2 + k_Fx_k y_k~,\quad\quad F=\{ F_{V}, A^{i}\}~,
\end{equation}
where $a_{F},b_{F},c_{F},d_{F},e_{F}$, and $k_F$ are free fit parameters. The parameterization is fitted to our lattice data: the results of the fit for $y_{k}\simeq0.3$ are shown by the red bands in Fig.~\ref{fig:4L_ReFinalFF}. The fit parameters are listed in the left panel of Tab.~\ref{tab:4L_fitpars}, while their correlations are
given in App.~\ref{app:4L_Corr}. 
\begin{table}[] 
\centering 
\begin{tabular}{|ccccccc|} \hline \hline &  &  &  & &  &  \\ [-2. ex] 
& $a_{F}$ & $b_{F}$ & $c_{F}$ & $d_{F}$ & $e_{F}$ & $k_{F}$ \\ [1. ex] \hline &  &  & & & &  \\ [-2. ex] 
$\mathrm{Re}[F_V]$ & $0.155(12)$ & $-0.182(45)$ & $-0.005(33)$ & $0.227(53)$ & $-0.027(27)$ & $0.036(31)$ \\[1.ex] \hline & & & & & & \\[-2.ex]$\mathrm{Re}[F_A]$ & $0.057(16)$ & $-0.055(74)$ & $-0.022(38)$ & $0.080(51)$ & $0.010(34)$ & $-0.017(61)$ \\[1.ex] \hline & & & & & & \\[-2.ex]$\mathrm{Re}[H_1]$ & $0.269(48)$ & $-0.19(11)$ & $0.002(64)$ & $0.298(90)$ & $-0.013(37)$ & $0.012(57)$ \\[1.ex] \hline & & & & & & \\[-2.ex]$\mathrm{Re}[H_2]$ & $0.19(11)$ & $-0.07(26)$ & $0.15(14)$ & $0.08(20)$ & $-0.154(89)$ & $0.07(14)$ \\[1.ex] \hline \hline\end{tabular} 
\begin{tabular}{|cccc|} \hline \hline &  &  &      \\ [-2. ex] 
& $\alpha_{F}^I$ & $\beta_{F}^I$ & $\gamma_{F}^I$  \\ [1. ex] \hline &  &  &    \\ [-2. ex] 
$\mathrm{Im}[F_V]$ & $-0.229(88)$ & $0.37(15)$ & $-0.041(90)$ \\[1.ex] \hline & & & \\[-2.ex]$\mathrm{Im}[F_A]$ & $-0.11(25)$ & $0.12(30)$ & $0.006(74)$ \\[1.ex] \hline & & &  \\[-2.ex]$\mathrm{Im}[H_1]$ & $-0.21(24)$ & $0.35(28)$ & $0.02(14)$  \\[1.ex] \hline & & &  \\[-2.ex]$\mathrm{Im}[H_2]$ & $0.05(16)$ & $-0.11(22)$ & $0.078(41)$  \\[1.ex] \hline \hline\end{tabular} 
\caption{Parameters describing the real (left table) and imaginary (right table) parts of the form factors $F_V$, $F_A$, $H_1$, and $H_2$ through the ansatz in Eqs.~(\ref{eq:4L_ansatzFF}) and~(\ref{eq:4L_ansatzImFF}), respectively.} 
\label{tab:4L_fitpars} 
\end{table}

To parameterize the imaginary parts of the form factors, , since they have larger errors, we adopt a
linear ansatz, after factoring out the non-analytic behaviour around threshold. The fit ansatz in this case is given by 
\begin{equation}
\label{eq:4L_ansatzImFF}
{\rm Im}[F(x_k, y_k)] =
\Theta \left(x_k - \frac{2m_{\pi}}{m_{K}} \right)\Phi(x_k)
\left(
\alpha_F^{I} + \beta_F^{I} x_k + \gamma_F^{I} y_k 
\right),~\qquad F=\{F_{V},A^{i}\}~, 
\end{equation}
where $\alpha_{F}^{I},\beta_{F}^{I}$ and $\gamma_{F}^{I}$ are free fit parameters, $\Theta(x)$ indicates the Heaviside step-function, while $\Phi(x_{k})$ is the P-wave phase-space factor
\begin{align}
\Phi(x_{k}) = \left[ 1- \left(\frac{2m_{\pi}}{m_{K}x_{k}}\right)^{2}\right]^{3/2}~.
\end{align}
The results of the fits are shown, for $y_{k}\simeq0.3$, by the red bands in Fig.~\ref{fig:4L_ImFinalFF}, while the values of the fit parameters are collected in the right panel of Tab.~\ref{tab:4L_fitpars}. The correlation matrix of the fit parameters is instead given in App.~\ref{app:4L_Corr}.

We have also explored replacing the polynomial part of the ansatz in Eq.~(\ref{eq:4L_ansatzImFF}) with a pole-like term describing the $\rho$-resonance contribution.
Within the current statistical uncertainties of our lattice results for
the imaginary parts, however, the pole and polynomial parameterizations
yield indistinguishable results.

\section{Conclusions}
\label{sec:conclusions}

In this work, we have presented for the first time, a first-principles non-perturbative lattice QCD determination of the real and imaginary parts
of the four structure-dependent form factors $F_V$, $H_1$, $F_A$, and $H_2$
describing the rare decay in the SM,
$K^- \to \ell^-\bar{\nu}_\ell \ell'^+\ell'^-$,
as functions of the photon momentum and virtuality over the full kinematical range.

Making use of the latest-generation ensembles produced by the Extended Twisted Mass Collaboration, we have computed all contributions to the form factors entering the decay amplitude of $K_{\ell2\ell'}$ at $O(\alpha_{\mathrm{em}}^2)$ (Fig.~\ref{Fig:Diagrams}), including the quark-disconnected contribution in which the photon is emitted by a sea quark (Diag.~(d)). The simulations have been performed directly at the physical pion mass, and a thorough assessment of the systematic uncertainties has been carried out. Finite-size effects have been investigated using three lattice spatial extents $L \in [3.8,,7.6]~\mathrm{fm}$, while discretization effects have been analyzed employing three lattice spacings $a \in [0.08,,0.057]~\mathrm{fm}$. Finally, we have discussed the errors associated with the isolation of the initial-state kaon, as well as those arising from the finite temporal extent of the lattice.

At the physical point, the $K_{\ell2\ell'}$ form factors cannot be determined by integrating the corresponding Euclidean correlation functions over Euclidean time at all values of the photon virtuality. Indeed, for photon virtualities $2m_\pi < \sqrt{k^2} < m_K$, the Wick rotation to Euclidean time is not possible and the form factors become complex. In this kinematical region, the real and imaginary parts of the form factors have been extracted using the Spectral Function Reconstruction  strategy~\cite{Frezzotti:2023nun}, combined with the Hansen--Lupo--Tantalo method~\cite{Hansen:2019idp}, which allows one to compute a regularized (smeared) version of the form factors by introducing a non-zero smearing parameter $\varepsilon$. A key advancement of this work has been an ad hoc implementation of the HLT method (see Secs.~\ref{sec:4L_model} and~\ref{sec:4L_FVsfrandAisfr}), with its parameters optimized for the problem at hand, together with the definition of improved estimators of the form factors to achieve better control of the crucial $\varepsilon \to 0$ limit.

All these aspects make the present work a substantial step forward with respect to previous exploratory lattice QCD calculations~\cite{Gagliardi:2022szw,Tuo:2021ewr}, which were performed on a single gauge ensemble and at heavier pion masses, $m_\pi > 0.3~\mathrm{GeV}$. This progress highlights the rapid advancement of lattice QCD and of its tools, which now enable calculations that were previously considered out of reach. The strategy developed in this work can be extended in the future to investigate analogous rare pseudoscalar-meson decays $P^- \to \ell^{-}\bar{\nu}_{\ell}\ell'^{+}\ell'^{-}$.

The main results of this work consist of the determination of the real and imaginary parts of the $K_{\ell2\ell'}$ form factors, shown as functions of the photon virtuality $x_k$ (in units of the kaon mass), at fixed photon momentum ($y_k = 2|\bs{k}|/m_K$), in Figs.~\ref{fig:4L_ReFinalFF} and~\ref{fig:4L_ImFinalFF}, respectively. We also fit their dependence on $x_k$ and $y_k$ using the ansätze in Eqs.~(\ref{eq:4L_ansatzFF}) and~(\ref{eq:4L_ansatzImFF}); the corresponding fit parameters are reported in Tab.~\ref{tab:4L_fitpars}, while their correlations are collected in App.~\ref{app:4L_Corr}.
In Fig.~\ref{fig:4L_ReFinalFF}, we additionally present a comparison of the real parts of the form factors with the previous exploratory ETMC determination~\cite{Gagliardi:2022szw} and with the ChPT predictions at $O(p^4)$. We find that the previous ETMC result is compatible within $1\sigma$ for $\mathrm{Re}[F_A]$, whereas all other form factors show differences larger than $2\sigma$, particularly in the region above the two-pion production threshold ($\sqrt{k^2} > 2m_\pi$). 

From a phenomenological perspective, the implications of our results will be explored in detail in the companion paper~\cite{DiPalma:pheno}. Our determination of the form factors enables a direct comparison between experimental measurements and Standard Model predictions for all four $K_{\ell2\ell'}$ decay channels, thereby providing a powerful probe of possible New Physics effects. At the same time, this work opens a new avenue for independent determinations of the CKM matrix element $\vert V_{us} \vert$ from rare decays.

\section{Acknowledgments}
\label{sec:akno}
We thank the ETMC for the most enjoyable collaboration. V.L., F.S., G.G., R.F., and N.T. have been supported by the Italian Ministry
of University and Research (MUR) and the European
Union (EU) – Next Generation EU, Mission 4, Component 1, PRIN 2022, CUP F53D23001480006. 
F.S. is supported by ICSC – Centro Nazionale di Ricerca in High Performance Computing, Big Data and Quantum Computing, funded by European Union - Next Generation EU and by Italian  Ministry of University and Research (MUR) project FIS\_00001556. We acknowledge support from the LQCD123, ENP, and SPIF Scientific Initiatives of
the Italian Nuclear Physics Institute (INFN).

The open-source packages tmLQCD~\cite{Jansen:2009xp,Abdel-Rehim:2013wba,Deuzeman:2013xaa,Kostrzewa:2022hsv}, LEMON~\cite{Deuzeman:2011wz}, DD-$\alpha$AMG~\cite{Frommer:2013fsa,Alexandrou:2016izb,Bacchio:2017pcp,Alexandrou:2018wiv}, QPhiX~\cite{joo2016optimizing,Schrock:2015gik} and QUDA~\cite{Clark:2009wm,Babich:2011np,Clark:2016rdz} have been used in the ensemble generation.

We gratefully acknowledge the ICSC - Centro Nazionale di Ricerca in High Performance Computing for providing computing time under the allocations RAC 1916318. We gratefully acknowledge CINECA for the provision of GPU time on Leonardo supercomputing facilities under the specific initiative INFN-LQCD123, and under project IscrB VITO-QCD, project IscrB SemBD, and project IscrB Hcee. We gratefully acknowledge EuroHPC Joint Undertaking for awarding us access to MareNostrum5 through the project EHPC-EXT-2024E01-031. The authors gratefully acknowledge the Gauss Centre for Supercomputing e.V. (www.gauss-centre.eu) for funding this project by providing computing time on the GCS Supercomputers SuperMUC-NG at Leibniz Supercomputing Centre. The authors acknowledge the Texas Advanced Computing Center (TACC) at The University of Texas at Austin for providing HPC resources (Project ID PHY21001). We gratefully acknowledge PRACE for awarding access to HAWK at HLRS within the project with Id Acid 4886. We acknowledge the Swiss National Supercomputing Centre (CSCS) and the EuroHPC Joint Undertaking for awarding this project access to the LUMI supercomputer, owned by the EuroHPC Joint Undertaking, hosted by CSC (Finland) and the LUMI consortium through the Chronos programme under project IDs CH17-CSCS-CYP. We acknowledge EuroHPC Joint Undertaking for awarding the project ID EHPC-EXT-2023E02-052 access to MareNostrum5 hosted by at the Barcelona Supercomputing Center, Spain.

\appendix

\section{Gounaris-Sakurai parametrization of the electromagnetic pion form factor }
\label{App:rhomunupipi_details}
In Sec.~\ref{sec:4L_model}, we introduce an effective model designed to capture the contribution of intermediate
$\pi^+\pi^-$ states to the spectral densities entering the SFR form factors
$F_{V,{\rm SFR}}(\varepsilon,x_k,y_k)$ and $A^{i}_{{\rm SFR}}(\varepsilon,x_k,y_k)$.
The starting point of our discussion is the contribution of the $\pi\pi$ states to the spectral density $\rho_{W;u}^{\mu\nu}$, defined in Eq.~(\ref{eq:4L_rhotwopions_model_revised}), which we report here for clarity 
\begin{align}
\label{eq:Apprhomunupipi}
\rho^{\mu\nu}_{W;\pi\pi}(E,\bs{k})
  \equiv -2\pi e_{u} \int 
      \frac{d\bs{p}_\pi}{(2\pi)^3\, 4 E_{\pi^+} E_{\pi^-}}\;
      \delta(E - E_{\pi^+} - E_{\pi^-})
   \bra{0} j^\mu_{I=1}(0) \ket{\pi\pi(\bs{k}, \bs{p}_\pi)}\,
   \bra{\pi\pi(\bs{k}, \bs{p}_\pi)} j^\nu_W(0) \ket{K^-(\bs{0})},
\end{align}
where $j^\mu_{I=1}$ is the $I=1$ component of the  electromagnetic current of quarks fields, defined in Eq.~(\ref{eq:jmuemIeq1}).
The matrix element of the electromagnetic current in Eq.~(\ref{eq:Apprhomunupipi}) can be parameterized as
\begin{equation}
\bra{0} j^\mu_{I=1}(0) \ket{\pi\pi(\bs{k}, \bs{p}_\pi)}
  = i \bigl(p^\mu_{\pi^{+}} - p^\mu_{\pi^{-}}\bigr)\, F_\pi(s),
\end{equation}
where $F_\pi(s)$ is the  timelike electromagnetic pion form factor. 
In this appendix, we provide an explicit expression for $F_\pi(s)$, modeling its dependence on $s=(p_{\pi^+} + p_{\pi^-})^2$ using the Gounaris--Sakurai (GS) parametrization~\cite{Gounaris:1968mw}.

In the GS framework, $F_\pi(s)$ is constructed under the assumption that the
$\pi\pi$ $P$-wave elastic scattering amplitude is dominated by the $\rho$
resonance. The form factor is written as
\begin{equation}
F_\pi(s) =
\frac{m_\rho^2 - A_{\pi\pi}(0)}
     {m_\rho^2 - s - A_{\pi\pi}(s)} \, ,
\end{equation}
where $A_{\pi\pi}(s)$ denotes the twice-subtracted $\pi\pi$
amplitude~\cite{Gounaris:1968mw}, given by
\begin{equation}
A_{\pi\pi}(s) =
h(m_\rho^2)
+ (s - m_\rho^2)\,\frac{h'(m_\rho^2)}{2m_\rho}
- h(s)
+ i\,\sqrt{s}\,\Gamma_{\rho\pi\pi}(s).
\label{eq:Apipi}
\end{equation}
Here, $\Gamma_{\rho\pi\pi}(s)$ is the energy-dependent decay width of the $\rho$
resonance into two pions. The running width can be expressed in terms of the
effective coupling $g_{\rho\pi\pi}$ as
\begin{equation}
\Gamma_{\rho\pi\pi}(s)
=
\frac{g_{\rho\pi\pi}^2}{6\pi}
\frac{k^3(s)}{s}, \quad \mathrm{with} \quad k(s) \equiv \sqrt{\frac{s}{4} - m_\pi^2}
\end{equation}
being the three-momentum of each pion in the center-of-mass frame of the $\rho$.
The coupling $g_{\rho\pi\pi}$ is fixed by the on-shell width,
$\Gamma_\rho \equiv \Gamma_{\rho\pi\pi}(m_\rho^2)$,
\begin{equation}
g_{\rho\pi\pi}
=
\sqrt{
\frac{48\pi\,m_\rho^2\,\Gamma_\rho}
     {(m_\rho^2 - 4m_\pi^2)^{3/2}}
     } \, .
\end{equation}
The functions $h(s)$ and $h'(s)$ appearing in Eq.~(\ref{eq:Apipi}) are defined as
\begin{align}
h(s) &=
\frac{2}{\pi}\,
\sqrt{s}\,\Gamma_{\rho\pi\pi}(s)\,
\ln\!\left(
\frac{\sqrt{s} + 2k(s)}{2m_\pi}
\right), \\[2mm]
h'(s) &=
\Gamma_{\rho\pi\pi}(s)\,
\frac{\sqrt{s}}{\pi\,k(s)}
\left[
1 +
\left(1 + \frac{2m_\pi^2}{s}\right)
\frac{\sqrt{s}}{k(s)}
\ln\!\left(
\frac{\sqrt{s} + 2k(s)}{2m_\pi}
\right)
\right].
\end{align}
By construction, the form factor is normalized to unity at $s=0$,
\begin{equation}
F_\pi(0) = 1,
\end{equation}
which follows from the analytic continuation of $A_{\pi\pi}(s)$ to $s=0$.
The two parameters of the GS model, namely the $\rho$-meson mass $m_\rho$ and
width $\Gamma_\rho$, are fixed to their physical values~\cite{ParticleDataGroup:2024cfk},
\begin{equation}
m_\rho = 0.775~\mathrm{GeV}, \qquad
\Gamma_\rho = 0.15~\mathrm{GeV},
\end{equation}
corresponding to an effective coupling
\begin{equation}
g_{\rho\pi\pi} \simeq 5.9 \, .
\end{equation}

\section{Evaluation of $F_{V,\mathrm{Wick}}(x_k, y_k)$ and $A^i_{\mathrm{Wick}}(x_k, y_k)$}
\label{app:4L_Wick}

\begin{figure}[]
    \centering
    \includegraphics[width=1.\columnwidth]{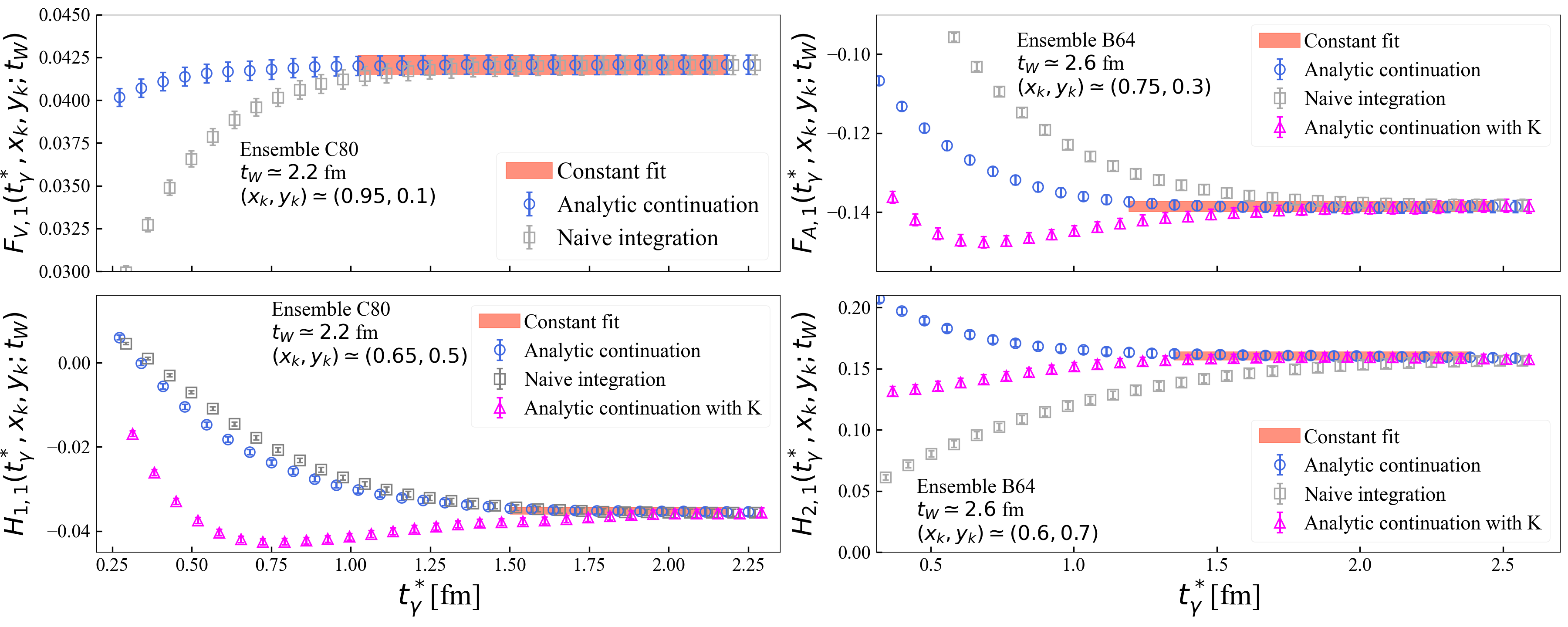}
    \caption{Quark-connected contributions to the form factors from the first
time ordering as functions of the intermediate time $t_\gamma^\ast$. The
results in the left (right) panels are obtained on the C80 (B64) ensemble with
$t_W \simeq 2.2~\mathrm{fm}$ ($t_W \simeq 2.6~\mathrm{fm}$), at different values
of the photon virtuality $x_k$ above the two-pion threshold and photon
momentum $y_k$. Blue circles, grey squares, and magenta triangles correspond to
different estimators (see text for details). Red bands represent the results of
constant fits to the blue circles. Data points have been slightly shifted
horizontally to improve readability.}
    \label{fig:4L_ABFFfirstpltx}
\end{figure}

\begin{figure}[]
    \centering
    \includegraphics[width=1.\columnwidth]{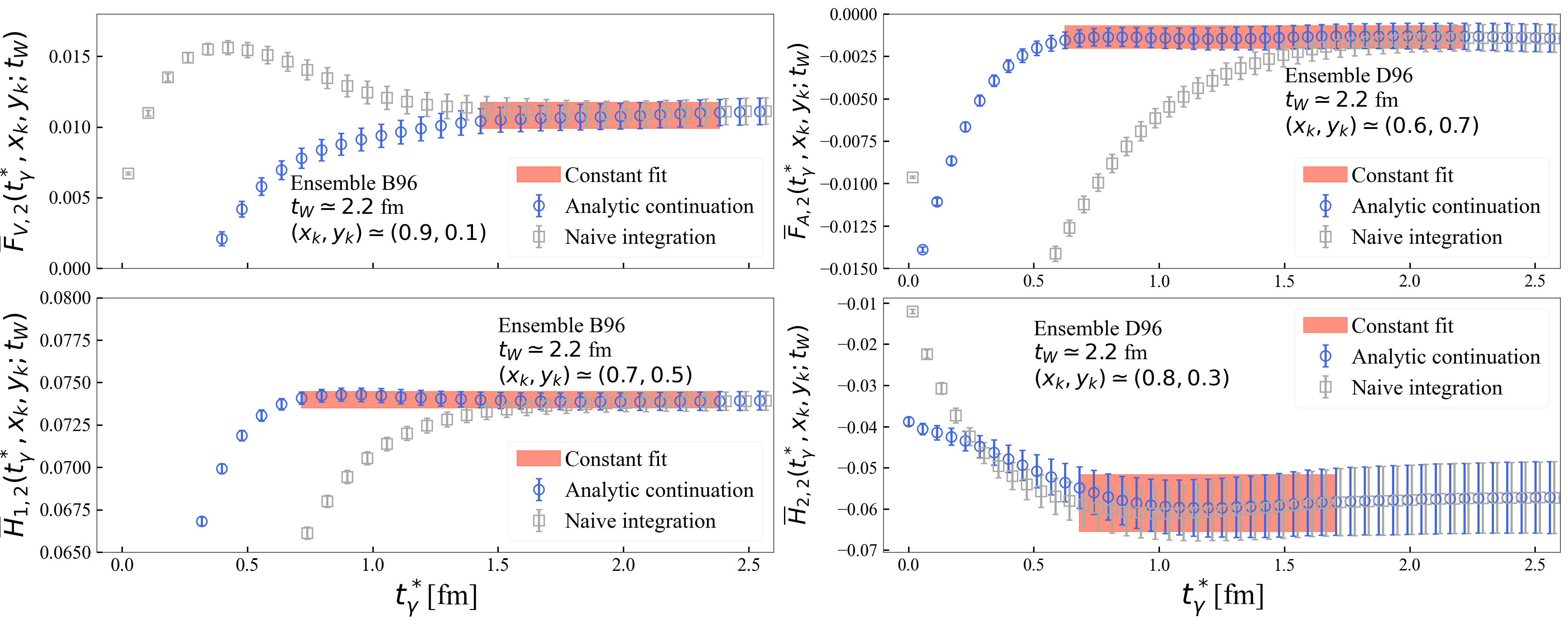}
    \caption{Quark-connected contributions to
$\bar{F}_{V,2}(x_k, y_k; t_W)$, $\bar{H}_{1,2}(x_k, y_k; t_W)$,
$\bar{F}_{A,2}(x_k, y_k; t_W)$, and $\bar{H}_{2,2}(x_k, y_k; t_W)$ as functions
of the intermediate time $t_\gamma^\ast$. The results in the right (left)
panels are obtained on the D96 (B96) ensemble with $t_W \simeq 2.2~\mathrm{fm}$
at different values of the photon virtuality $x_k$ and momentum $y_k$. Blue
circles and gray squares correspond to different estimators (see text for
details). Red bands represent constant fits to the blue circles. Data points
have been slightly shifted horizontally to improve readability.}
    \label{fig:4L_ABFFsecondpltx}
\end{figure}

\begin{figure}[]
    \centering
    \includegraphics[width=1.\columnwidth]{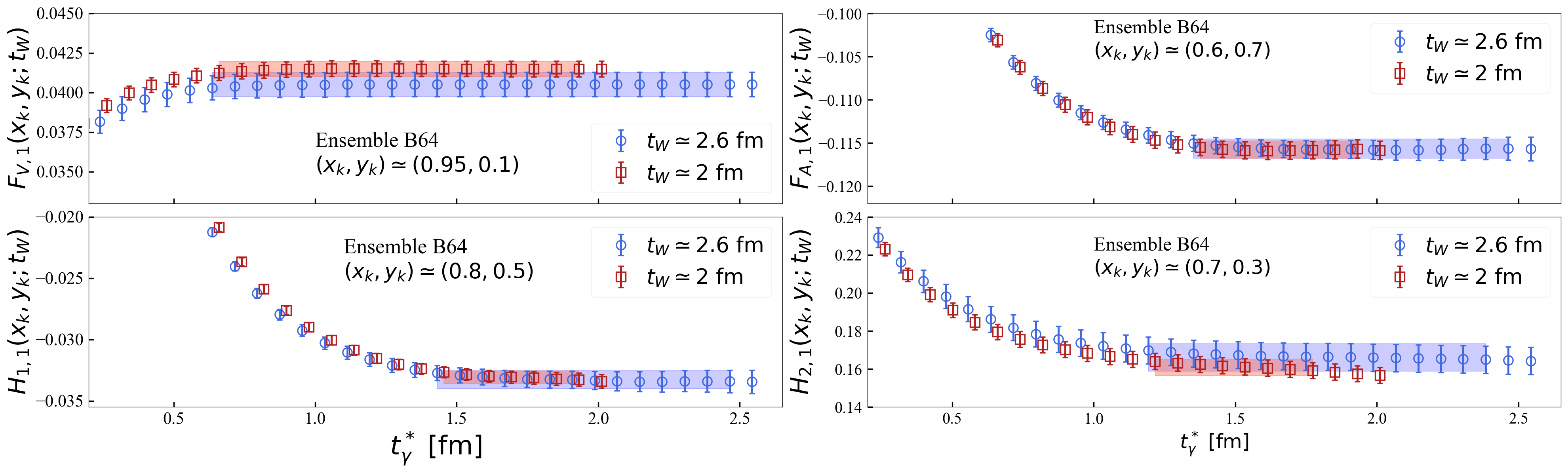}
    \caption{Quark-connected contribution to the form factors from the first
time ordering. Results are shown for the B64 ensemble using $t_W \simeq
2.0~\mathrm{fm}$ (red squares) and $t_W \simeq 2.6~\mathrm{fm}$ (blue circles).
Coloured bands represent constant fits to the corresponding data points. Data
points have been slightly shifted horizontally for clarity.}
    \label{fig:4L_ABFFfirstTOtwdep}
\end{figure}

\begin{figure}[]
    \centering
    \includegraphics[width=1.\columnwidth]{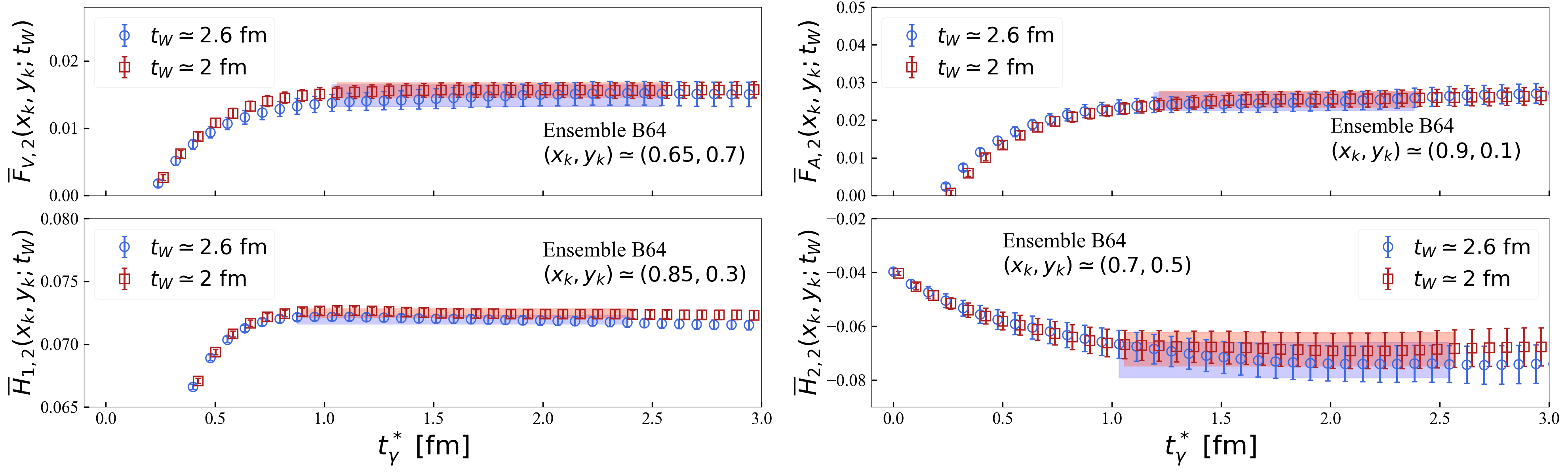}
    \caption{Quark-connected contributions to
$\bar{F}_{V,2}(x_k, y_k; t_W)$, $\bar{H}_{1,2}(x_k, y_k; t_W)$,
$\bar{F}_{A,2}(x_k, y_k; t_W)$, and $\bar{H}_{2,2}(x_k, y_k; t_W)$ on the B64
ensemble, using $t_W \simeq 2.0~\mathrm{fm}$ (red squares) and $t_W \simeq
2.6~\mathrm{fm}$ (blue circles). Coloured bands represent constant fits to the
corresponding data points. Data points have been slightly shifted horizontally
for clarity.}
    \label{fig:4L_ABFFsecondTOtwdep}
\end{figure}

\begin{figure}[]
    \centering
    \includegraphics[width=1.\columnwidth]{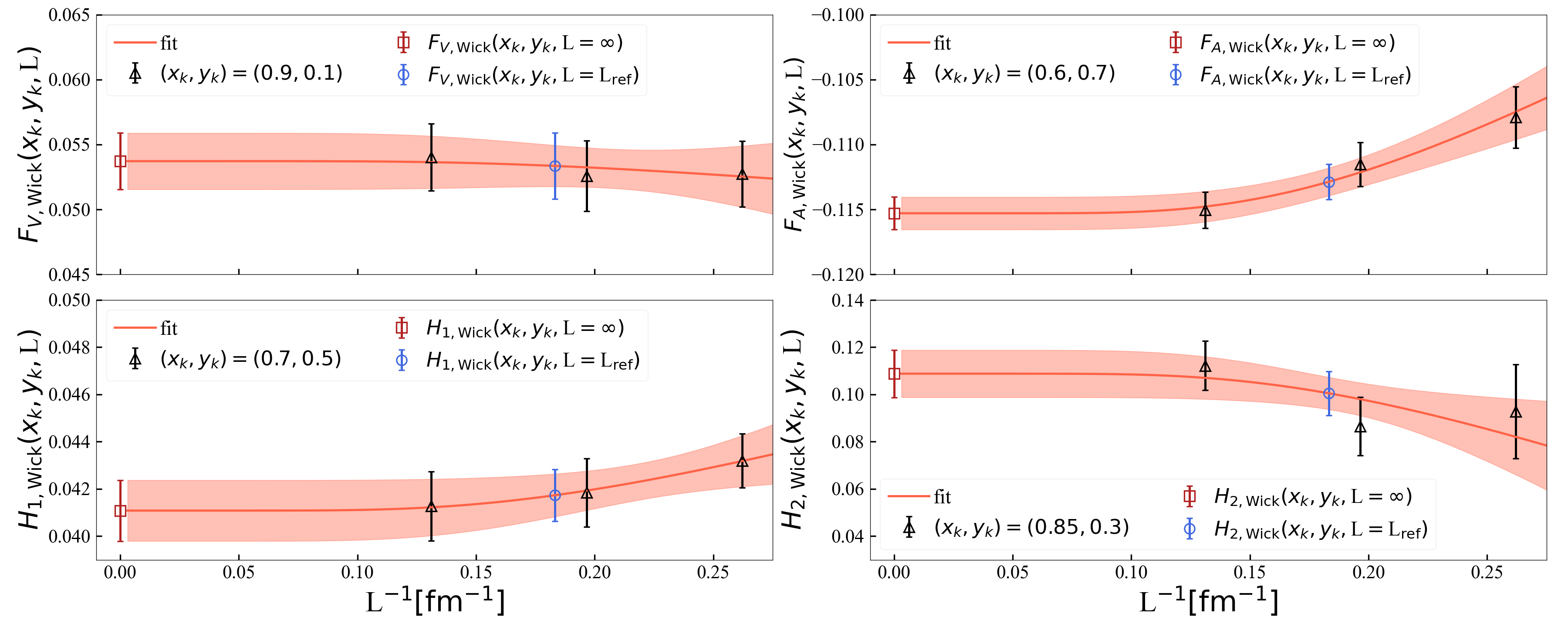}
    \caption{Dependence of the quark-connected contributions to
$F_{V,\mathrm{Wick}}(x_k, y_k,L)$, $H_{1,\mathrm{Wick}}(x_k, y_k,L)$,
$F_{A,\mathrm{Wick}}(x_k, y_k,L)$, and $H_{2,\mathrm{Wick}}(x_k, y_k,L)$
on the inverse lattice spatial extent, $L^{-1}$. The black triangles show the
values obtained on the B48, B64, and B96 ensembles, ordered from larger to
smaller $L^{-1}$. The red bands represent exponential fits based on the ansatz
in Eq.~(\ref{eq:4L_FSE_ansatz}). The red squares indicate the corresponding
extrapolations to the infinite-volume limit at non-zero lattice spacing. The
blue circles denote the form factors interpolated to the reference spatial
extent $L_{\mathrm{ref}} \simeq 5.46~\mathrm{fm}$, matching the volume of the
C80 and D96 ensembles.}
    \label{fig:4L_ABVfit}
\end{figure}

\begin{figure}[]
    \centering
    \includegraphics[width=1.\columnwidth]{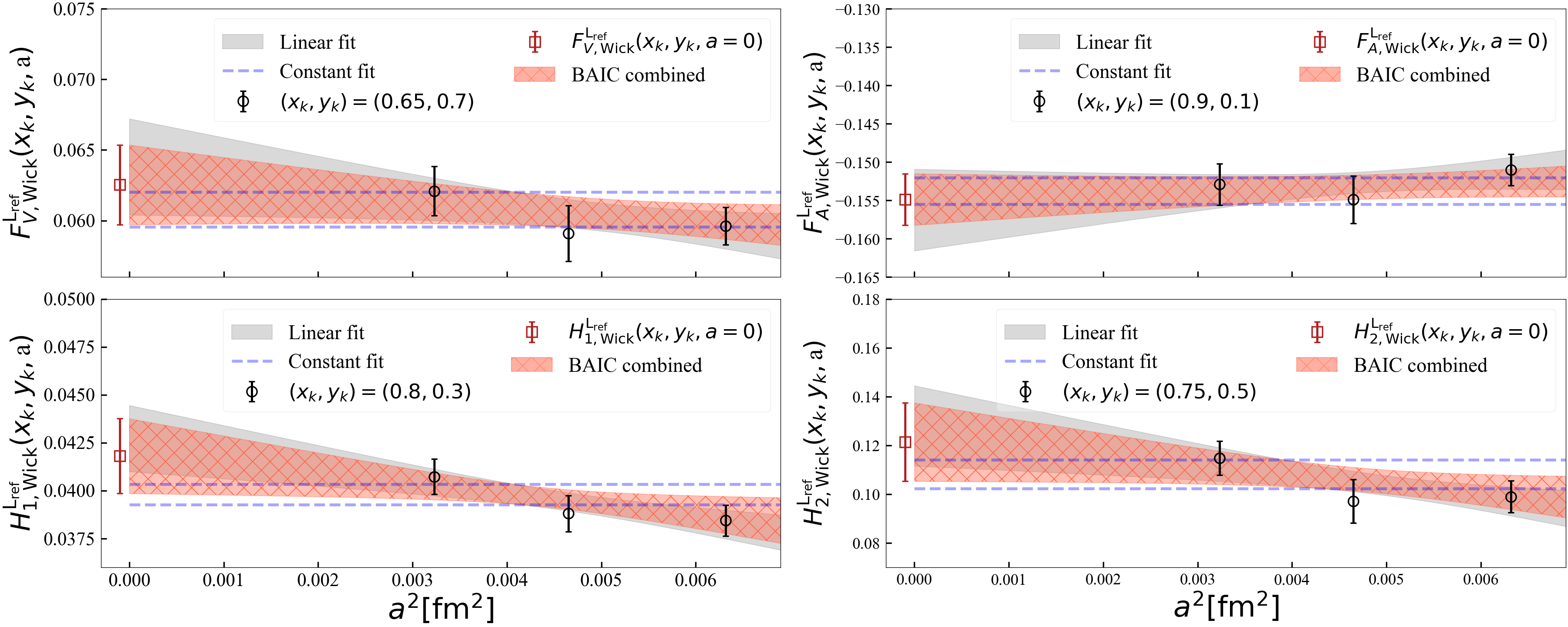}
    \caption{Continuum-limit extrapolation of the quark-connected contributions
to $F_{V,\mathrm{Wick}}(x_k, y_k,a)$, $H_{1,\mathrm{Wick}}(x_k, y_k,a)$,
$F_{A,\mathrm{Wick}}(x_k, y_k,a)$, and $H_{2,\mathrm{Wick}}(x_k, y_k,a)$
at fixed lattice spatial extent $L_{\mathrm{ref}} \simeq 5.46~\mathrm{fm}$. The
black circles correspond, from finer to coarser lattice spacing, to the D96,
C80, and B ensembles. Grey bands show the linear fit to all three points, while
the are between the blue lines correspond to a constant fit to the two finest lattice spacings.
The meshed red bands represent the BAIC-weighted combination of the two fits.
Red squares mark the continuum-limit extrapolated values at $L_{\mathrm{ref}}$.}
    \label{fig:4L_ABafit}
\end{figure}

\begin{figure}[]
    \centering
    \includegraphics[width=1.\columnwidth]{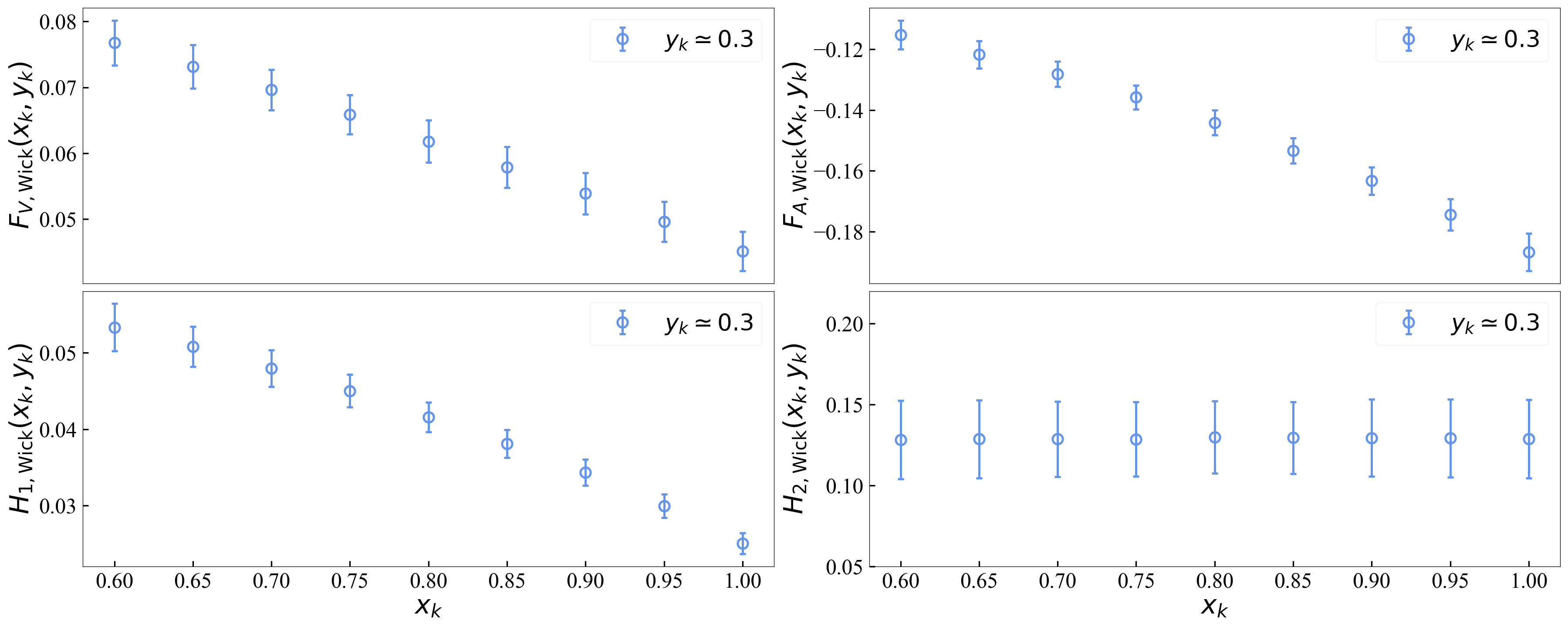}
    \caption{Quark-connected contribution to the form factors above the two-pion threshold from the terms that can be estimated by integrating the relevant correlation functions over Euclidean time (blue circles) as functions of the photon virtuality, at fixed photon momentum $y_k\simeq 0.3$.}
    \label{fig:4L_AboveWick}
\end{figure}

In this appendix, we discuss the evaluation of the quark-connected contributions
to the real parts of the form factors above the two-pion threshold that do not
require the SFR method,  and can be computed by integrating the relevant differential form factors over Euclidean time: $F_{\mathrm{Wick}}(x_k, y_k)$, with $F = \{F_V, A^i\}$ (see  Eqs.~(\ref{eq:FV_Wick_def}) and~(\ref{eq:FA_wick})). \\ \\ 
The analysis of these contributions follows exactly the same steps as the
analysis of the contributions below the two-pion threshold, presented in
Sec.~\ref{Sec:belowth}, where all the details can be found. We start by decomposing the form factors
into contributions from the two time orderings:
\begin{equation}
\label{eq:4L_appFvWick}
F_{\mathrm{Wick}}(x_k, y_k; t_W)
 = F_{1}(x_k, y_k; t_W) + \bar{F}_{2}(x_k, y_k; t_W), \quad F \in \{F_{V}, A^i\},
\end{equation}
where $\bar{F}_{2}(x_k, y_k; t_W)$ collects the contribution from the second
time ordering that can be obtained by integrating the differential form factors over positive Euclidean time ($t_\gamma > 0$). For the vector form factor, this term is given by  
\begin{equation}
\label{eq:4L_appFV2bar}
\bar{F}_{V,2}(x_k, y_k; t_W)
 \equiv F_{V,2;s}(x_k,y_k;t_W)
+\int_{0}^{\infty}dt_\gamma\,e^{-E_\gamma t_\gamma}\,
\delta F_{V;u}(t_\gamma,x_k,y_k;t_W),
\end{equation}
while for the axial form factors reads
\begin{align}
\label{eq:4L_appAi2bar}
\bar{A}^i_{2}(x_k, y_k; t_W)
\equiv A^{i}_{2;s}(x_{k},y_{k};t_{W})
+
\int_{0}^{\infty} dt_{\gamma}~
e^{-E_{\gamma}t_{\gamma}}\,
\delta \overline{A}^{i}_{u}(t_{\gamma},x_{k},y_{k};t_{W})
- \int_{0}^{\infty} dt_{\gamma}~\delta A^{i}_{{\rm pt};u}(t_{\gamma}, x_{k},y_{k};t_{W}).
\end{align}
The first step of the analysis consists in computing separately, on each gauge
ensemble, the two contributions in Eqs.~(\ref{eq:4L_appFvWick}), and in controlling the effects of truncating the
integrals over $t_\gamma$ as was done in Sec.~\ref{Sec:4L_tgammaintegral}. \\ \\ 
Fig.~\ref{fig:4L_ABFFfirstpltx} shows the quark-connected contribution from
the first time ordering to the form factors.
As in Fig.~\ref{fig:4L_FFfirstpltx}, the blue circles are obtained using the
estimators defined in Eqs.~(\ref{eq:4L_contFV1_new}) and~(\ref{eq:esti_impro_axial}), the
grey squares correspond to naive integrations of the lattice data, and the magenta triangles are
obtained by performing the analytic continuation to large and negative times using directly the $K$ meson with momentum $-\bs{k}$ as an asymptotic state. The red bands represent constant
fits to the blue circles and provide our estimators of
$F_{V,1}(x_k, y_k; t_W)$ and $A^i_{1}(x_k, y_k; t_W)$. \\ \\ 
Fig.~\ref{fig:4L_ABFFsecondpltx} shows the quark-connected contributions to
$\bar{F}_{V,2}(x_k, y_k; t_W)$ and $\bar{A}^i_{2}(x_k, y_k; t_W)$.
The blue circles are obtained using the asymptotic approximations in
Eqs.~(\ref{eq:FV2_new}) and~(\ref{eq:FA2_est}) for the contribution from the strange quark $A_{2;s}^i$, while the gray squares correspond to
naive integrations. \\ \\ 
The dependence on $t_W$ is analyzed following the same methodology as in
Sec.~\ref{Sec:4L_tweff}. Fig.~\ref{fig:4L_ABFFfirstTOtwdep} shows a comparison
of the quark-connected contributions from the first time ordering computed on
the B64 ensemble at two values of $t_W$, namely $t_W \simeq 2.0~\mathrm{fm}$ and
$t_W \simeq 2.6~\mathrm{fm}$.
The analogous comparison for $\bar{F}_{V,2}(x_k, y_k; t_W)$ and
$\bar{A}^i_{2}(x_k, y_k; t_W)$ is presented in
Fig.~\ref{fig:4L_ABFFsecondTOtwdep}.
As in the kinematic region below the two-pion threshold, no significant
dependence on $t_W$ is observed. A systematic uncertainty associated with the
choice of fixed $t_W$ is assigned to
$F_{V,\mathrm{Wick}}(x_k, y_k; t_W)$ and $A^i_{\mathrm{Wick}}(x_k, y_k; t_W)$
according to Eq.~(\ref{eq:4L_Systtw}). A total of 460 such systematic errors are
estimated: 80\% of them are smaller than the corresponding statistical
uncertainty, 98\% are smaller than twice the statistical uncertainty, and none
exceeds $2.9$ times the statistical uncertainty. \\ \\ 
We now turn to the evaluation of finite-size and discretization effects. Fig.~\ref{fig:4L_ABVfit} shows the
dependence of the form factors on the inverse spatial extent, $L^{-1}$, using
results from the B48, B64, and B96 ensembles (black triangles).
Finite-size effects are estimated by fitting the data with the ansatz of
Eq.~(\ref{eq:4L_FSE_ansatz}); the resulting fitted bands are shown in red. The
values interpolated to the reference volume $L_{\mathrm{ref}} \simeq
5.46~\mathrm{fm}$, at which the continuum limit is taken, are indicated by blue
circles. The finite-size effects are found to be small and well described by the
exponential model employed. \\ \\ 
The continuum limit is then performed following the procedure described in
Sec.~\ref{Sec:4L_Continuum}. The results at fixed spatial extent
$L_{\mathrm{ref}}$ are shown in Fig.~\ref{fig:4L_ABafit}. A linear fit (grey
bands) is carried out for $F_{V,\mathrm{Wick}}(x_k, y_k, a)$ and
$A^i_{\mathrm{Wick}}(x_k, y_k, a)$ using the results at the three lattice
spacings (black points). 
A constant fit (area between blue lines) is then performed using
only the two finest lattice spacings. The two extrapolations are combined using
BAIC, and the result is shown as a meshed red band. \\ \\ 
Finally, the infinite-volume limit is obtained by adding the finite-size
corrections, estimated previously following Eq.~(\ref{eq:4L_finite-VCorr}), to the continuum-limit results
(red squares in Fig.~\ref{fig:4L_ABafit}). 
Fig.~\ref{fig:4L_AboveWick} shows $F_{V,\mathrm{Wick}}(x_k, y_k)$ and $A^{i}_{\mathrm{Wick}}(x_k, y_k)$, already extrapolated to infinite volume and to the continuum-limit, as
functions of the photon virtuality at fixed photon momentum, $y_k \simeq 0.3$. 
In Sec.~\ref{Sec:4L_finalFF} we combine our determination of 
$F_{V,{\rm Wick}}(x_k,y_k)$ and $A^{i}_{\rm Wick}(x_k,y_k)$ 
with the SFR form factors discussed in Sec.~\ref{Sec:4L_aboveth} 
to obtain the full vector and axial form factors above the $\pi\pi$ threshold.

\section{Determination of the kaon's mean square charge radius $\langle r_{K}^{2}\rangle$}
\label{app:kaon_radius}

In this appendix we describe the determination of the kaon's mean square charge radius
$\langle r^{2}_{K}\rangle$, which enters the kaon–pole contribution in
Eqs.~\ref{eq:4L_CmunuAK_new}–\ref{eq:FK_expanded}. Since the subtraction of the
kaon pole is applied only to the quark–connected contribution, we restrict the
evaluation of $\langle r^{2}_{K}\rangle$ to the connected diagrams.

The starting point of the calculation is the three–point correlation function
\begin{align}
\label{eq:C3_kaon_lat}
C_{3K}^{\mu}(t,\bs{k};t_{J}) =
\sum_{\bs{x},\bs{y}} e^{i\bs{k}(\bs{x}-\bs{y})}
\langle 0| P_{K}(t,\bs{x}) J_{\rm em}^{\mu}(t_{J},\bs{y}) P_{K}^{\dagger}(0) |0\rangle ,
\end{align}
where only the quark–connected contribution is retained.
$P_{K}^{\dagger}$ denotes the same smeared kaon interpolating operator used in
the calculation of the hadronic tensor and discussed in
Sec.~\ref{sec:Lattice_setup}.

As discussed in Sec.~\ref{sec:Lattice_setup}, on the lattice the spatial momentum
$\bs{k}$ is not injected through a Fourier transform, which would restrict the
accessible momenta to integer multiples of $2\pi/L$, but instead through twisted
boundary conditions. This allows us to reach very small momenta and therefore
probe directly the slope of the form factor near $q^{2}=0$.

After amputating the initial and final kaon states the correlator becomes
\begin{align}
C_{K}(t,\bs{k};t_{J}) \equiv
\frac{4E_{K}(\bs{k})m_{K}}{Z_{K}(\bs{k})Z_{K}(\bs{0})}
\, e^{t_{J}(m_{K}-E_{K}(\bs{k}))}
\, e^{E_{K}(\bs{k})t}
\, C^{0}_{3K}(t,\bs{k};t_{J}) ,
\end{align}
where $E_{K}(\bs{k})$ denotes the energy of a kaon state
with momentum $-\bs{k}$ and 
\begin{align}
Z_{K}(\bs{k}) = \langle K^{-}(-\bs{k})| P_{K}^{\dagger}(0) |0\rangle .
\end{align}
We determine both $E_{K}$ and $Z_{K}(\bs{k})$ from the kaon two–point correlation function
with momentum $-\bs{k}$ from an analysis of the effective-mass
plateau.

\begin{figure}
    \centering
    \includegraphics[width=0.5\linewidth]{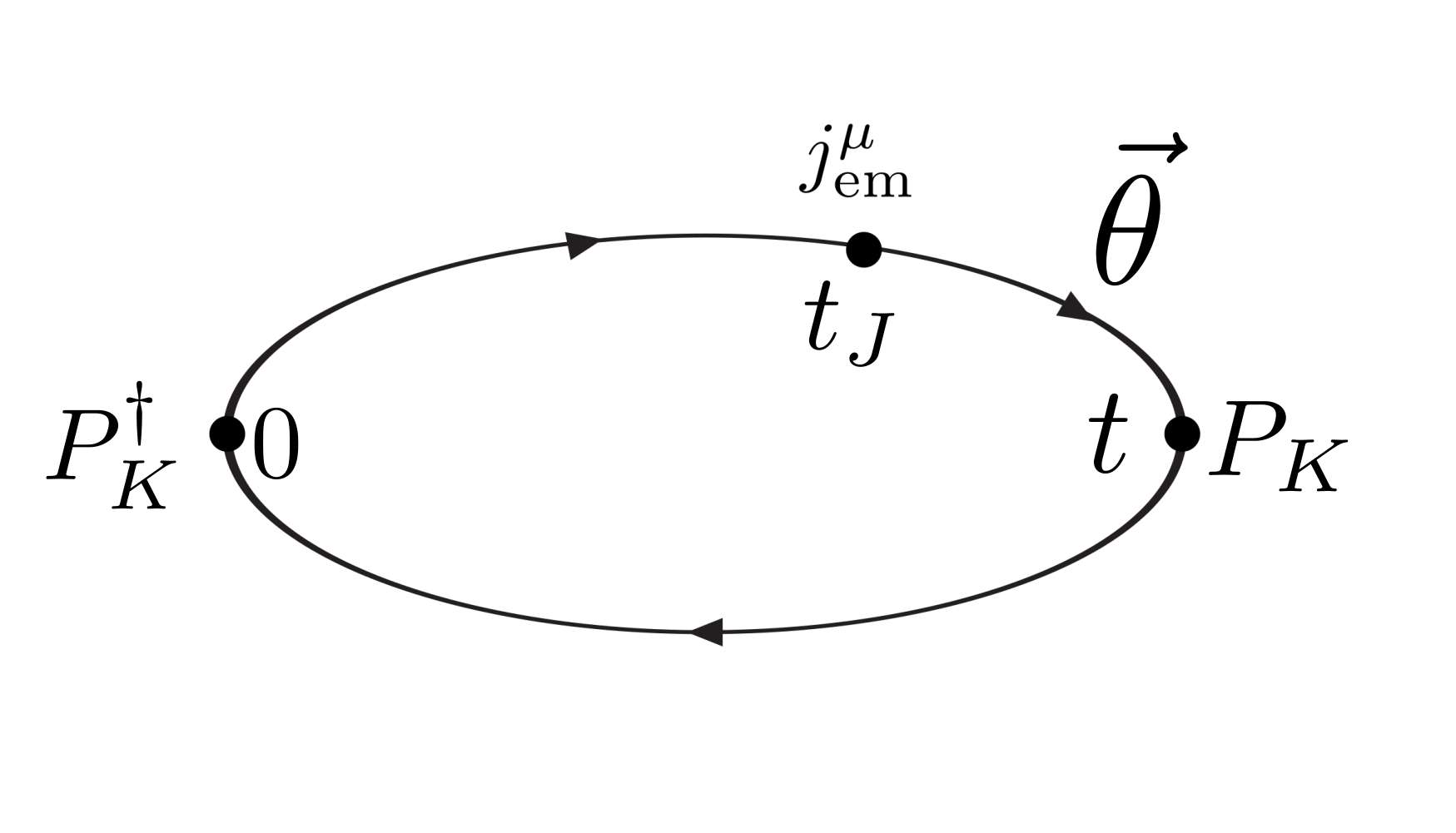}
    \caption{Schematic illustration of the setup used in the evaluation of
    the correlation function $C_{3K}^{\mu}(t,\bs{k};t_{J})$ in
    Eq.~(\ref{eq:C3_kaon_lat}). The zero-momentum kaon interpolator is placed at
    Euclidean time $0$, the electromagnetic current is inserted on a quark line
    at a fixed time $t_{J}$, and the three–point function is computed for all
    sink times $t$ where the kaon with momentum $-\bs{k}$ is annihilated. The
    momentum $\bs{k}$ is injected through twisted boundary conditions,
    $|\bs{k}|=\theta2\pi/L$.}
    \label{fig:kaon_radius}
\end{figure}

A schematic illustration of the setup used to compute the correlation function
in Eq.~(\ref{eq:C3_kaon_lat}) is shown in Fig.~\ref{fig:kaon_radius}. The kaon
source at rest is placed at Euclidean time $0$, the electromagnetic current is
inserted at a fixed time $t_{J}$ on a quark line (either the anti-up or strange
quark line), and the three–point function is computed for all sink times $t$ at
which the kaon with momentum $-\bs{k}$ is annihilated.

From the correlator $C_{K}(t,\bs{k};t_{J})$ we construct an estimator for the
kaon electromagnetic form factor,
\begin{align}
F^{K}_{\rm em}(t,\bs{k};t_{J})
\equiv
\frac{C_{K}(t,\bs{k};t_{J})}{m_{K}+E_{K}(\bs{k})}~,
\end{align}
which approaches the physical form factor in the limits
$t_{J}\to\infty$ and $t-t_{J}\to\infty$, specifically
\begin{align}
F^{K}_{\rm em}(q_{K}^{2}) =
\lim_{\substack{t_{J}\to\infty\\ t-t_{J}\to\infty}}
F^{K}_{\rm em}(t,\bs{k};t_{J}) ~,\qquad
q_{K}^{2} = -2m_{K}(E_{K}(\bs{k})-m_{K})
          = -|\bs{k}|^{2} + O(|\bs{k}|^{4})~.
\end{align}
The kaon charge radius is defined from the slope of the electromagnetic form
factor at vanishing momentum transfer,
\begin{align}
\langle r^{2}_{K}\rangle =
6\,\frac{d F_{\rm em}^{K}(q_{K}^{2})}{dq_{K}^{2}}
\bigg|_{q_{K}^{2}=0}.
\end{align}

To access this derivative we construct the following discrete estimator,
\begin{align}
\label{eq:radius_estimator}
r^{2}_{K}(t,\bs{k};t_{J}) \equiv
-\frac{3}{|\bs{k}|^{2}}
\left(
F_{\rm em}^{K}(t,\bs{k};t_{J})
+
F_{\rm em}^{K}(t,-\bs{k};t_{J})
-
2F_{\rm em}^{K}(t,\bs{0};t_{J})
\right).
\end{align}
In the limits $t_{J},t-t_{J}\to\infty$ and $|\bs{k}|\to0$, this estimator
converges to the kaon mean square radius $\langle r^{2}_{K}\rangle$.

It is essential that the left-hand side of Eq.~\ref{eq:radius_estimator} be evaluated
in a fully correlated way, using the same set of gauge configurations and
stochastic sources for all the terms on the right-hand side.

\begin{figure}
    \centering
    \includegraphics[width=0.7\linewidth]{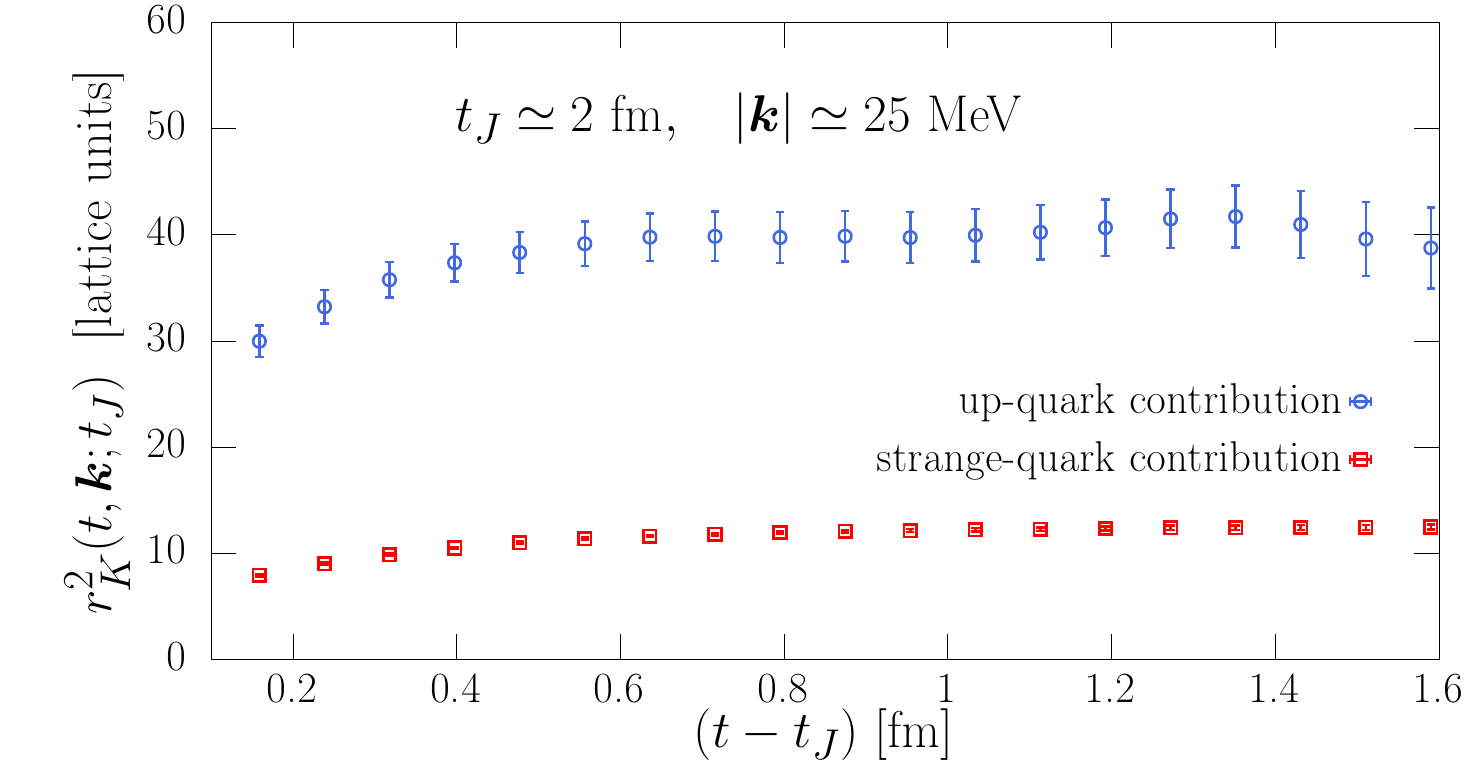}
    \caption{Estimator $r^{2}_{K}(t,\bs{k};t_{J})$ defined in
    Eq.~\ref{eq:radius_estimator} on the B64 ensemble for
    $t_{J}\simeq 2~{\rm fm}$ and $|\bs{k}|\simeq 25~{\rm MeV}$.
    Blue circles correspond to the contribution obtained inserting the
    electromagnetic current on the anti-up quark line, while red squares
    correspond to the insertion on the strange quark line.}
    \label{fig:estimator_square_radius}
\end{figure}

\begin{figure}
    \centering
    \includegraphics[width=0.7\linewidth]{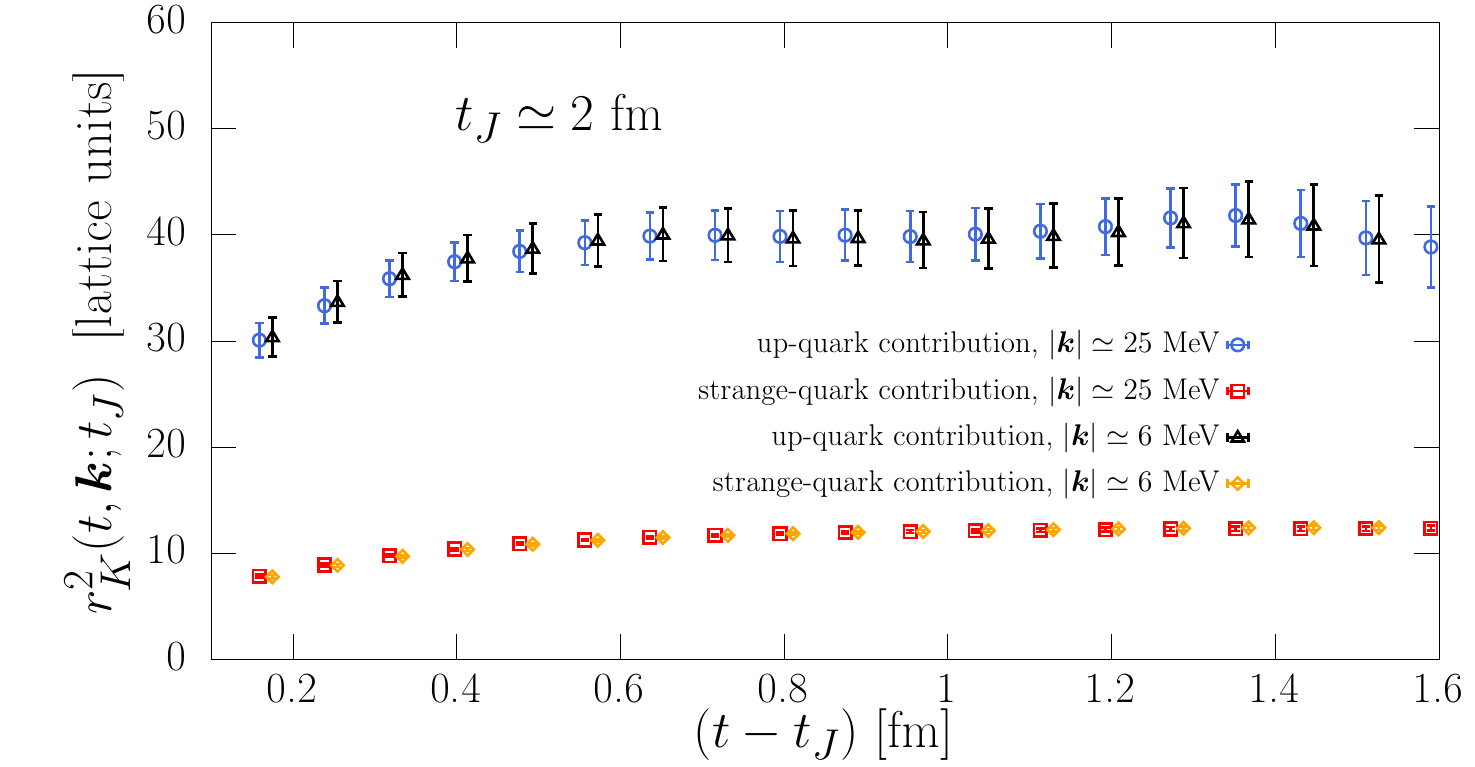}
    \caption{Check of the small-momentum limit of the estimator
    $r^{2}_{K}(t,\bs{k};t_{J})$ on the B64 ensemble.
    Results obtained with $|\bs{k}|\simeq 25~{\rm MeV}$ are compared with those
    obtained using a smaller momentum $|\bs{k}|\simeq 6~{\rm MeV}$.
    Within uncertainties the two determinations are consistent, indicating that
    the $|\bs{k}|\to0$ limit has been reached.}
    \label{fig:rk_eps}
\end{figure}

\begin{figure}
    \centering
    \includegraphics[width=0.7\linewidth]{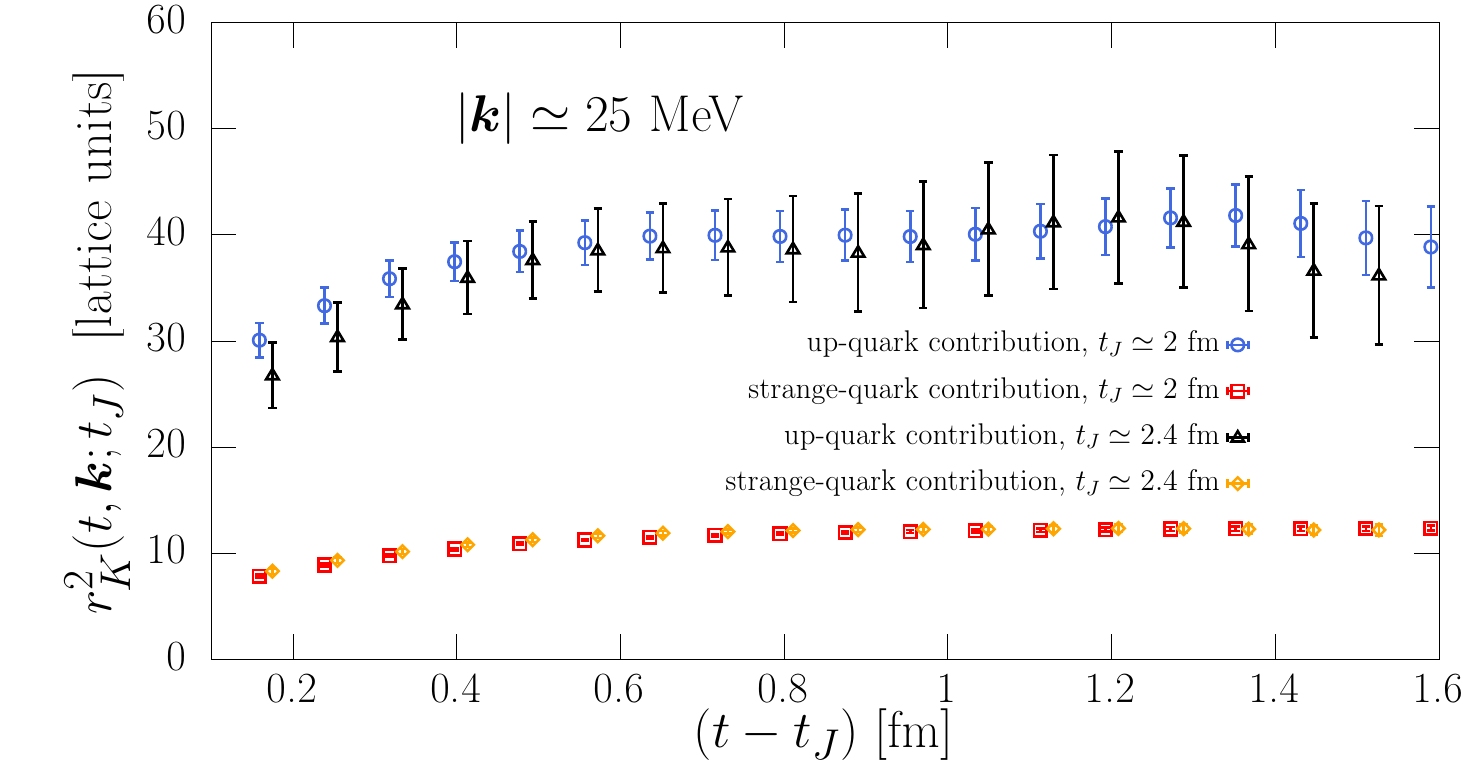}
    \caption{Check of finite-$t_{J}$ effects in the estimator
    $r^{2}_{K}(t,\bs{k};t_{J})$ on the B64 ensemble.
    Results obtained with $t_{J}\simeq 2~{\rm fm}$ are compared with those
    obtained using a larger value $t_{J}\simeq 2.4~{\rm fm}$.
    Within statistical uncertainties no dependence on $t_{J}$ is
    observed, indicating that finite-$t_{J}$ effects negligible.}
    \label{fig:rk_tj}
\end{figure}

The resulting estimator $r^{2}_{K}(t,\bs{k};t_{J})$ obtained on the B64 ensemble
is shown in Fig.~\ref{fig:estimator_square_radius}. We compute the estimator for
all values of $t$ at a fixed current insertion time, $t_{J}\gtrsim 2~{\rm fm}$.
Blue circles correspond to the up–quark contribution, obtained by retaining only
the up–quark component of the electromagnetic current $J^{\mu}_{\rm em}$, while
red squares represent the strange–quark contribution. Their sum gives the full
kaon square radius. We employ a small momentum $|\bs{k}|\simeq 25~{\rm MeV}$,
ensuring that the estimator has effectively reached the $|\bs{k}|\to0$ limit.
This has been verified explicitly on the B64 ensemble by repeating the
calculation with a smaller momentum $|\bs{k}| \simeq 6~{\rm MeV}$ (see
Fig.~\ref{fig:rk_eps}). Moreover, we have verified that finite-$t_{J}$ effects
are negligible within our statistical precision by considering a second, larger
value $t_{J}\simeq 2.4~{\rm fm}$ on the B64 ensemble (see Fig.~\ref{fig:rk_tj}).

As is evident from Fig.~\ref{fig:estimator_square_radius}, the up–quark
contribution, being significantly noisier than the strange–quark contribution,
dominates the final uncertainty. The precision of our current lattice
calculation is about $4\%$, which is fully satisfactory for determining the
kaon-pole contribution in Eq.~(\ref{eq:4L_CmunuAK_new}) with percent-level
accuracy.

A constant fit is performed in the region where a clear plateau is observed.
The resulting values of the kaon's mean square radius at the three lattice
spacings considered in this work are reported in
Tab.~\ref{tab:rsqKZem}.

\section{Correlation matrices of the parameters entering the form factor's effective parameterization.}
\label{app:4L_Corr}
The correlation matrices of the parameters describing the ansatz of the real and imaginary parts of the form factors in Eqs.~(\ref{eq:4L_ansatzFF}) and~(\ref{eq:4L_ansatzImFF}) are collected in Tabs.~\ref{tab:4L_Recorr} and~\ref{tab:4L_Imcorr}, respectively. 
\begin{table}[t] 
\centering 
\begin{tabular}{|ccccccc|} \hline \hline &  &  &  & &  &  \\ [-2. ex] 
& $a_{V}$ & $b_{V}$ & $c_{V}$ & $d_{V}$ & $e_{V}$ & $k_V$ \\ [1. ex] \hline &  &  & & & &  \\ [-2. ex] 
$a_V$  & $1.0$ & $-0.55$ & $-0.56$ & $0.34$ & $0.33$ & $0.39$\\[1.ex] \hline & & & & & & \\[-2.ex]$b_V$  & $-0.55$ & $1.0$ & $-0.22$ & $-0.87$ & $0.3$ & $-0.37$\\[1.ex] \hline & & & & & & \\[-2.ex]$c_V$  & $-0.56$ & $-0.22$ & $1.0$ & $0.34$ & $-0.92$ & $-0.09$\\[1.ex] \hline & & & & & & \\[-2.ex]$d_V$  & $0.34$ & $-0.87$ & $0.34$ & $1.0$ & $-0.27$ & $-0.1$\\[1.ex] \hline & & & & & & \\[-2.ex]$e_V$  & $0.33$ & $0.3$ & $-0.92$ & $-0.27$ & $1.0$ & $-0.22$\\[1.ex] \hline & & & & & & \\[-2.ex]$k_V$  & $0.39$ & $-0.37$ & $-0.09$ & $-0.1$ & $-0.22$ & $1.0$\\[1.ex] \hline \hline\end{tabular} 
\begin{tabular}{|ccccccc|} \hline \hline &  &  &  & &  &  \\ [-2. ex] 
& $a_{F_A}$ & $b_{F_A}$ & $c_{F_A}$ & $d_{F_A}$ & $e_{F_A}$ & $k_{F_A}$ \\ [1. ex] \hline &  &  & & &  & \\ [-2. ex] 
$a_{F_A}$  & $1.0$ & $-0.46$ & $-0.7$ & $0.39$ & $0.46$ & $0.26$\\[1.ex] \hline & & & & & & \\[-2.ex]$b_{F_A}$  & $-0.46$ & $1.0$ & $-0.23$ & $-0.71$ & $0.38$ & $-0.76$\\[1.ex] \hline & & & & & & \\[-2.ex]$c_{F_A}$  & $-0.7$ & $-0.23$ & $1.0$ & $0.05$ & $-0.92$ & $0.29$\\[1.ex] \hline & & & & & & \\[-2.ex]$d_{F_A}$  & $0.39$ & $-0.71$ & $0.05$ & $1.0$ & $0.01$ & $0.12$\\[1.ex] \hline & & & & & & \\[-2.ex]$e_{F_A}$  & $0.46$ & $0.38$ & $-0.92$ & $0.01$ & $1.0$ & $-0.57$\\[1.ex] \hline & & & & & & \\[-2.ex]$k_{F_A}$  & $0.26$ & $-0.76$ & $0.29$ & $0.12$ & $-0.57$ & $1.0$\\[1.ex] \hline \hline\end{tabular} 
\begin{tabular}{|ccccccc|} \hline \hline &  &  &  & & &    \\ [-2. ex] 
& $a_{H_1}$ & $b_{H_1}$ & $c_{H_1}$ & $d_{H_1}$ & $e_{H_1}$ & $k_{H_1}$ \\ [1. ex] \hline &  &  & & & &  \\ [-2. ex] 
$a_{H_1}$  & $1.0$ & $-0.94$ & $-0.72$ & $0.8$ & $0.28$ & $0.82$\\[1.ex] \hline & & & & & & \\[-2.ex]$b_{H_1}$  & $-0.94$ & $1.0$ & $0.55$ & $-0.95$ & $-0.17$ & $-0.67$\\[1.ex] \hline & & & & & & \\[-2.ex]$c_{H_1}$  & $-0.72$ & $0.55$ & $1.0$ & $-0.38$ & $-0.81$ & $-0.83$\\[1.ex] \hline & & & & & & \\[-2.ex]$d_{H_1}$  & $0.8$ & $-0.95$ & $-0.38$ & $1.0$ & $0.09$ & $0.46$\\[1.ex] \hline & & & & & & \\[-2.ex]$e_{H_1}$  & $0.28$ & $-0.17$ & $-0.81$ & $0.09$ & $1.0$ & $0.45$\\[1.ex] \hline & & & & & & \\[-2.ex]$k_{H_1}$  & $0.82$ & $-0.67$ & $-0.83$ & $0.46$ & $0.45$ & $1.0$\\[1.ex] \hline \hline\end{tabular} 
\begin{tabular}{|ccccccc|} \hline \hline &  &  &  & &  &  \\ [-2. ex] 
& $a_{H_2}$ & $b_{H_2}$ & $c_{H_2}$ & $d_{H_2}$ & $e_{H_2}$ & $k_{H_2}$ \\ [1. ex] \hline &  &  & & &   &\\ [-2. ex] 
$a_{H_2}$  & $1.0$ & $-0.88$ & $-0.7$ & $0.73$ & $0.26$ & $0.62$\\[1.ex] \hline & & & & & & \\[-2.ex]$b_{H_2}$  & $-0.88$ & $1.0$ & $0.46$ & $-0.9$ & $-0.08$ & $-0.66$\\[1.ex] \hline & & & & & & \\[-2.ex]$c_{H_2}$  & $-0.7$ & $0.46$ & $1.0$ & $-0.31$ & $-0.83$ & $-0.51$\\[1.ex] \hline & & & & & & \\[-2.ex]$d_{H_2}$  & $0.73$ & $-0.9$ & $-0.31$ & $1.0$ & $0.06$ & $0.33$\\[1.ex] \hline & & & & & & \\[-2.ex]$e_{H_2}$  & $0.26$ & $-0.08$ & $-0.83$ & $0.06$ & $1.0$ & $0.1$\\[1.ex] \hline & & & & & & \\[-2.ex]$k_{H_2}$  & $0.62$ & $-0.66$ & $-0.51$ & $0.33$ & $0.1$ & $1.0$\\[1.ex] \hline \hline\end{tabular} 
\caption{Correlations between the parameters of Eq.~(\ref{eq:4L_ansatzFF}), describing the real part of $F_V$(upper left), $F_A$(upper right), $H_1$(lower left), and $H_2$(lower right).} 
\label{tab:4L_Recorr} 
\end{table} 

\begin{table}[] 
\centering 
\begin{tabular}{|cccc|} \hline \hline &  &  &      \\ [-2. ex] 
& $\alpha_{V}^I$ & $\beta_{V}^I$ & $\gamma_{V}^I$ \\ [1. ex] \hline &  &  &    \\ [-2. ex] 
$\alpha_V$  & $1.0$ & $-0.87$ & $-0.38$\\[1.ex] \hline & & &  \\[-2.ex]$\beta_V$  & $-0.87$ & $1.0$ & $0.07$\\[1.ex] \hline & & &  \\[-2.ex]$\gamma_V$  & $-0.38$ & $0.07$ & $1.0$\\[1.ex] \hline \hline\end{tabular} 
\begin{tabular}{|cccc|} \hline \hline &  &  &      \\ [-2. ex] 
& $\alpha_{F_A}$ & $\beta_{F_A}$ & $\gamma_{F_A}$  \\ [1. ex] \hline &  &  &    \\ [-2. ex] 
$\alpha_{F_A}$  & $1.0$ & $-0.99$ & $-0.23$\\[1.ex] \hline & & &  \\[-2.ex]$\beta_{F_A}$  & $-0.99$ & $1.0$ & $0.16$\\[1.ex] \hline & & &  \\[-2.ex]$\gamma_{F_A}$  & $-0.23$ & $0.16$ & $1.0$\\[1.ex] \hline \hline\end{tabular} 
\begin{tabular}{|cccc|} \hline \hline &  &  &      \\ [-2. ex] 
& $\alpha_{H_1}$ & $\beta_{H_1}$ & $\gamma_{H_1}$ \\ [1. ex] \hline &  &  &    \\ [-2. ex] 
$\alpha_{H_1}$  & $1.0$ & $-0.98$ & $-0.49$\\[1.ex] \hline & & &  \\[-2.ex]$\beta_{H_1}$  & $-0.98$ & $1.0$ & $0.37$\\[1.ex] \hline & & &  \\[-2.ex]$\gamma_{H_1}$  & $-0.49$ & $0.37$ & $1.0$\\[1.ex] \hline \hline\end{tabular} 
\begin{tabular}{|cccc|} \hline \hline &  &  &     \\ [-2. ex] 
& $\alpha_{H_2}$ & $\beta_{H_2}$ & $\gamma_{H_2}$ \\ [1. ex] \hline &  &  &    \\ [-2. ex] 
$\alpha_{H_2}$  & $1.0$ & $-0.99$ & $-0.14$\\[1.ex] \hline & & &  \\[-2.ex]$\beta_{H_2}$  & $-0.99$ & $1.0$ & $0.12$\\[1.ex] \hline & & &  \\[-2.ex]$\gamma_{H_2}$  & $-0.14$ & $0.12$ & $1.0$\\[1.ex] \hline \hline\end{tabular} 
\caption{Correlations between the parameters of Eq.~(\ref{eq:4L_ansatzImFF}), describing the imaginary part of $F_V$, $F_A$, $H_1$, and $H_2$, from left to right respectively.} 
\label{tab:4L_Imcorr} 
\end{table}

\clearpage

\bibliographystyle{JHEP}
\bibliography{biblio}
\end{document}